\documentclass[letterpaper,12pt]{article}
\usepackage[margin=22mm]{geometry}
\usepackage{amsmath}
\usepackage{amsfonts}
\usepackage{amssymb}
\usepackage{verbatim}
\usepackage{subfigure}
\usepackage{caption}
\usepackage{setspace}
\renewcommand{\baselinestretch}{1.5}

\usepackage{color,bm}
\usepackage{graphicx}
\usepackage{array}
\usepackage{float}
\usepackage{afterpage}
\usepackage{arydshln}
\usepackage{titlesec}
\usepackage{blkarray}
\usepackage{mathtools}
\usepackage{pgfplots}
\usepackage{tikz}
\usetikzlibrary{bayesnet}
\usepackage{blkarray}
\usepackage{multirow}
\usepackage[ruled,vlined]{algorithm2e}
\usepackage{booktabs}
\usepackage{enumitem}
\usepackage{xr}
\usepackage{lscape}

\usepackage{hyperref} 
\hypersetup{
	colorlinks=true,
	linkcolor=blue,        
	citecolor=blue,        
	filecolor=magenta,     
	urlcolor=blue
}
\usepackage{setspace}

\newtheorem{proposition}{Proposition}

\tikzset{
  LabelStyle/.style = { rectangle, rounded corners, draw,
                        minimum width = 2em, fill = yellow!50,
                        text = red, font = \bfseries },
  VertexStyle/.append style = { inner sep=2pt,
                                font = \bfseries},
  EdgeStyle/.append style = {->,  left} }

\immediate\write18{texcount -tex -sum  \jobname.tex > \jobname.wordcount.tex}

\usepackage[sort]{natbib}
\usepackage{authblk}

\providecommand{\keywords}[1]
{
  \small	
  \textbf{\textit{Keywords---}} #1
}

\title{
A Note on Improving Variational Estimation for Multidimensional Item Response Theory
}
\author{ Chenchen Ma$^1$, Jing Ouyang$^1$, Chun Wang$^2$ and Gongjun Xu$^1$\\
    \small\em$^1$\;Department of Statistics, University of Michigan\\
 \small\em$^2$\;College of Education, University of Washington}
\date{} 

\begin{document}
\maketitle

\pagestyle{plain}
\setcounter{page}{1}
\pagenumbering{arabic}

\renewcommand{\baselinestretch}{1.3} 
\begin{abstract}
\normalsize
Survey instruments and assessments are frequently used in many domains of social science. When the constructs that these assessments try to measure become multifaceted, multidimensional item response theory (MIRT) provides a unified framework and convenient statistical tool for item analysis, calibration, and scoring. However, the computational challenge of estimating MIRT models prohibits its wide use because many of the extant methods can hardly provide results in a realistic time frame when the number of dimensions, sample size, and test length are large. Instead, variational estimation methods, such as Gaussian Variational Expectation Maximization (GVEM) algorithm, have been recently proposed to solve the estimation challenge by providing a fast and accurate solution. However, results have shown that variational estimation methods may produce some bias on discrimination parameters during confirmatory model estimation, and this note proposes an importance weighted version of GVEM (i.e., IW-GVEM) to correct for such bias under MIRT models. We also use the adaptive moment estimation method to update the learning rate for gradient descent automatically. Our simulations show that IW-GVEM can effectively correct bias with  modest increase of computation time, compared with GVEM.  The proposed method may also shed light on improving the variational estimation for other psychometrics models.
\end{abstract} 

\renewcommand{\baselinestretch}{1.5}

\section{Introduction}
\normalsize
\label{sec-intro}

Developing, refining, and validating survey questionnaires that measure target latent traits such as personality or cognitive abilities has always been a core agenda in education and psychology, and this focus is also extended to health measurement and culminates in a multi-decade initiative on patient-reported outcome measures. Psychometric methods and tools are an integral part of achieving this focus. When the constructs that these assessments try to measure become increasingly complex, multidimensional item response theory (MIRT), also known as item factor analysis, provides a unified framework and convenient statistical tool for item analysis, calibration, and scoring. However, the increasing scale and complexity of survey designs, especially in large-scale assessments (LSA), require MIRT models with many latent factors. For instance, the English Language Proficiency Assessment for the 21st Century (ELPA21) across two gradebands consists of 8 domain-level traits measured by more than 600 items \citep{cresst2017English}. The existing computational algorithms for fitting high-dimensional MIRT models are insufficient to navigate the massive amount of assessment data, reflected by excessively long computation time and unstable estimation results.

MIRT provides a powerful tool for enriching the information gained in educational assessment \citep{hartig2009multidimensional}. For instance, cognitive instructional psychology considers ``science knowledge'' and ``mathematical ability'' as highly differentiated theoretical constructs that consist of both basic facts and skills as well as deeper or higher order understanding \citep{kupermintz1995dedication,hamilton1995enhancing}. As another example, the 2003 assessment framework of PISA \citep{oecd2003pisa} contains a hierarchy of ability dimensions with general ``knowledge and skills'' at the highest level, followed by reading, math, science, and problem solving. Then at the lowest level are the sub-domains such as ``space and shape'', ``change and relationships'', and ``quantity'' nested within math. Hence, dimensions on different levels vary in their degree of generality and abstraction. Oftentimes, the highest level represents a broad competency level, whereas lower levels represent narrower and more specific abilities. If the intention is to model both the overall and lower-level abilities simultaneously, the model will be high dimensional \citep{briggs2003introduction}.

Even though the research and development in statistics and psychometrics have provided increasingly sophisticated measurement models to better assess constructs in social sciences, the practice still lags behind \citep{cai2018improving}. Unidimensional IRT models continue to dominate the current applications in many domains. One reason is that when the number of items, sample size, and the number of dimensions are all large, the current computational algorithms for MIRT estimation may not be powerful enough to produce results in a reasonable time frame (or ever) \citep{cresst2017English}.  For instance, due to the large number of students and items within each gradeband, the operational analysis approach used for ELPA21 is a two-step approach: in the first step, a unidimensional IRT model is fitted to the item response data for each domain subtest to obtain item parameter estimates; then in the second step, a restricted hierarchical model (i.e., testlet model, \citeauthor{wainer2007testlet},\citeyear{wainer2007testlet}; \citeauthor{gibbons1992full},\citeyear{gibbons1992full}; \citeauthor{cai2011generalized},\citeyear{cai2011generalized}) is fitted to estimate the correlations between the four domains \citep{thissen2013using}. Such a two-step process has two limitations: (1) the item parameter calibration errors are ignored in the second step, and (2) the restricted hierarchical model is only an approximation to the independent-cluster MIRT model.
Various full-information methods have been proposed to deal with the computational challenge, which are listed below with pros and cons. The list is by no means exhaustive, but it includes some of the most popular methods that are available in commercial software packages or R packages.\footnote{The limited-information method such as weighted least squares is not reviewed here as it handles high-dimensional models very differently, and it cannot handle missing data very well.}
\begin{enumerate}
\item Adaptive Gaussian quadrature.  Compared to the regular Gauss-Hermite quadrature (e.g., \citeauthor{bock1981marginal}, \citeyear{bock1981marginal}), even though the number of quadrature points per dimension is reduced, the total number of quadrature points still increases exponentially with the number of dimensions. Moreover, an extra step is needed to compute the posterior mode and variance of latent factors in each iteration, which adds additional computation costs \citep{pinheiro1995approximations}.
\item Monte Carlo techniques. This family of methods include, for instance, the Monte Carlo EM algorithm \citep{mcculloch1997maximum,wang2015mixture}, stochastic EM algorithm \citep{von2010stochastic, zhang2020improved}, or Metropolis-Hastings Robbins-Monro algorithm \cite{cai2010high, cai2010metropolis}.  These methods circumvent intractable integrations by sampling from the posterior distributions; however, they may still computationally intensive   for complicated high-dimensional models.
Fully Bayesian estimation methods, such as Markov chain Monte Carlo (MCMC; \citeauthor{albert1992bayesian},  \citeyear{albert1992bayesian}; \citeauthor{patz1999applications},  \citeyear{patz1999applications}) can also be considered in this category. The Bayesian approach is also computationally costly as it needs a long chain to converge for complex models,   though it is preferable with smaller sample sizes. 
\item Analytic dimension reduction.  For models assuming certain conditional independence among factors (such as the bi-factor models), the conditional independence relations can be used to partition the joint space of all latent variables into smaller subsets. {\color{black}As a result, brute force numerical integration over the joint latent space can be replaced by a sequence of integrations over smaller subsets of latent variables, which helps reduce the computation burden dramatically. This strategy to deal with high-dimensional integration challenges is known as analytic dimension reduction~\citep{cai2011generalized, gibbons1992full, rijmen2008latent}.} One limitation, though, is that the algebraic manipulations of the likelihood of a specific model might become very complicated, and they differ for different models (e.g., \citeauthor{cai2011generalized}, \citeyear{cai2011generalized}; \citeauthor{gibbons1992full}, \citeyear{gibbons1992full}). Hence, there is no universal rule that applies to any model.
\item Laplace approximation. This method is based on second-order Taylor expansion of the log-integrand around its mode \citep{lindstrom1988newton} such that the high-dimensional integral becomes tractable. {\color{black} This method is a classical and popularly used method for generalized linear mixed-effects models (GLMM), and it is available in many software packages, such as the ``lem4” R package \citep{bates2014fitting}.} However, this approximation may not be accurate when the dimension increases to 3 or higher,   the sample size is small \citep{jeon2017variational}, or the likelihood function is skewed.
\end{enumerate}

 {\color{black} Besides the full-information methods above, a recent constraint joint maximum likelihood estimation (CJMLE) was proposed by~\cite{chenlizhang}, which is more computationally efficient than many marginal maximum likelihood methods, and the estimator has the theoretical guarantee to be consistent under high-dimensional settings. Extending CJMLE, the singular value decomposition (SVD) based estimator was proposed by~\cite{zhang2020note}, which further improves the performance of CJMLE. These joint maximum likelihood methods enjoy the low computational cost but sacrifice the flexibility of latent factors by treating them as fixed effects. {\color{black}For instance, it would be hard conceptually to generalize the algorithm to a multiple-group condition in which unbiased estimation of group-specific population distributions is often needed than estimation of individual person's latent trait as a fixed effect.} }

In light of the limitations of the above-mentioned methods, variational estimation methods that leverage advances in statistical and machine learning  have recently gained increasing interests in psychometrics \citep{jeon2017variational,cho2021gaussian,cho2022qestimation}. { \color{black} Among numerous variational estimation methods, \cite{rijmen2013fitting} was one of the first to use a variational estimation technique for MIRT models that approximates the likelihood function by a computationally tractable lower bound, but it only studied MIRT models with discrete latent factors. Later, a wide range of studies on variational methods were conducted for the estimation of more complex models~\citep{hui2017variational, natesan2016bayesian}. 
Recently \cite{jeon2017variational} proposed variational maximization-maximization (VMM) algorithm for the generalized linear mixed models (GLMMs), which outperforms Laplace approximation with a small sample size. 
However, they rely on some iterative numerical algorithms to attain the solutions in each maximization step, resulting in a slow speed in running the algorithm. 
To further increase computational efficiency, many researchers brought up variational autoencoder (VAE), a deep learning based variational method to tackle the estimation problems in MIRT models~\citep{curi2019interpretable, wu2020variational}. Extending from VAE, the importance-weighted VAE (IW-VAE) is developed and exhibits competitive performances to other estimation methods~\citep{urban2021deep, liu2022estimating} {\color{black} at large sample sizes.} 
However, the two IW-VAE methods 
lack theoretical support for the consistency of estimators. 
{\color{black} In addition, although they are powerful in handling large-scale data, their performances in small to medium-sample data may not be as well (see supplementary materials for more details).} \cite{cho2021gaussian,cho2022qestimation} proposed a  Gaussian Variational Expectation-Maximization (GVEM) algorithm, which has shown to be computationally fast and produces comparable and sometimes more accurate parameter estimates than the MH-RM algorithm {\color{black} and than the CJMLE method} in high-dimensional exploratory item factor analysis models (i.e., M2PL and M3PL in  \citeauthor{cho2021gaussian}, \citeyear{cho2021gaussian}). {\color{black} Moreover, \cite{cho2021gaussian} proved that the estimated parameters from GVEM algorithm are consistent under the high-dimensional setting.}
However, {\color{black}
we found that directly applying the GVEM algorithm in confirmatory MIRT models would generate relatively large bias on discrimination parameters, especially when the correlations among factors are high and the sample size is not large (please see Section~\ref{sec-simulation studies} for the detailed simulation results).} Such a bias issue happens commonly to variational  estimation   for various statistical models \citep{bishop2006pattern}. 


To correct the bias in the variational algorithms for MIRT models, we propose an importance weighted GVEM algorithm (denoted as IW-GVEM hereafter) {\color{black}, which is an extension of GVEM algorithm by performaning additional steps after GVEM convergence}. The primary idea is to use an importance weighted variational inference technique to create a tighter variational lower bound to the target, otherwise intractable, marginal likelihood. Because the variational lower-bound proposed in \cite{cho2021gaussian,cho2022qestimation} is replaced by a weighted average based on importance sampling \citep{domke2018importance}, the desirable closed-form solution in the M-step is no longer applicable. Instead, we propose to use \emph{Adam} \citep{kingma2014adam}, a popular algorithm for first-order gradient-based optimization. This computationally efficient algorithm updates the objective function stochastically based on adaptive estimates of lower-order moments, and it is especially well-suited for large data and complex models.
{\color{black} Moreover, different from the IW-VAE methods rooted in deep neural network models where substantial theoretical works on the consistency of the estimators remain to be done, our proposed IW-GVEM is a more transparent method that comes with theoretical guarantees
under the high-dimensional
setting. }

In what follows, this note   briefly describes the M2PL model and the original GVEM algorithm and then introduces the IW-GVEM algorithm in Section 2, followed by a comprehensive simulation study in Section 3. We end the paper with discussions and future directions.

\section{Methods}
\label{sec-model}

\subsection{M2PL}
\label{sec-m2pl}
Multidimensional 2PL model is one of the most widely used MIRT models in practice \citep{reckase2009multidimensional}. With M2PL, the item response function of the $i$th individual to the $j$th item is modeled by
	\begin{equation}\label{2PL}
	P(Y_{ij}=1 \mid \boldsymbol{\theta}_i)= \frac{\exp(\boldsymbol{a}_j^\top\boldsymbol{\theta}_i-b_j)}{1+\exp(\boldsymbol{a}_j^\top\boldsymbol{\theta}_i-b_j)},
	\end{equation}
where $Y_{ij}$ for $i=1,...,N$ and $j=1,...,J$ is a binary response, $\boldsymbol{a}_j$ denotes a $K$-dimensional vector of item discrimination parameters for item $j$, and $b_j$ specifies the corresponding difficulty level with item difficulty parameter as $b_j/\|\boldsymbol{a}_j\|_2$.  Following notations in \cite{cho2021gaussian}, we use $\boldsymbol{Y}_i$  to denote the response vector of the $i$th subject, and $\boldsymbol{\theta}_i$ to denote the latent trait vector of the $i$th subject.
We write $A = (\boldsymbol{\alpha_j}, j=1,\dots, J)$ and $B = (b_j, j = 1, \dots, J)$. For model identification, oftentimes the means and variances of $\boldsymbol{\theta}$ are fixed as   zeros and ones, respectively,  and   the covariance (which is actually correlation) of  $\boldsymbol{\theta}$   is freely estimated.

\subsection{GVEM}
Let $\boldsymbol{\Delta}=({\boldsymbol{A},\boldsymbol{B},\boldsymbol{\rho}})$ denote the set of unknown parameters for M2PL, where $\boldsymbol{\rho}$ denotes the correlations of $\boldsymbol{\theta}$. As discussed, the population means of $\boldsymbol{\theta}$ {\color{black} are} fixed at 0, and the population variances are fixed at 1. The correlations among $\boldsymbol{\theta}$'s can be freely estimated. Then the {\color{black} log-marginal likelihood} of responses $\boldsymbol{Y}$ is
\begin{equation}
	 l(\boldsymbol{\Delta}\mid \boldsymbol{Y})= \sum_{i=1}^N \log P(\boldsymbol{Y}_{i}\mid \boldsymbol{\Delta})=\sum_{i=1}^N\log\int \prod_{j=1}^J P(Y_{ij}\mid \boldsymbol{\Delta},\boldsymbol{\theta}_i) \phi(\boldsymbol{\theta}_i)d\boldsymbol{\theta}_i,
\label{mml}
\end{equation}
where $\phi$ denotes a $K$-dimensional Gaussian distribution of $\boldsymbol{\theta}$ with mean 0 and covariance $\Sigma_{\boldsymbol{\theta}}$. It is the potentially high-dimensional integration in Equation \eqref{mml} that makes direct maximization of the log-marginal likelihood computationally prohibitive.
{\color{black} The log-likelihood of response $\bm{Y}$ has an equivalent form
\begin{equation*}
     l(\boldsymbol{\Delta} \mid \boldsymbol{Y})=\sum_{i=1}^N \int_{\bm{\theta}_i} \log P(\boldsymbol{Y}_{i}\mid \boldsymbol{\Delta}) \times q_i (\bm{\theta}_i) d \bm{\theta}_i,
\end{equation*}
where $q_i (\bm{\theta}_i)$ can be any probability density function satisfying $\int_{\bm{\theta}_i} q_i (\bm{\theta}_i) d \bm{\theta}_i = 1$. }

The main idea behind variational inference is to approximate the intractable integral in Equation \eqref{mml} with a computationally feasible form, known as the evidence lower bound \citep[ELBO;][]{ormerod2010explaining,blei2017variational}. {\color{black} Because $P(\boldsymbol{Y}_{i}\mid \boldsymbol{\Delta}) = P(\boldsymbol{Y}_{i}, \bm{\theta}_i \mid \boldsymbol{\Delta})/ P(\bm{\theta}_i \mid \boldsymbol{Y}_{i}, \boldsymbol{\Delta})$, we write $l(\boldsymbol{\Delta} \mid \boldsymbol{Y})$ as 
\begin{eqnarray*}
    l(\boldsymbol{\Delta} \mid \boldsymbol{Y}) &=& \sum_{i=1}^N \int_{\bm{\theta}_i} \log \frac{P(\boldsymbol{Y}_{i}, \bm{\theta}_i \mid \boldsymbol{\Delta})}{P(\bm{\theta}_i \mid \boldsymbol{Y}_{i}, \boldsymbol{\Delta})} \times q_i (\bm{\theta}_i) d \bm{\theta}_i \\
    &=&
    \sum_{i=1}^N \int_{\bm{\theta}_i} \log \frac{P(\boldsymbol{Y}_{i}, \bm{\theta}_i \mid \boldsymbol{\Delta})q_i (\bm{\theta}_i)}{P(\bm{\theta}_i \mid \boldsymbol{Y}_{i}, \boldsymbol{\Delta})q_i (\bm{\theta}_i)} \times q_i (\bm{\theta}_i) d \bm{\theta}_i \\
      &=& \sum_{i=1}^N \int_{\bm{\theta}_i} \log \frac{P(\boldsymbol{Y}_{i}, \bm{\theta}_i \mid \boldsymbol{\Delta})}{q_i (\bm{\theta}_i)} \times q_i (\bm{\theta}_i) d \bm{\theta}_i + KL \{q_i (\bm{\theta}_i) \mid  
 P(\bm{\theta}_i \mid \boldsymbol{Y}_{i}, \boldsymbol{\Delta}) \},
\end{eqnarray*}
where $KL \{q_i (\bm{\theta}_i) \mid  
 P(\bm{\theta}_i \mid \boldsymbol{Y}_{i}, \boldsymbol{\Delta}) \} = \int_{\bm{\theta}_i} \log \frac{q_i (\bm{\theta}_i) }{P(\bm{\theta}_i \mid \boldsymbol{Y}_{i}, \boldsymbol{\Delta})} \times q_i (\bm{\theta}_i) d \bm{\theta}_i$ is non-negative. This is because 
 \begin{eqnarray*}
   - KL \{q_i (\bm{\theta}_i) \mid  
 P(\bm{\theta}_i \mid \boldsymbol{Y}_{i}, \boldsymbol{\Delta}) \}&=&  \int_{\bm{\theta}_i} \log \frac{P(\bm{\theta}_i \mid \boldsymbol{Y}_{i}, \boldsymbol{\Delta})}{q_i (\bm{\theta}_i) } \times q_i (\bm{\theta}_i) d \bm{\theta}_i \\
 & \leqslant & \int_{\bm{\theta}_i} \left( \frac{P(\bm{\theta}_i \mid \boldsymbol{Y}_{i}, \boldsymbol{\Delta})}{q_i (\bm{\theta}_i) }-1\right) \times q_i (\bm{\theta}_i) d \bm{\theta}_i \\
 & \leqslant & \int_{\bm{\theta}_i} P(\bm{\theta}_i \mid \boldsymbol{Y}_{i}, \boldsymbol{\Delta}) d \bm{\theta}_i - \int_{\bm{\theta}_i} q_i (\bm{\theta}_i)  d \bm{\theta}_i \\&=& 1- 1 = 0
 \end{eqnarray*}
Therefore, we have a lower bound of log-likelihood that 
\begin{eqnarray}
     l(\boldsymbol{\Delta} \mid \boldsymbol{Y}) &\geqslant &\sum_{i=1}^N \int_{\bm{\theta}_i} \log \frac{P(\boldsymbol{Y}_{i}, \bm{\theta}_i \mid \boldsymbol{\Delta})}{q_i (\bm{\theta}_i)} \times q_i (\bm{\theta}_i) d \bm{\theta}_i \nonumber \\
     &=& \sum_{i=1}^N E_{q_i(\boldsymbol{\theta}_i)}\bigg{[}\log \frac{P(\boldsymbol{Y}_{i},\boldsymbol{\theta}_i \mid  \boldsymbol{\Delta} )}{q_i(\boldsymbol{\theta}_i)}\bigg{]} =: ELBO,
\label{elbo}
\end{eqnarray}
where the last term $\sum_{i=1}^N E_{q_i(\boldsymbol{\theta}_i)}\bigg{[}\log \frac{P(\boldsymbol{Y}_{i},\boldsymbol{\theta}_i \mid  \boldsymbol{\Delta} )}{q_i(\boldsymbol{\theta}_i)}\bigg{]}$ is the ELBO for $l(\boldsymbol{\Delta}|\boldsymbol{Y})$ in Equation \eqref{mml}. Maximizing the   log-marginal likelihood is then approximated by maximizing ELBO, and $q_i(\boldsymbol{\theta}_i)$, the variational distribution, needs to be carefully chosen to minimize the gap between the log-marginal likelihood and its ELBO.

    }


 The key is to find $q_i(\boldsymbol{\theta}_i)$ so that ELBO approximates the marginal likelihood $l(\boldsymbol{\Delta}|\boldsymbol{Y})$ as close as possible. Note that when $q_i(\boldsymbol{\theta}_i)$ is the posterior density of $\boldsymbol{\theta}_i$, i.e., $q_i(\boldsymbol{\theta}_i)=P(\boldsymbol{\theta}_i\mid \boldsymbol{Y}_{i}, \boldsymbol{\Delta} )$, maximizing ELBO is equivalent to \cite{bock1981marginal}'s marginal maximum likelihood/expectation-maximization (MML/EM) algorithm. Instead, as the choice of $q_i(\boldsymbol{\theta}_i)$ determines the computational cost and success of the algorithm,
\cite{cho2021gaussian, cho2022qestimation} proposed a  choice of $q_i(\boldsymbol{\theta}_i)$ that satisfied two criteria: (1) it is easy to maximize, and (2) it approximates the true {\color{black} log-marginal likelihood} well.
Due to the independence of the students’ responses in general IRT models, $q_i(\boldsymbol{\theta}_i)$ is selected for each individual separately. Specifically, under M2PL, the joint distribution of $\boldsymbol{\theta}_i$ and $\boldsymbol{Y}_i$ is,
\begin{align}
	\nonumber
	&\quad \log P(Y_{i}, \boldsymbol{\theta}_i \mid \boldsymbol{\alpha}, \bm{b},  {\color{black} \bm{\rho}}) \\
	\label{eq-joint_p}
	& = \sum_{j=1}^J \Big\{ Y_{ij} (\boldsymbol{\alpha}_j^\top\boldsymbol{\theta}_i - b_j) + \log \frac{1}{1+\exp(\boldsymbol{\alpha}_j^\top\boldsymbol{\theta}_i - b_j)}
	\Big\} + \log \phi_{\boldsymbol{\theta}}(\boldsymbol{\theta}_i) \\
	\nonumber
	& \geq \sum_{j=1}^J \log \frac{e^{\xi_{ij}}}{1+e^{\xi_{ij}}} + \sum_{j=1}^J Y_{ij}(\boldsymbol{\alpha}_j^\top\boldsymbol{\theta}_i - b_j) + \sum_{j=1}^J \frac{b_j -\boldsymbol{\alpha}_j^\top\boldsymbol{\theta}_i - \xi_{ij}}{2} \\
	\nonumber
	& \quad - \sum_{j=1}^J \eta(\xi_{ij})\{(b_j -\boldsymbol{\alpha}_j^\top\boldsymbol{\theta}_i)^2 - \xi_{ij}^2\} + \log \phi_{\boldsymbol{\theta}}(\boldsymbol{\theta}_i)\\
	& := l(Y_i, \boldsymbol{\theta}_i \mid \boldsymbol{\alpha}, \bm{b},  {\color{black} \bm{\rho}}, \xi_{ij}),
\label{varilower}
\end{align}
where $\xi_{ij}$ is the variational parameter for the $i$th subject, which will be updated iteratively in the M-step of GVEM, and {\color{black}$\eta (\xi_{ij}) = (2 \xi_{i, j})^{-1}[e^{\xi_{i, j}} /(1+e^{\xi_{i, j}})-1 / 2]$}. The derivation is as follows. 
Because the difficulty of handling the marginal distribution of $P(\boldsymbol{Y}_i)$ mostly comes from the logistic sigmoid function, which makes the integration over $\theta$ not a closed form in the E-step. As a result, \cite{cho2021gaussian}   used a local variational approximation method \citep{jordan2004graphical}. {\color{black} Denote $x_{ij} = b_j - \boldsymbol{\alpha}_j^\top\boldsymbol{\theta}_i$, the local variational method gives the following variational lower bound for the sigmoid function:
\begin{eqnarray*}
    \frac{1}{1+\exp(\boldsymbol{\alpha}_j^\top\boldsymbol{\theta}_i - b_j)} & = & \frac{\exp (x_{ij})}{1+\exp (x_{ij}) } \\
    &= &   \max_{\xi_{ij}} \frac{\exp(\xi_{ij})}{ 1+\exp(\xi_{ij}) } \exp \left\{ \frac{x_{ij} - \xi_{ij} }{2} - \eta(\xi_{ij} ) (x_{ij}^2 - \xi_{ij}^2) \right\} \\
    & \geqslant &  \frac{\exp(\xi_{ij})}{ 1+\exp(\xi_{ij}) } \exp \left\{ \frac{x_{ij} - \xi_{ij} }{2} - \eta(\xi_{ij} ) (x_{ij}^2 - \xi_{ij}^2) \right\},
\end{eqnarray*}
and by applying the above lower bound to Equation~\eqref{eq-joint_p}, we get Equation \eqref{varilower}, which provides a variational lower bound for $ \log P(Y_{i}, \boldsymbol{\theta}_i \mid \boldsymbol{\alpha}, \bm{b},  {\color{black} \bm{\rho}})$. }

By variational inference theory, we can show that the variational distributions  $q_i(\boldsymbol{\theta}_i)$  (for $i=1,…, N$) that minimize the distances between the lower bound and the joint distribution follow a Gaussian distribution with closed-form mean and variance,
  i.e.,
  $ q_i(\boldsymbol{\theta}_i) \sim N(\boldsymbol{\theta}_i \mid \mu_i,\Sigma_i )$
	where the mean parameter of the normal distribution is
	\begin{eqnarray}\label{update_mu}
		\mu_i ={\Sigma_i}  \times \sum_{j=1}^J  \Big \{  2 \eta (\xi_{i,j} ) b_j  +Y_{ij}  -   \frac{1}{2}    \Big\} \boldsymbol{\alpha}_j 
		 \label{eq-mu}
	\end{eqnarray}
and the covariance matrix is  
	\begin{eqnarray} \label{update_Sigma}
		\big(  \Sigma_i \big)^{-1}  =   \big( {\Sigma_{\boldsymbol{\theta}}} \big)^{-1} + 2\sum_{j=1}^J\eta(\xi_{i,j} ) \boldsymbol{\alpha}_j  \boldsymbol{\alpha}_j^\top.
		\label{eq-sigma}
	\end{eqnarray}
In the confirmatory model estimation, we update population covariance matrix $\boldsymbol{\Sigma}_{\boldsymbol{\theta}}$ by
\begin{equation}
\boldsymbol{\Sigma}_{\boldsymbol{\theta}}=\frac{1}{N}\sum_{i=1}^N (\boldsymbol{\Sigma}_i+\boldsymbol{\mu}_i\boldsymbol{\mu}_i^\top).
\end{equation}
But because we need to fix the diagonal elements of  $\boldsymbol{\Sigma}_{\boldsymbol{\theta}}$ during estimation to fix the scale, we propose to rescale $\boldsymbol{\Sigma}_{\boldsymbol{\theta}}$ after the M-step converges, i.e., 
\begin{equation*}
\boldsymbol{\Sigma}_{\boldsymbol{\theta}}^{*}=((\sqrt{diag(\boldsymbol{\Sigma}_{\boldsymbol{\theta}})})^{-1})^\top \boldsymbol{\Sigma}_{\boldsymbol{\theta}}(\sqrt{diag(\boldsymbol{\Sigma}_{\boldsymbol{\theta}})})^{-1},
\end{equation*}
and the discrimination parameter needs to be rescaled accordingly, i.e., $\boldsymbol{\alpha}_j^*=\boldsymbol{\alpha}_j\sqrt{diag(\boldsymbol{\Sigma}_{\boldsymbol{\theta}})}$. For the exploratory analysis, $\boldsymbol{\Sigma}_{\boldsymbol{\theta}}$ is fixed at an identity matrix during estimation, and a post-hoc rotation will then produce proper non-zero correlations. In the following, we assume that the GVEM algorithm has converged and we fix the variational parameter $\xi_{ij}$ as the final estimates. {\color{black} In other words, we do not update $\xi_{ij}$ in the later iterative steps and $\xi_{ij}$ is fixed at the initialization GVEM step in Algorithm~\ref{algo:iw-gvem-m2pl}.}

\subsection{Importance Sampling}
\label{sec-is}
{\color{black}Referring back to the basic idea underlying variational inference, i.e., the ELBO for log-likelihood of response $ l(\boldsymbol{\Delta} \mid \boldsymbol{Y})$ in the inequality~\eqref{elbo}, it can be seen that a tighter lower bound is attained when $R\equiv  {P(\boldsymbol{Y}_{i},\boldsymbol{\theta}_i \mid  \boldsymbol{\Delta} )}/{q_i(\boldsymbol{\theta}_i)}$ around its mean $P(\boldsymbol{Y}_{i}\mid \boldsymbol{\Delta})$.} Therefore we can consider different random variables with the same mean that are more concentrated. For example, we can draw {\color{black}$M$} i.i.d. samples from $q(\boldsymbol{z})$, and average the estimates as in importance sampling (IS):
\begin{equation}
			R_M = \frac{1}{M}\sum_{m=1}^M R_m = \frac{1}{M} \sum_{m=1}^M \frac{p( \boldsymbol{x},  \boldsymbol{z}_m)}{ q( \boldsymbol{z}_m)}, \  \boldsymbol{z}_m \sim q(\cdot).
\end{equation}
This lead to a tighter ``importance weighted ELBO" (IW-ELBO) on $\log P ( \boldsymbol{x})$,
\begin{equation}
			\text{IW-ELBO}_M =  E_{q(\boldsymbol{Z})}\Big[  \log \frac{1}{M} \sum_{m=1}^M \frac{p( \boldsymbol{z}_m,  \boldsymbol{x})}{q( \boldsymbol{z}_m)} \Big] := \mathcal{L}_M( \boldsymbol{x}).
\label{iw-elbo}
\end{equation}
It is shown that $\mathcal{L}_M( \boldsymbol{x})$ converge to $\log p( \boldsymbol{x})$ as $M$ goes to infinity \citep{burda2015importance}, which is summarized  in the following result.
\begin{proposition}
\label{prop:converge}
			For all $M$, the lower bounds satisfy
			\begin{equation*}
				\log p( \boldsymbol{x}) \geq \mathcal{L}_{M+1} \geq \mathcal{L}_M.
			\end{equation*}
			Moreover, if $p( \boldsymbol{x},  \boldsymbol{z}) / q( \boldsymbol{z} |  \boldsymbol{x})$ is bounded, then $\mathcal{L}_M$ approaches $\log p( \boldsymbol{x})$ as $M$ goes to infinity.
\end{proposition}
Motivated by this result, we  use the importance sampling method   and calculate the derivatives of   $\mathcal{L}_M$ to further perform gradient based optimization.
		Specifically, denote $w_m = p( \boldsymbol{x},  \boldsymbol{z}_m) / q( \boldsymbol{z}_m) $, then the derivatives of   $\mathcal{L}_M$ with respect to $\boldsymbol{\theta}$ are
		\begin{align}
			\nonumber
			\nabla_{\boldsymbol{\theta}}\mathcal{L}_M ( \boldsymbol{x}) & = \nabla_{\boldsymbol{\theta}} {E}_{q( \boldsymbol{Z} )}\Big[  \log \frac{1}{M} \sum_{m=1}^M w_m \Big]\\
			\nonumber
			& =    {E}_{q( \boldsymbol{Z})} \Big[  \nabla_{\boldsymbol{\theta}}  \log \frac{1}{M} \sum_{m=1}^M w_m \Big]\\
			\nonumber
			& =   {E}_{q( \boldsymbol{Z})} \Big[\sum_{m=1}^M \tilde{w}_m  \nabla_{\boldsymbol{\theta}}   \log w_m \Big],
		\end{align}
		where $\tilde{w}_m = w_m / \sum_{m'=1}^M w_{m'}$ and
\begin{equation}
			\nabla_{\boldsymbol{\theta}}   \log w_m = \nabla_{\boldsymbol{\theta}}   \log p( \boldsymbol{x},  \boldsymbol{z}_m) - \nabla_{\boldsymbol{\theta}}   \log q( \boldsymbol{z}_m).
\label{eq-gradient}
\end{equation}

\subsection{IW-GVEM}
The primary idea of IW-GVEM is to replace Equation \eqref{elbo} with importance weighted ELBO as in Equation \eqref{iw-elbo}. That is, for each $i = 1,\dots, N$, we draw $M$ samples from $q_i(\boldsymbol{\theta}_i)$ for $S$ times:
		\[
		\boldsymbol{\theta}_i^{(s, m)} \sim q_i(\boldsymbol{\theta}_i), \text{ for } s = 1, \dots, S, m = 1, \dots, M.
		\]
Define $w_i^{(s, m)} = p(Y_i, \boldsymbol{\theta}_i^{(s,m)}) / q_i(\boldsymbol{\theta}_i^{(s,m)})$, where $p(Y_i, \boldsymbol{\theta}_i^{(s,m)}) = P(Y_i, \boldsymbol{\theta}_i^{(s,m)} \mid \boldsymbol{\alpha},  \boldsymbol{b}, {\color{black} \bm{\rho}})$ as in equation \eqref{eq-joint_p},
		and $q_i(\boldsymbol{\theta}_i^{(s,m)}) \sim N(\boldsymbol{\theta}_i^{(s,m)} \mid \mu_i, \Sigma_i)$, then $\mathcal{L}_M ( \boldsymbol{Y})$ can be approximated by
		\[
		\mathcal{L}_M ( \boldsymbol{Y}) \approx \sum_{i=1}^N \Big( \frac{1}{S}\sum_{s=1}^S\big[ \log \frac{1}{M}\sum_{m=1}^M w_i^{(s, m)} \big] \Big).
		\]
		Note $w_i^{(s,m)}$ is a function of parameters $(\xi_i, \boldsymbol{\alpha},  \boldsymbol{b} , {\color{black} \bm{\rho}})$.

To learn parameters, we use a stochastic gradient ascent method, which needs to calculate the gradients of $\mathcal{L}_M( \boldsymbol{Y})$.
		Based on equation \eqref{eq-gradient}, the gradients can be approximated by 
		\[
			\nabla_{\boldsymbol{\alpha}}\mathcal{L}_M ( \boldsymbol{Y}) \approx \sum_{i=1}^N \Big(
			\frac{1}{S}\sum_{s=1}^S
			\sum_{m=1}^M \Tilde{w}_i^{(s,m)} \nabla_{\boldsymbol{\alpha}}  \big[ \log P(Y_i, \boldsymbol{\theta}_i^{(s,m)} \mid \boldsymbol{\alpha},  \boldsymbol{b},  {\color{black} \bm{\rho}}) 
			- \nabla_{\boldsymbol{\alpha}}   \log q_i(\boldsymbol{\theta}_i^{(s,m)} \mid Y_i)
			\big]
			\Big),
		\]
		where $\Tilde{w}_i^{(s,m)} = w_i^{(s,m)} / \sum_{m'=1}^M w_i^{(s,m')}$.
		Note that $q_i(\boldsymbol{\theta}_i^{(s,m)} \mid Y_i)$ does not depend on the parameters in the current iteration.
		Therefore, we only need to calculate $\Tilde{w}_i^{(s,m)}$ and $\nabla_{\boldsymbol{\alpha}} P(Y_i, \boldsymbol{\theta}_i^{(s,m)} \mid \boldsymbol{\alpha},  \boldsymbol{b},  {\color{black} \bm{\rho}})$. 
		Similarly we can calculate $\nabla_{ \boldsymbol{b}}\mathcal{L}_M ( \boldsymbol{Y})$ and $\nabla_{\Sigma_{\boldsymbol{\theta}}}\mathcal{L}_M ( \boldsymbol{Y})$.
Specifically, we have
\begin{eqnarray}
\nonumber
\nabla_{\boldsymbol{\alpha}_j} \log P(Y_i, \boldsymbol{\theta}_i^{(s,m)} \mid \boldsymbol{\alpha},  \boldsymbol{b},\boldsymbol{\rho})
			& =& \Tilde{w}_i^{(s,m)} \nabla_{\boldsymbol{\alpha}_j}  \big[ \log P(Y_i, \boldsymbol{\theta}_i^{(s,m)} \mid \boldsymbol{\alpha},  \boldsymbol{b},  {\color{black} \bm{\rho}}) 
			\big]\\
			& =&\Tilde{w}_i^{(s,m)} \big[
			\big(Y_{ij} - 1 + \frac{1}{1 + \exp(\boldsymbol{\alpha}_j^\top \boldsymbol{\theta}_i^{(s,m)} - b_j)}\big) \boldsymbol{\theta}_i^{(s,m)}
			\big]\label{eq-grad-alpha}, \\
\nonumber
\nabla_{b_j} \log P(Y_i, \boldsymbol{\theta}_i^{(s,m)} \mid \boldsymbol{\alpha},  \boldsymbol{b},\boldsymbol{\rho}) 
			 & =& \Tilde{w}_i^{(s,m)} \nabla_{b_j}  \big[ \log P(Y_i, \boldsymbol{\theta}_i^{(s,m)} \mid \boldsymbol{\alpha},  \boldsymbol{b},  {\color{black} \bm{\rho}}) 
			\big]\\
& =& \Tilde{w}_i^{(s,m)} \big[
			1 - Y_{ij} - \frac{1}{1 + \exp(\boldsymbol{\alpha}_j^\top \boldsymbol{\theta}_i^{(s,m)} - b_j)}
			\big]			
\label{eq-grad-b}, \\
\nonumber
			\nabla_{\Sigma_{\boldsymbol{\theta}}} \log P(Y_i, \boldsymbol{\theta}_i^{(s,m)} \mid \boldsymbol{\alpha},  \boldsymbol{b},\boldsymbol{\rho}) 
			& =& \Tilde{w}_i^{(s,m)} \nabla_{\Sigma_{\boldsymbol{\theta}}}  \big[ \log P(Y_i, \boldsymbol{\theta}_i^{(s,m)} \mid \boldsymbol{\alpha},  \boldsymbol{b},  {\color{black} \bm{\rho}}) 
			\big] \\
& =& \Tilde{w}_i^{(s,m)} \big[
			\frac{1}{2}\Sigma_{\boldsymbol{\theta}} - \frac{1}{2} \boldsymbol{\theta}_i^{(s,m)} (\boldsymbol{\theta}_i^{(s,m)})^\top
						\big].\label{eq-grad-sigma}
\end{eqnarray}	
\noindent To summarize, in the $(t+1)th$ iteration, we perform the following:

\begin{enumerate}
	\item For $i=1, \dots, N$, draw $M$ samples from $q_i (\boldsymbol{\theta}_i)$ {\color{black}for $S$ times}.
	\item Calculate $w_i^{(s, m)} = P(Y_i, \boldsymbol{\theta}_i^{(s,m)}\mid \boldsymbol{\alpha}, \boldsymbol{b},  {\color{black} \bm{\rho}}) / q_i(\boldsymbol{\theta}_i^{(s,m)})$ and $\Tilde{w}_i^{(s,m)} = w_i^{(s,m)} / \sum_{m'=1}^M w_i^{(s,m')}$.
	\item Calculate the gradients according to equations \eqref{eq-grad-alpha}, \eqref{eq-grad-b} and \eqref{eq-grad-sigma}.
\end{enumerate}

Proper learning rate scheduling is important in gradient-based algorithms.
In this work, we apply the \textbf{Ada}ptive \textbf{m}oment estimation (Adam) method \citep{kingma2014adam}, which has been extensively used in deep learning research and applications, to adjust the learning rate in our training process.
In Adam, we compute individual adaptive learning rates for each parameter from estimates of the first and second moments of the gradients.
Specifically in the $t$th iteration, we calculate exponential moving averages of the gradient (denoted as $\boldsymbol{v}_t$) and the squared gradient (denoted as $\boldsymbol{s}_t$) with exponential decay rates $\beta_1$ and $\beta_2$ respectively.
The moving averages can be seen as estimates of the first and second moments of the gradients.
Then we correct these biased exponential moving averages by $1 - \beta_1^t$ and $1 - \beta_2^t$ respectively and update parameters using standardized gradients.
The concrete steps of generic Adam are provided below, where $\boldsymbol{g}_t$ is the gradient (corresponding to that in equations \eqref{eq-grad-alpha}, \eqref{eq-grad-b} and \eqref{eq-grad-sigma}, respectively) in the $t$th iteration:
\begin{enumerate}
	\item $\boldsymbol{v}_t = \beta_1 \boldsymbol{v}_{t-1} + (1-\beta_1)\boldsymbol{g}_t$ (update biased first moment estimate)
	\item ${\color{black} \boldsymbol{r}_t} = \beta_2 {\color{black} \boldsymbol{r}_{t-1}} + (1-\beta_2)\boldsymbol{g}_t^2$ (update biased second moment estimate)
	\item $\hat{\boldsymbol{v}}_t = \boldsymbol{v}_t / (1 - \beta_1^t),\ {\color{black} \hat{\boldsymbol{r}}_t} = {\color{black} \boldsymbol{r}_t} / (1 - \beta_2^t)$ (compute bias-corrected moment estimates)
	\item $\hat{\boldsymbol{g}}_t = \eta \hat{\boldsymbol{v}_t} /(\sqrt{{\color{black} \hat{\boldsymbol{r}}_t}} + \epsilon)$, where $\eta$ is learning rate (update the final gradient)
\end{enumerate}

\noindent With this, the proposed Importance-Weighted Gaussian Variational EM (IW-GVEM) algorithm is summarized in Algorithm \ref{algo:iw-gvem-m2pl}. {\color{black} For the choice of hyperparameters, we follow the suggestions in \cite{kingma2014adam} and adopt the default setting that $\beta_1 = 0.9$ and $\beta_2 = 0.999$. Empirically in our simulation studies, for better convergence performance, we let the learning rate of $\bm{\Sigma}_{\theta}$ to be $0.1\eta$ while the learning rate for $\bm{a}$ and $\bm{b}$ to be $\eta$, and we search for an optimal learning rate  $\eta$ with the maximum ELBO over a list $\{0.01, 0.05, 0.1, 0.5\}$. Lastly, we set $\epsilon = 0.001$. 

In terms of convergence criteria, we evaluate the Euclidean norm of the difference between the estimated parameters of the current step and those of the previous step. When the difference is less than a certain tolerance value, the algorithm is stopped. For our simulation studies, in obtaining the initial model parameter using the GVEM algorithm, we reach convergence at $(l+1)$th iteration if $\| \bm{\alpha}_{GV}^{l+1} - \bm{\alpha}_{GV}^{l}\|_2 + \| \bm{b}_{GV}^{t+1} - \bm{b}_{GV}^{t}\|_2 +  \| \bm{\Sigma}_{\theta, GV}^{l+1} - \bm{\Sigma}_{\theta, GV}^{l}\|_2 \leqslant 0.0001$. In IW-GVEM, we reach convergence at $(t+1)$th iteration when $\max\{ \| \bm{\alpha}^{t+1} - \bm{\alpha}^{t}\|_2, \| \bm{b}^{t+1} - \bm{b}^{t}\|_2, \| \bm{\Sigma}_{\theta}^{t+1} - \bm{\Sigma}_{\theta}^{t}\|_2 \} \leqslant 0.0001$ or the iteration stops when it reaches certain maximum iteration number.} 

\begin{algorithm}[ht!]
\caption{IW-GVEM for M2PL}
\label{algo:iw-gvem-m2pl}
\SetKwInOut{Input}{Input}
\SetKwInOut{Output}{Output}

\KwData{Binary response matrix $ \boldsymbol{Y}\in \{0, 1\}^{N\times J}$.}

Run GVEM algorithm and obtain $ \boldsymbol{\mu}_{i, \text{GV}}$, $ \boldsymbol{\Sigma}_{i,\text{GV}}$, $\boldsymbol{\alpha}_{\text{GV}}$, $ \boldsymbol{b}_{\text{GV}}$, $ \boldsymbol{\Sigma}_{{\theta},\text{GV}}$, and $\xi_{ij}$. These values will serve as initial values for IW-GVEM.

Set hyper-parameters $S$, $M$ for importance sampling, and $\beta_1$, $\beta_2$, $\eta$ and $\epsilon$ for Adam.


Set $ \boldsymbol{v}^{(0)}_{ \boldsymbol{\alpha}_j} =  \boldsymbol{0}$, $ \boldsymbol{v}^{(0)}_{ \boldsymbol{b}_j} =  \boldsymbol{0}$, $ \boldsymbol{v}^{(0)}_{ \boldsymbol{\Sigma}_{ \boldsymbol{\theta}}} =  \boldsymbol{0}$, $ \boldsymbol{r}^{(0)}_{ \boldsymbol{\alpha}_j} =  \boldsymbol{0}$, $ \boldsymbol{r}^{(0)}_{ \boldsymbol{b}_j} =  \boldsymbol{0}$, $ \boldsymbol{r}^{(0)}_{ \boldsymbol{\Sigma}_{ \boldsymbol{\theta}}} =  \boldsymbol{0}$.

\While{not converged}{

In the $t$-th iteration,

\For{$i \in [N]$}{
	draw $M$ samples from $q_i ( \boldsymbol{\theta}_i) = N( \boldsymbol{\theta}_i \mid  \boldsymbol{\mu}_{i,\text{GV}},  \boldsymbol{\Sigma}_{i,\text{GV}})$ {\color{black}for $S$ times}.
}

\For{$i \in [N],\ s \in [S]$ and $m \in [M]$}{
	\medskip
	$w_i^{(s, m)} = p\big(  \boldsymbol{Y}_i,  \boldsymbol{\theta}_i^{(s,m)}\mid  \boldsymbol{\alpha},  \boldsymbol{b},  {\color{black} \bm{\rho}} \big)\  / \ q_i \big( \boldsymbol{\theta}_i^{(s,m)}\big)$,\quad
	$\Tilde{w}_i^{(s,m)} = w_i^{(s,m)} / \sum_{m'=1}^M w_i^{(s,m')}$.	
}

\For{$j \in [J]$}{
	\medskip
	$ \boldsymbol{g}_{ \boldsymbol{\alpha}_j} = \sum_{i=1}^N \Big( \frac{1}{S} \sum_{s=1}^S \sum_{m=1}^M \Tilde{w}_i^{(s,m)} \big[
			Y_{ij} - 1 + 1 / \big(1 + \exp\{ \boldsymbol{\alpha}_j^\top  \boldsymbol{\theta}_i^{(s,m)} - b_j\}\big)\big]  \boldsymbol{\theta}_i^{(s,m)}\Big)$,
			
	\medskip
	$ \boldsymbol{g}_{b_j} = \sum_{i=1}^N \Big( \frac{1}{S} \sum_{s=1}^S \sum_{m=1}^M \Tilde{w}_i^{(s,m)} \big[
			1 - Y_{ij} - 1 / \big(1 + \exp\{ \boldsymbol{\alpha}_j^\top  \boldsymbol{\theta}_i^{(s,m)} - b_j\}\big)\big] \Big)$.
	
}

$ \boldsymbol{g}_{ \boldsymbol{\Sigma}_{ \boldsymbol{\theta}}} = \sum_{i=1}^N \Big( \frac{1}{S} \sum_{s=1}^S \sum_{m=1}^M \Tilde{w}_i^{(s,m)} \big[
			 \boldsymbol{\Sigma}_{ \boldsymbol{\theta}} -  \boldsymbol{\theta}_i^{(s,m)} ( \boldsymbol{\theta}_i^{(s,m)})^\top
						\big] / 2\Big)$.

\medskip
\For{$j \in [J]$}{
	\medskip
	$ \boldsymbol{v}^{(t)}_{ \boldsymbol{\alpha}_j} = \beta_1  \boldsymbol{v}^{(t-1)}_{ \boldsymbol{\alpha}_j} + (1-\beta_1) \boldsymbol{g}_{ \boldsymbol{\alpha}_j}$,
	\quad
	$ \boldsymbol{r}^{(t)}_{ \boldsymbol{\alpha}_j} = \beta_2  \boldsymbol{r}^{(t-1)}_{ \boldsymbol{\alpha}_j} + (1-\beta_2) \boldsymbol{g}_{ \boldsymbol{\alpha}_j} \textcolor{red}{\cdot} \boldsymbol{g}_{ \boldsymbol{\alpha}_j} $,
	
	\medskip
	$ \boldsymbol{v}^{(t)}_{ \boldsymbol{\alpha}_j} =  \boldsymbol{v}^{(t)}_{ \boldsymbol{\alpha}_j} / (1 - \beta_1^{t})$,
	\quad
	$ \boldsymbol{r}^{(t)}_{ \boldsymbol{\alpha}_j} =  \boldsymbol{r}^{(t)}_{ \boldsymbol{\alpha}_j} / (1 - \beta_2^{t})$,
	
	\medskip
	$ \boldsymbol{v}^{(t)}_{ \boldsymbol{b}_j} = \beta_1  \boldsymbol{v}^{(t-1)}_{ \boldsymbol{b}_j} + (1-\beta_1) \boldsymbol{g}_{ \boldsymbol{b}_j}$,
	\quad
	$ \boldsymbol{r}^{(t)}_{ \boldsymbol{b}_j} = \beta_2  \boldsymbol{v}^{(t-1)}_{ \boldsymbol{b}_j} + (1-\beta_2) \boldsymbol{g}_{ \boldsymbol{b}_j} \cdot  \boldsymbol{g}_{ \boldsymbol{b}_j}$,
	
	\medskip
	$ \boldsymbol{v}^{(t)}_{ \boldsymbol{b}_j} =  \boldsymbol{v}^{(t)}_{ \boldsymbol{b}_j} / (1 - \beta_1^{t})$,
	\quad
	$ \boldsymbol{r}^{(t)}_{ \boldsymbol{b}_j} =  \boldsymbol{r}^{(t)}_{ \boldsymbol{b}_j} / (1 - \beta_2^{t})$.
}

\medskip

$ \boldsymbol{v}^{(t)}_{ \boldsymbol{\Sigma}_{ \boldsymbol{\theta}}} = \beta_1  \boldsymbol{v}^{(t-1)}_{ \boldsymbol{\Sigma}_{ \boldsymbol{\theta}}} + (1-\beta_1) \boldsymbol{g}_{ \boldsymbol{\Sigma}_{ \boldsymbol{\theta}}}$,
\quad
$ \boldsymbol{r}^{(t)}_{ \boldsymbol{\Sigma}_{ \boldsymbol{\theta}}} = \beta_2  \boldsymbol{r}^{(t-1)}_{ \boldsymbol{\Sigma}_{ \boldsymbol{\theta}}} + (1-\beta_2) \boldsymbol{g}_{ \boldsymbol{\Sigma}_{ \boldsymbol{\theta}}} \textcolor{red}{\cdot} \boldsymbol{g}_{ \boldsymbol{\Sigma}_{ \boldsymbol{\theta}}} $,

\medskip
$ \boldsymbol{v}^{(t)}_{ \boldsymbol{\Sigma}_{ \boldsymbol{\theta}}} =  \boldsymbol{v}^{(t)}_{ \boldsymbol{\Sigma}_{ \boldsymbol{\theta}}} / (1 - \beta_1^{t})$,
\quad
$ \boldsymbol{r}^{(t)}_{ \boldsymbol{\Sigma}_{ \boldsymbol{\theta}}} =  \boldsymbol{r}^{(t)}_{ \boldsymbol{\Sigma}_{ \boldsymbol{\theta}}} / (1 - \beta_2^{t})$.

\medskip
\For{$j \in [J]$}{
	\medskip
	$\hat{ \boldsymbol{g}}_{ \boldsymbol{\alpha}_j}  = \eta  \boldsymbol{v}^{(t)}_{ \boldsymbol{\alpha}_j} \ \big/ \ \big(\sqrt{ \boldsymbol{r}^{(t)}_{ \boldsymbol{\alpha}_j}} + \epsilon \big)$, \quad
	$\hat{ \boldsymbol{\alpha}}_j^{(t)} = \hat{ \boldsymbol{\alpha}}_j^{(t-1)} + \hat{ \boldsymbol{g}}_{ \boldsymbol{\alpha}_j}$,
	
	\medskip
	$\hat{ \boldsymbol{g}}_{ \boldsymbol{b}_j}  = \eta  \boldsymbol{v}^{(t)}_{ \boldsymbol{b}_j} \ \big/ \ \big(\sqrt{ \boldsymbol{r}^{(t)}_{ \boldsymbol{b}_j}} + \epsilon \big)$, \quad
	$\hat{ \boldsymbol{b}}_j^{(t)} = \hat{ \boldsymbol{b}}_j^{(t-1)} + \hat{ \boldsymbol{g}}_{ \boldsymbol{b}_j}$.
}

$\hat{ \boldsymbol{g}}_{ \boldsymbol{\Sigma}_{ \boldsymbol{\theta}}}  = \eta  \boldsymbol{v}^{(t)}_{ \boldsymbol{\Sigma}_{ \boldsymbol{\theta}}} \ \big/ \ \big(\sqrt{ \boldsymbol{r}^{(t)}_{ \boldsymbol{\Sigma}_{ \boldsymbol{\theta}}}} + \epsilon \big)$, \quad
	$\hat{ \boldsymbol{\Sigma}}_{ \boldsymbol{\theta}}^{(t)} = \hat{ \boldsymbol{\Sigma}}_{ \boldsymbol{\theta}}^{(t-1)} + \hat{ \boldsymbol{g}}_{ \boldsymbol{\Sigma}_{ \boldsymbol{\theta}}}$.
}

\Output{$\hat{ \boldsymbol{\alpha}}, \ \hat{ \boldsymbol{b}}$ and $\hat{ \boldsymbol{\Sigma}}_{ \boldsymbol{\theta}}$.}

\end{algorithm}{}

\section{Simulation Studies}
\label{sec-simulation studies}
\subsection{Design}
\label{sec-simulation}
We conducted comprehensive simulation studies to evaluate the performance of the proposed method under various manipulated conditions.
We follow similar designs as in \cite{cho2021gaussian} and consider different settings: (1) sample size: $N$ = 200 or 500; (2) number of domains:  $K$ = 2 or 5; (3) test length:  $J$ = 30 if $K$ = 2 or  $J$ = 55 if $K$ = 5; (4) both within and between multidimensional structures; (5) factor correlations: low correlation $r\sim \text{unif}(0.1, 0.3)$ or high correlation $r \sim \text{unif}(0.5, 0.7)$; and (6) confirmatory or exploratory analysis.

Similar to \cite{cho2021gaussian}, for the between-item multidimensional structure, we had equal numbers of items loaded on each factor.
For the within-item multidimensional structure, when $K=2$, about one third of the items were loaded onto the first, or the second, or both factors respectively.
In the cases where $K=5$, there were about one-third of the items loaded onto one, two, or three factors respectively.
For the model parameters, we simulated the item discrimination parameters $\alpha_{j,k}$ from uniform distribution on $[1,2]$, and difficulty parameter $b_j$ from the standard normal distribution.
We generated the latent traits $\boldsymbol{\theta}_j$ from multivariate normal distribution $N(\boldsymbol{0}, \boldsymbol{\Sigma}_{\boldsymbol{\theta}})$, where the diagonal elements of $\boldsymbol{\Sigma}_{\boldsymbol{\theta}}$ were all 1 and off-diagonal elements were generated from uniform distributions.
Specifically, in high-correlation settings, the uniform distribution was set to be $\text{unif}(0.5, 0.7)$, whereas in the low-correlation settings we set it to be $\text{unif}(0.1, 0.3)$.

For evaluation, we compared the bias and Root Mean Squared Errors (RMSEs) of model parameters, as well as computation time between GVEM and IW-GVEM. For exploratory analysis, we did a promax rotation after model convergence, and compared the rotated parameters to the true values \citep{cho2022qestimation}.
For IW-GVEM, we first ran GVEM algorithm to get initial estimates of model parameters, and then ran several gradient descent steps using importance sampling to correct the bias.
To select a proper initial learning rate for the gradient algorithm, we first sampled a set of data aside based on the GVEM estimates.
After we got model parameter estimates using importance sampling,
we calculated the lower bound as in our objective function based on the previously sampled data set, and chose the learning rate corresponding to the largest lower bound.
In the simulation studies, we set $S$ and $M$ to be 10.
Our empirical experiments have shown that increasing $S$ and $M$ did not result in significant improvements and 10 was large enough for the simulation settings.
The results were averaged over 100 repetitions.

\subsection{Results}
\label{sec-results}

Figures \ref{fig:bias-k2-confirm} and   \ref{fig:rmse-k2-confirm} present the bias and RMSE of confirmatory M2PL model when $K=2$. {\color{black} Note that in confirmatory analysis, there are discrimination parameters specified to be zeros. These zero-constrained terms are excluded in the bias and RMSE computation.} The two separately colored boxes represent the distribution of respective criteria across 100 replications from IW-GVEM (denoted as ``IS'' in the figure) and the original GVEM algorithm. As shown, GVEM already performs well by producing close to 0 bias for $\boldsymbol{b}$ and $\boldsymbol{\Sigma}_{\theta}$. It is the discrimination parameter, $\boldsymbol{\alpha}$, that has a non-ignorable bias. The IW-GVEM algorithm effectively corrects such bias on $\boldsymbol{\alpha}$ across all conditions without deteriorating the estimation accuracy of other parameters. And because the bias is corrected, the RMSE of $\boldsymbol{\alpha}$ is also smaller consistently compared to that from GVEM, whereas again, there is no appreciable difference between IW-GVEM and GVEM in terms of RMSE on $\boldsymbol{b}$ and $\boldsymbol{\Sigma}_{\theta}$. Sorting through the manipulated conditions, it is ``within-item'' multidimensional structure in combination with high factor correlation tends to yield larger RMSE for both methods and all parameters.

Figures \ref{fig:bias-k5-confirm} and \ref{fig:rmse-k5-confirm} present the bias and RMSE of confirmatory M2PL model when  $K=5$.
 The trend observed from the $K=2$ condition continues to hold here. That is, IW-GVEM can correct bias on $\boldsymbol{\alpha}$ effectively and hence also brings down its RMSE, whereas bias on the other parameters are already close to 0 from both methods and their RMSE's are also comparable. Increasing the number of dimensions certainly makes the model estimation harder to converge, and the estimates are also more variable, especially for $\boldsymbol{b}$ and $\boldsymbol{\Sigma}_{\theta}$, as reflected by wider boxes for those parameters in Figure \ref{fig:bias-k5-confirm}.

Figures \ref{fig:bias-k2-explore} to \ref{fig:rmse-k5-explore} presents the results from exploratory estimation condition, in the same order as before. For the exploratory M2PL model estimation, GVEM generally performs well and the bias on $\boldsymbol{\alpha}$ is already small to begin with. This is consistent with the results reported in literature \citep{cho2021gaussian, cho2022qestimation}. {\color{black} Even so,  under all settings, the RMSEs of IW-GVEM are still smaller than or equal to that of GVEM. IW-GVEM can still further bring down the bias of $\boldsymbol{\alpha}$ to near 0 for most cases. The exceptional case when the bias of $\bm{\alpha}$ from IW-GVEM is larger than the bias from GVEM is for the ``within item, correlation is high" condition. This case is the most difficult case where the items were loaded on factors via a more complicated setting and the correlations among factors are relatively high. Nonetheless, this special case has overall good estimation performance as the estimation bias from IW-GVEM is still close to the bias from GVEM, and the RMSE from IW-GVEM is lower than the RMSE from GVEM.} In addition, when $K=2$ the bias of $\boldsymbol{\Sigma}_{\theta}$ appears to depart from 0 and IW-GVEM does not correct for such bias, although the RMSE of $\boldsymbol{\Sigma}_{\theta}$ is kept small across the board. The bias of $\boldsymbol{\Sigma}_{\theta}$ gets closer to 0 when $K$ increases and when the factor correlation is low. Because in the exploratory estimation mode, specific types of rotations will affect resulting factor correlations, the bias in $\boldsymbol{\Sigma}_{\theta}$ estimation is less of a concern. {\color{black} Although the increase in the number of dimensions $K$ could lead to a more complicated model and bring challenges to parameter estimation, the increase in test length, on the other hand, improves the estimation accuracy of parameters. Specifically, at $K = 5$, we use test length $J = 55$ which is greater than $J = 30$ at $K = 2$. This increase in test length explains the results that the biases at $K =5$ are closer to 0 than that at $K =2$ for some cases.} Overall, the results from GVEM and IW-GVEM are very close.

Table \ref{time1} presents the computation time for confirmatory M2PL estimation under both GVEM and IW-GVEM algorithms. Understandably, IW-GVEM takes longer time under all conditions because both the important sampling step and the gradient descent optimization are time consuming compared to closed-form updates in GVEM. Unsurprisingly, Both methods need longer time for larger sample sizes. It is more interesting to note that, other things being equal, when the multidimensional structure is ``within-item'', GVEM almost doubles (when $K=2$) or sometimes even triples (when $K=5$) the computation time compared to the ``between-item'' condition. But for IW-GVEM, the computation time is rather stable across these two multidimensional structures. Similarly, high correlation among factors is known to be more challenging, hence computation time increases by about 50\% or more for GVEM from low to high correlation conditions, but the computation time of IW-GVEM seems to be unaffected. These all suggest that IW-GVEM is better suited for more complex models. The same patterns remain for the exploratory M2PL estimation, as shown in Table \ref{time2}, although exploratory analysis in general takes longer time than confirmatory analysis, simply because   more parameters are needed  to be updated simultaneously.



\begin{figure}[htbp!]
    \centering
    \subfigure{
        \includegraphics[width=2.5in]{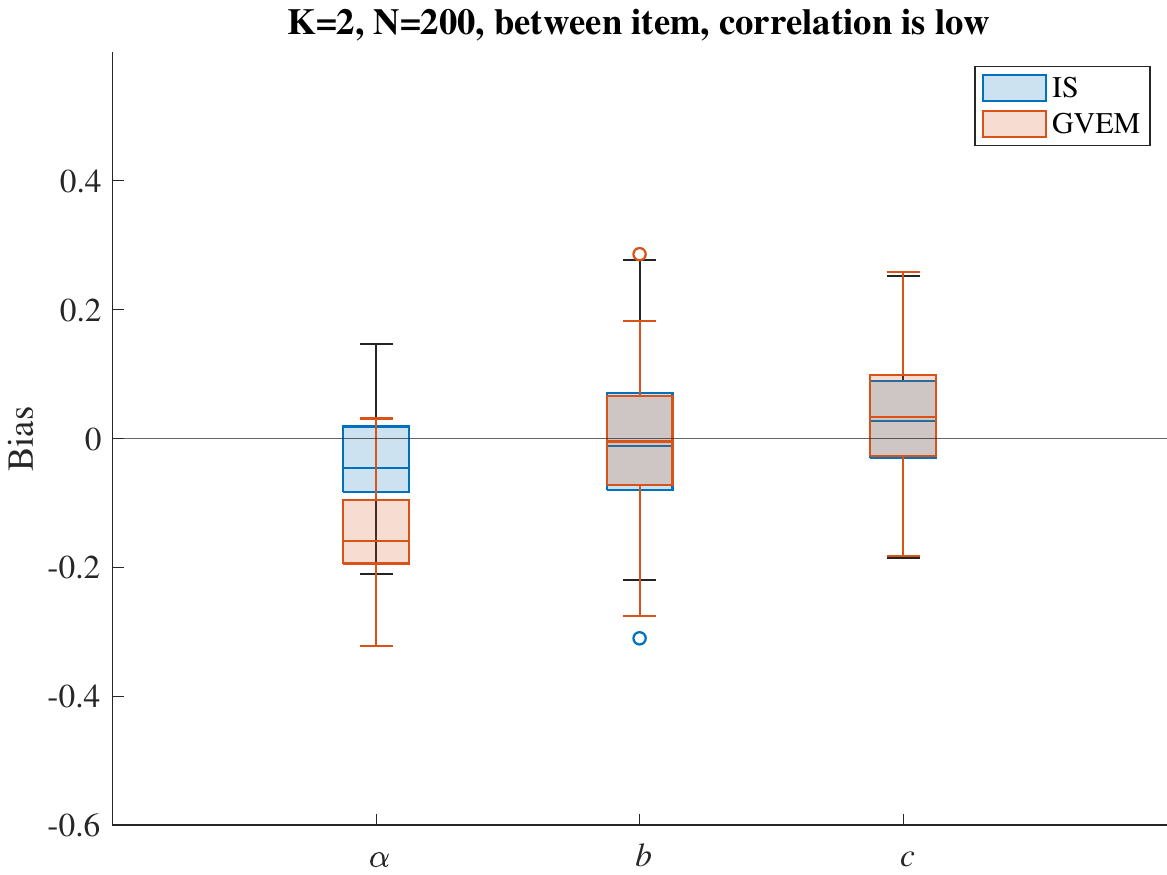}}
    \hspace{0.6in}
    \subfigure{
        \includegraphics[width=2.5in]{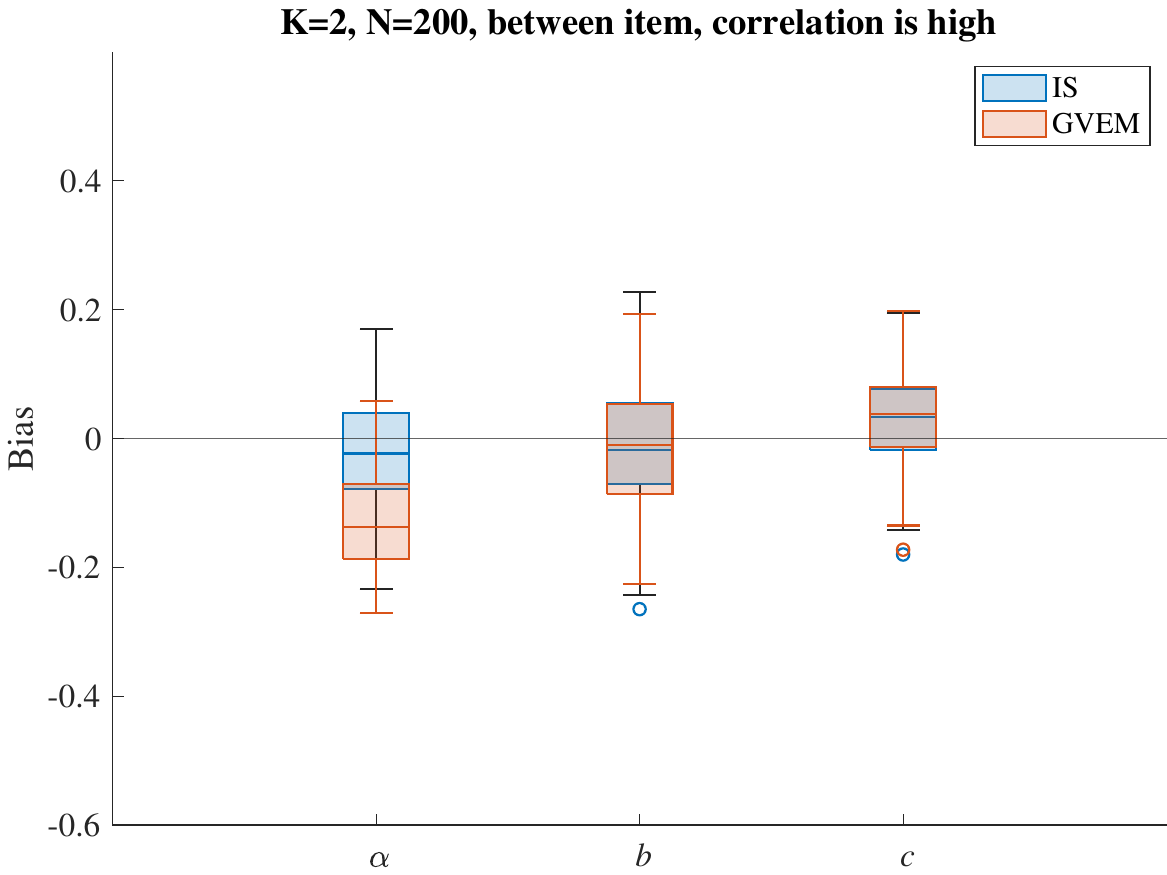}
        }
    \\
    \subfigure{
        \includegraphics[width=2.5in]{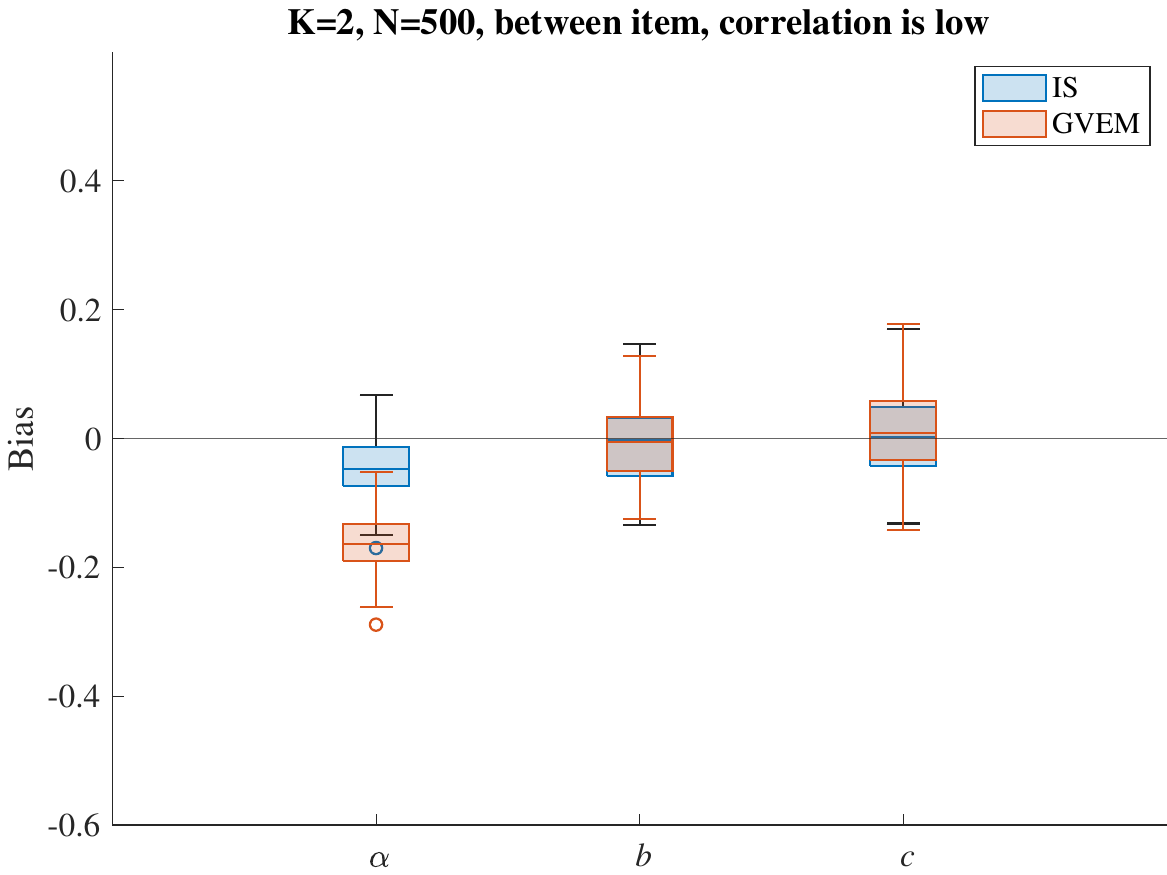}}
    \hspace{0.6in}
    \subfigure{
        \includegraphics[width=2.5in]{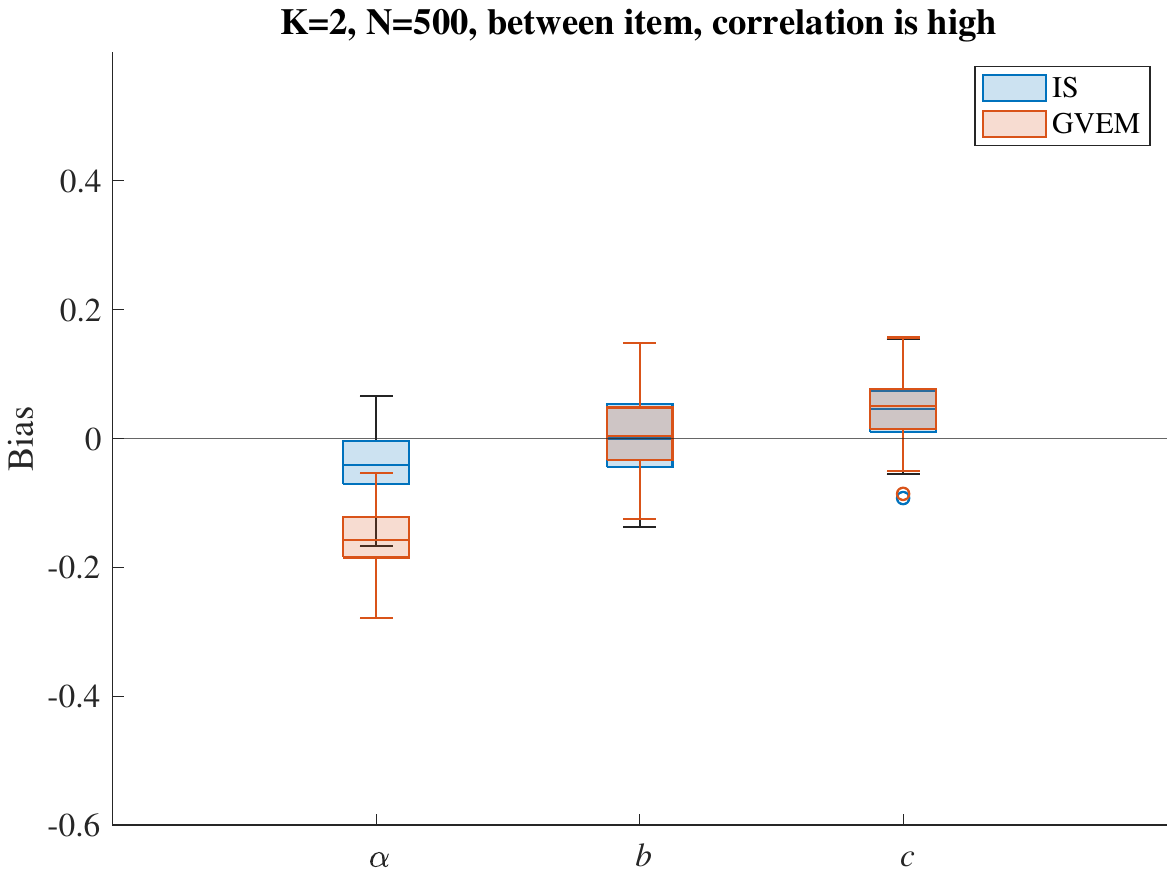}}
    \\
    \subfigure{
        \includegraphics[width=2.5in]{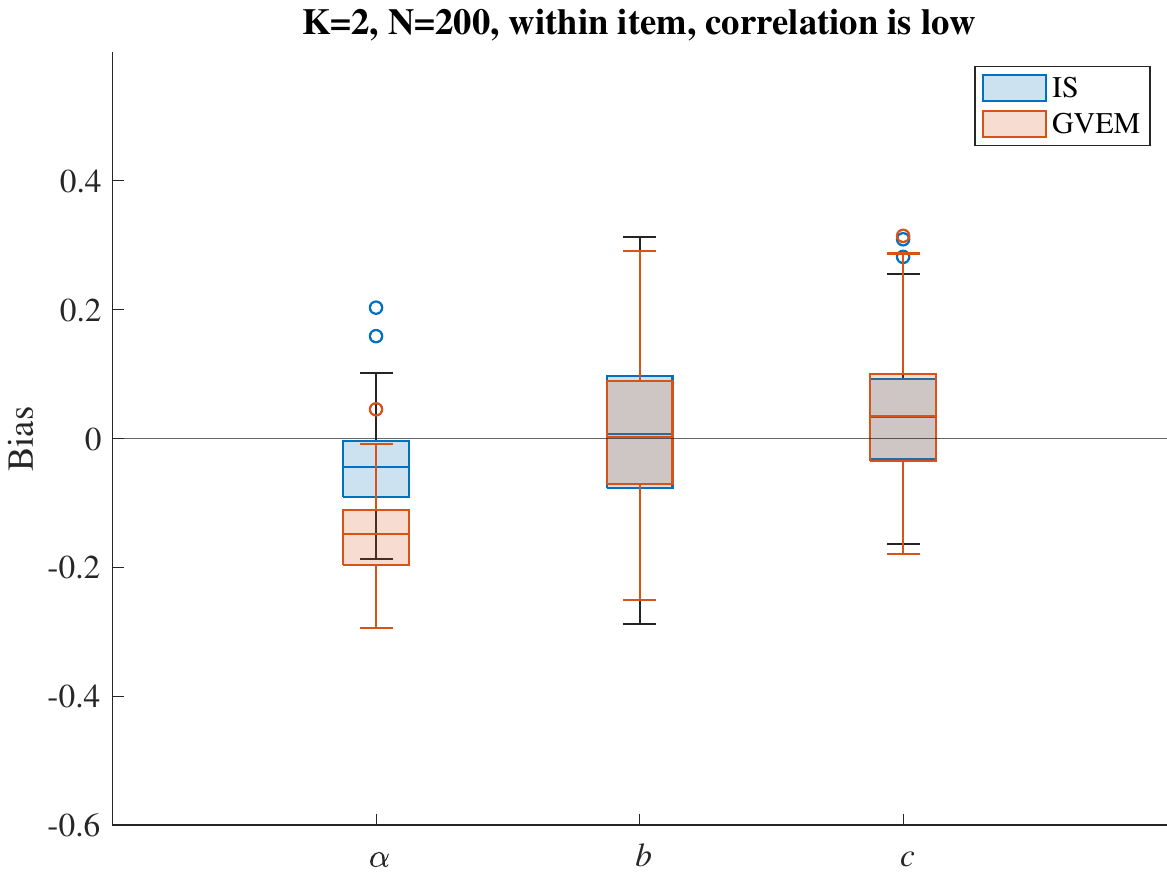}}
    \hspace{0.6in}
    \subfigure{
        \includegraphics[width=2.5in]{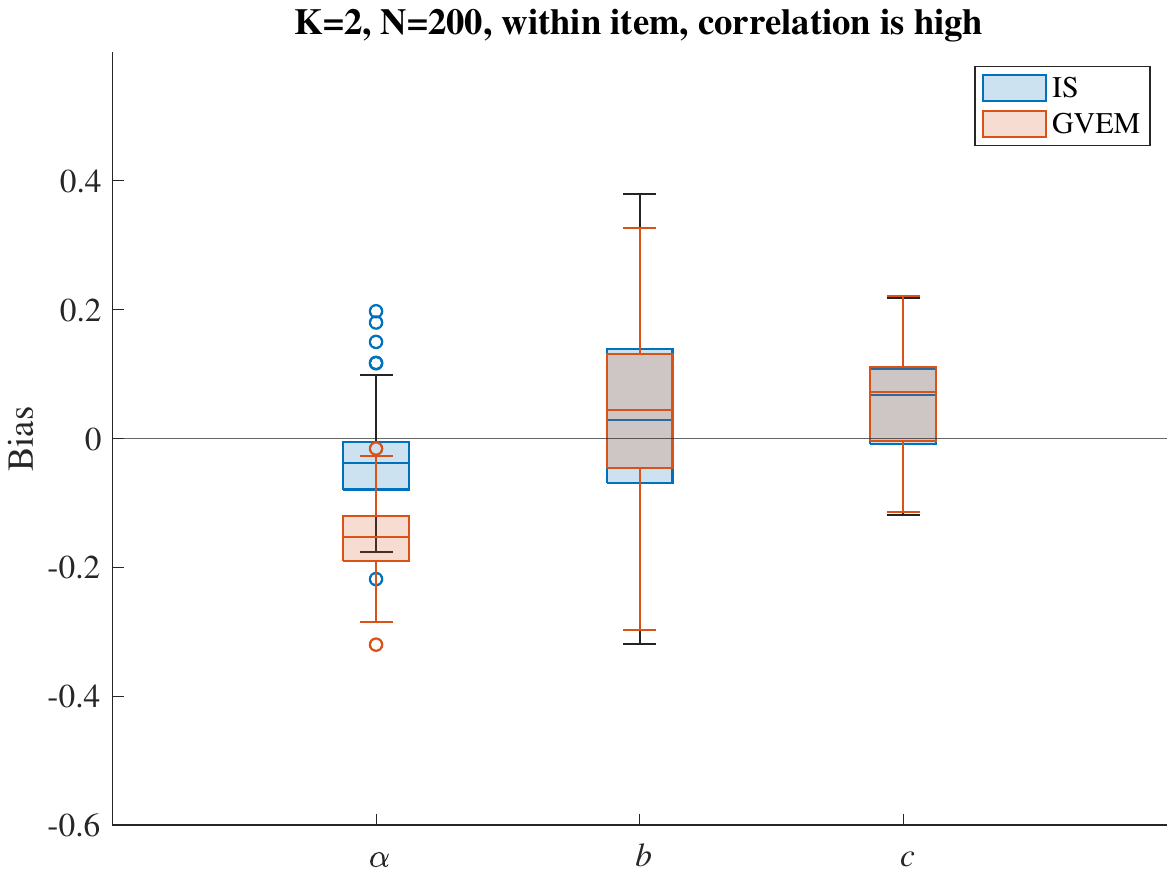}}
    \\
    \subfigure{
        \includegraphics[width=2.5in]{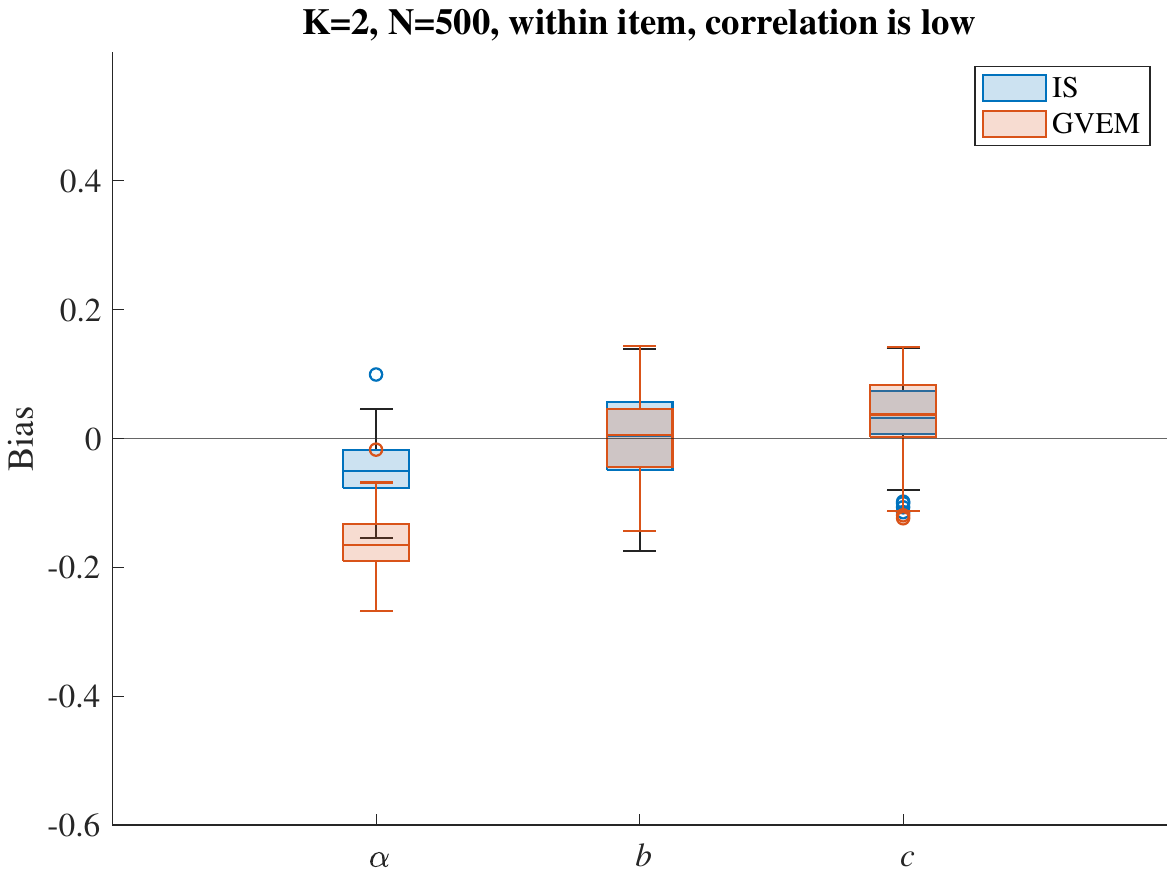}}
    \hspace{0.6in}
    \subfigure{
        \includegraphics[width=2.5in]{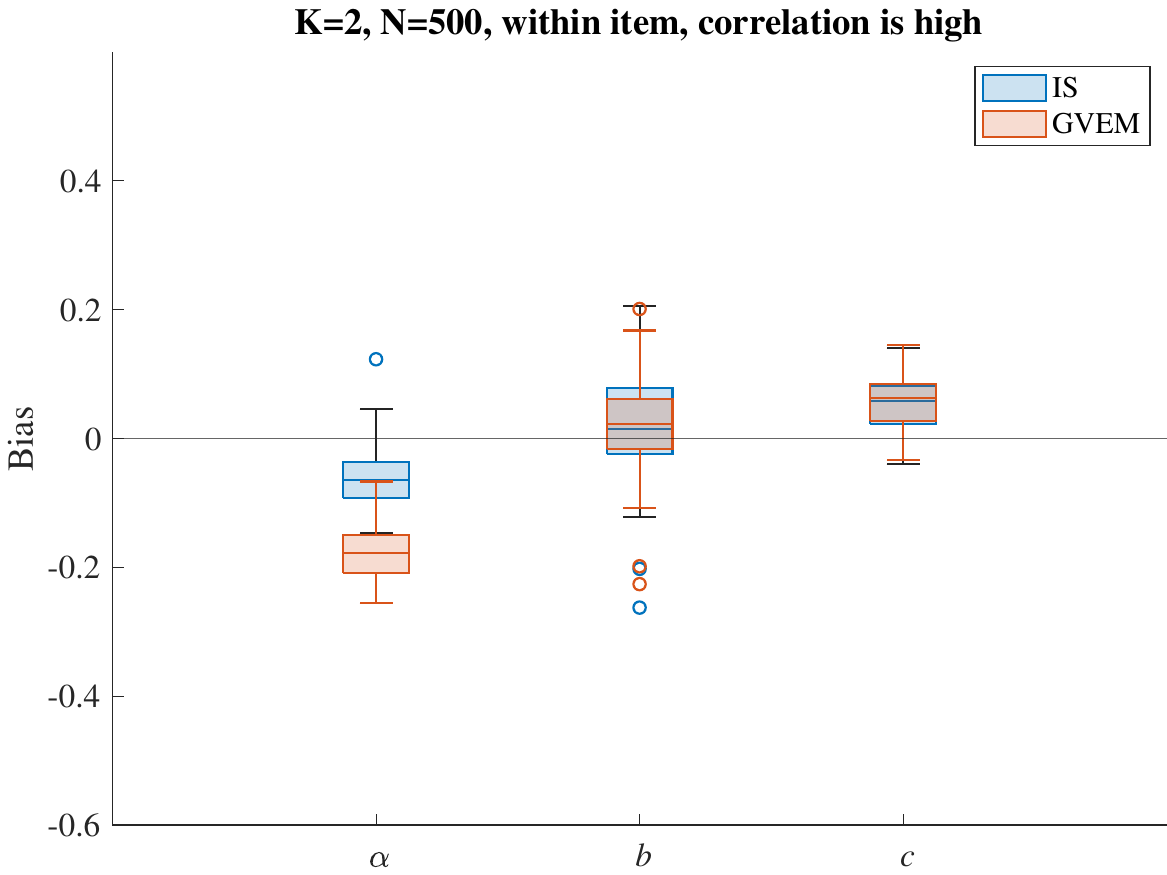}}
    \\
    \caption{Bias for $K=2$ under confirmatory analysis 
    }
    \label{fig:bias-k2-confirm}
\end{figure}

\newpage 

\begin{figure}[ht!]
    \centering
    \subfigure{
        \includegraphics[width=2.5in]{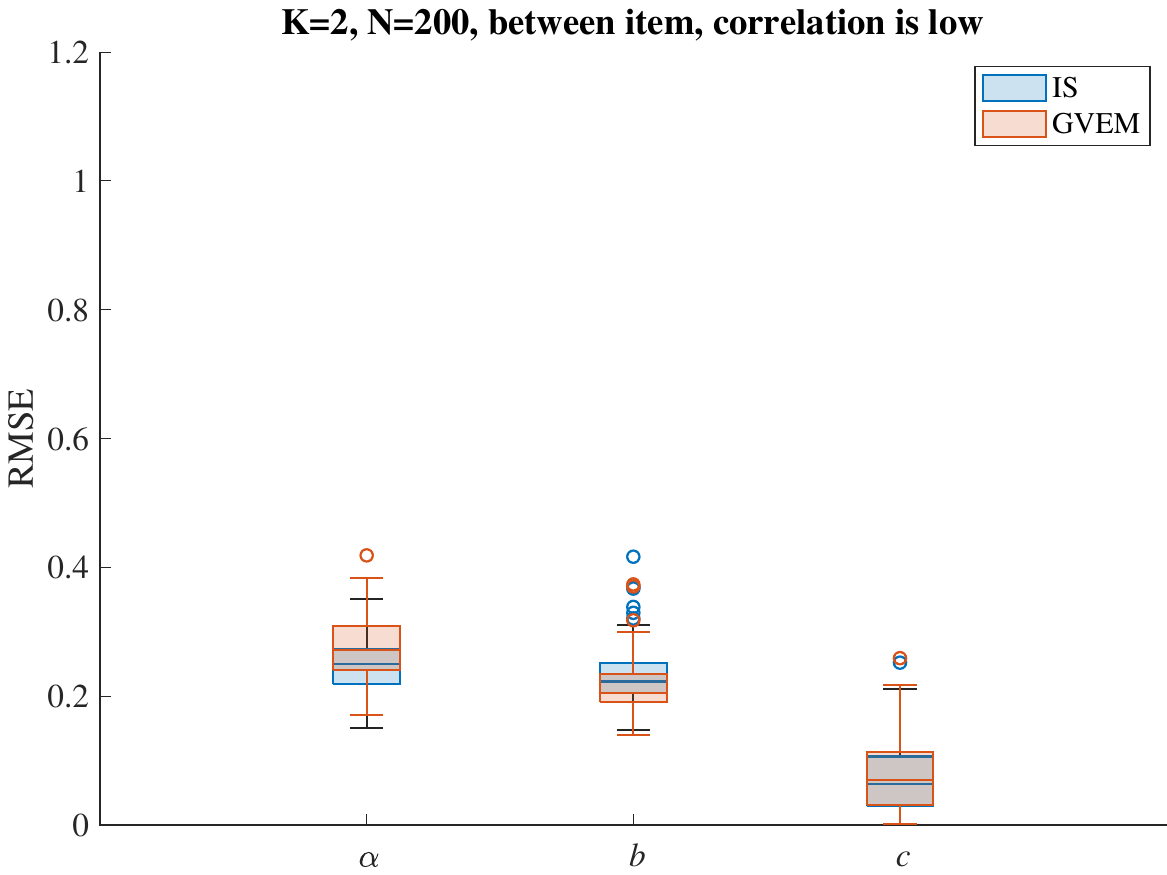}}
    \hspace{0.6in}
    \subfigure{
        \includegraphics[width=2.5in]{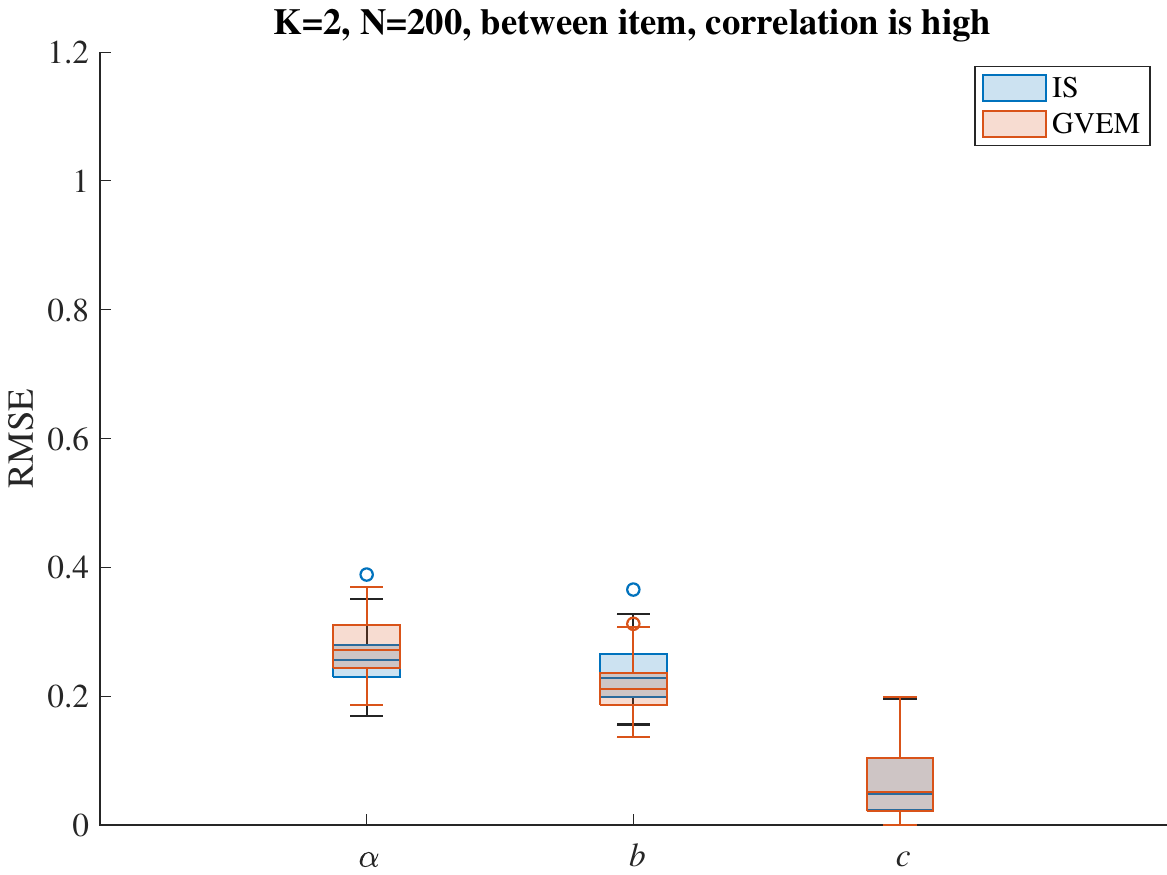}}
    \\
    \subfigure{
        \includegraphics[width=2.5in]{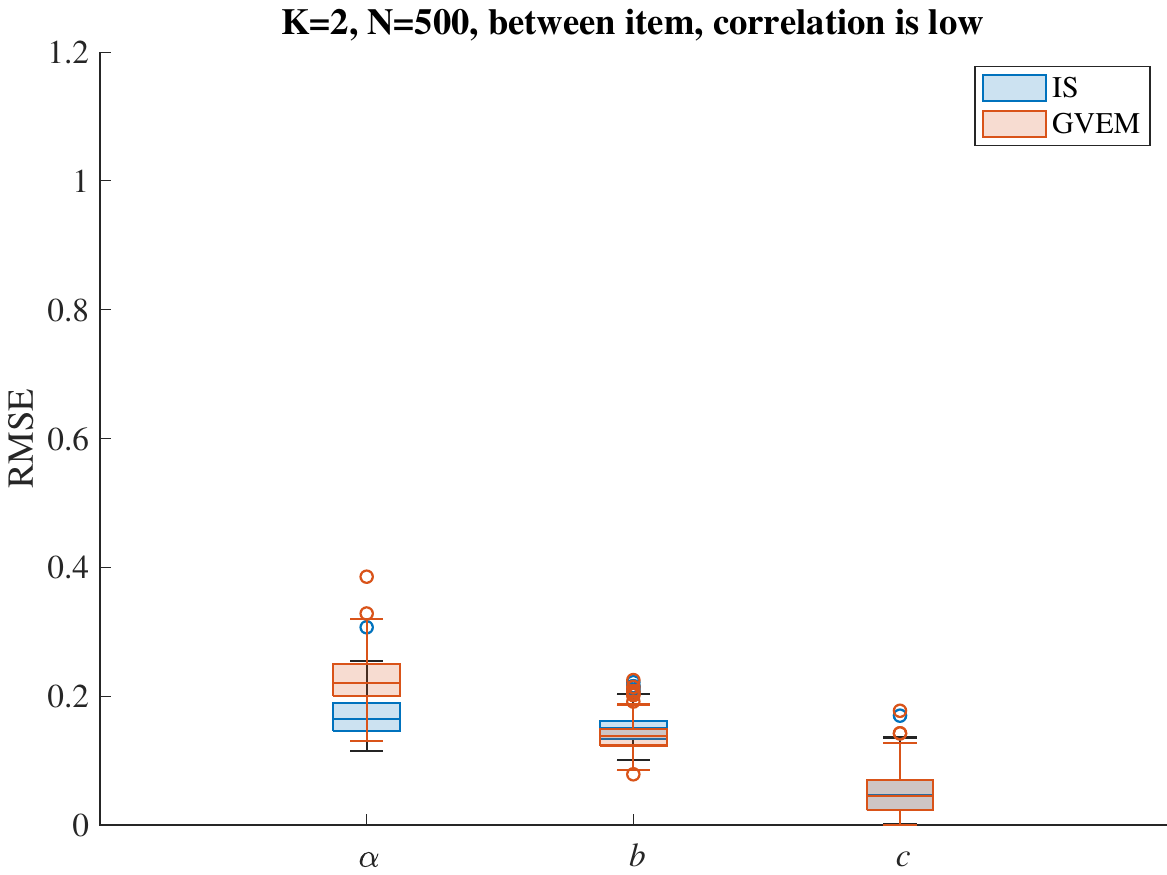}}
    \hspace{0.6in}
    \subfigure{
        \includegraphics[width=2.5in]{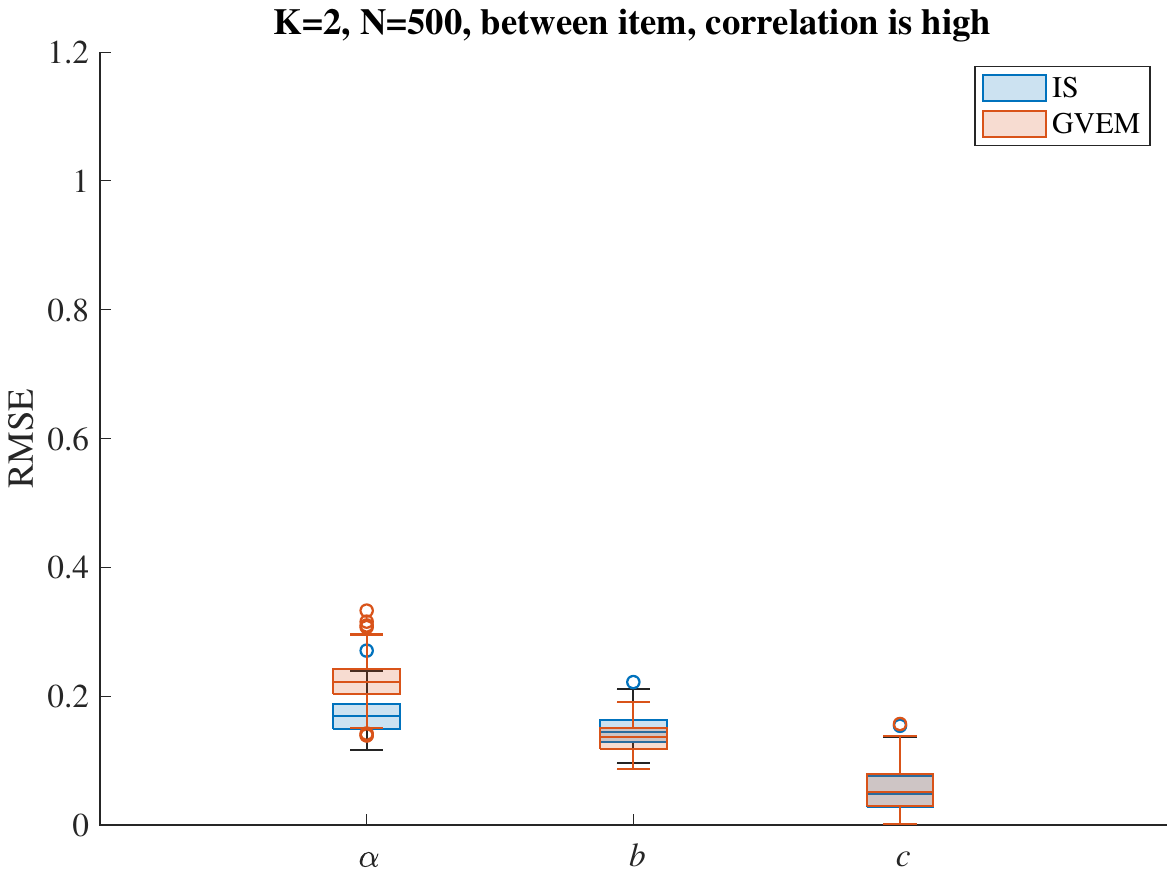}}
    \\
    \subfigure{
        \includegraphics[width=2.5in]{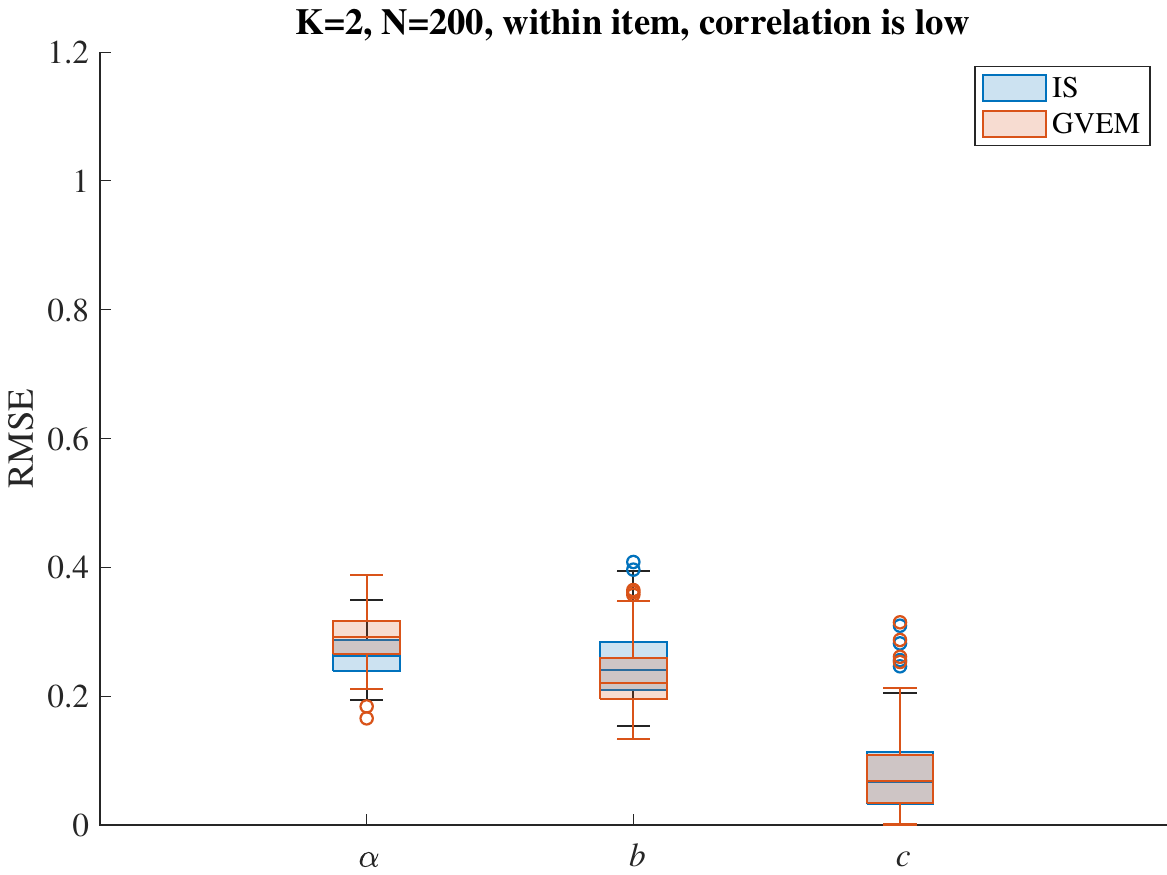}}
    \hspace{0.6in}
    \subfigure{
        \includegraphics[width=2.5in]{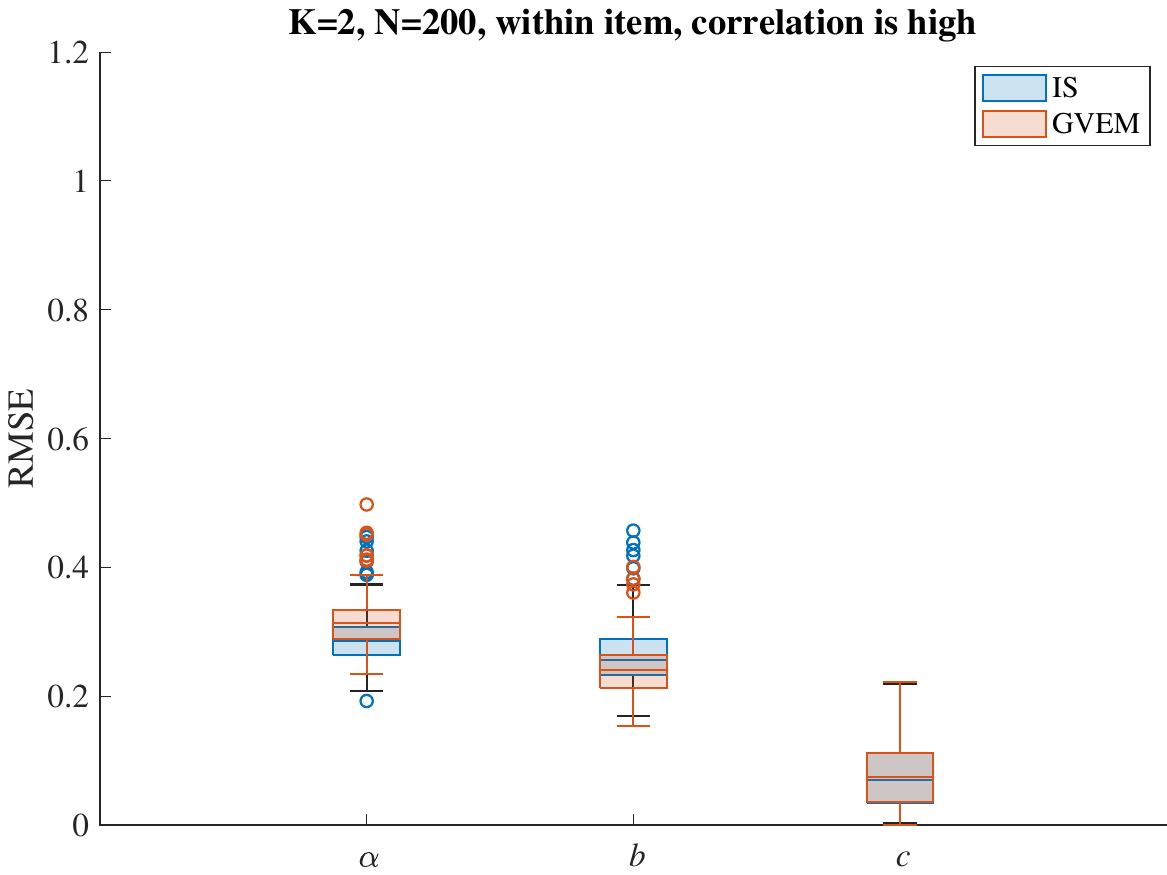}}
    \\
    \subfigure{
        \includegraphics[width=2.5in]{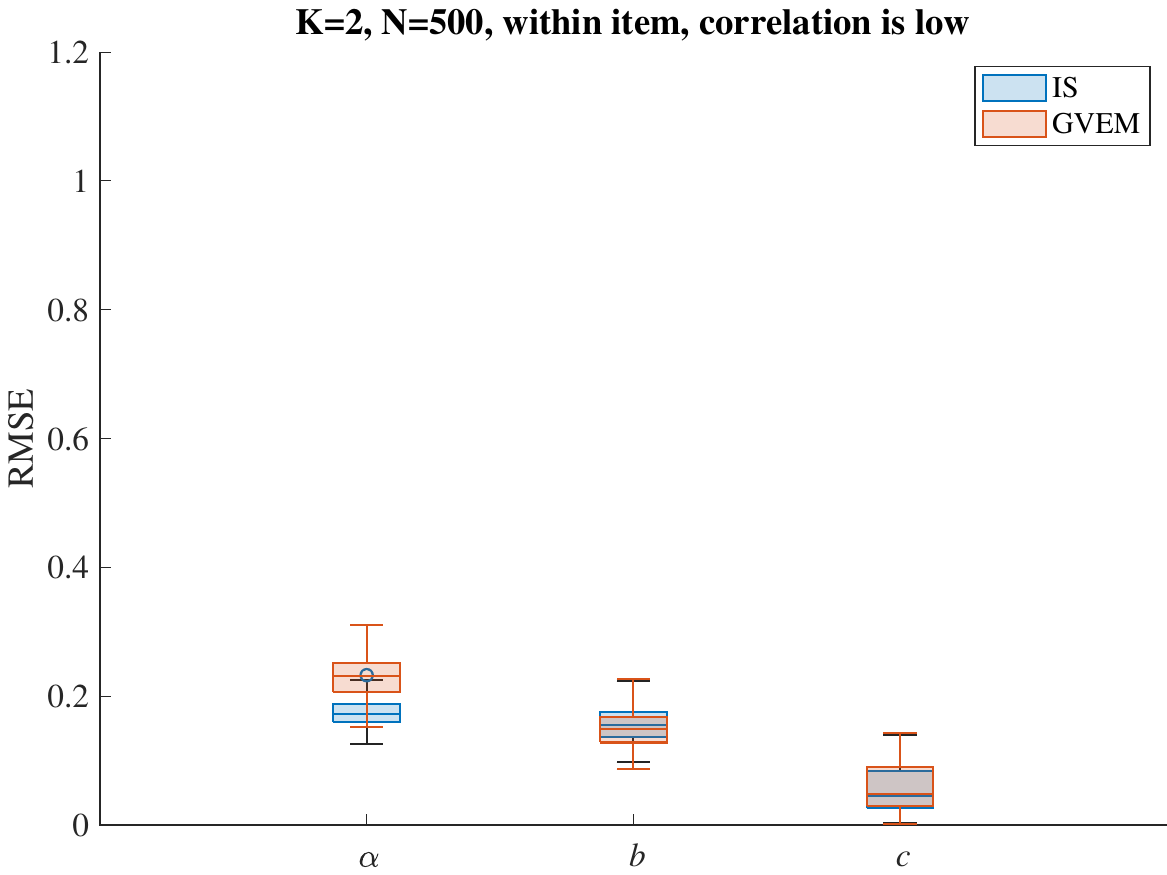}}
    \hspace{0.6in}
    \subfigure{
        \includegraphics[width=2.5in]{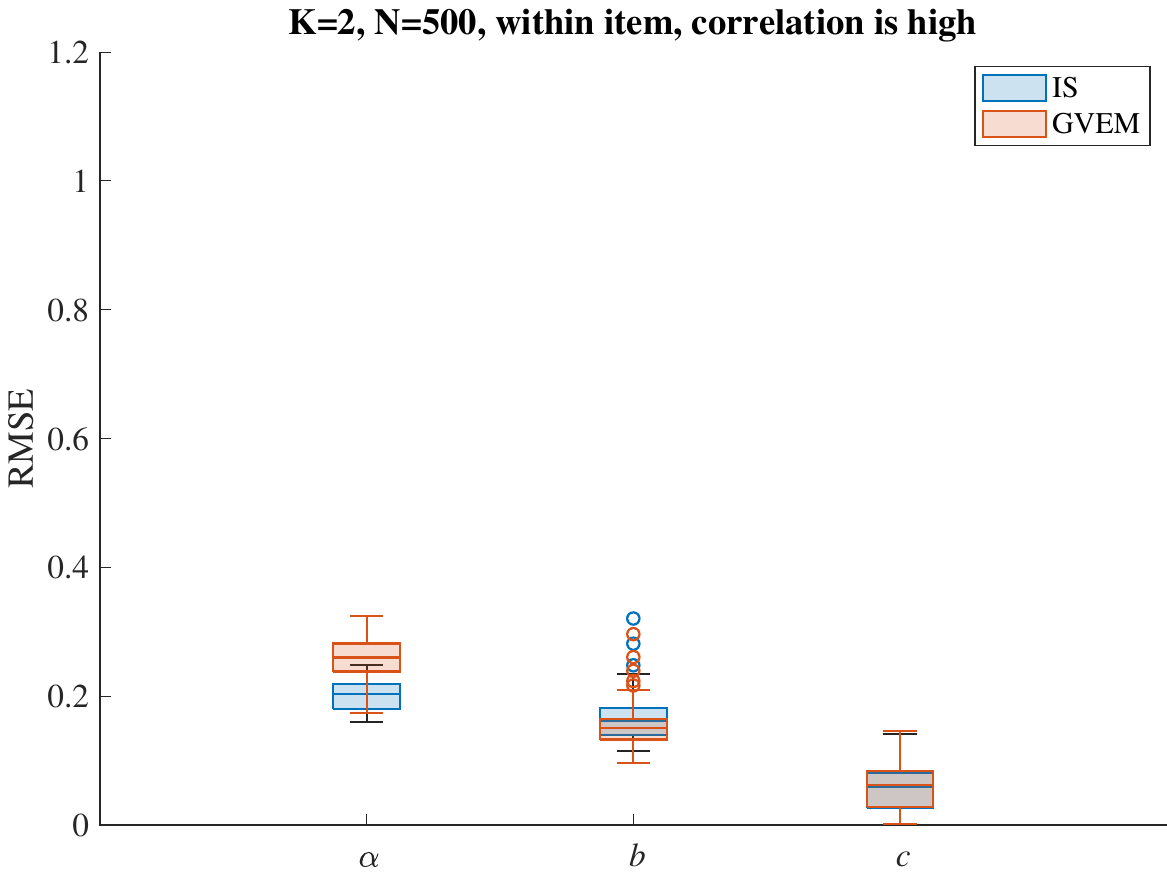}}
    \\
    \caption{RMSE for   $K=2$ under confirmatory analysis
    }
    \label{fig:rmse-k2-confirm}
\end{figure}

\newpage 

\begin{figure}[ht!]
    \centering
    \subfigure{
        \includegraphics[width=2.5in]{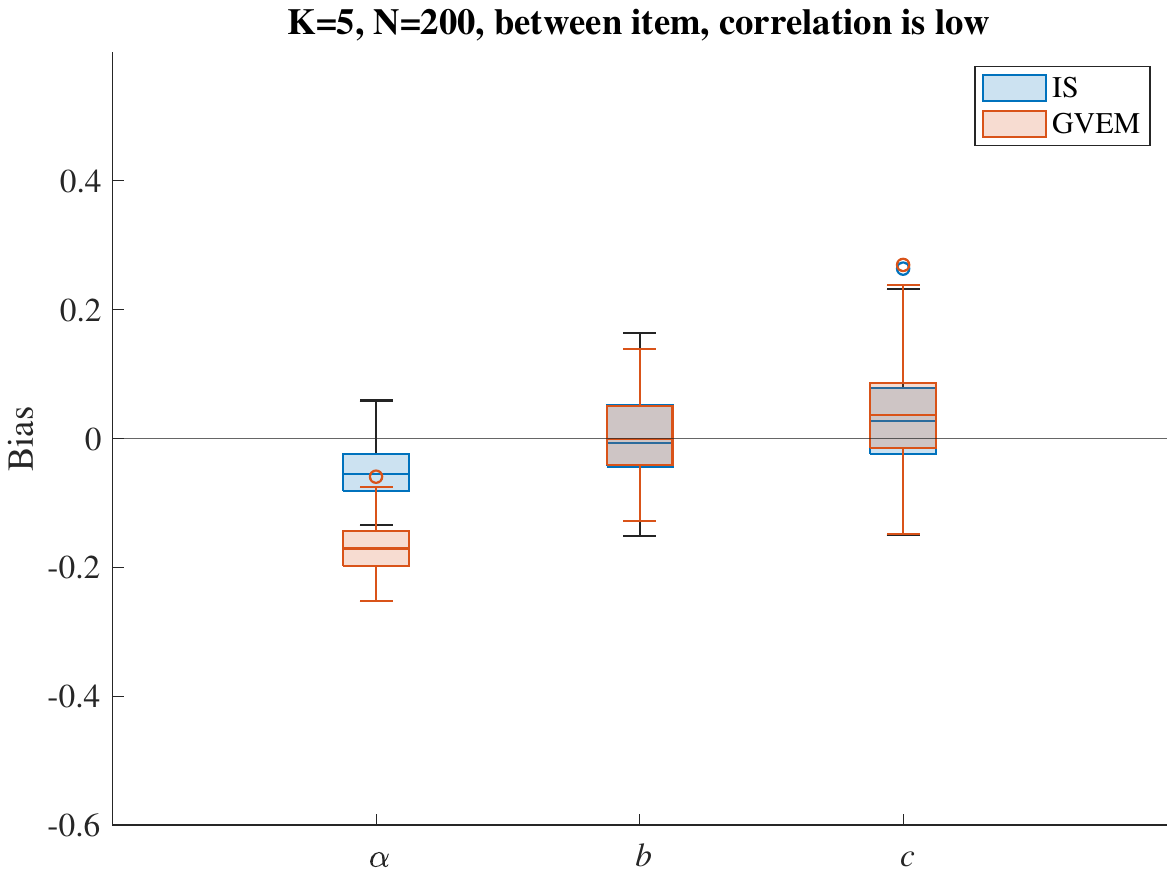}}
    \hspace{0.6in}
    \subfigure{
        \includegraphics[width=2.5in]{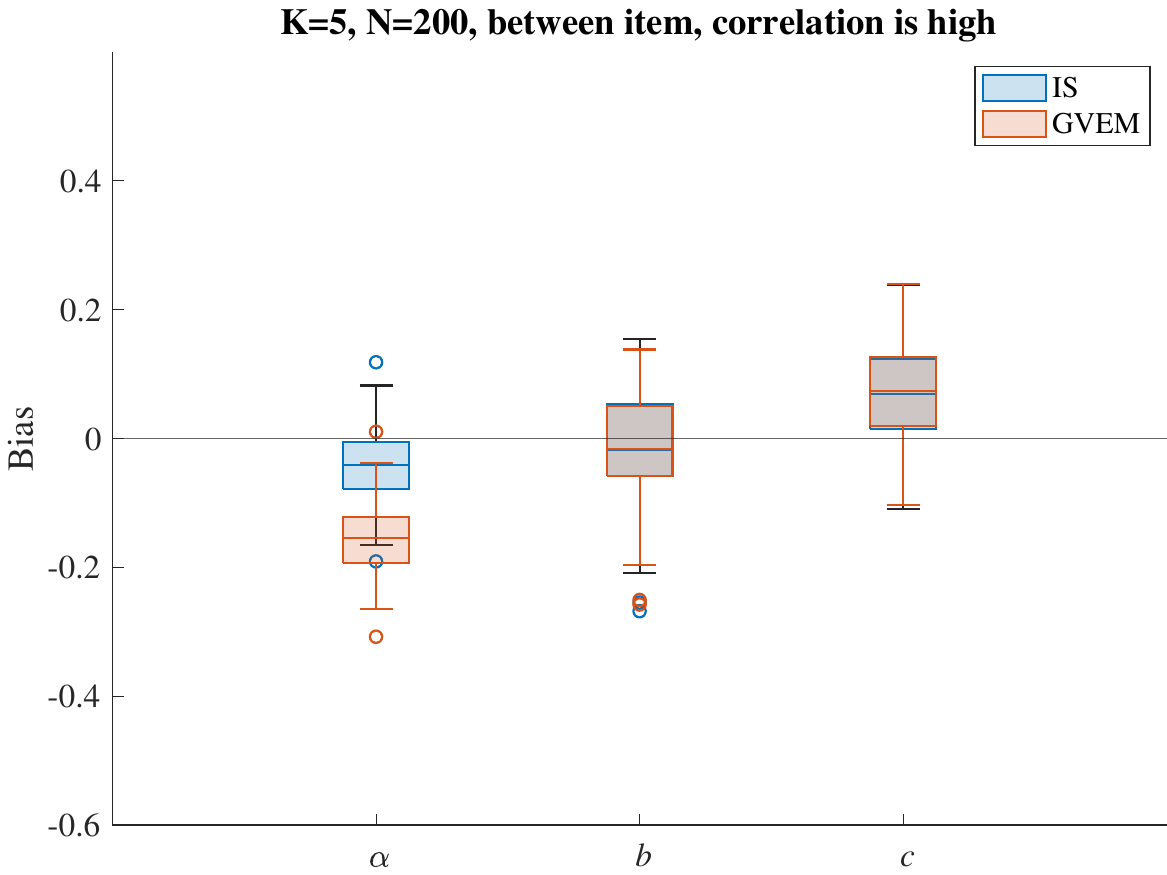}}
    \\
    \subfigure{
        \includegraphics[width=2.5in]{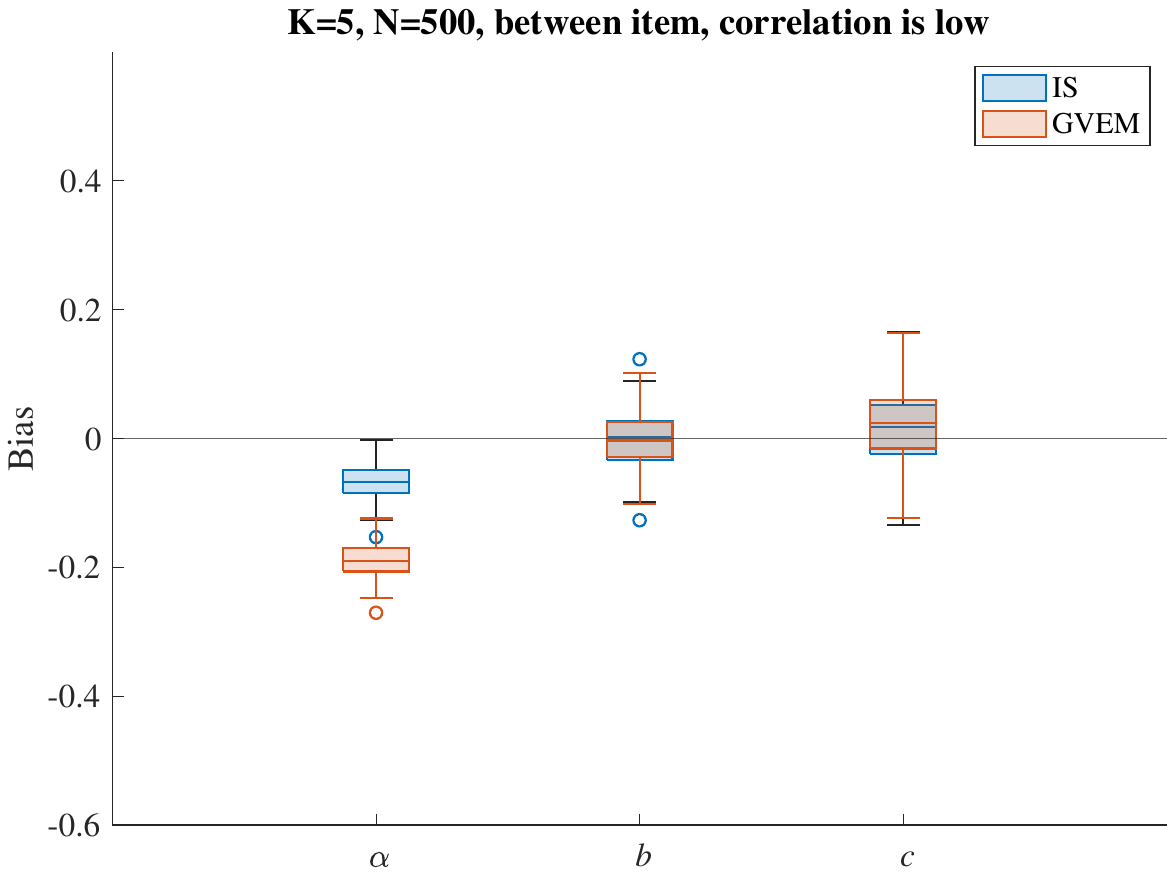}}
    \hspace{0.6in}
    \subfigure{
        \includegraphics[width=2.5in]{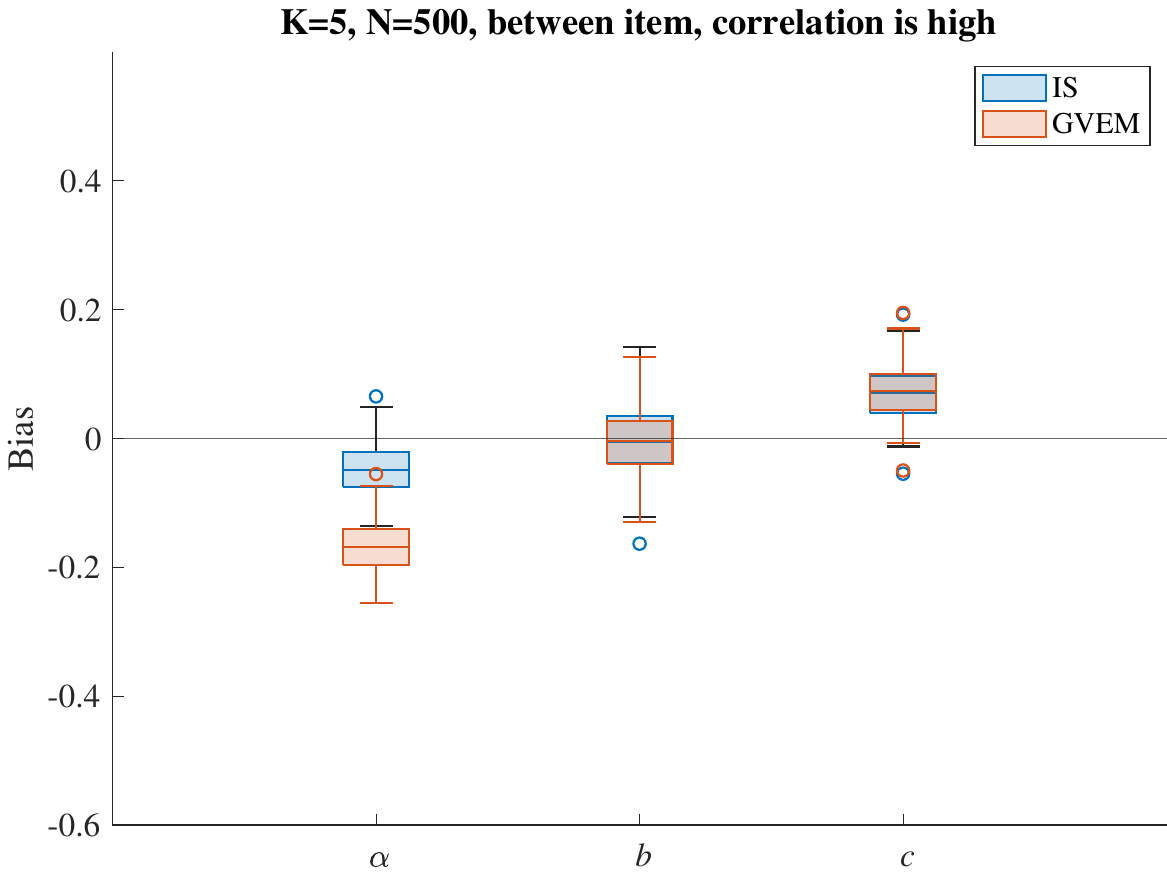}}
    \\
    \subfigure{
        \includegraphics[width=2.5in]{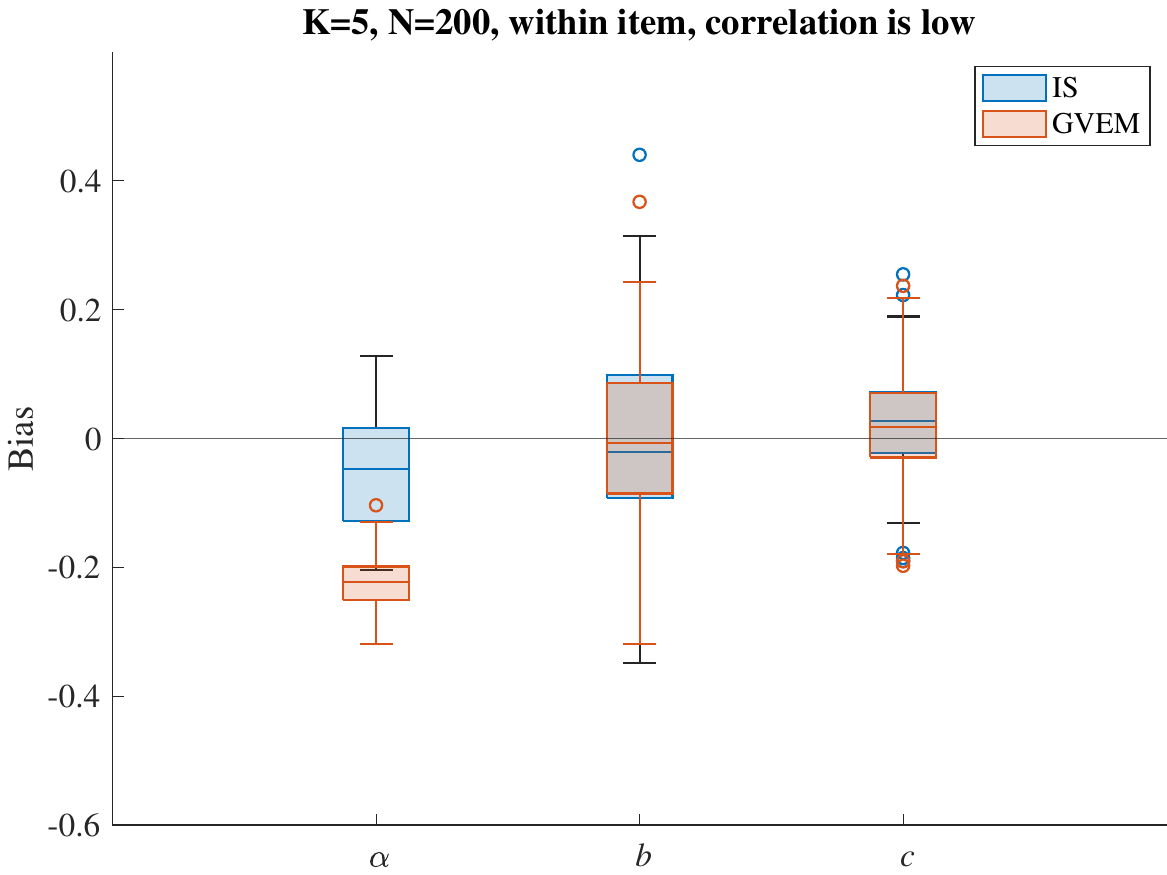}}
    \hspace{0.6in}
    \subfigure{
        \includegraphics[width=2.5in]{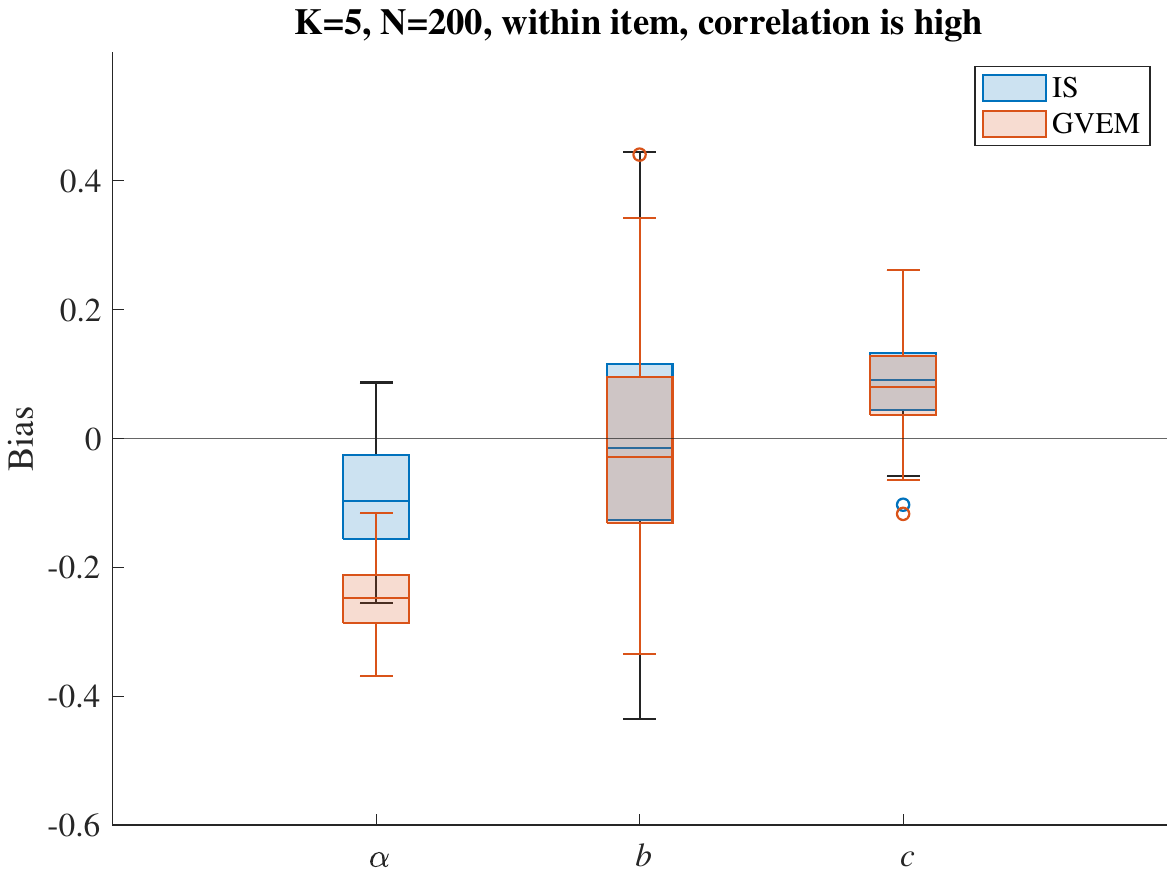}}
    \\
    \subfigure{
        \includegraphics[width=2.5in]{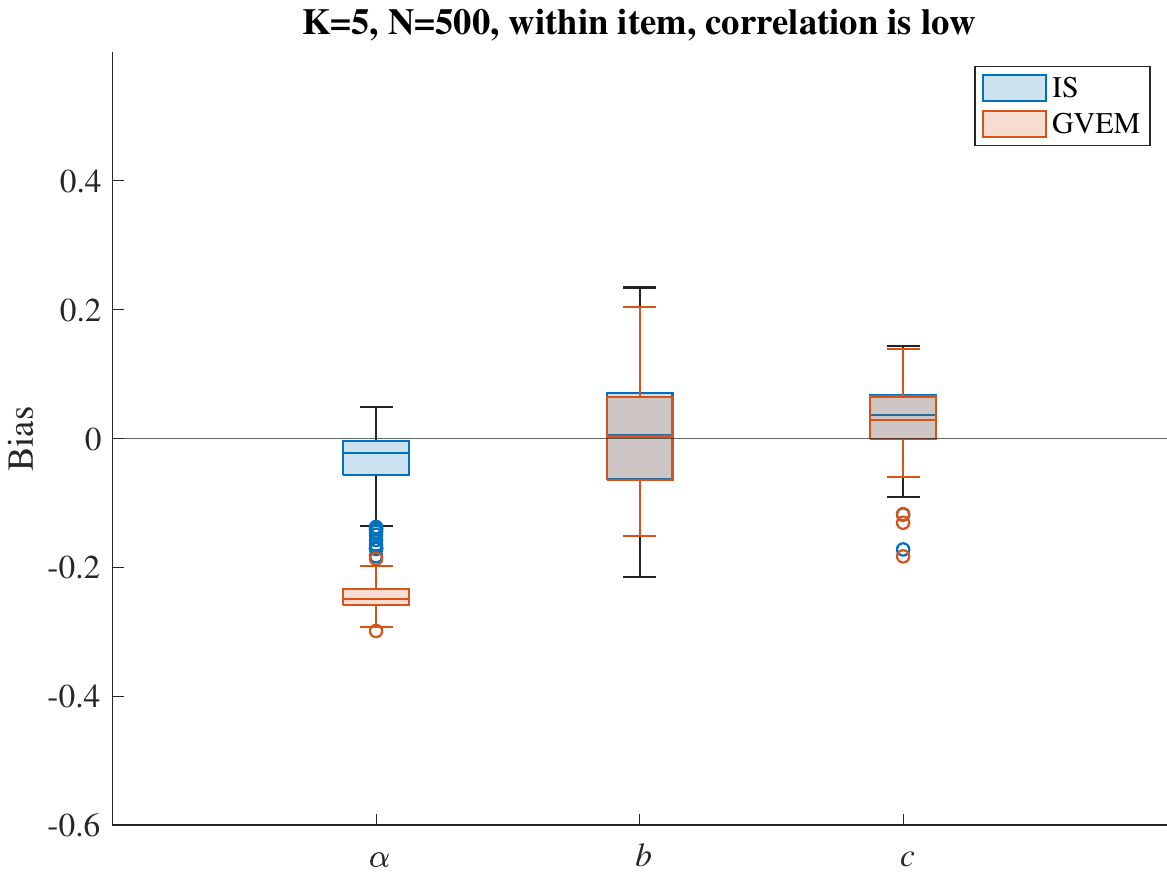}}
    \hspace{0.6in}
    \subfigure{
        \includegraphics[width=2.5in]{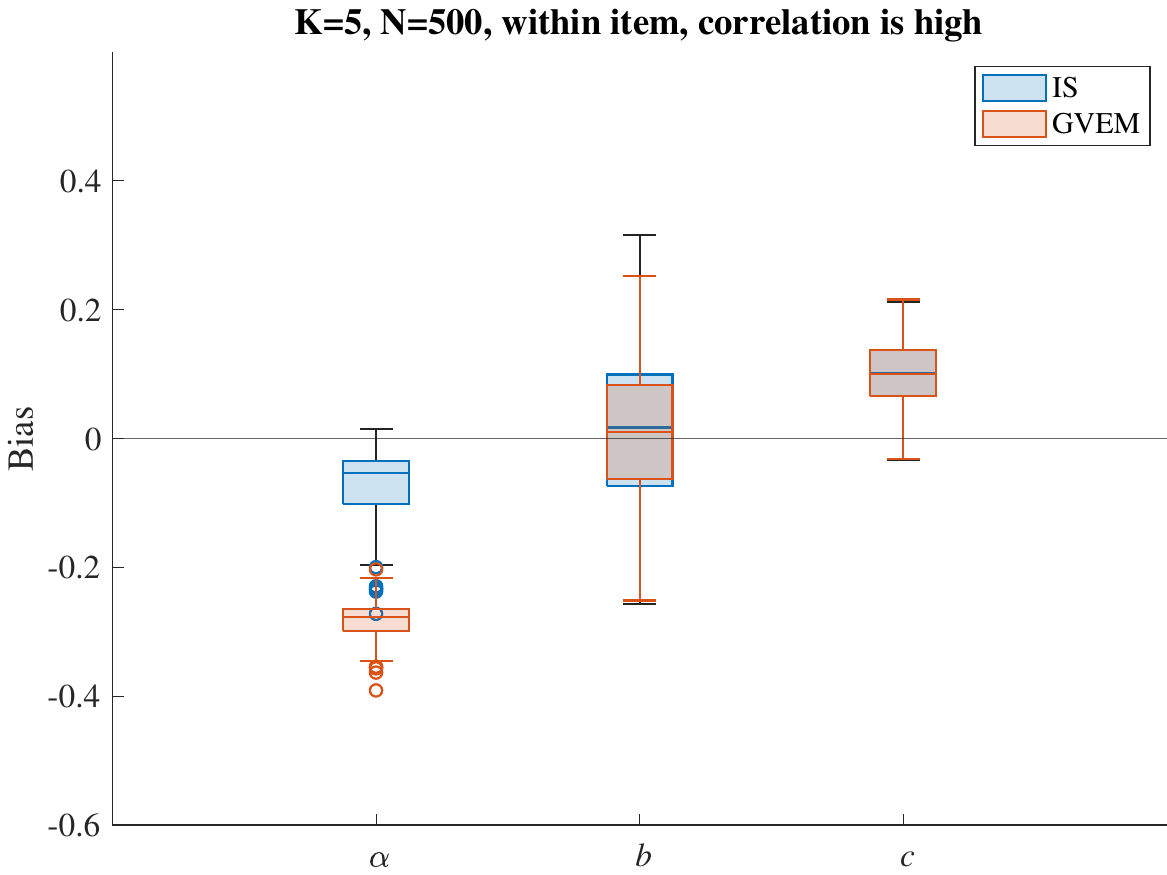}}
    \\
    \caption{Bias for   $K=5$ under confirmatory analysis
    }
    \label{fig:bias-k5-confirm}
\end{figure}

\newpage 

\begin{figure}[ht!]
    \centering
    \subfigure{
        \includegraphics[width=2.5in]{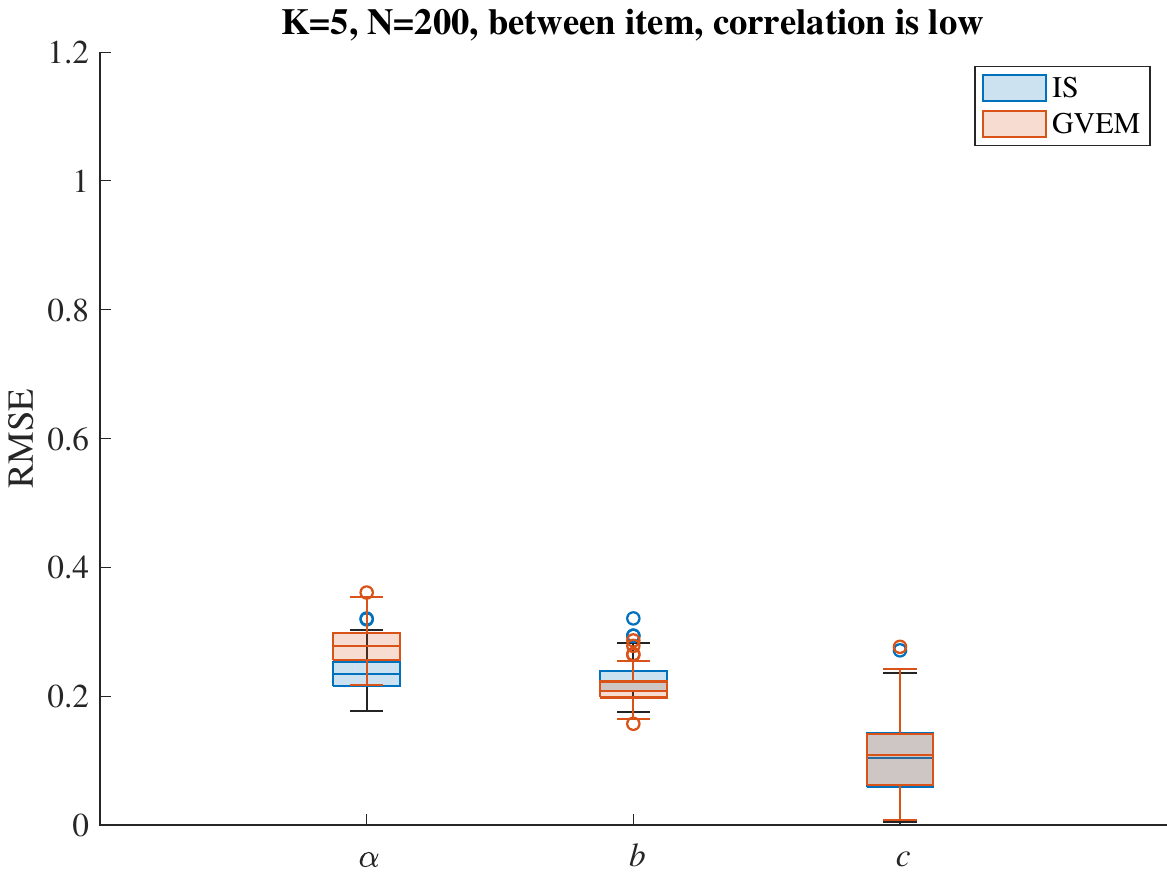}}
    \hspace{0.6in}
    \subfigure{
        \includegraphics[width=2.5in]{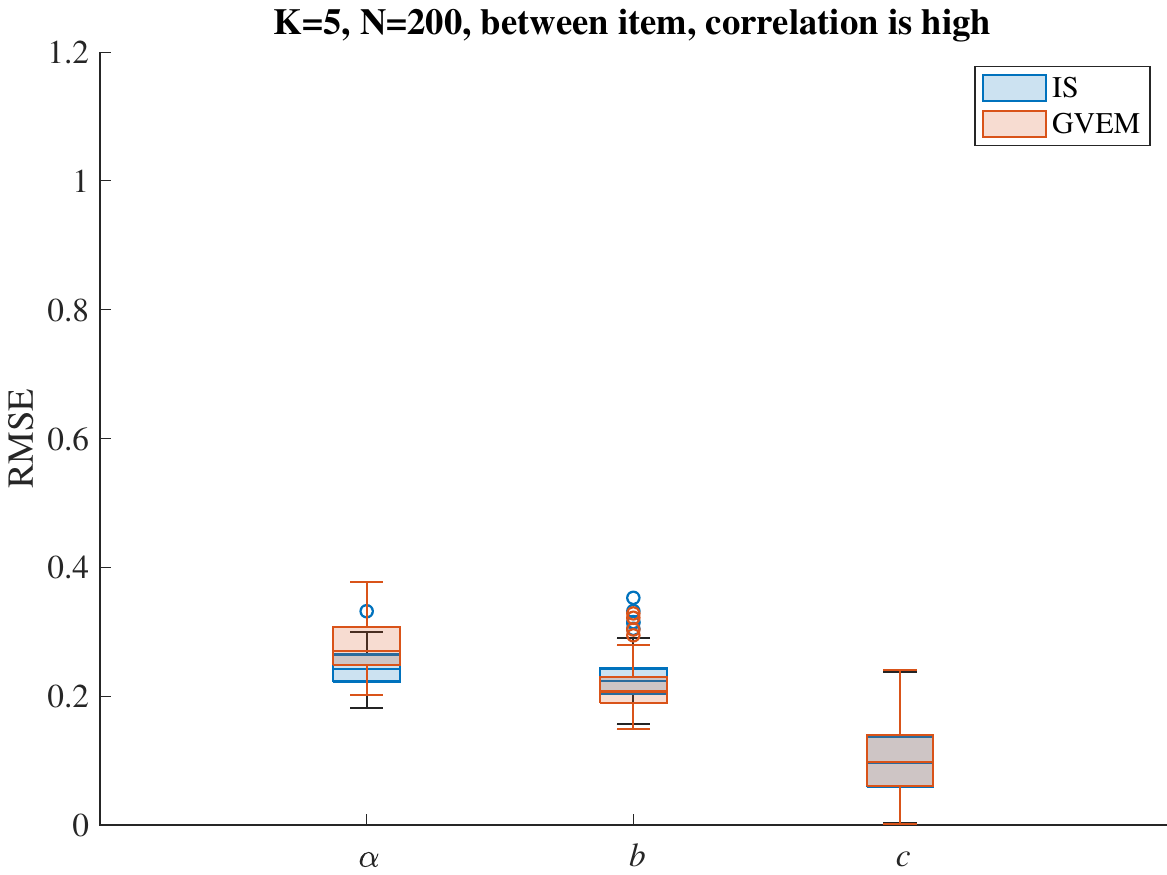}}
    \\
    \subfigure{
        \includegraphics[width=2.5in]{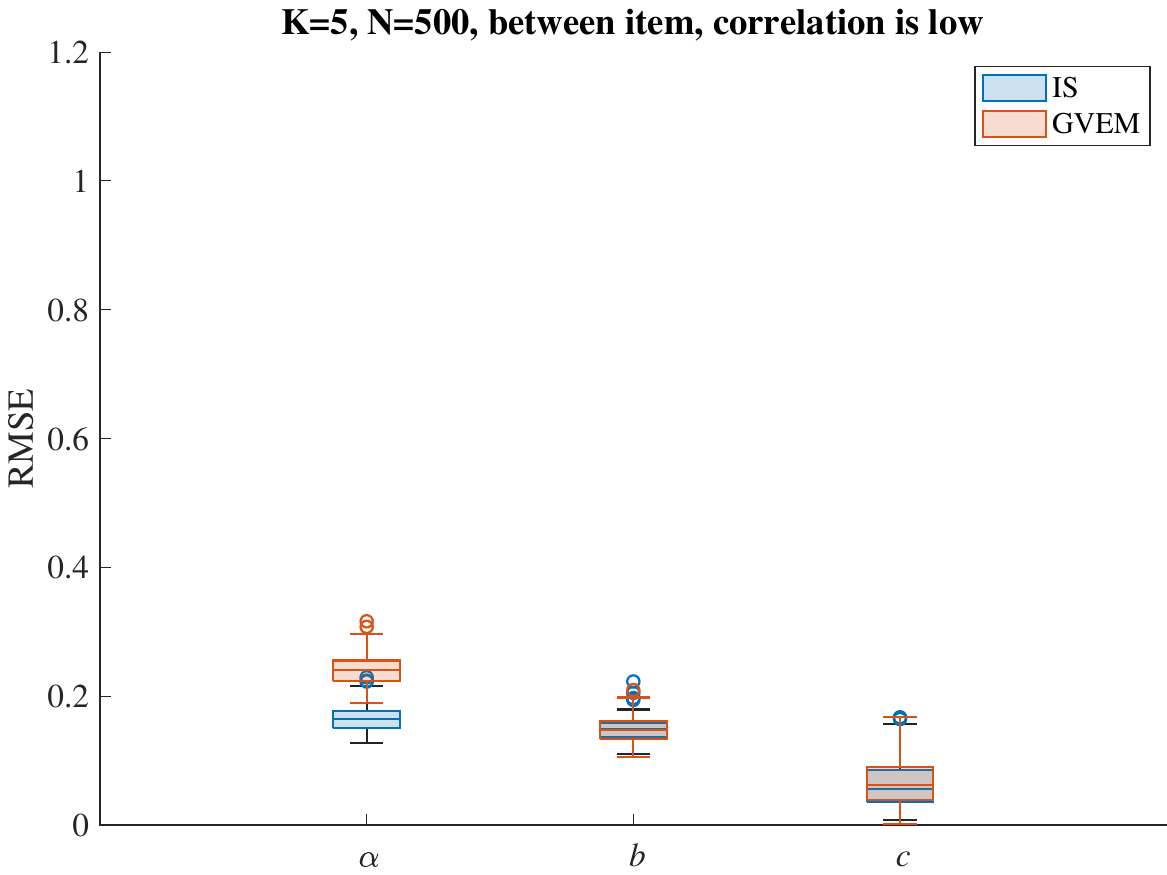}}
    \hspace{0.6in}
    \subfigure{
        \includegraphics[width=2.5in]{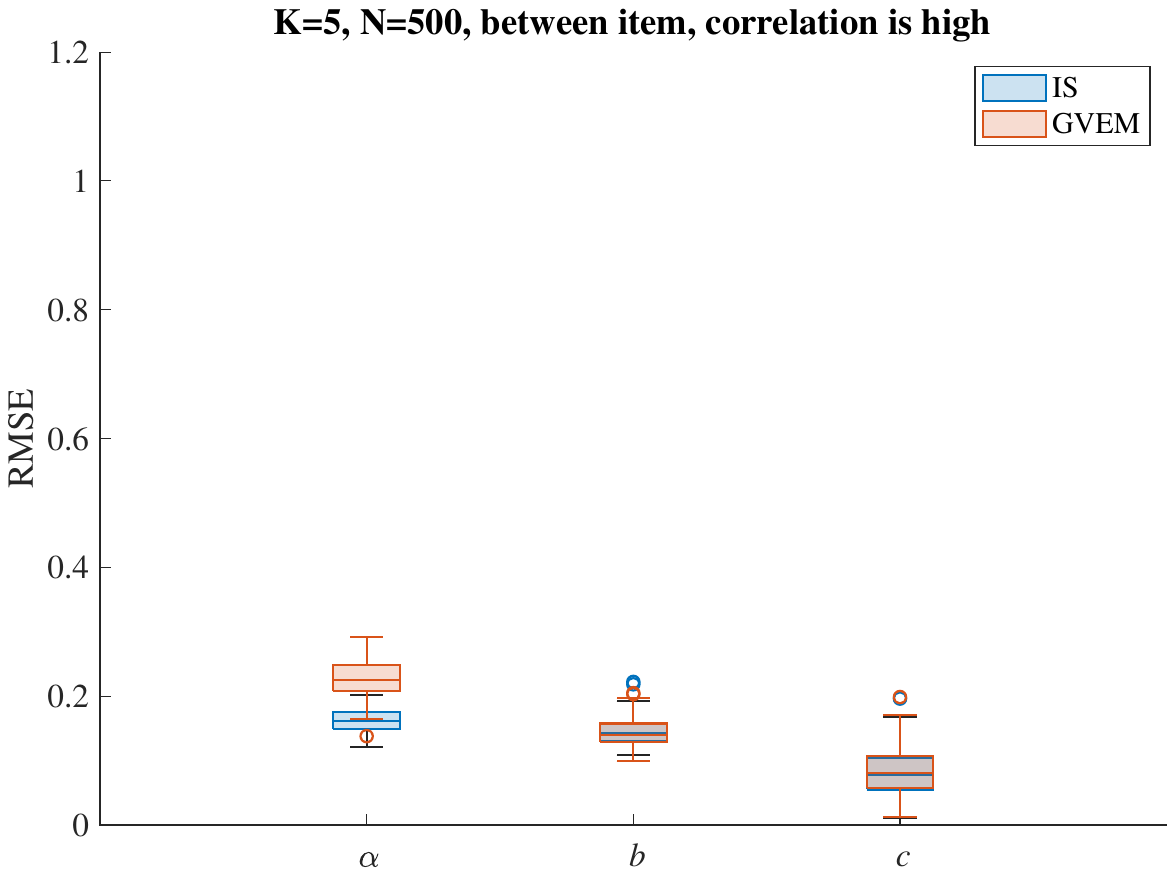}}
    \\
    \subfigure{
        \includegraphics[width=2.5in]{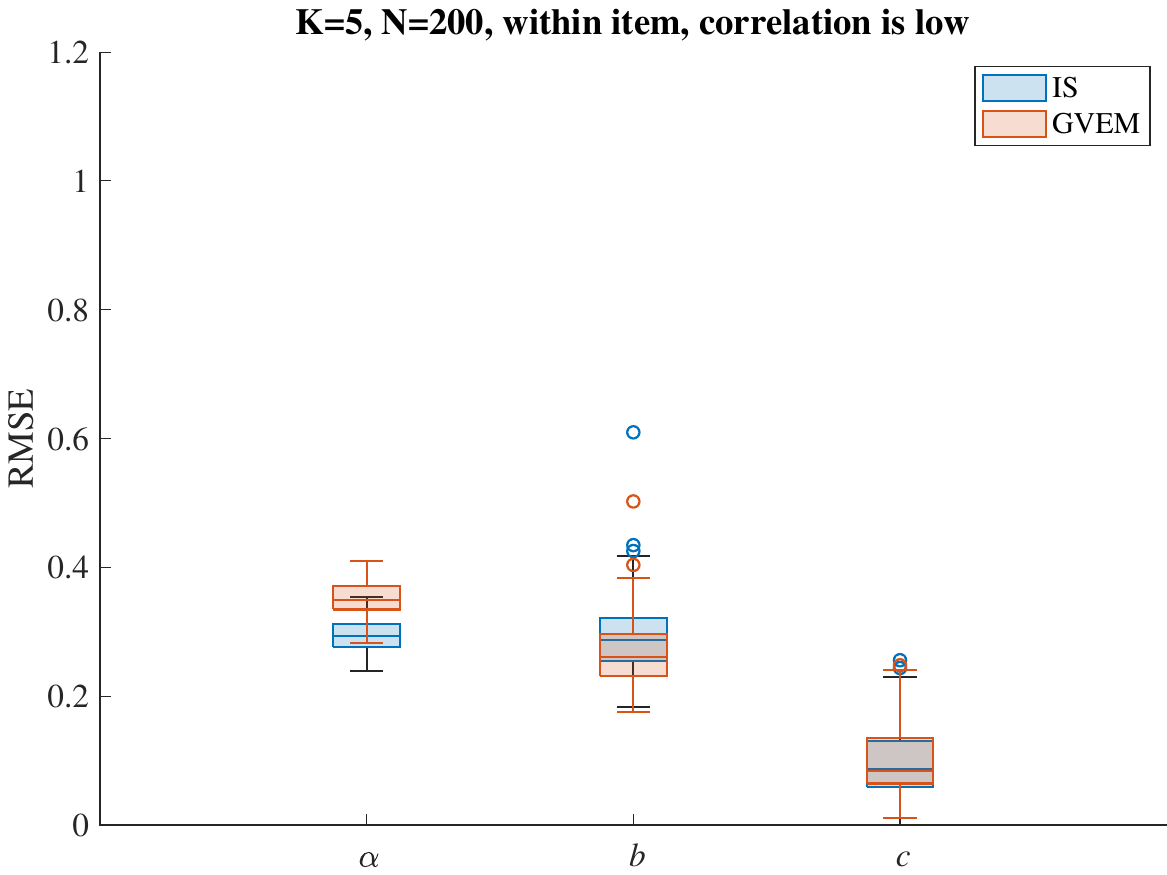}}
    \hspace{0.6in}
    \subfigure{
        \includegraphics[width=2.5in]{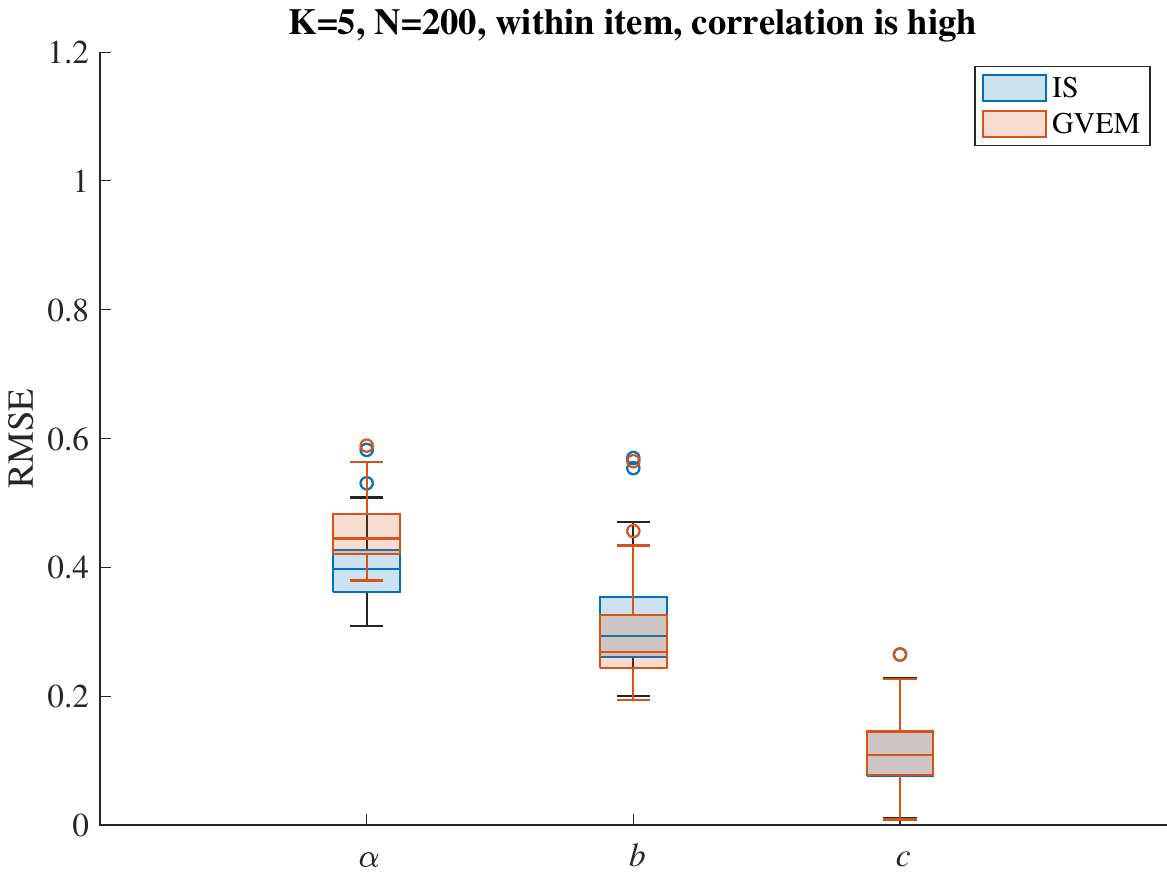}}
    \\
    \subfigure{
        \includegraphics[width=2.5in]{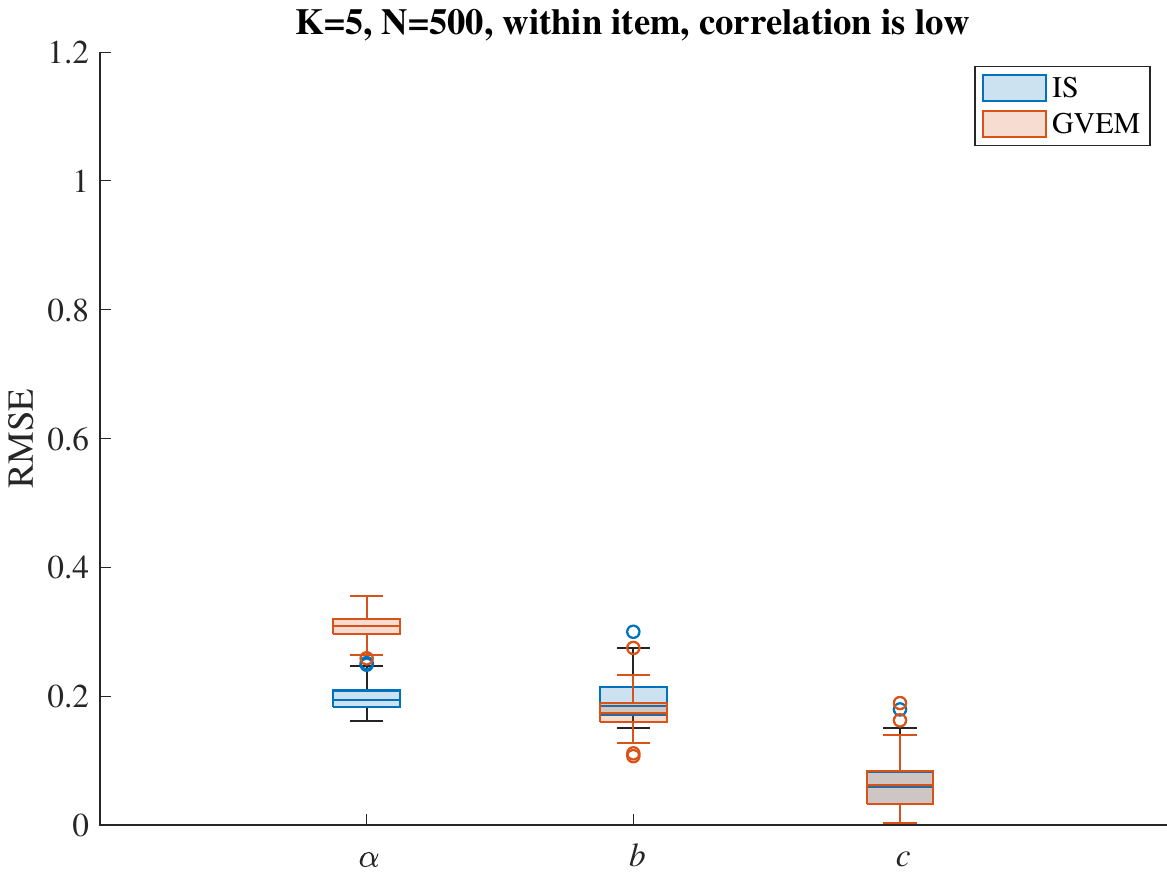}}
    \hspace{0.6in}
    \subfigure{
        \includegraphics[width=2.5in]{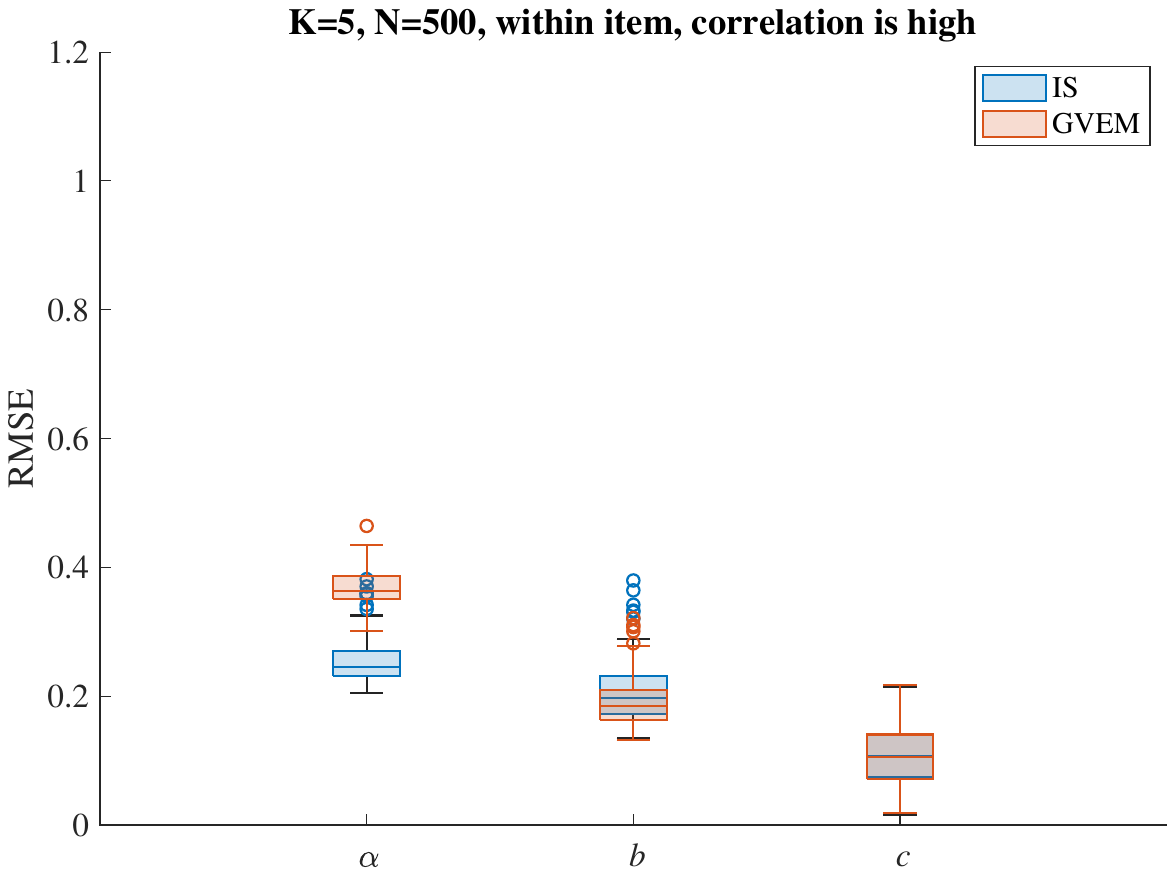}}
    \\
    \caption{RMSE for   $K=5$ under confirmatory analysis
    }
    \label{fig:rmse-k5-confirm}
\end{figure}


\newpage 

\begin{figure}[ht!]
    \centering
    \subfigure{
        \includegraphics[width=2.5in]{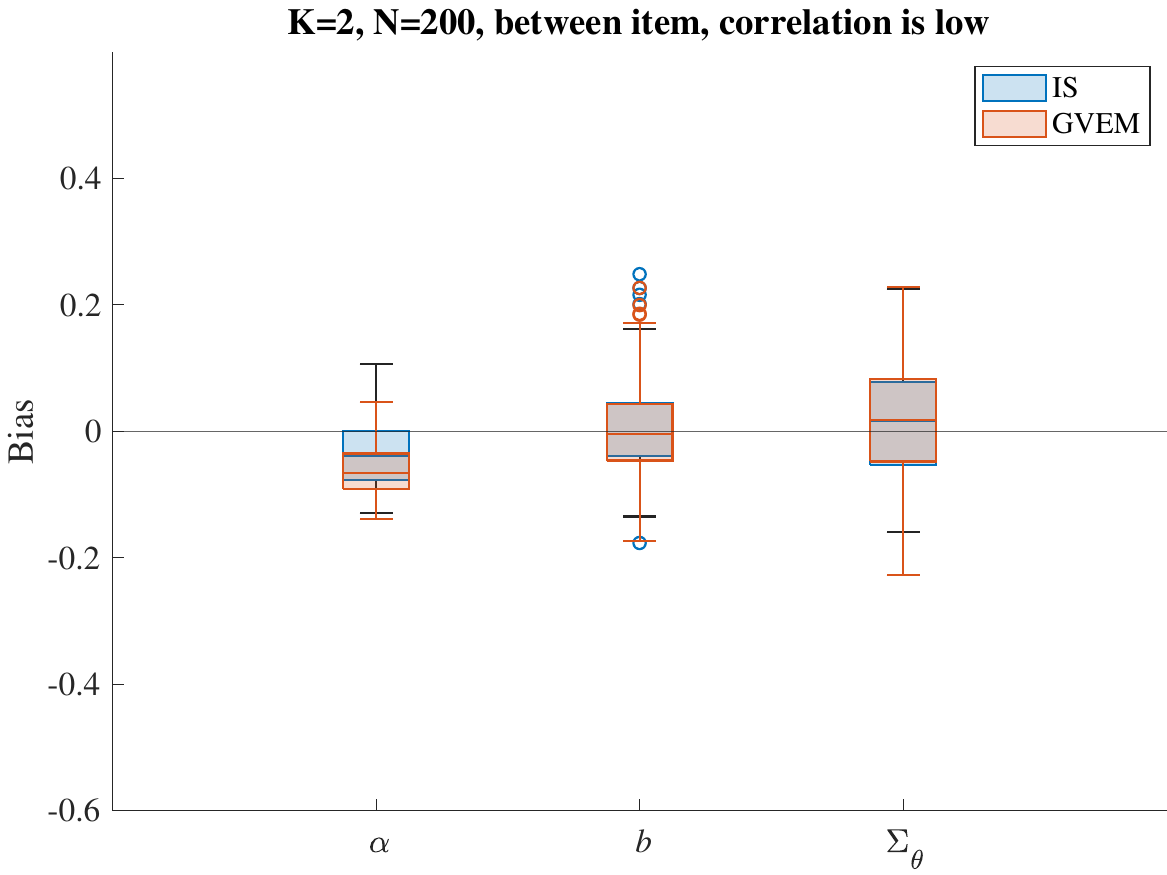}}
    \hspace{0.6in}
    \subfigure{
        \includegraphics[width=2.5in]{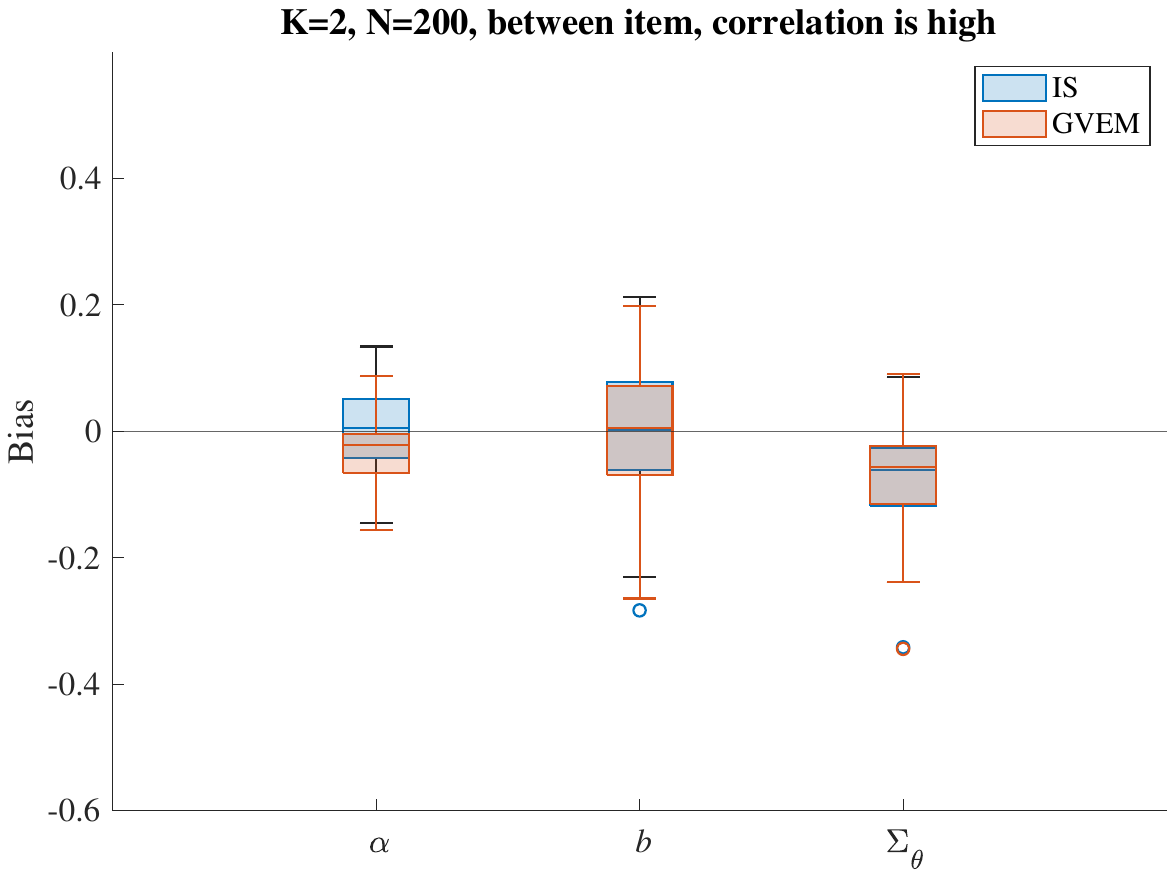}}
    \\
    \subfigure{
        \includegraphics[width=2.5in]{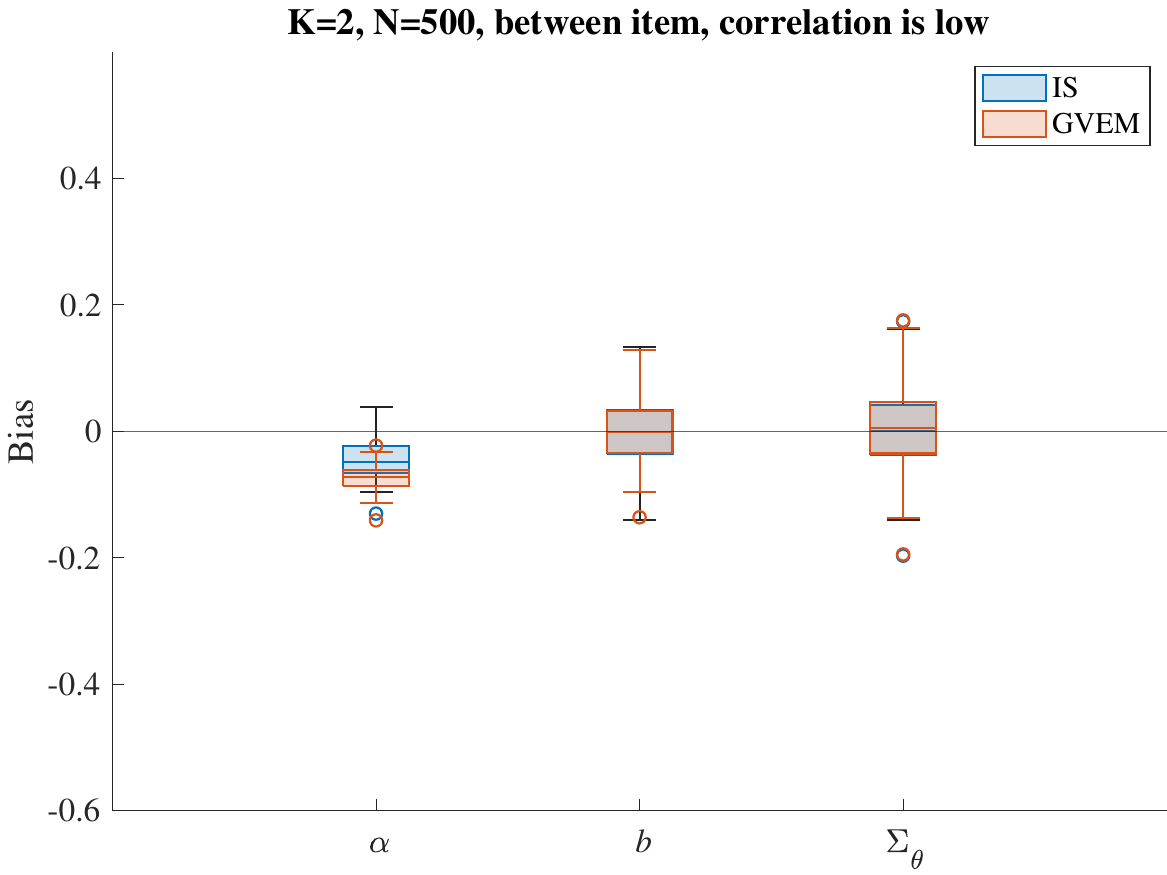}}
    \hspace{0.6in}
    \subfigure{
        \includegraphics[width=2.5in]{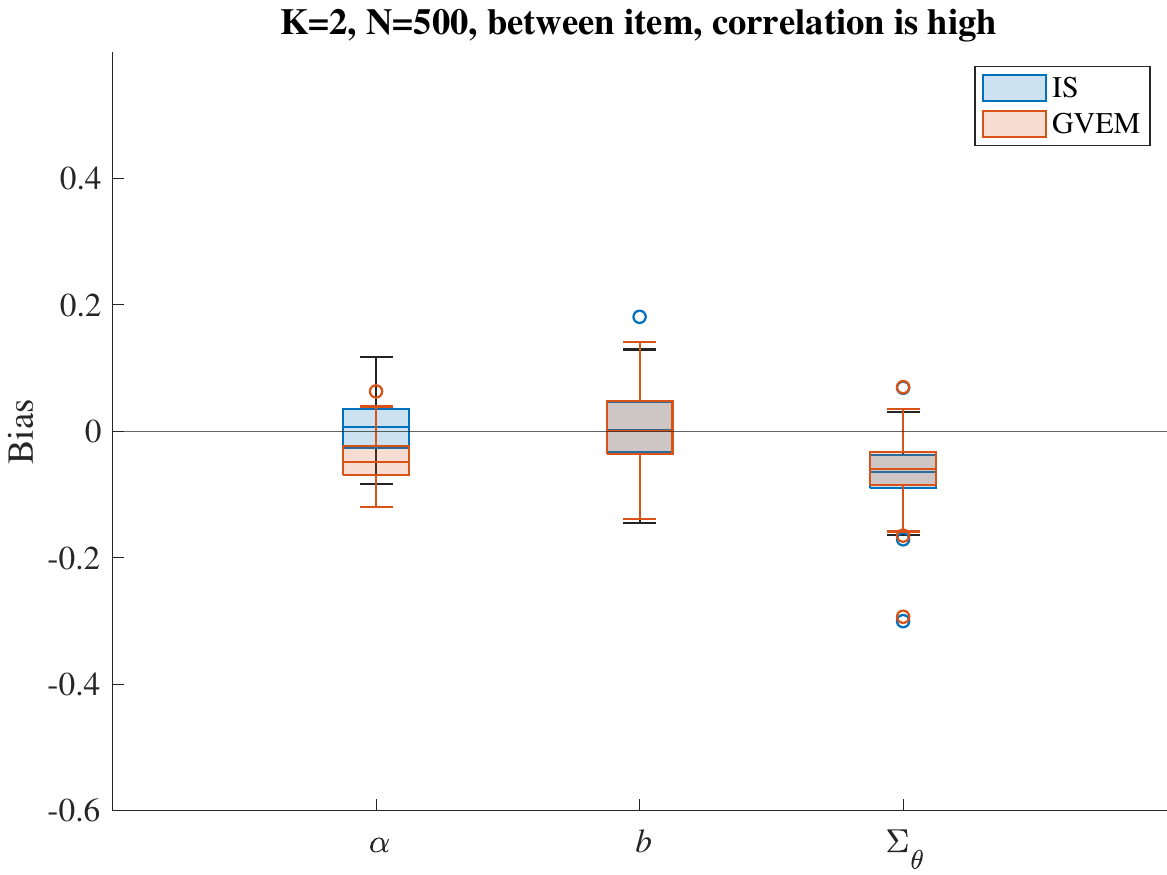}}
    \\
    \subfigure{
        \includegraphics[width=2.5in]{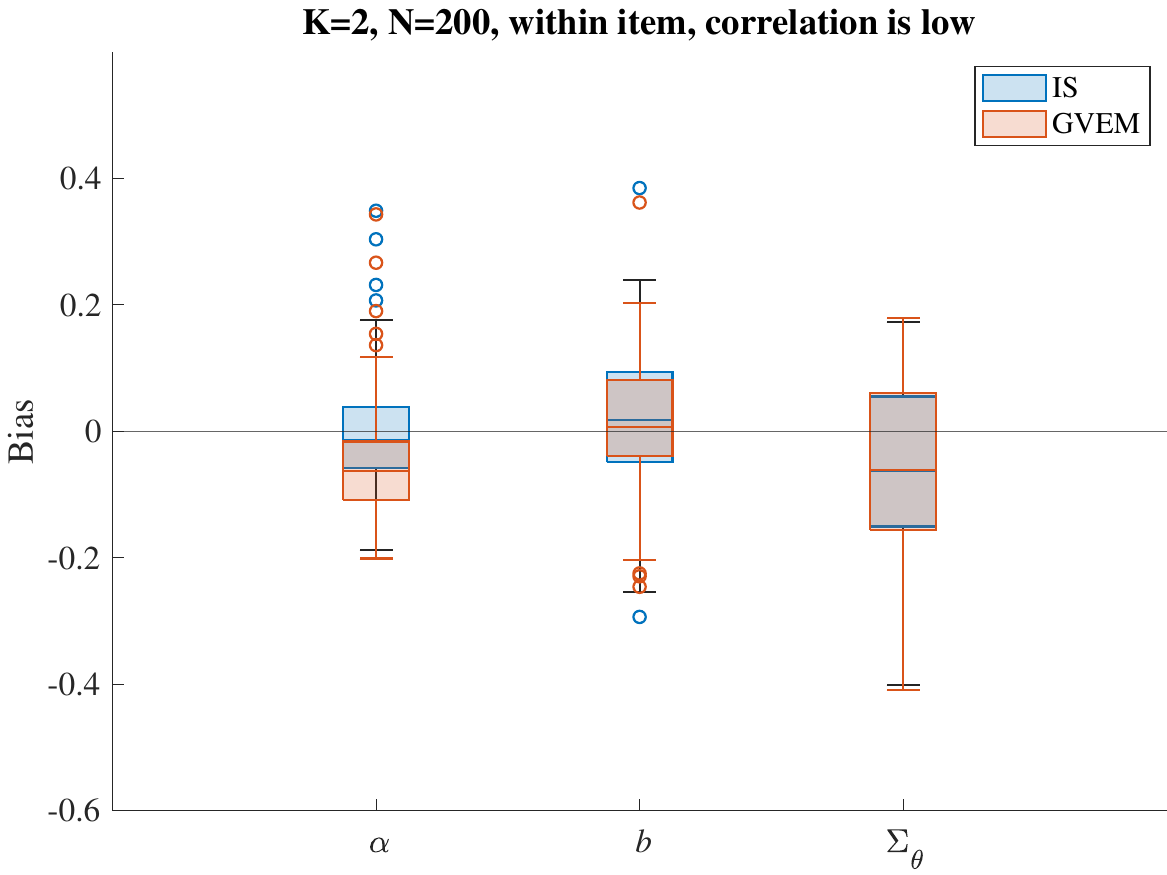}}
    \hspace{0.6in}
    \subfigure{
        \includegraphics[width=2.5in]{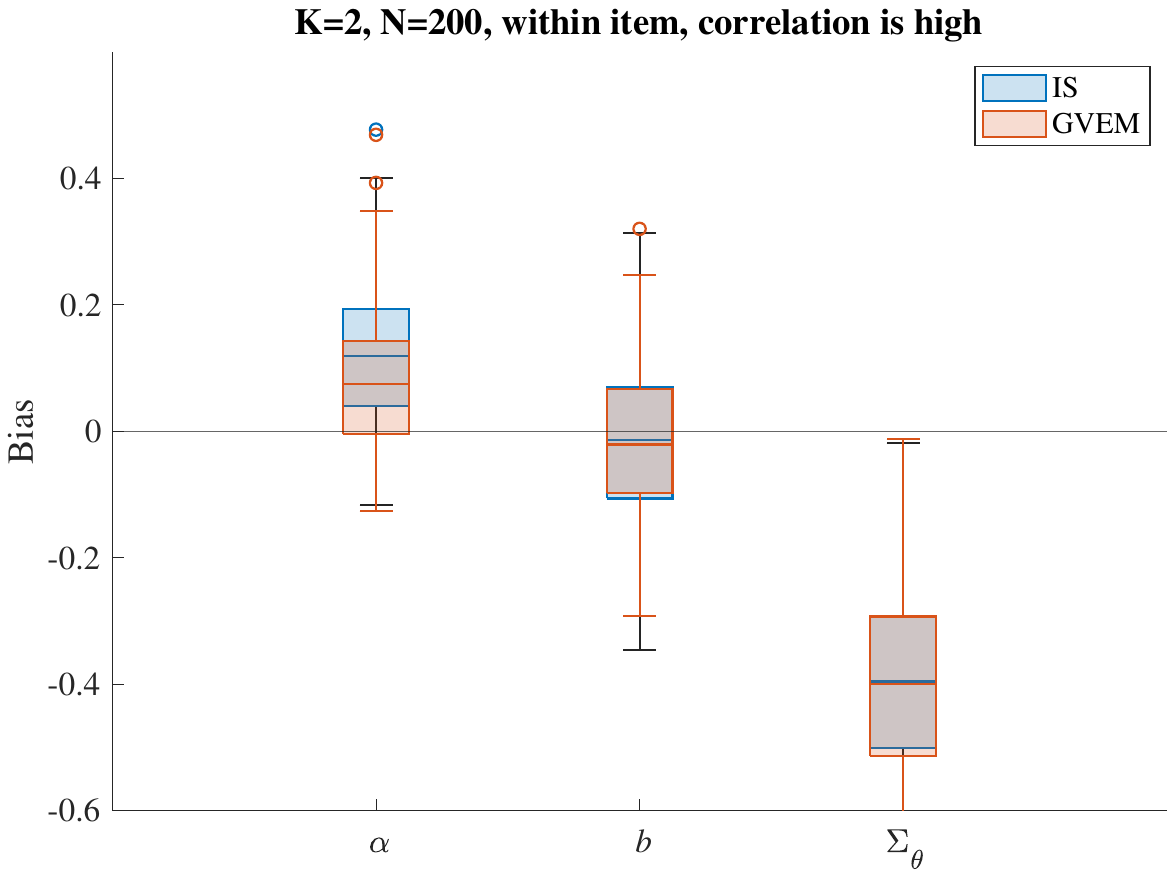}}
    \\
    \subfigure{
        \includegraphics[width=2.5in]{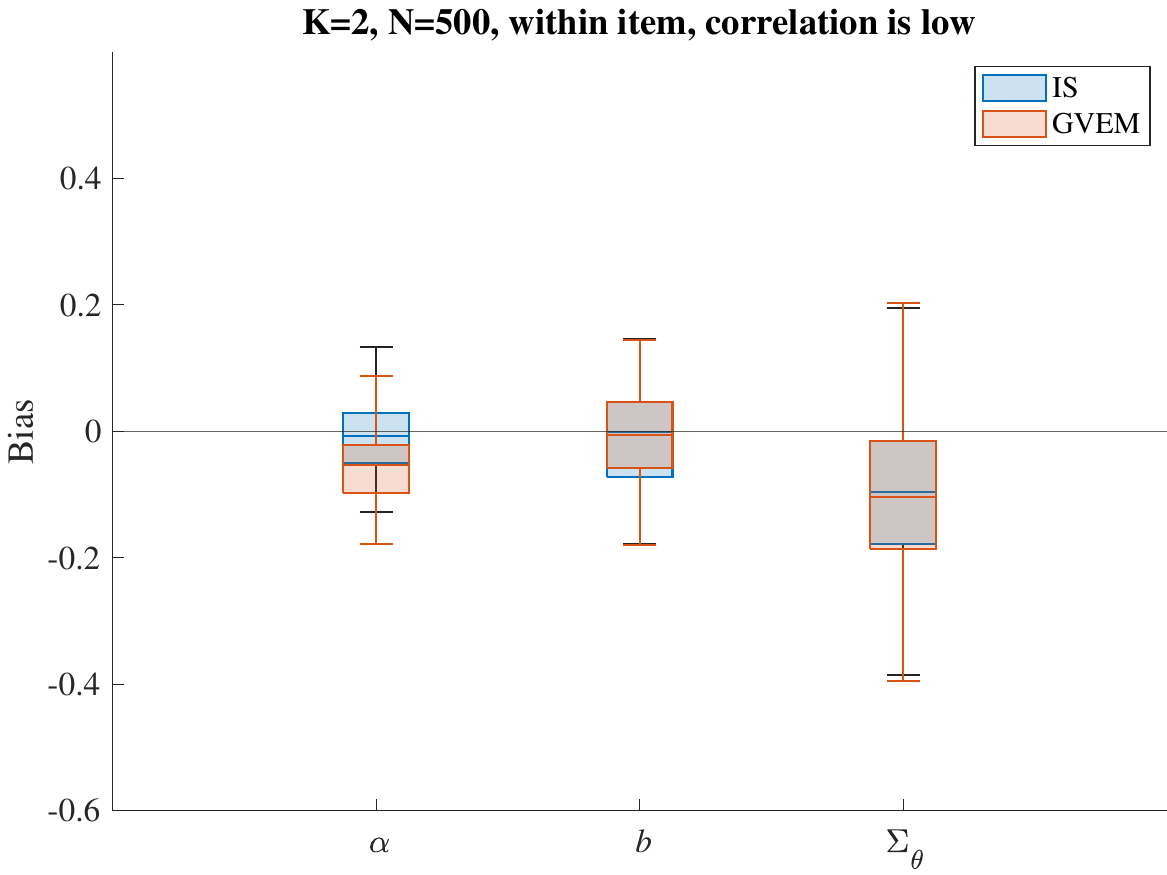}}
    \hspace{0.6in}
    \subfigure{
        \includegraphics[width=2.5in]{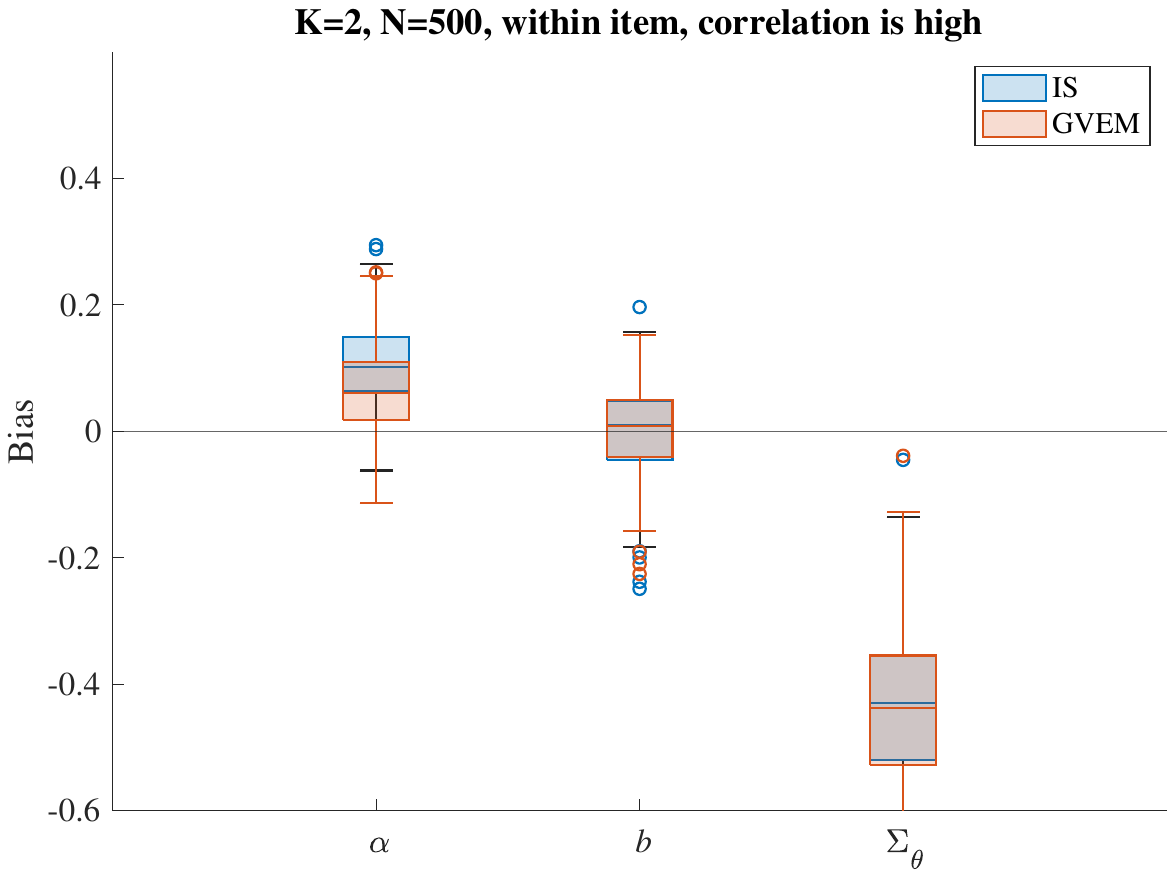}}
    \\
    \caption{Bias for $K=2$ under exploratory analysis   }
    \label{fig:bias-k2-explore}
\end{figure}

\newpage 

\begin{figure}[ht!]
    \centering
    \subfigure{
        \includegraphics[width=2.5in]{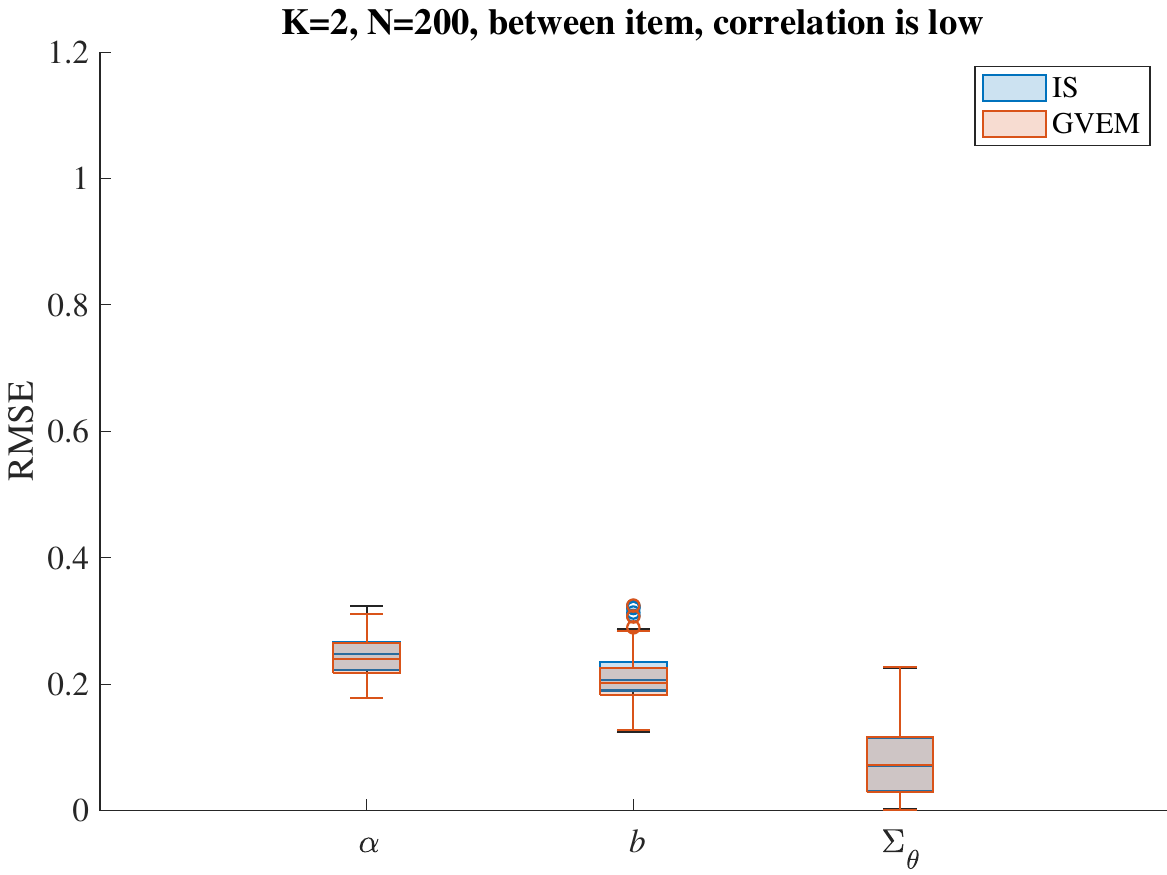}}
    \hspace{0.6in}
    \subfigure{
        \includegraphics[width=2.5in]{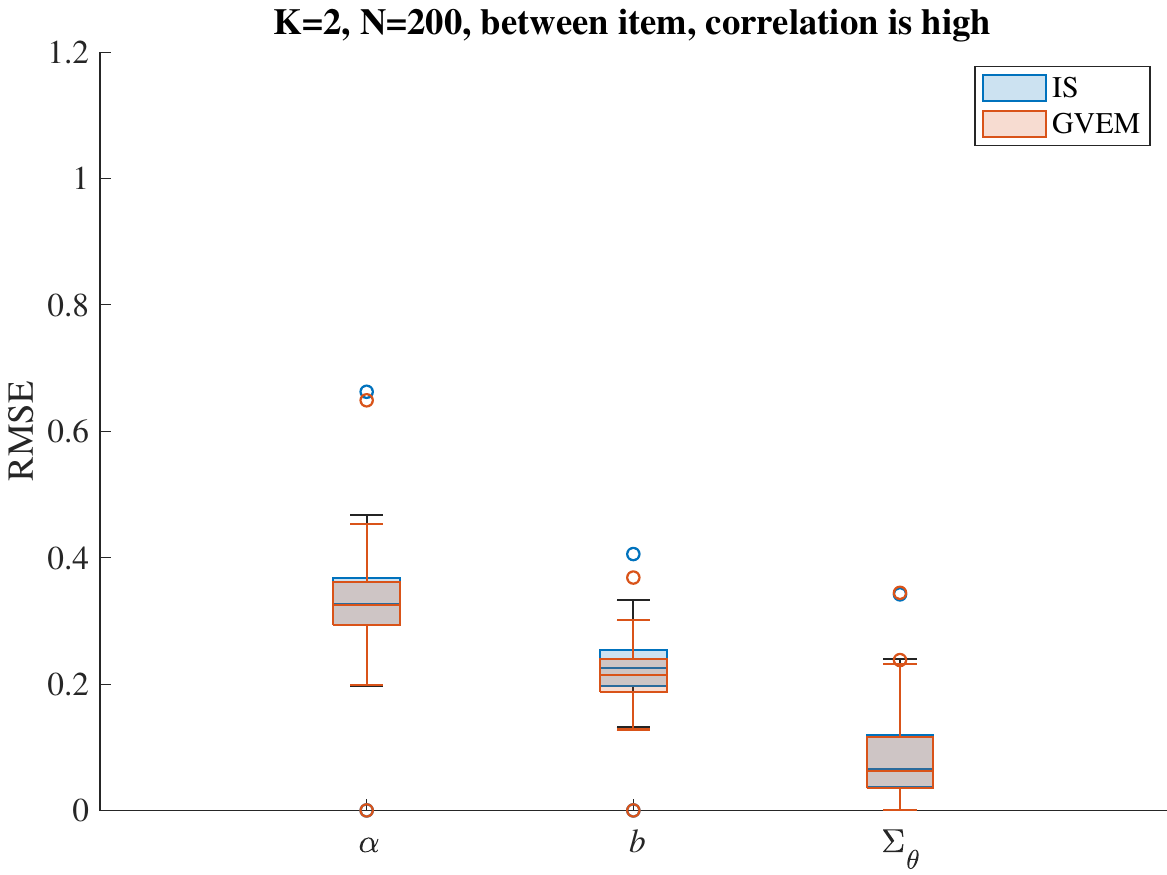}}
    \\
    \subfigure{
        \includegraphics[width=2.5in]{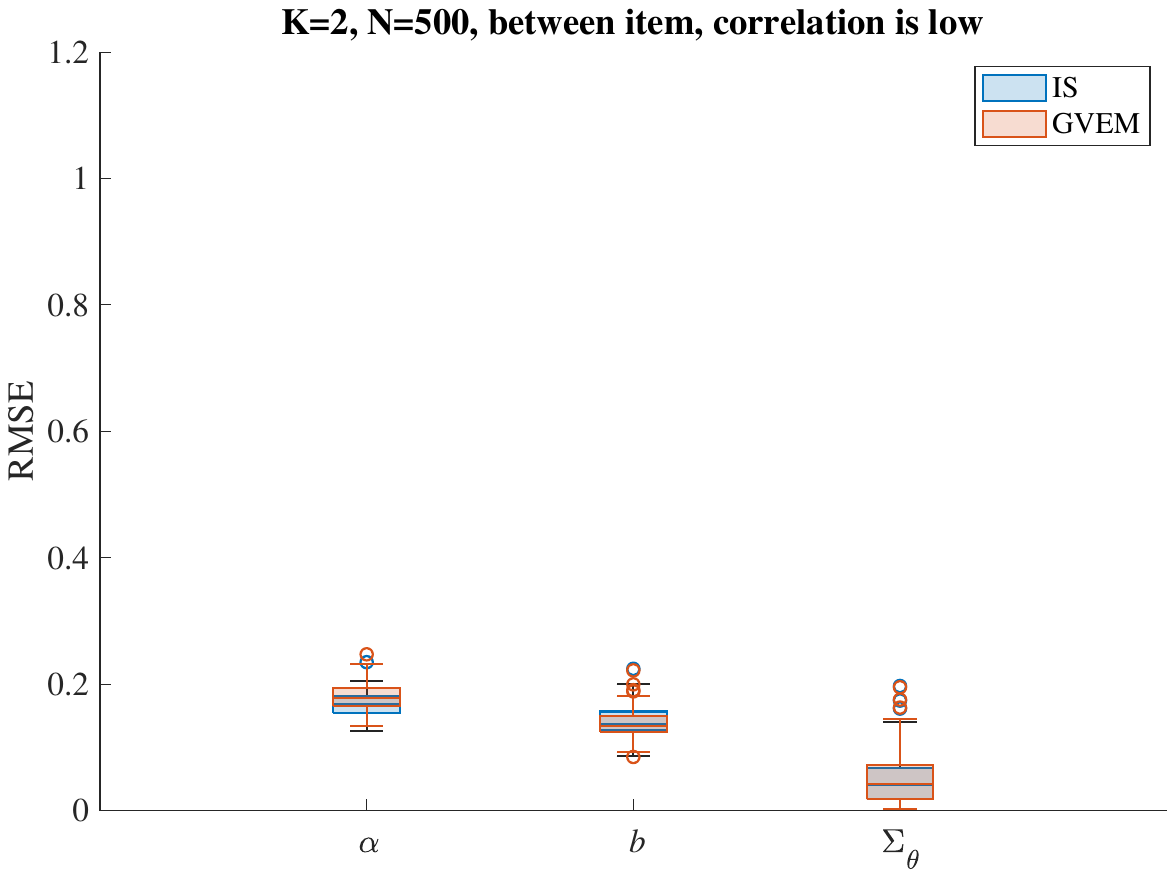}}
    \hspace{0.6in}
    \subfigure{
        \includegraphics[width=2.5in]{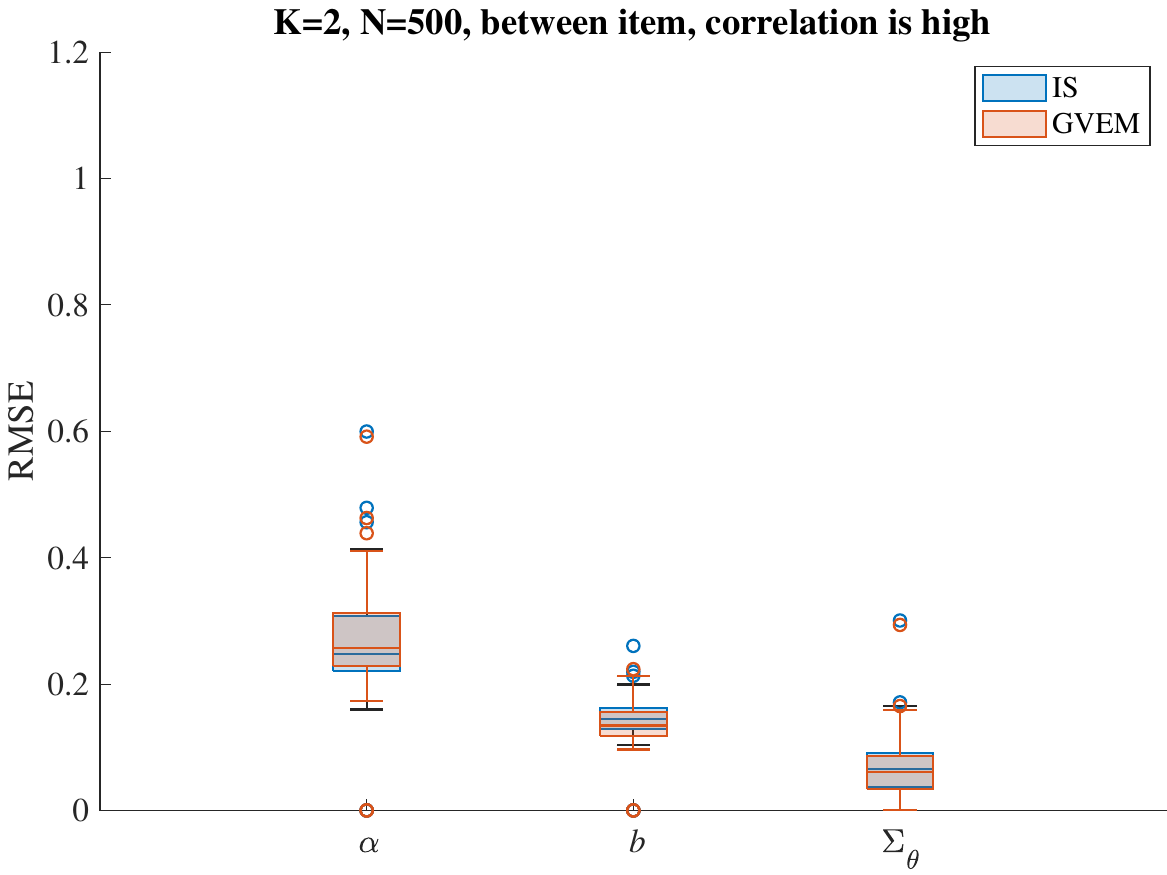}}
    \\
    \subfigure{
        \includegraphics[width=2.5in]{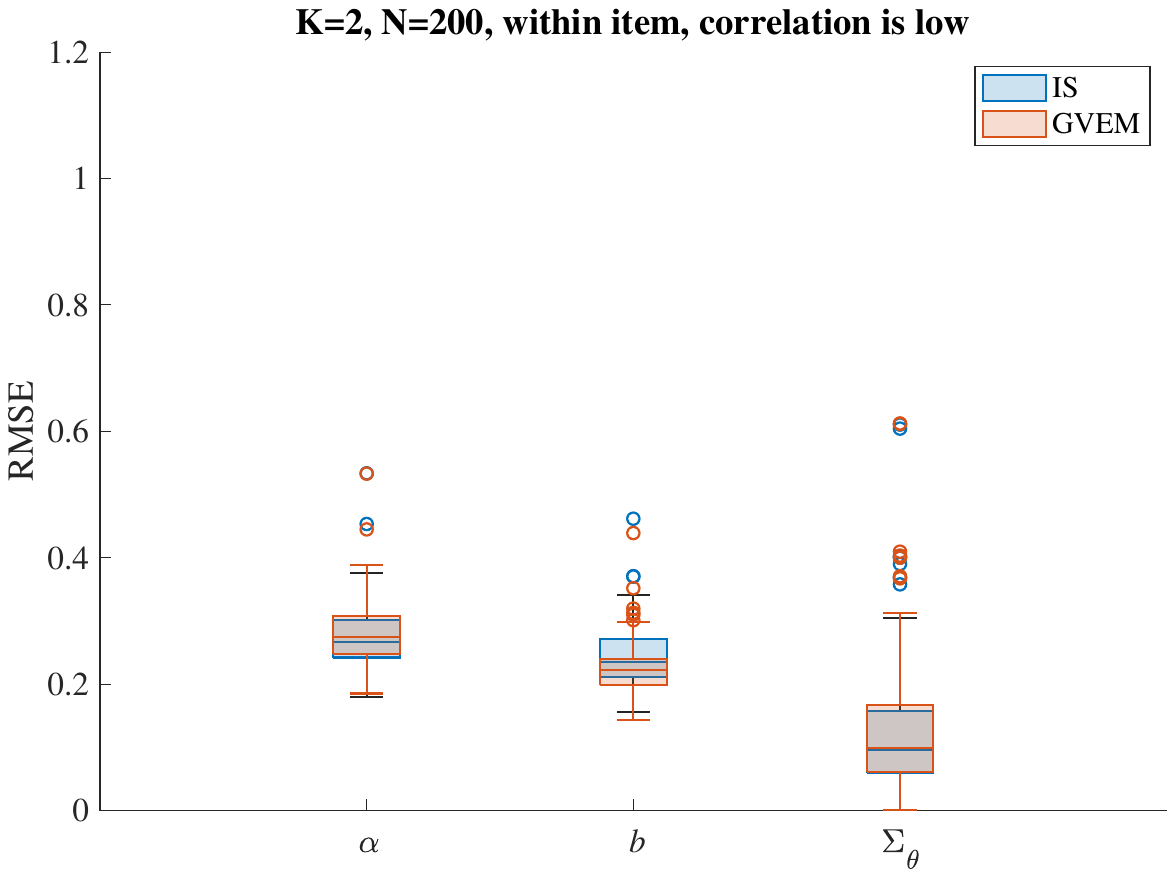}}
    \hspace{0.6in}
    \subfigure{
        \includegraphics[width=2.5in]{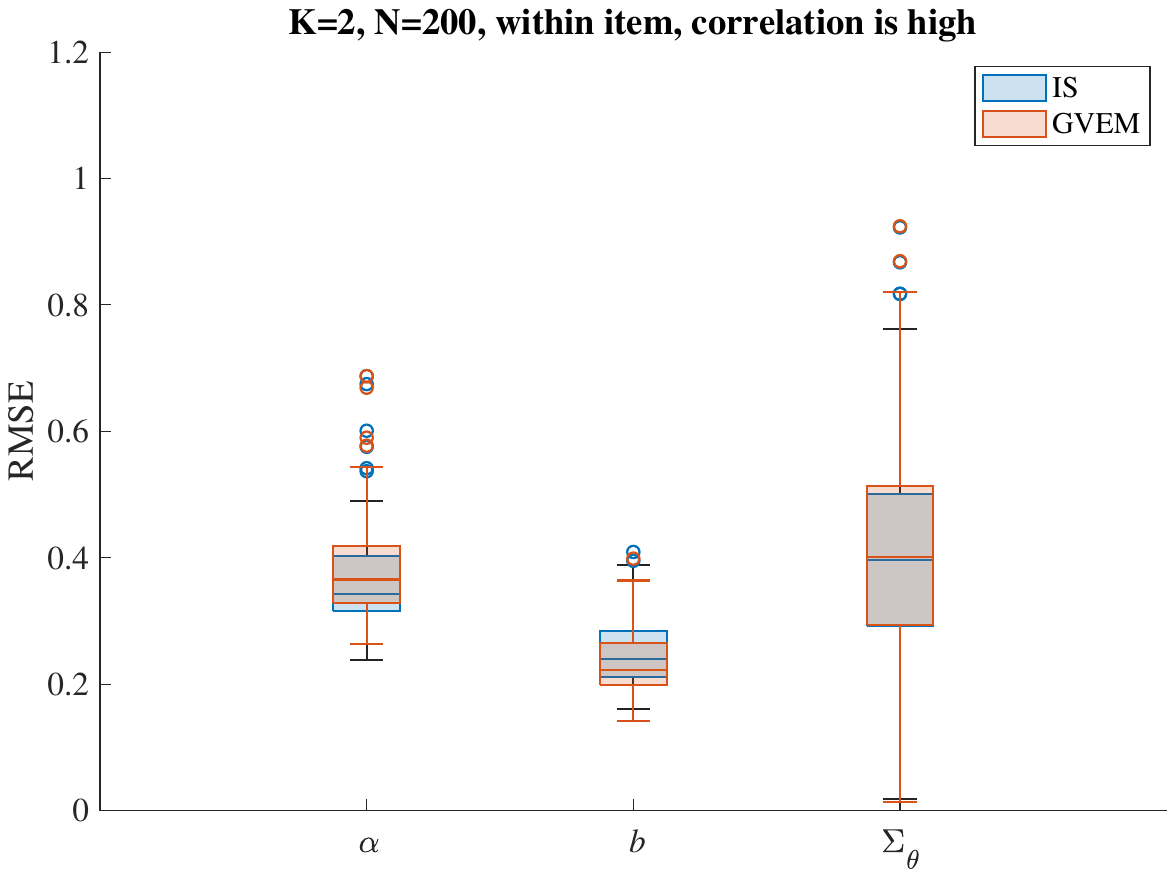}}
    \\
    \subfigure{
        \includegraphics[width=2.5in]{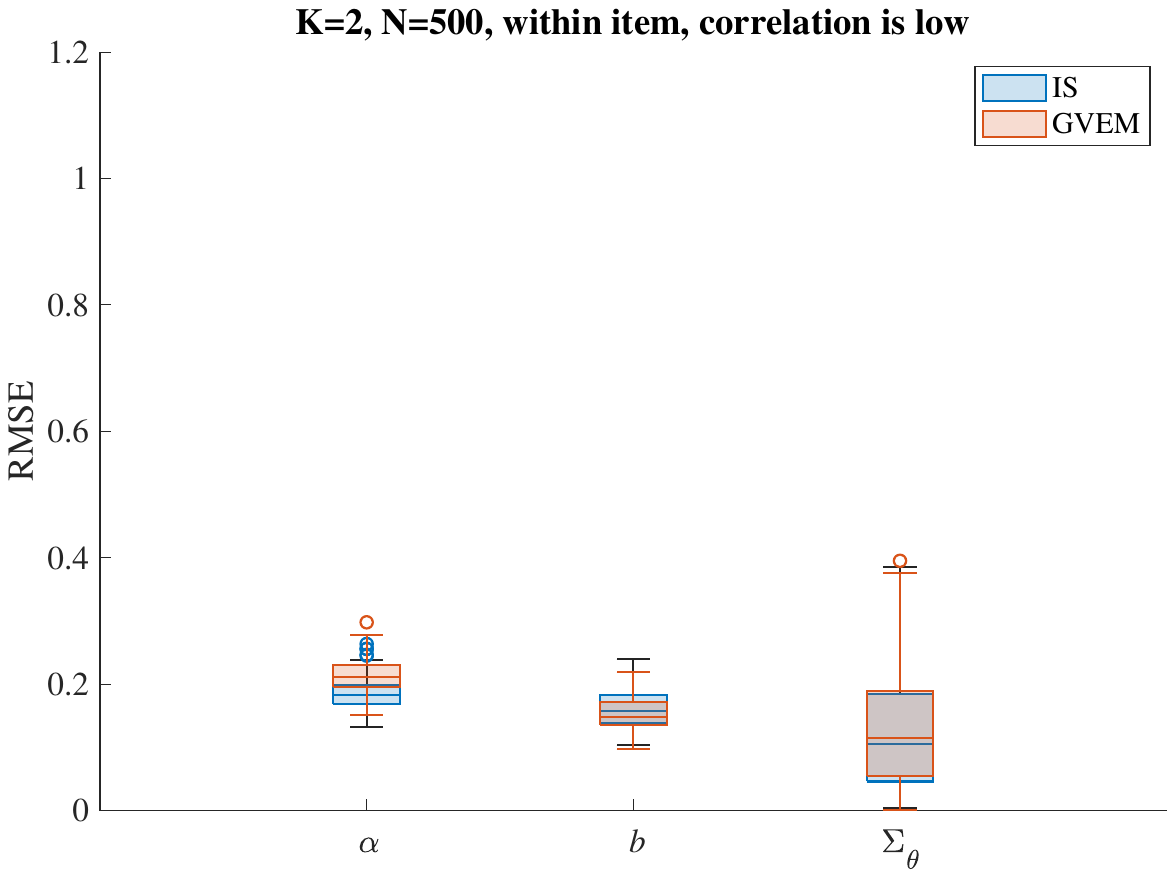}}
    \hspace{0.6in}
    \subfigure{
        \includegraphics[width=2.5in]{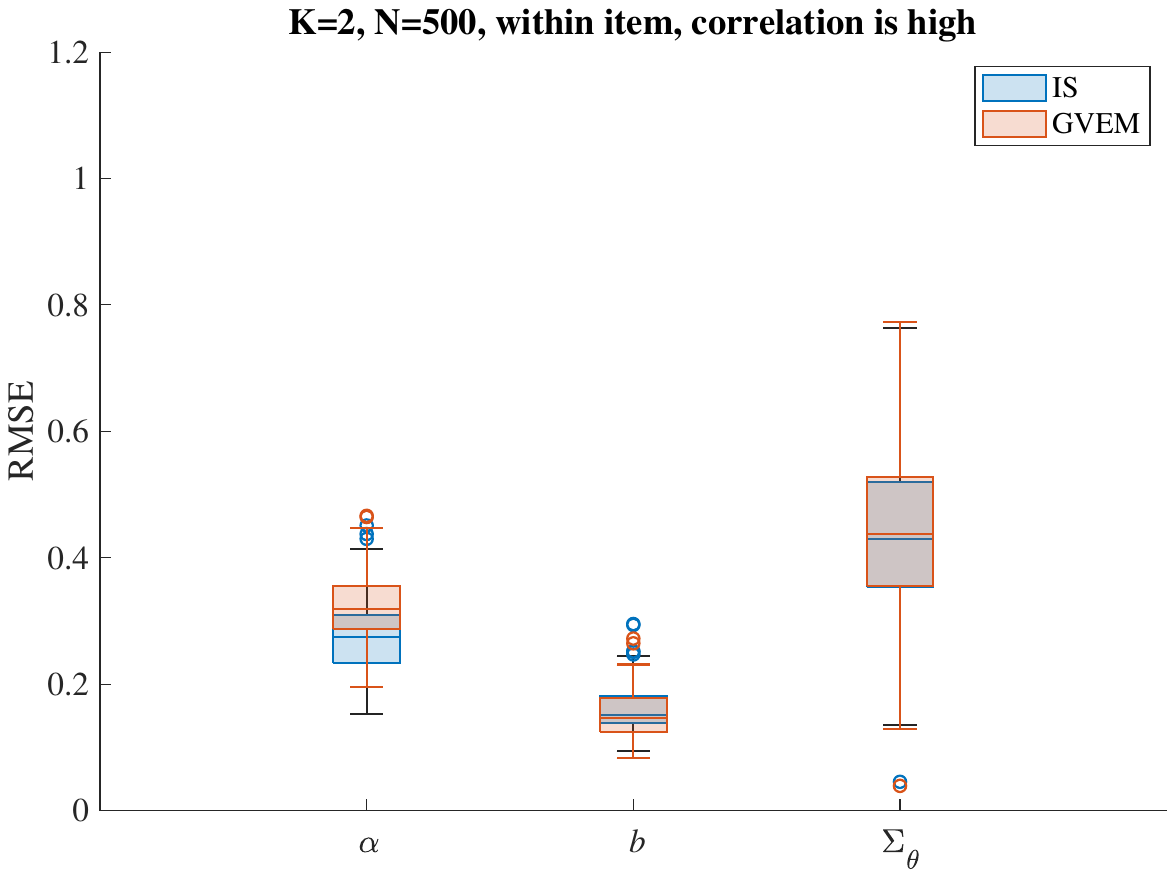}}
    \\
    \caption{RMSE for   $K=2$ under exploratory analysis
    }
    \label{fig:rmse-k2-explore}
\end{figure}

\newpage 

\begin{figure}[ht!]
    \centering
    \subfigure{
        \includegraphics[width=2.5in]{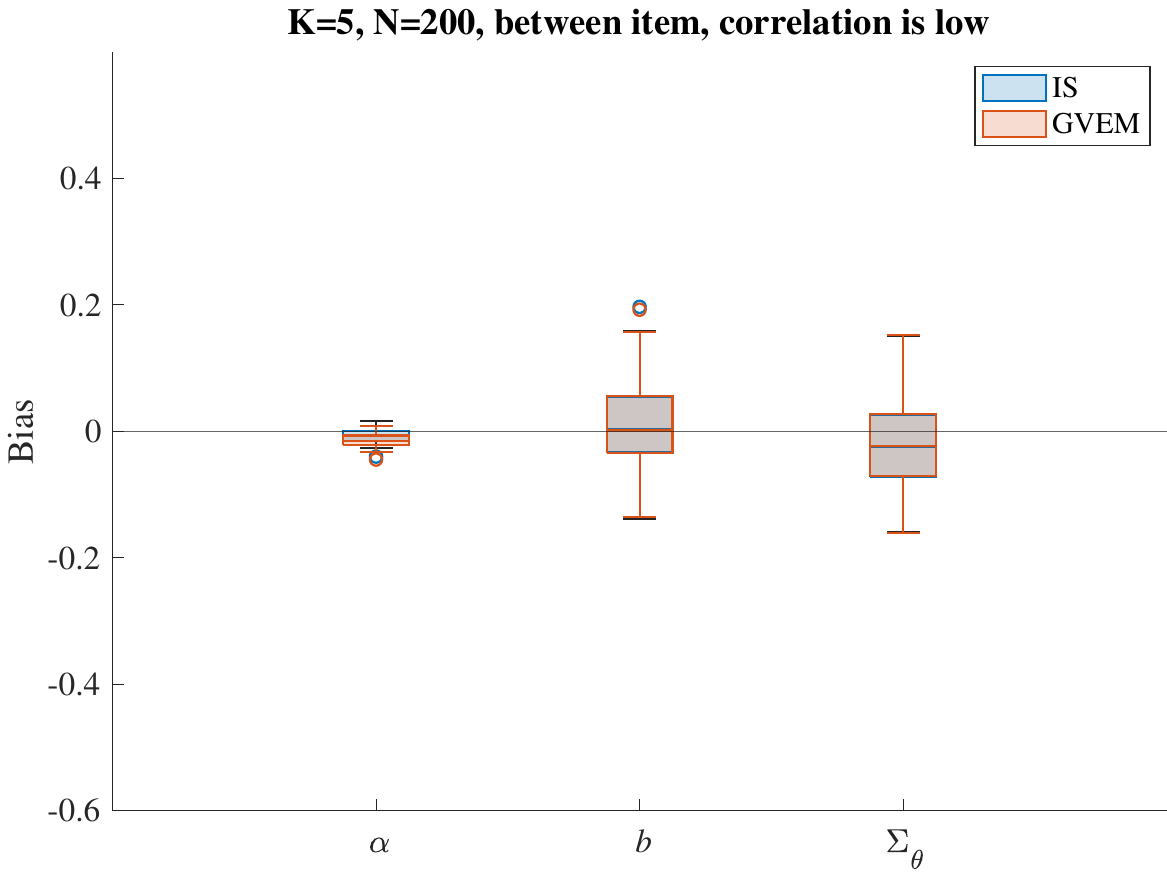}}
    \hspace{0.6in}
    \subfigure{
        \includegraphics[width=2.5in]{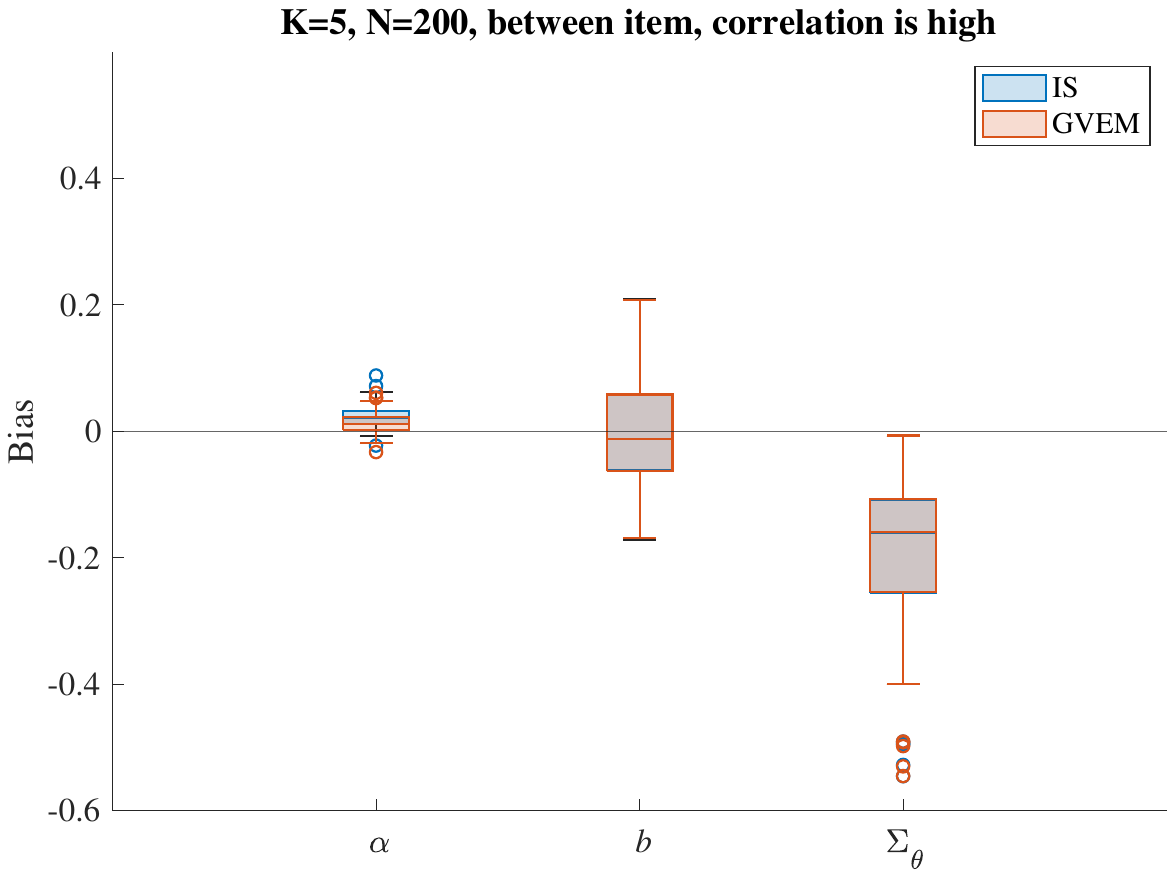}}
    \\
    \subfigure{
        \includegraphics[width=2.5in]{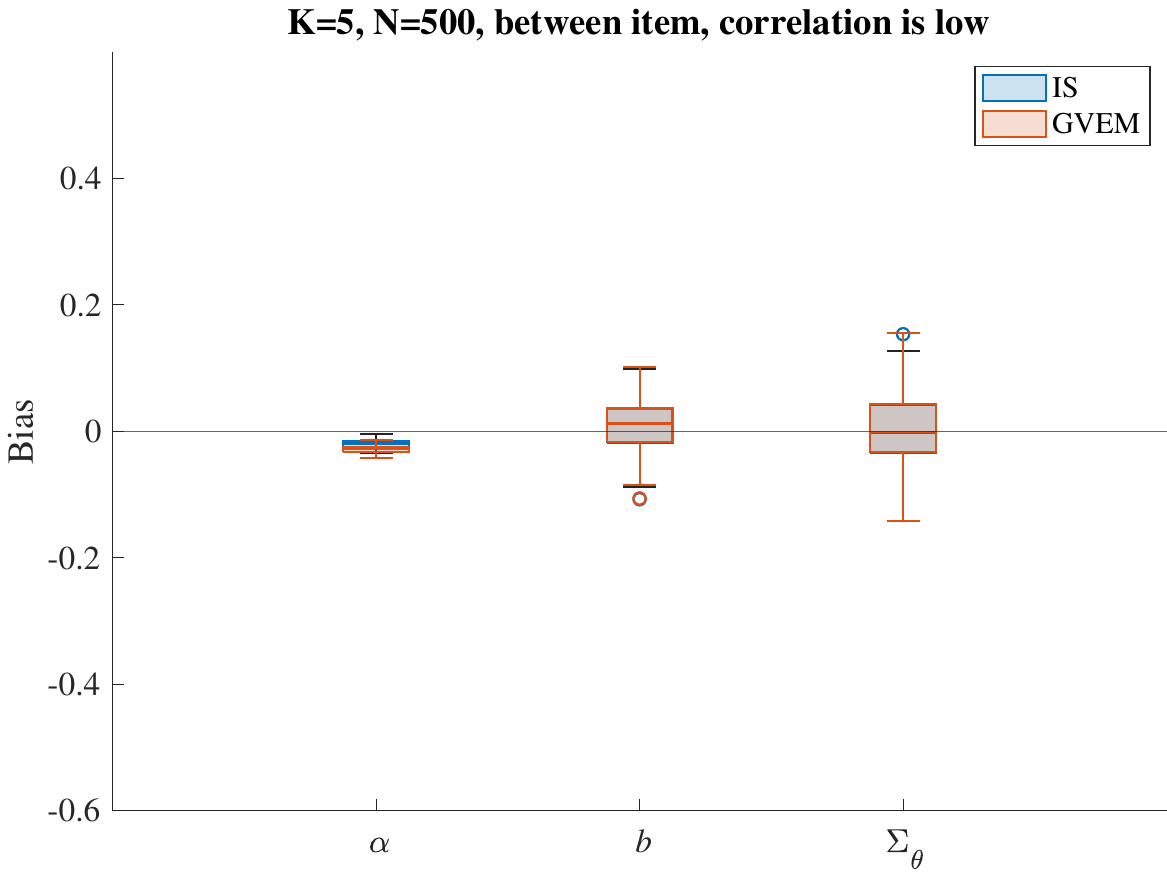}}
    \hspace{0.6in}
    \subfigure{
        \includegraphics[width=2.5in]{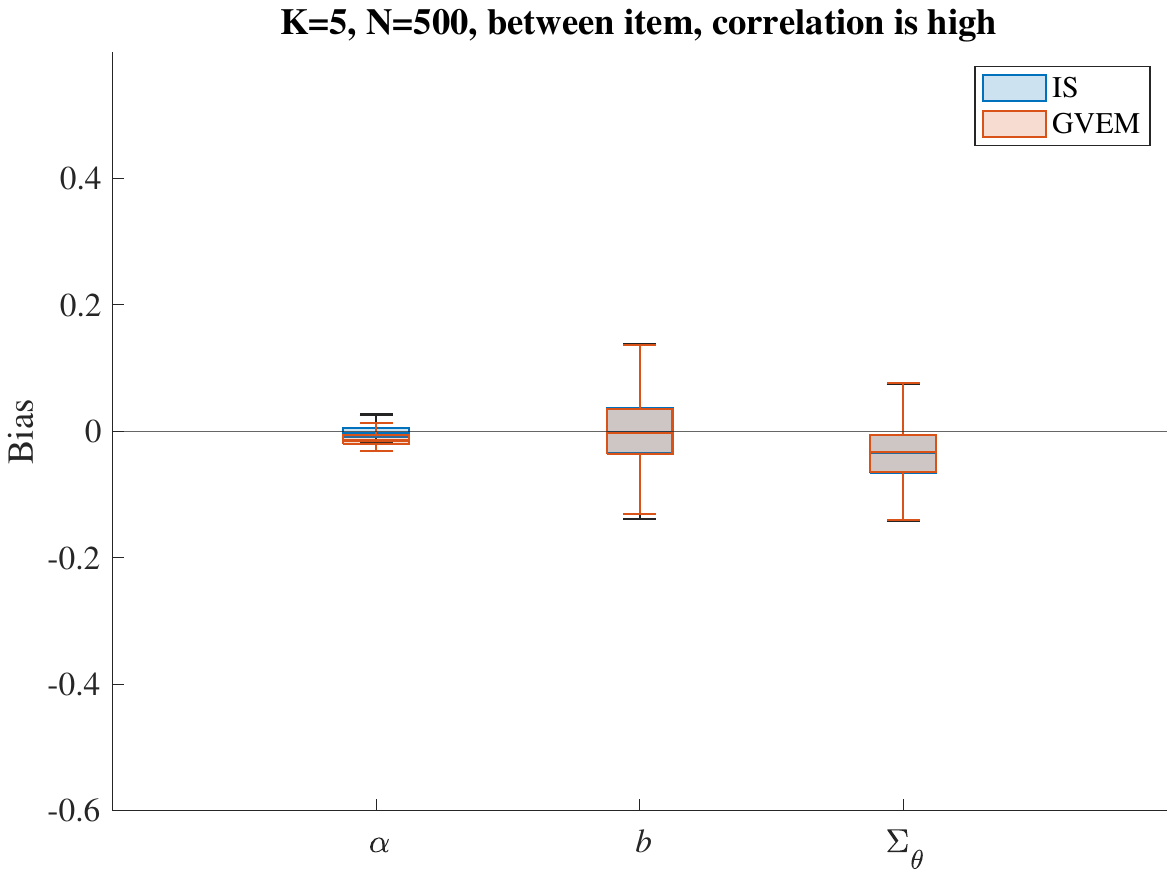}}
    \\
    \subfigure{
        \includegraphics[width=2.5in]{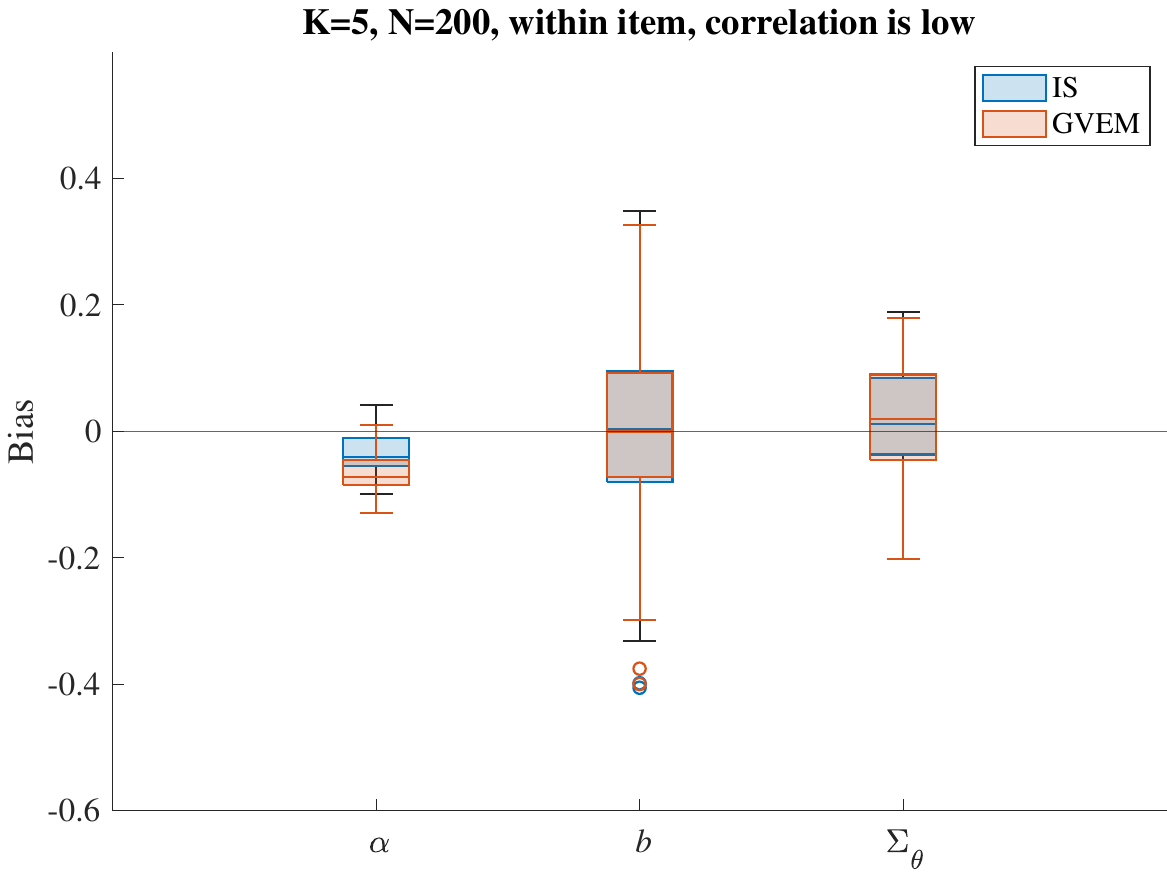}}
    \hspace{0.6in}
    \subfigure{
        \includegraphics[width=2.5in]{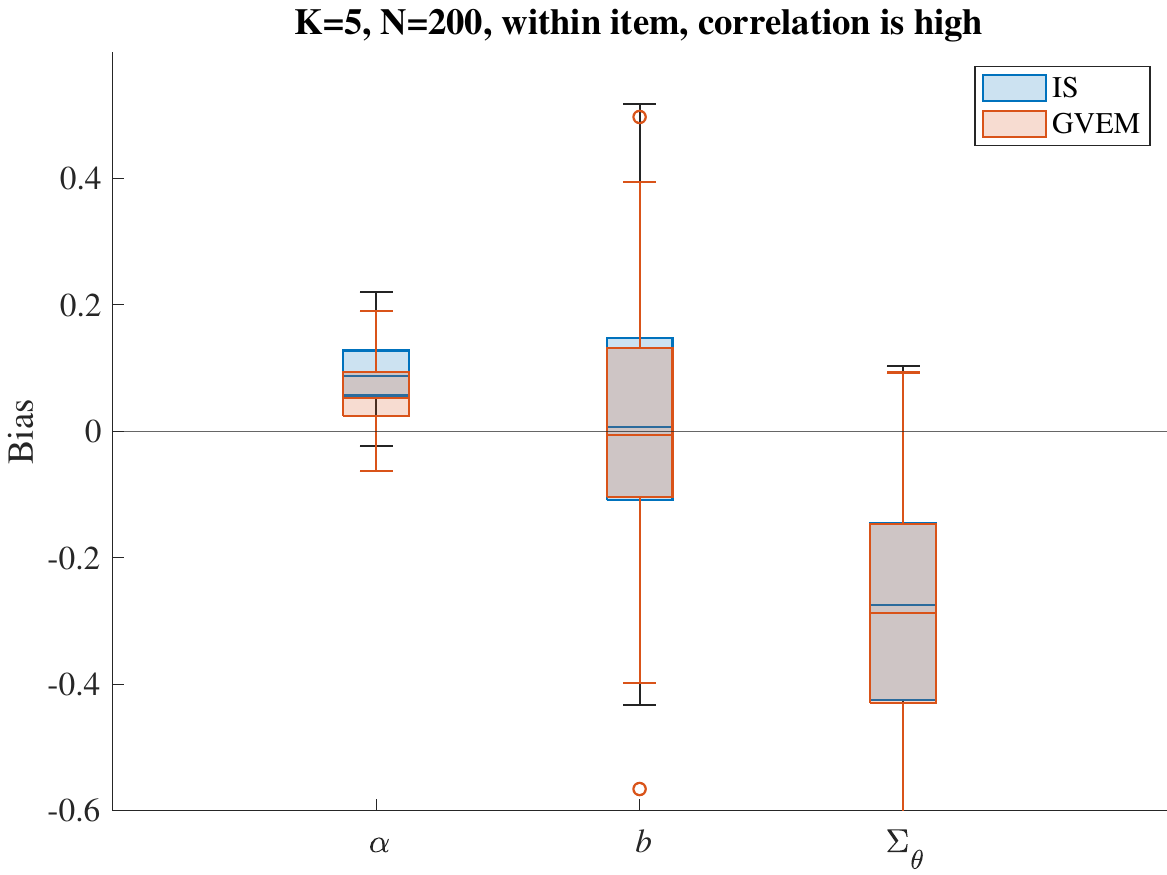}}
    \\
    \subfigure{
        \includegraphics[width=2.5in]{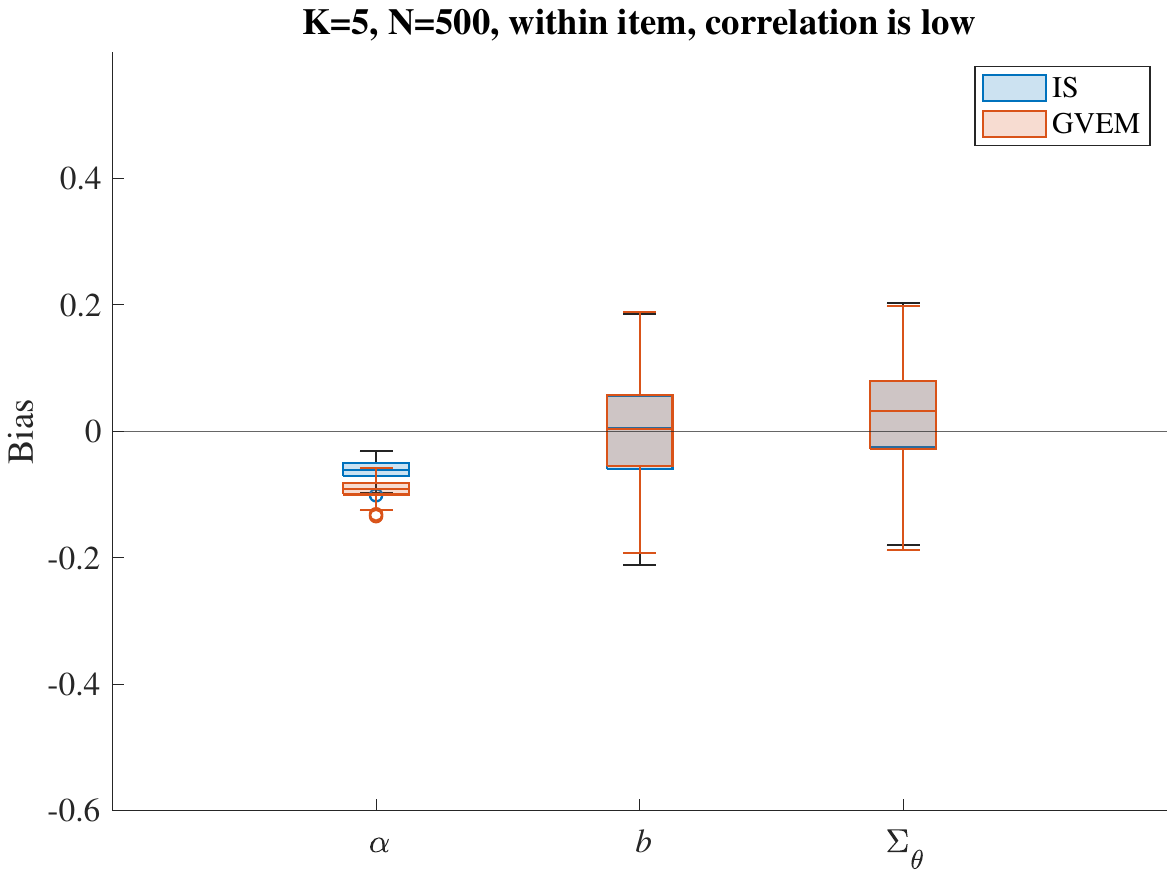}}
    \hspace{0.6in}
    \subfigure{
        \includegraphics[width=2.5in]{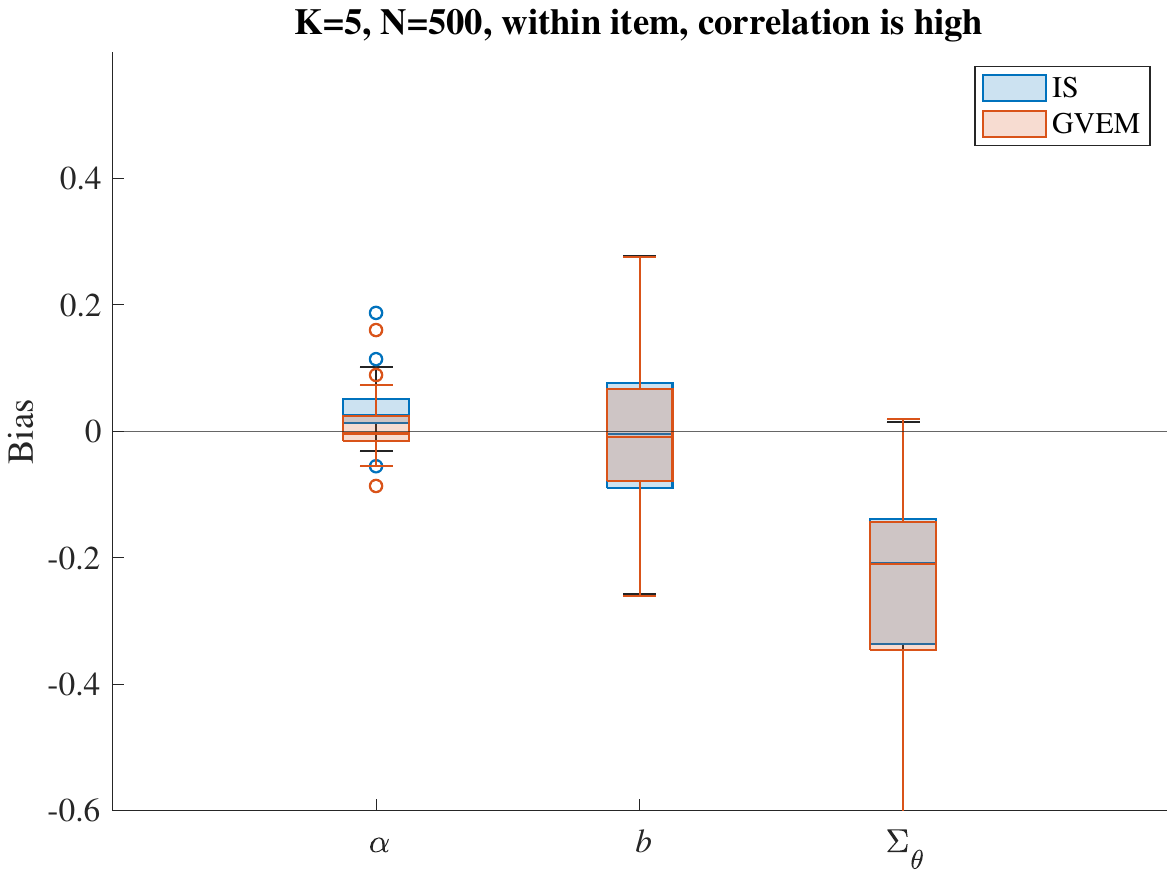}}
    \\
    \caption{Bias for   $K=5$ under exploratory analysis}
    \label{fig:bias-k5-explore}
\end{figure}

\newpage 

\begin{figure}[ht!]
    \centering
    \subfigure{
        \includegraphics[width=2.5in]{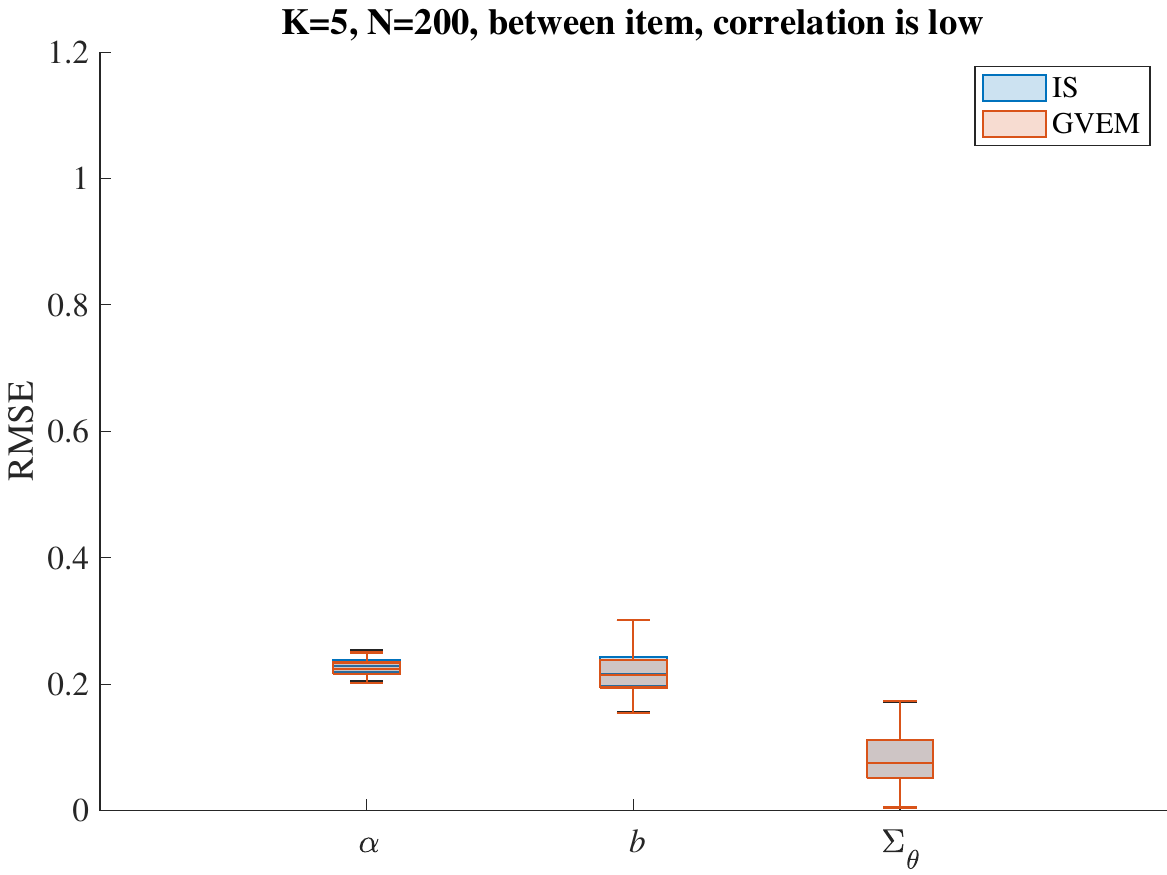}}
    \hspace{0.6in}
    \subfigure{
        \includegraphics[width=2.5in]{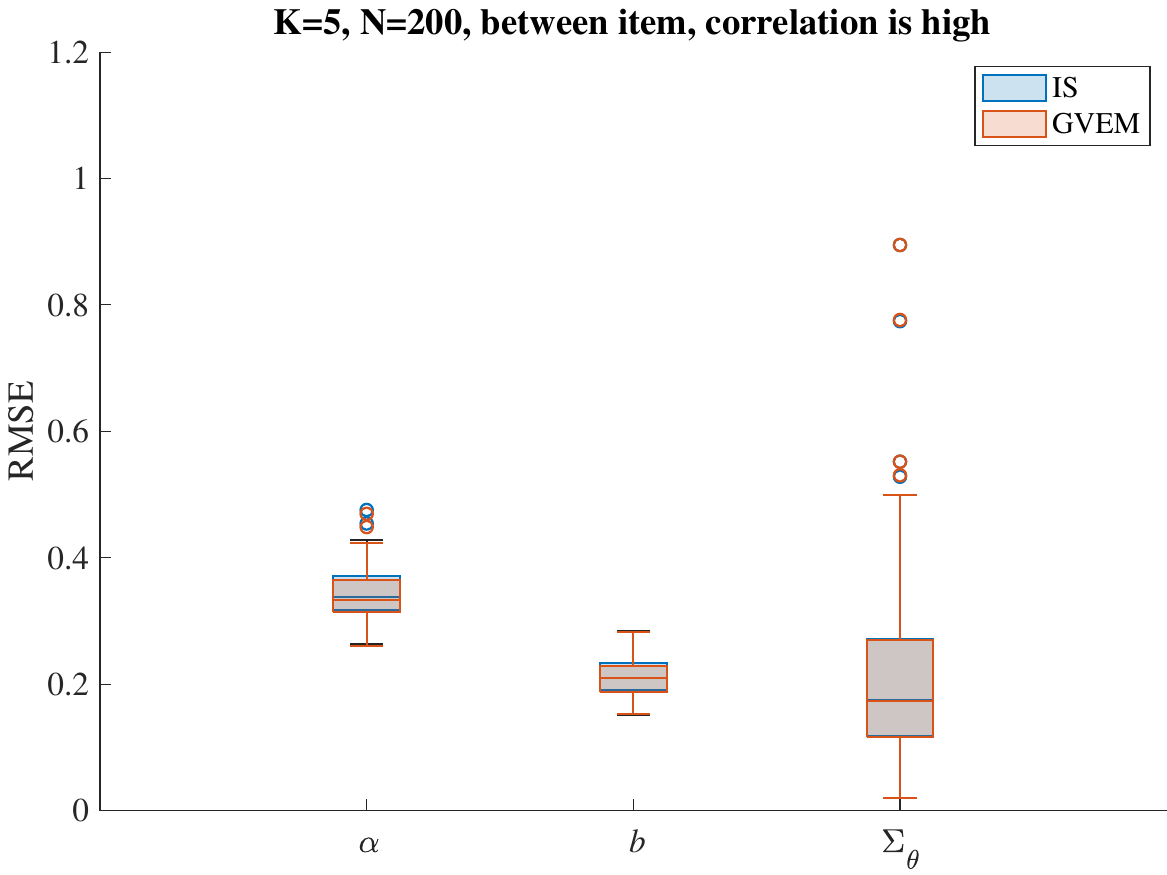}}
    \\
    \subfigure{
        \includegraphics[width=2.5in]{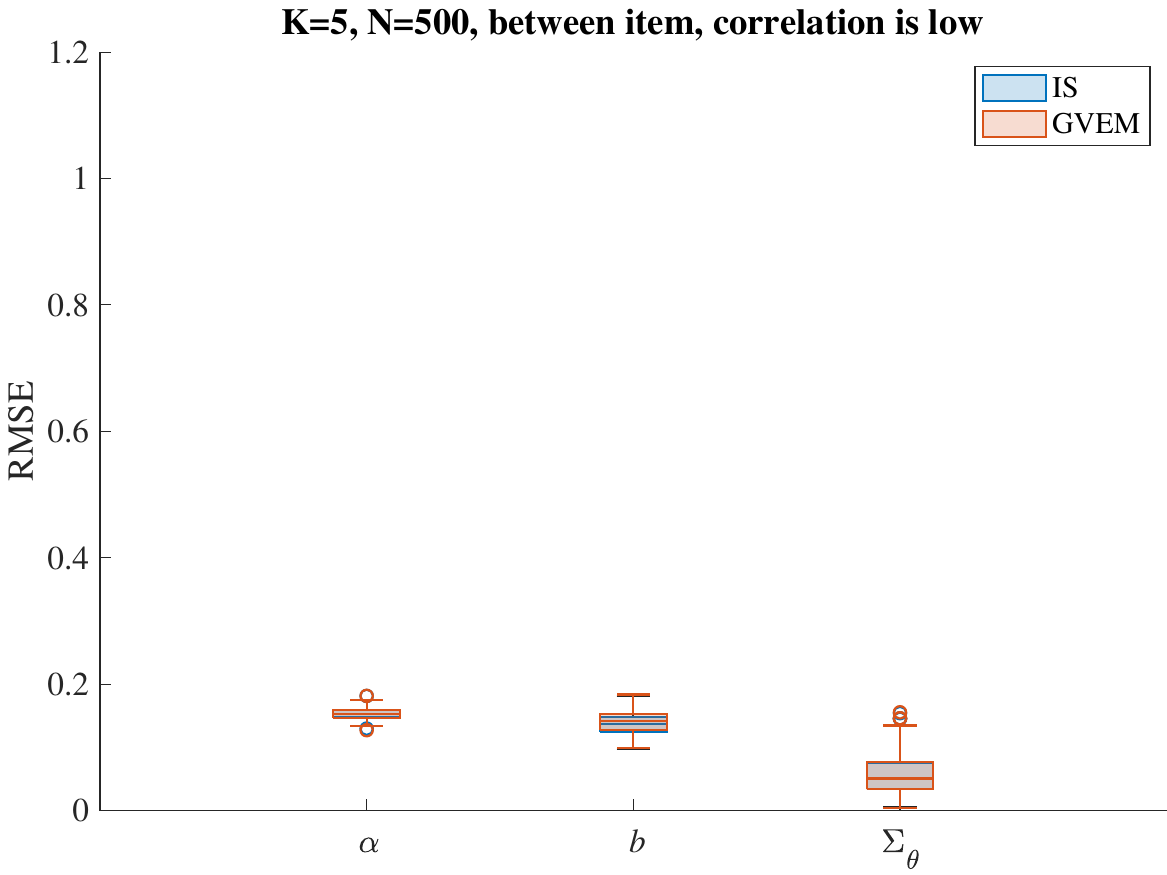}}
    \hspace{0.6in}
    \subfigure{
        \includegraphics[width=2.5in]{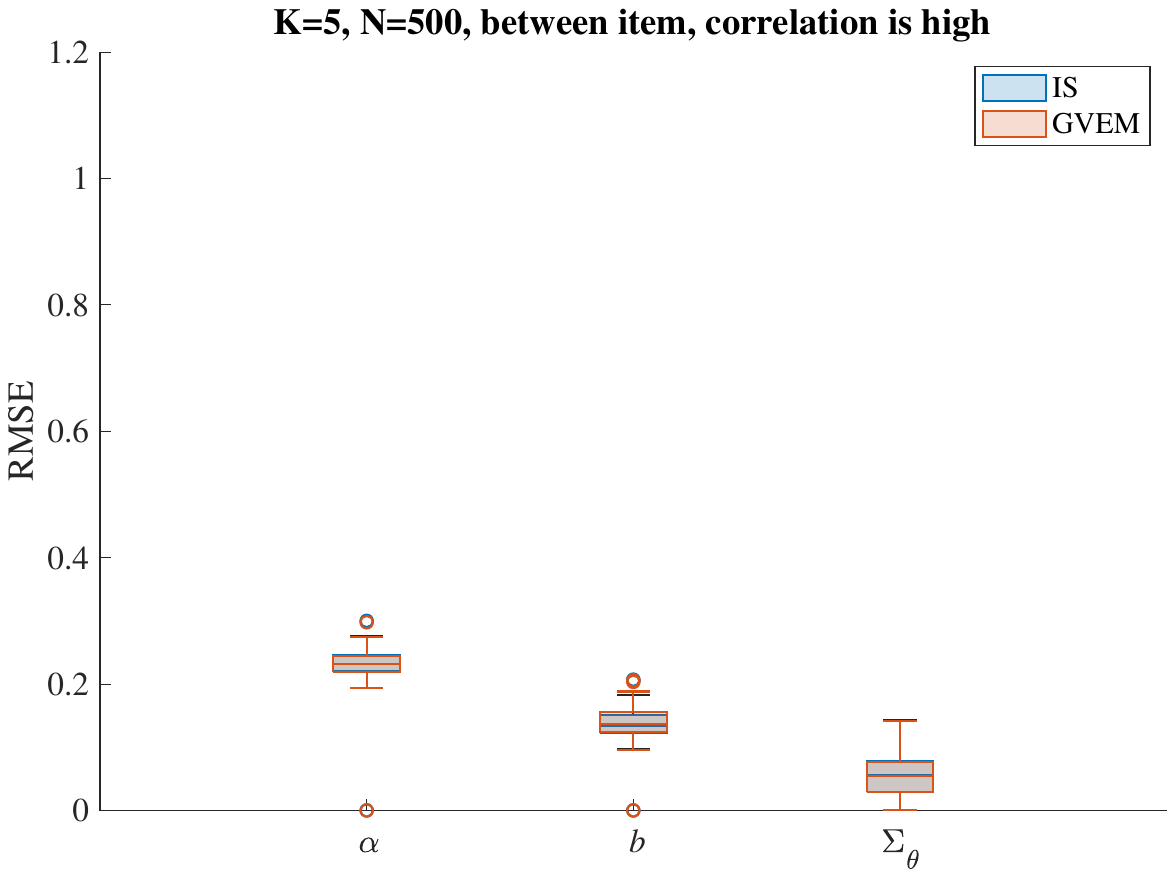}}
    \\
    \subfigure{
        \includegraphics[width=2.5in]{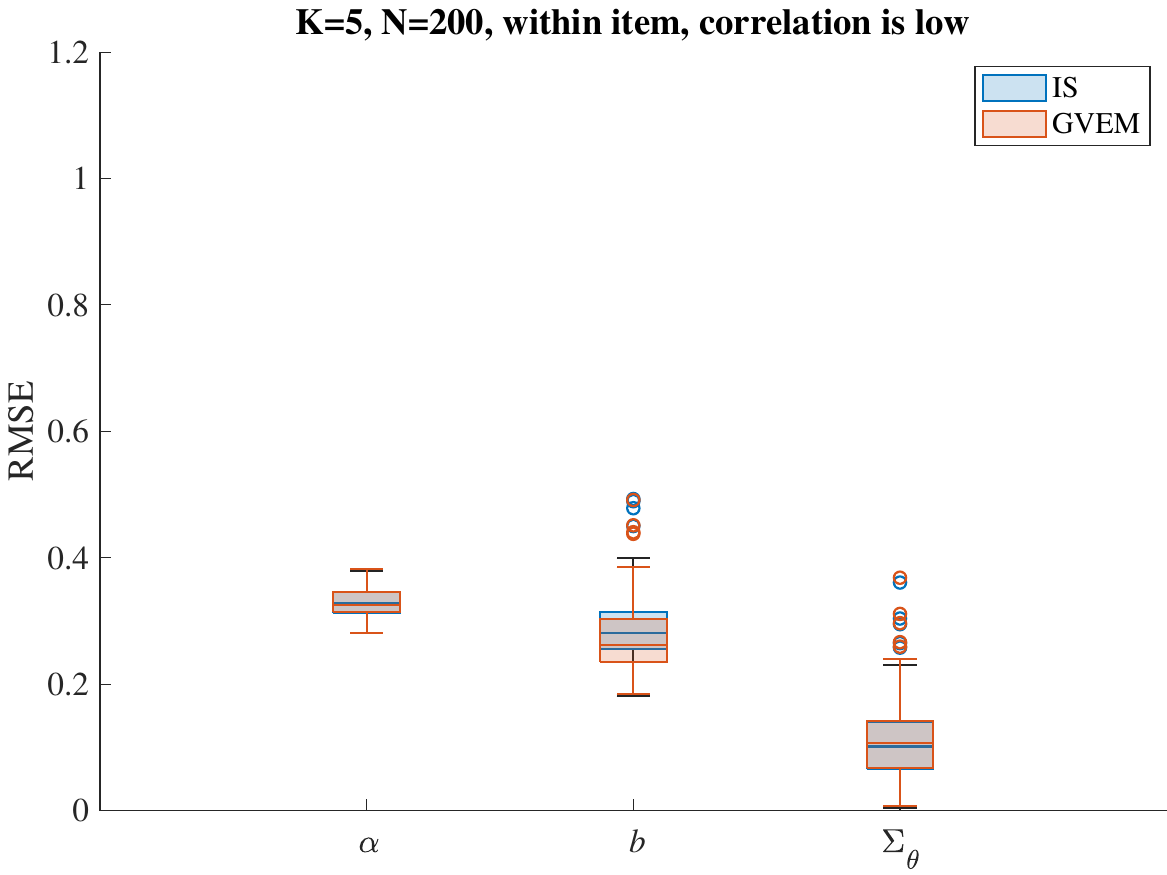}}
    \hspace{0.6in}
    \subfigure{
        \includegraphics[width=2.5in]{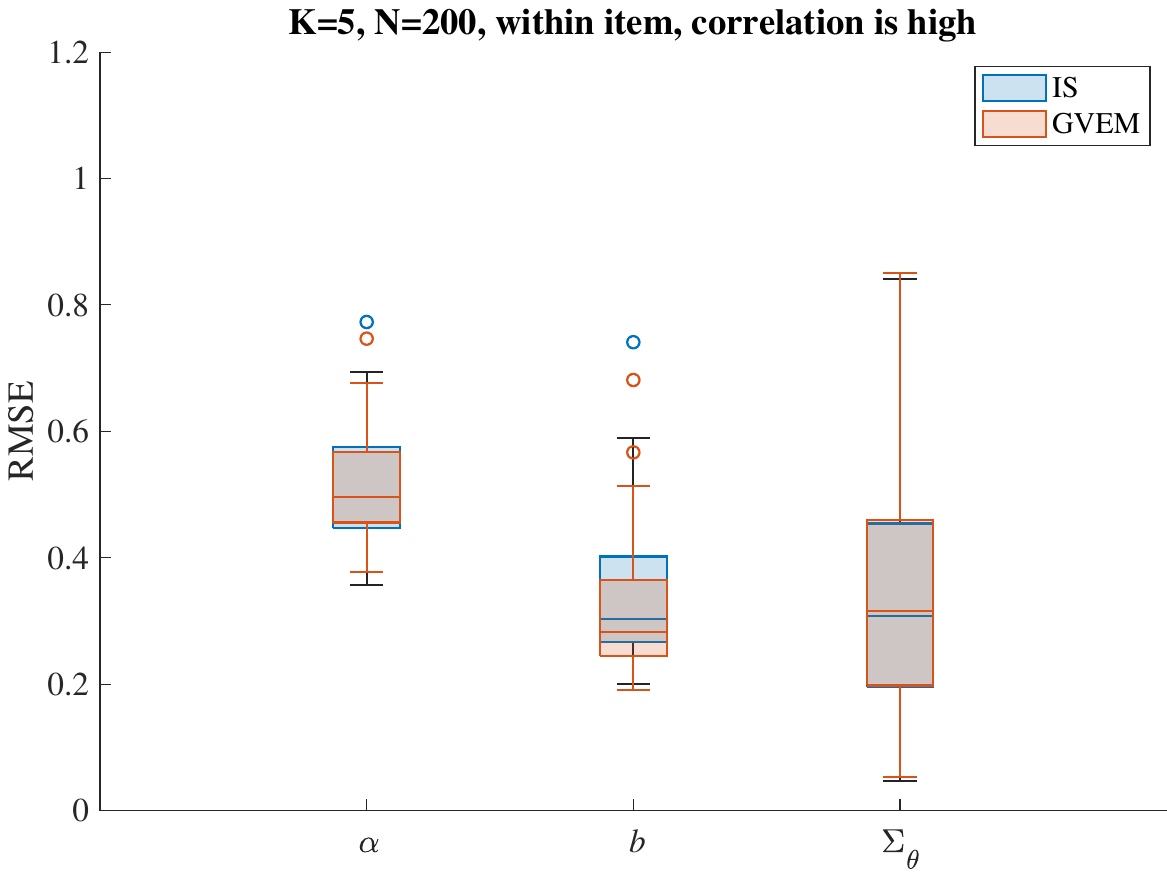}}
    \\
    \subfigure{
        \includegraphics[width=2.5in]{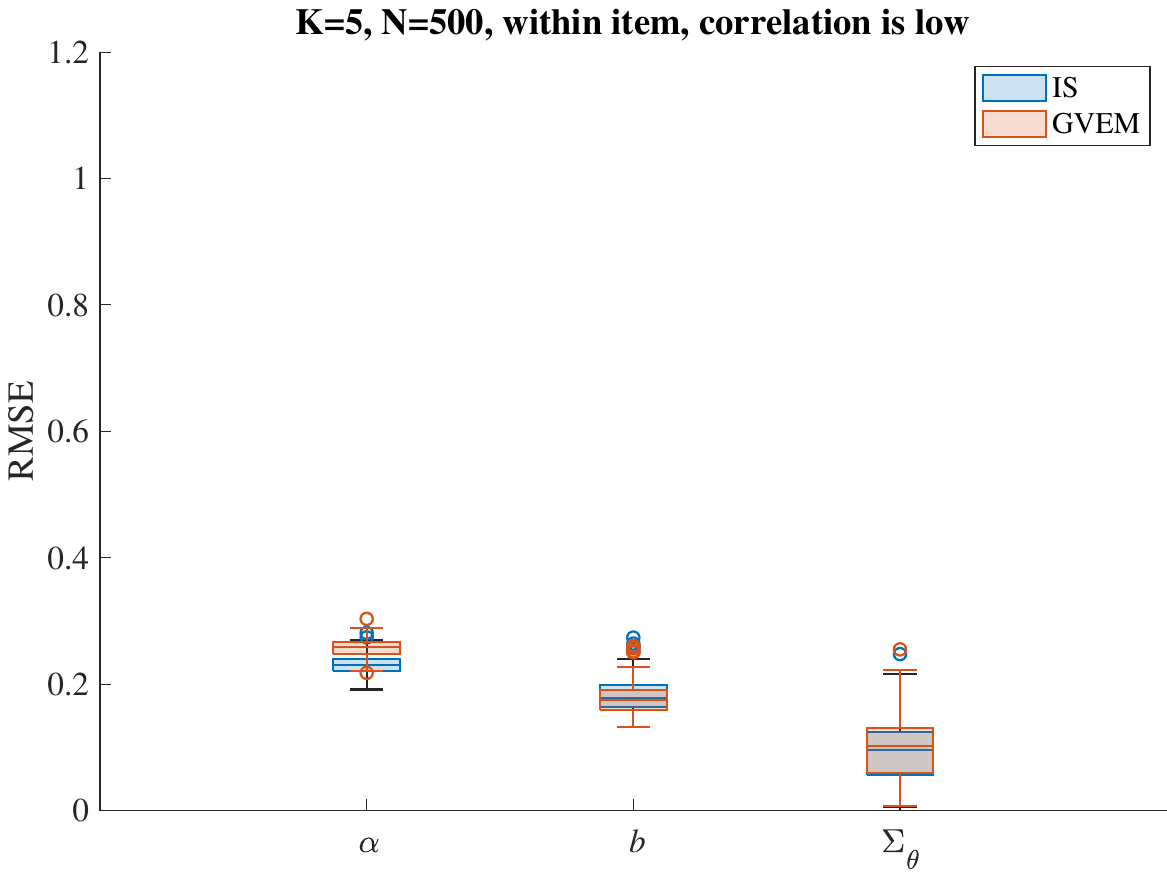}}
    \hspace{0.6in}
    \subfigure{
        \includegraphics[width=2.5in]{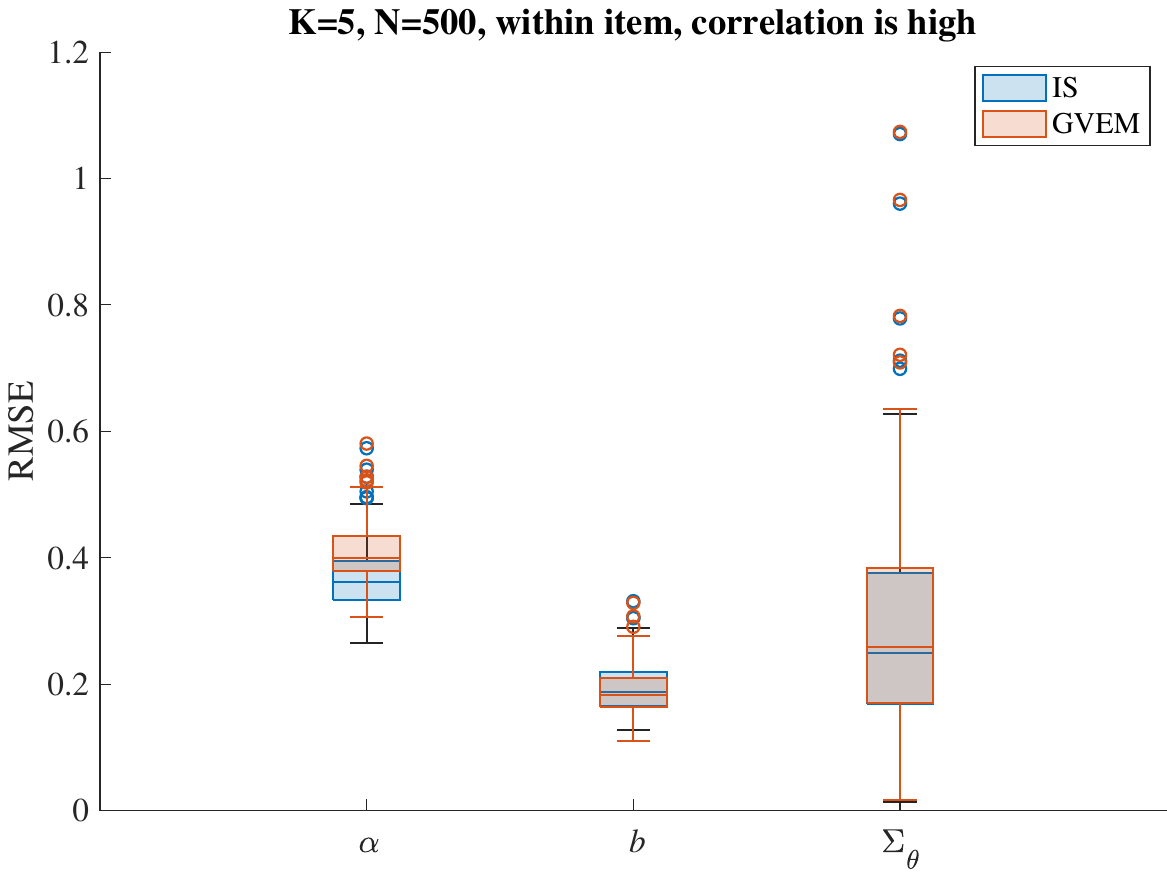}}
    \\
    \caption{RMSE for  $K=5$ under exploratory analysis
    }
    \label{fig:rmse-k5-explore}
\end{figure}

 \begin{table}[htp!]
\centering
\caption{Computation time (seconds) for the confirmatory M2PL estimation}
\begin{tabular}{c|c|c|cccccc}
\hline
 \multicolumn{1}{l|}{N} & \multicolumn{1}{|l|}{ r}    & \multicolumn{1}{|l|}{ Model}    
 &  \multicolumn{2}{c}{K=2}     &  \multicolumn{2}{c}{K=5}   \\\cline{4-9}
 	  &  &  & GVEM & IW-GVEM    & GVEM & IW-GVEM   \\ \cline{1-9}
 	\multirow{4}{0.3in}{200} &\multirow{2}{0.3in}{Low} & Between 
	& 0.68 & 2.88   & 1.15 & 11.31    \\ 
 	& & Within & 1.17 & 2.89  & 4.96 & 11.59    \\\cline{2-9}
    &\multirow{2}{0.3in}{High}  & Between & 1.06 & 2.98    & 2.33 & 12.81    \\ 
 	& & Within & 1.52 & 2.89 &   11.61 &   15.44  \\\cline{1-9}
 	\multirow{4}{0.3in}{500} &\multirow{2}{0.3in}{Low} & Between & 1.52 & 6.88    & 2.29 & 33.98    \\ 
 	& & Within & 2.51 & 6.95 &  10.80 & 35.52    \\\cline{2-9}
    &\multirow{2}{0.3in}{High}  & Between & 1.90 & 7.04    & 3.70 & 33.49    \\ 
 	& & Within & 3.35 & 6.95    & 21.62 & 34.45    \\\cline{1-9}
	\hline
\end{tabular}
\label{time1}
\end{table}

\begin{table}[htp!]
\centering
\caption{Computation time (seconds) for the exploratory M2PL estimation}
\begin{tabular}{c|c|c|cccccc}
\hline
 \multicolumn{1}{l|}{N}  & \multicolumn{1}{|l|}{ r} & \multicolumn{1}{|l|}{ Model}    &  \multicolumn{2}{c}{K=2}     &  \multicolumn{2}{c}{K=5}   \\\cline{4-9}
 	  &  &  & GVEM & IW-GVEM  &    GVEM & IW-GVEM   \\\cline{1-9}
 	\multirow{4}{0.3in}{200} &\multirow{2}{0.3in}{Low} & Between 
	& 0.93 & 2.00    &6.17&25.67    \\ 
 	& & Within & 1.10 & 2.01    &11.18&25.75    \\\cline{2-9}
    &\multirow{2}{0.3in}{High}  & Between & 1.13 & 2.39    & 12.53& 26.27    \\ 
 	& & Within & 1.38 & 2.03    &20.55 &26.18    \\\cline{1-9}
 	\multirow{4}{0.3in}{500} &\multirow{2}{0.3in}{Low} & Between 
	&  2.10 & 5.02   &13.42&68.10    \\ 
 	& & Within & 2.51 & 4.98    &24.16&68.04    \\\cline{2-9}
    &\multirow{2}{0.3in}{High}  & Between 
    & 2.66 & 5.91   &21.60&69.43    \\ 
 	& & Within & 3.39 & 5.40  &42.27&68.69    \\ \cline{1-9}
	\hline
\end{tabular}
\label{time2}
\end{table}

\section{Discussion}
\label{sec-diss}
In this note, we proposed an importance weighted version of GVEM to correct its bias on the $\boldsymbol{\alpha}$ estimates in the confirmatory M2PL models. Because the evidence lower bound (ELBO), a key component of variational inference, is derived based on Jensen's inequality, the ELBO will approximate the log-marginal distribution (i.e., $\log P(\boldsymbol{X})$) more closely when $R\equiv  {P(\boldsymbol{X},\boldsymbol{Z})}/{q(\boldsymbol{Z})}$ is more concentrated around its mean $P(\boldsymbol{X})$. Hence, the primary idea of IW-GVEM is to replace $R$ with its sample mean by drawing i.i.d. samples from variational distribution $q(\boldsymbol{z})$. In so doing, we achieve a tighter bound of Jensen's inequality, but at the slight cost of computational efficiency. The added computation time is mainly due to sampling in the E-step and gradient descent in the M-step. From our simulation results, the bias correction is  effective for confirmatory models and the extra computation time is acceptable because even with additional computational cost, the total time is still short. In fact, the time increase from GVEM to IW-GVEM is at a slow rate in that the time ratio between the two methods is smaller for more complex models (i.e., $K=5$, within-item multidimensional structure, and high correlations). Note that for exploratory M2PL models, the original GVEM is still recommended because it already produces almost unbiased results and hence importance sampling seems unnecessary, although it does not introduce any undesirable bias either. {\color{black} Theoretically, \cite{cho2021gaussian} proved that the estimated factor loading matrix and estimated latent factor from the GVEM algorithm is consistent as $N \rightarrow \infty$ and $J \rightarrow \infty$. The proposed IW-GVEM algorithm is based on the GVEM estimation, 
hence with consistent initial GVEM estimators, the final estimators from the IW-GVEM algorithm also have the theoretical guarantee to be consistent in the high-dimensional setting. Moreover, compared to ELBO in GVEM, the importance-weighted ELBOs are greatly improved after importance sampling. In finite-sample simulations, importance-weighted ELBOs at  $M = 5, 10, 50$, and $100$ are all larger than  ELBO from GVEM and converge as $M$ increases  (See Appendix~\ref{sec:additional simulation study}).
}

In IW-GVEM, we propose to use the adaptive moment estimation method to automatically update the learning rate on the fly. Our preliminary results showed that the Adam algorithm performs better than fixed learning rate. Further, we also evaluated the effect of Monte Carlo sample size (i.e., $S=10, 50, 100$) and sample size for the importance sampling step (i.e., $M=10, 50$) and noted essentially the same results. Hence, we set {\color{black} $S=10$} and $M=10$ in our simulation study, which explains the only modest increase in computation time.

Aside from GVEM, another recently proposed fast algorithm for high-dimensional IRT estimation is the joint maximum likelihood estimation \citep {chenlizhang}. This method treats the latent abilities as fixed effect parameters instead of random variables. Although this approach is innovative and their algorithm appears to produce accurate parameter estimates efficiently, the interpretation of person parameters is different such that caution needs to be exercised when one intends to generalize findings to a certain population. Plus, treating each individual as a separate fixed effect is, at the conceptual level, hard to justify when generalizing M2PL to a multiple-group MIRT model. This is because the goal of a multiple-group extension is to allow for unbiased marginal estimation of group-specific population distributions.

Instead, the GVEM method can be generalized to multiple-group MIRT in a more straightforward fashion. Our other study exploring multiple-group GVEM for differential item functioning detection (DIF) reveals that it can very well detect uniform DIF, but the power of detecting DIF on discrimination parameter is low. This is likely due to the estimation bias on $\boldsymbol{\alpha}$ from GVEM in the confirmatory model estimation, and hence the IW-GVEM will likely improve detection of the non-uniform DIF, in particular the DIF on discrimination parameters. Our study can also be extended in other directions. For instance, like in \cite{cho2021gaussian}, the IW-GVEM can be extended to M3PL models. Moreover, the current IW-GVEM algorithm does not automatically output standard error of item parameter estimates, and hence future studies may consider combining it with the supplemented EM algorithm \citep{caisem, chenwang} to produce accurate SE estimates. In addition to MIRT, the proposed method may   also shed light on improving the performance of the variational estimation for other psychometric models, such as  generalized linear mixed models \citep{jeon2017variational} and  cognitive diagnosis models \citep{yamaguchi2020variational,yamaguchi2020variational1}. 

\subsection*{Competing interest and data availability statement}
 The authors declare that they have no conflict of interest.   The simulation code and datasets generated during the current study are available 
 {\color{black} at https://github.com/jingoystat/A-Note-on-Improving-Variational-Estimation-for-Multidimensional-Item-Response-Theory}.

\renewcommand{\baselinestretch}{1.25}

\newpage
\appendix

\section*{Supplementary Material}
{\color{black} 
\section{Additional Comparitive Studies}

\subsection{Comparing IW-GVEM with Importance-Weighted Variational Bayesian Method}

In recent literature, researchers also proposed importance-weighted variational Bayesian (IW-VB) methods for the estimation of MIRT models. In particular, \citeauthor{urban2021deep} (\citeyear{urban2021deep}) and ~\citeauthor{liu2022estimating} (\citeyear{liu2022estimating}) proposed to use importance-weighted variational autoencoder (IW-VAE) for exploratory factor analysis. This method is a deep learning based variational method and is computationally fast in large data sets.
Although IW-VB methods handle large-scale data with high computational efficiency, their performances at {\color{black} relatively small-sized and medium-sized data are not competitive.
 While MCMC could be an alternative method for small samples, in situations with small to medium sample sizes, our variational method is faster and more competitive   than MCMC.}


In this section, we provide additional finite sample simulation results to show that 
our method outperforms the IW-VB methods in {small to medium samples}. To illustrate it, we compare our proposed IW-GVEM method and IW-VB method by~\cite{liu2022estimating} at $N = 200$, $N = 500$ {\color{black} and $N = 1000$}. Because their method focuses only on exploratory MIRT, we will compare the performance of our method (denoted as ``IS" in the figure) to IW-VB for exploratory analysis. The simulation settings follow the same settings as in Section~\ref{sec-simulation}. The results are presented in Figures~\ref{fig:bias-k2-between-explore compare vb}--\ref{fig:rmse-k5-within-explore compare vb}. From the results, we see the biases of IW-GVEM are closer to 0 than the IW-VB method under all simulation settings. The RMSEs of our proposed method are substantially smaller than the IW-VB in \cite{liu2022estimating}. 



\newpage 
\begin{figure}[ht!]
    \centering
    \subfigure{
        \includegraphics[width=2.5in]{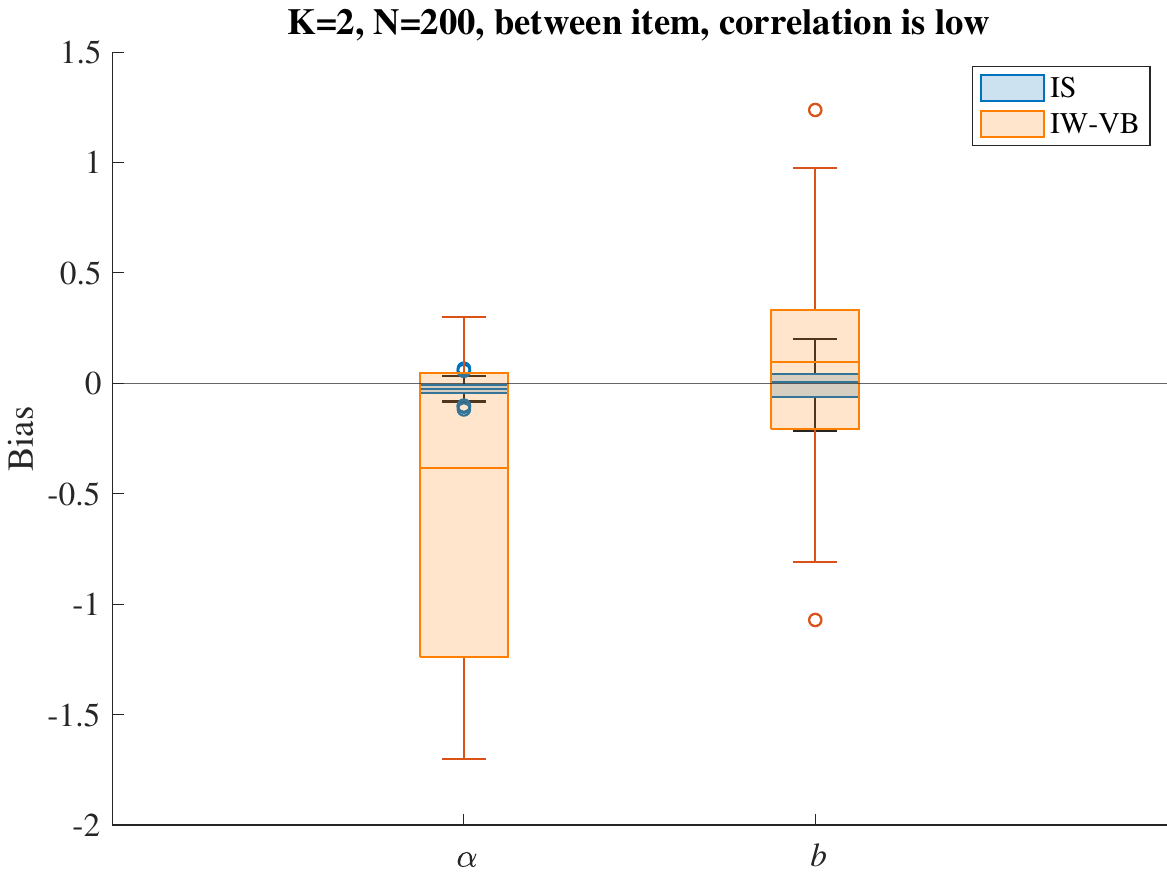}}
    \hspace{0.6in}
    \subfigure{
        \includegraphics[width=2.5in]{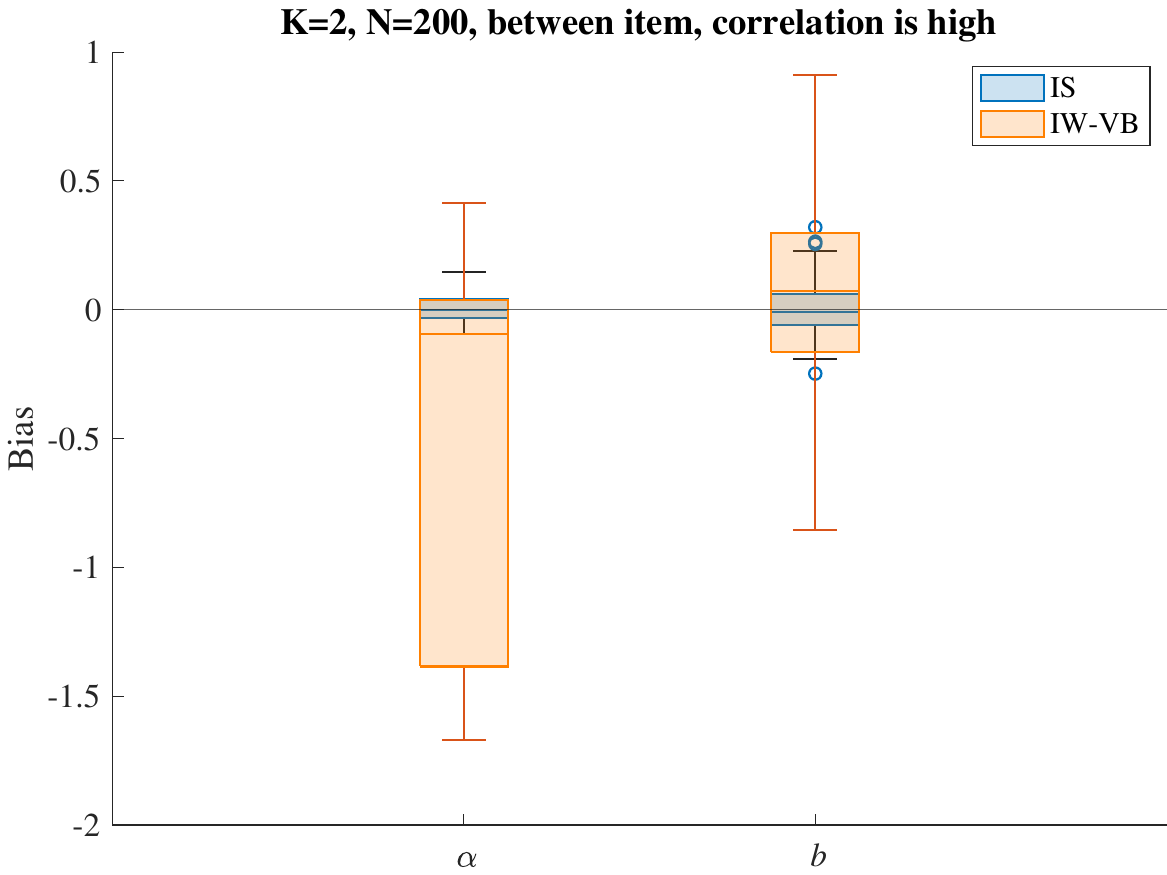}}
    \\
    \subfigure{
        \includegraphics[width=2.5in]{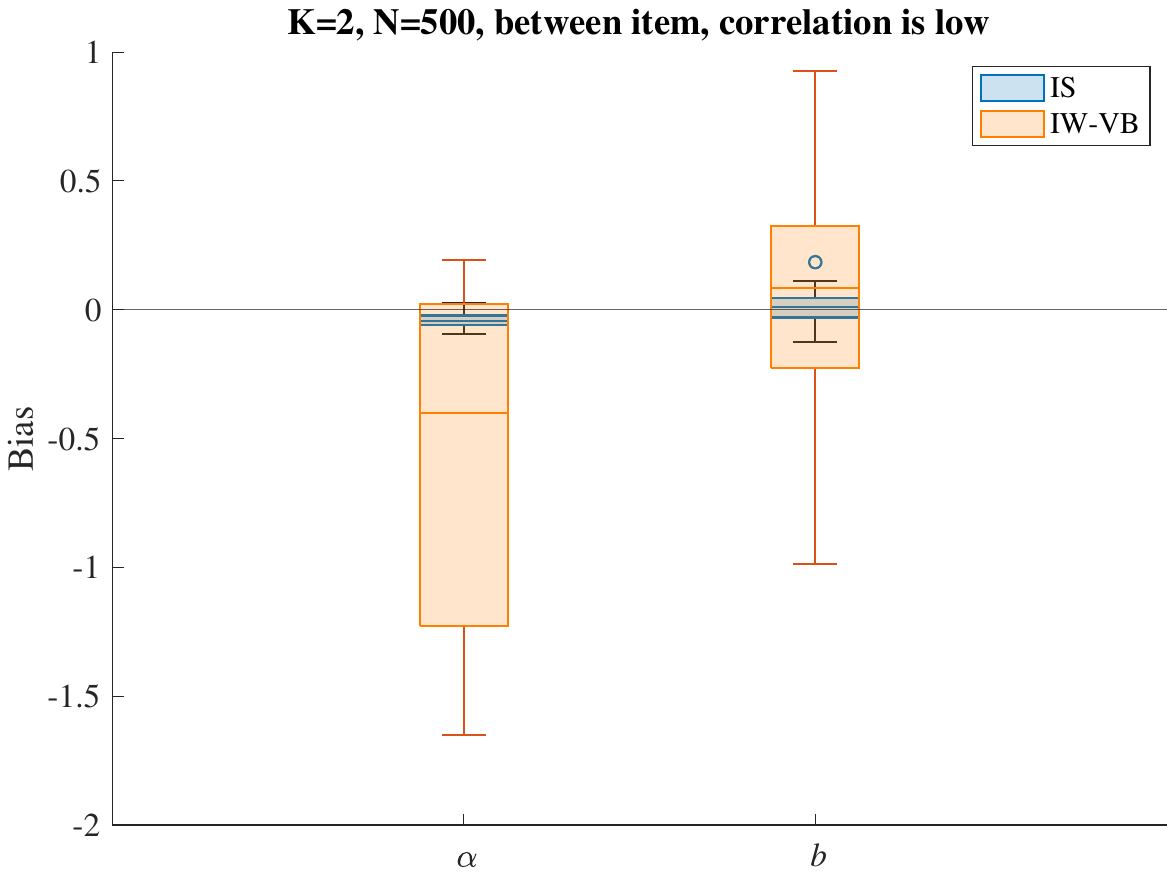}}
    \hspace{0.6in}
    \subfigure{
        \includegraphics[width=2.5in]{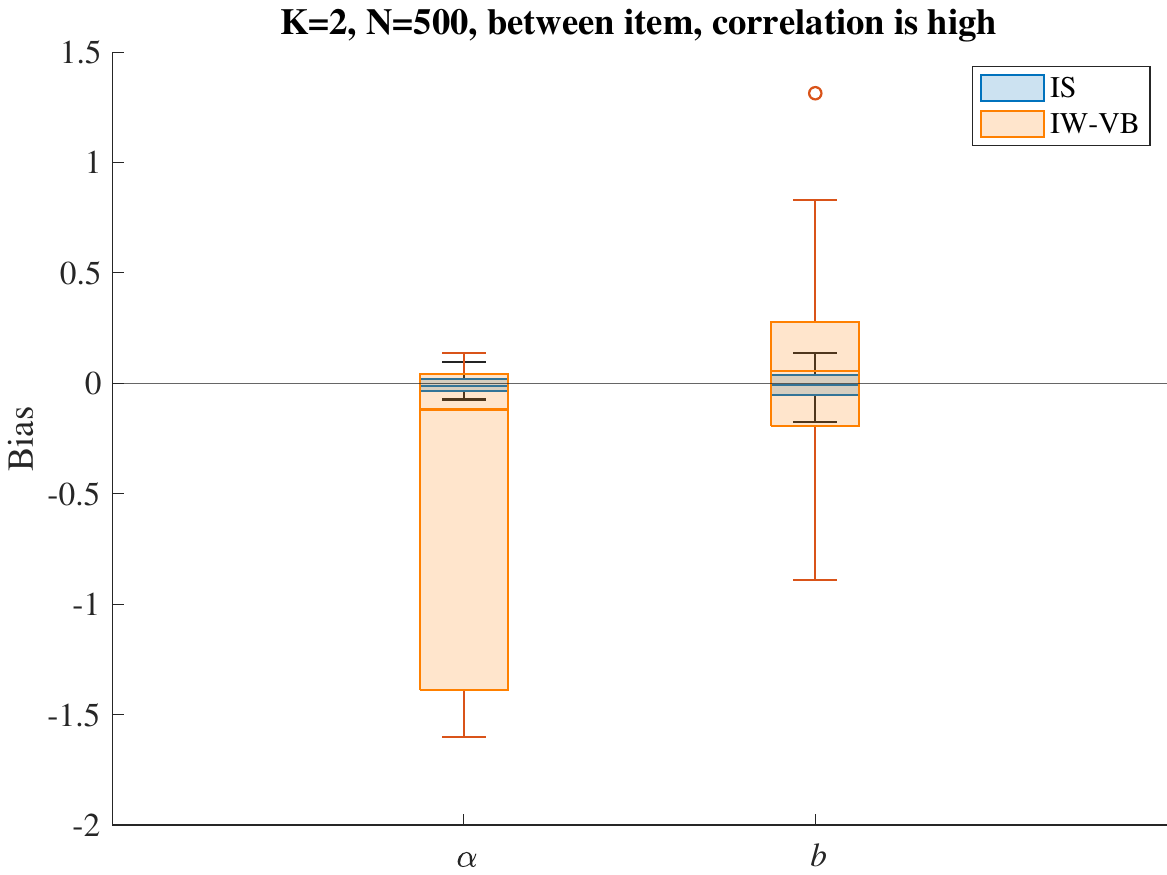}}
    \\
   \subfigure{
        \includegraphics[width=2.5in]{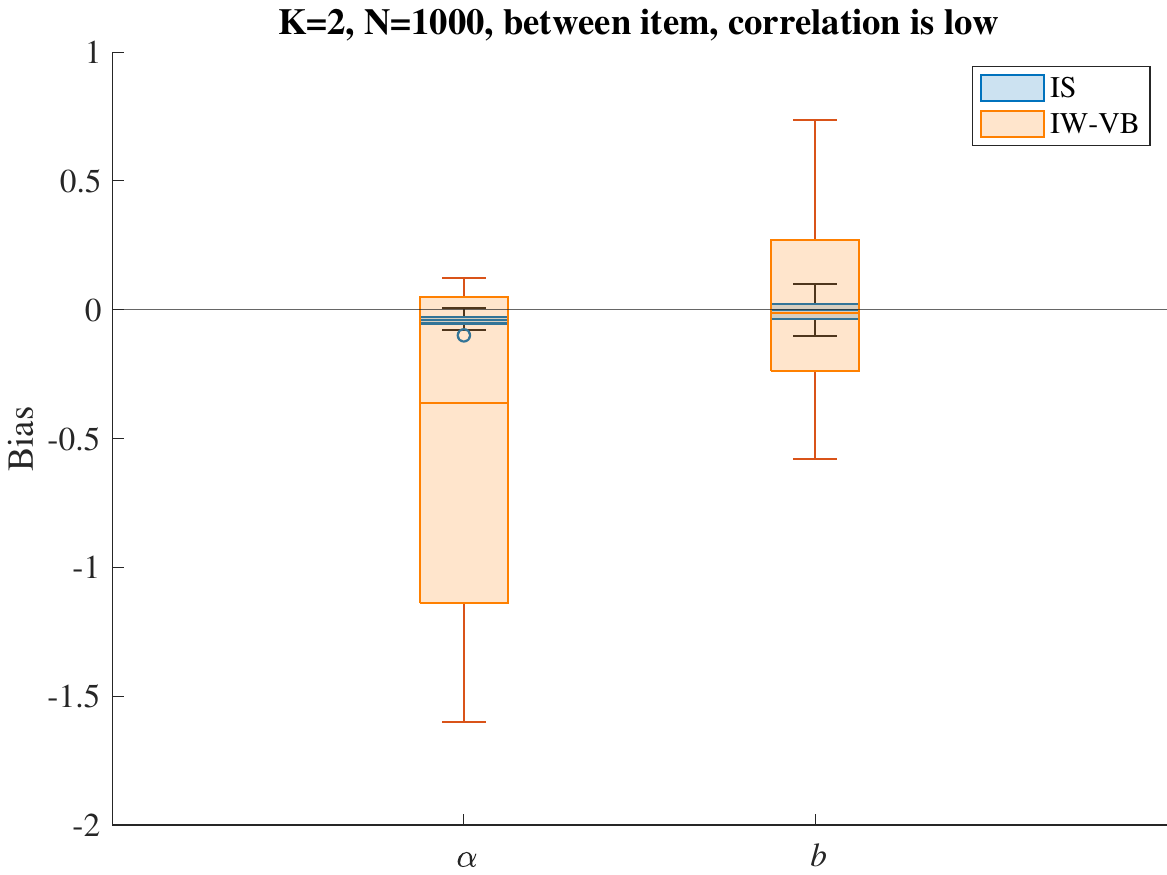}}
    \hspace{0.6in}
    \subfigure{
        \includegraphics[width=2.5in]{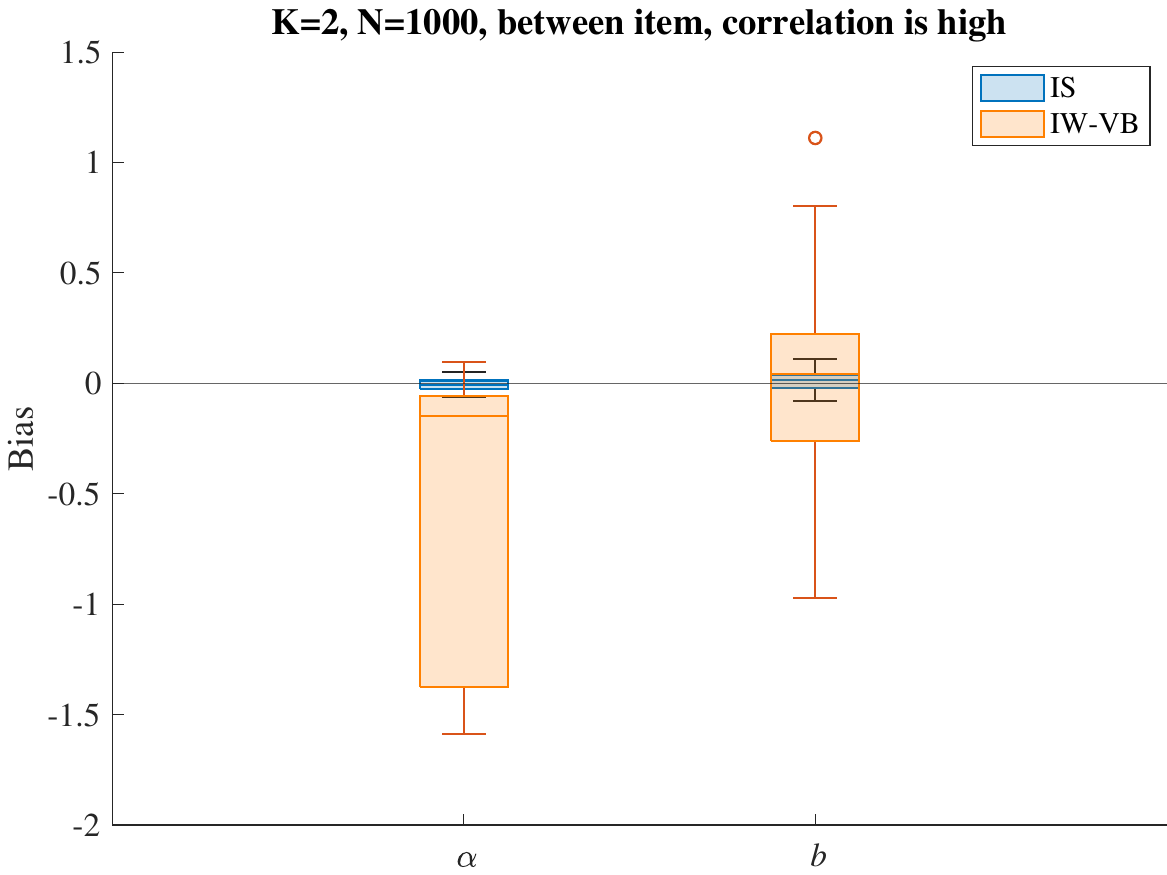}}
    \\
    \caption{Bias for $K=2$ between item under exploratory analysis   }
    \label{fig:bias-k2-between-explore compare vb}
\end{figure}

\newpage 
\begin{figure}[ht!]
    \centering
    \subfigure{
        \includegraphics[width=2.5in]{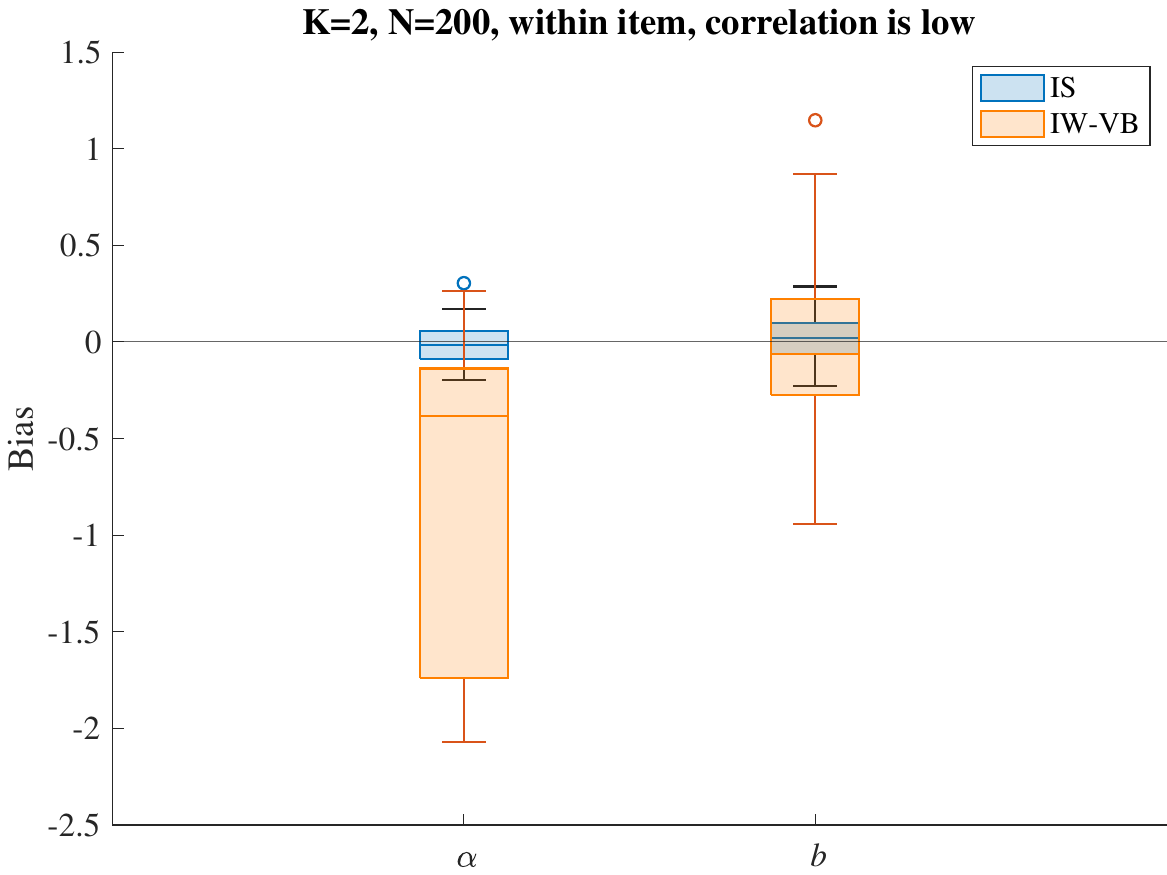}}
    \hspace{0.6in}
    \subfigure{
        \includegraphics[width=2.5in]{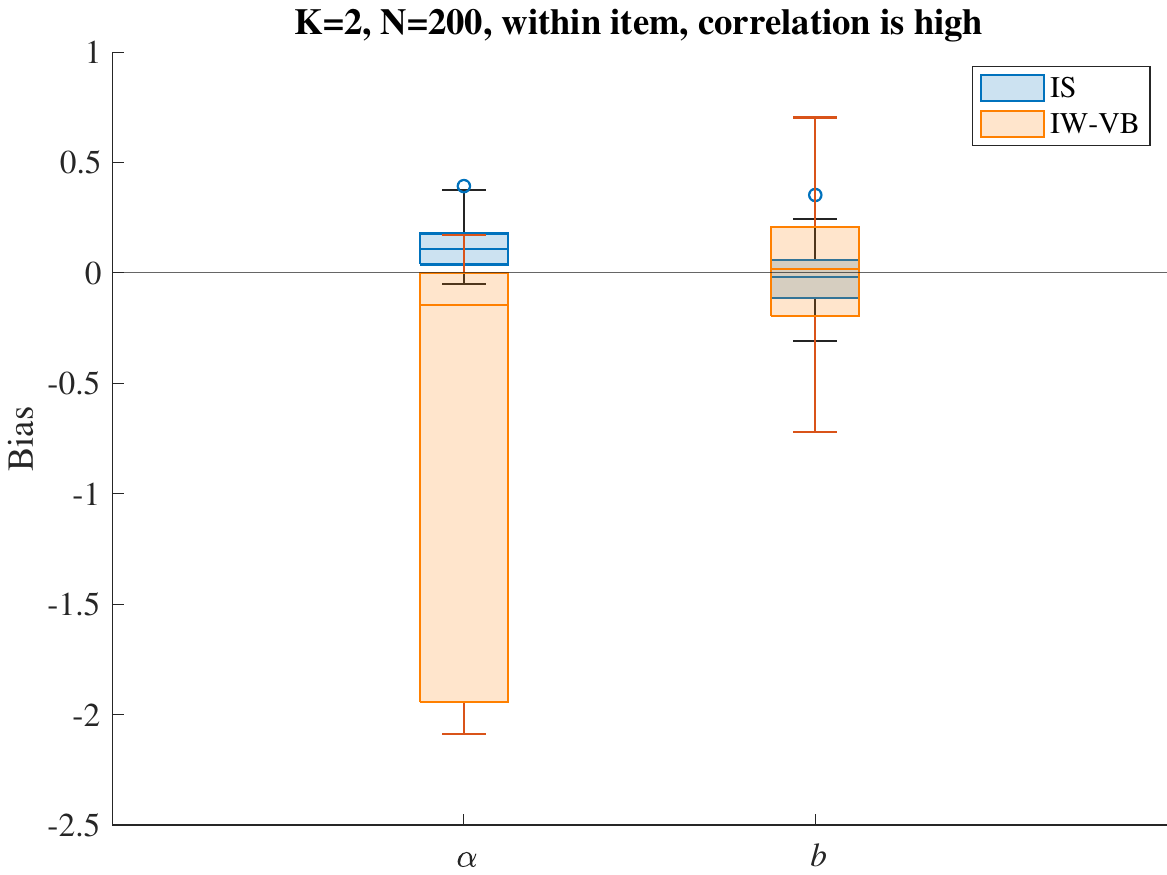}}
    \\
    \subfigure{
        \includegraphics[width=2.5in]{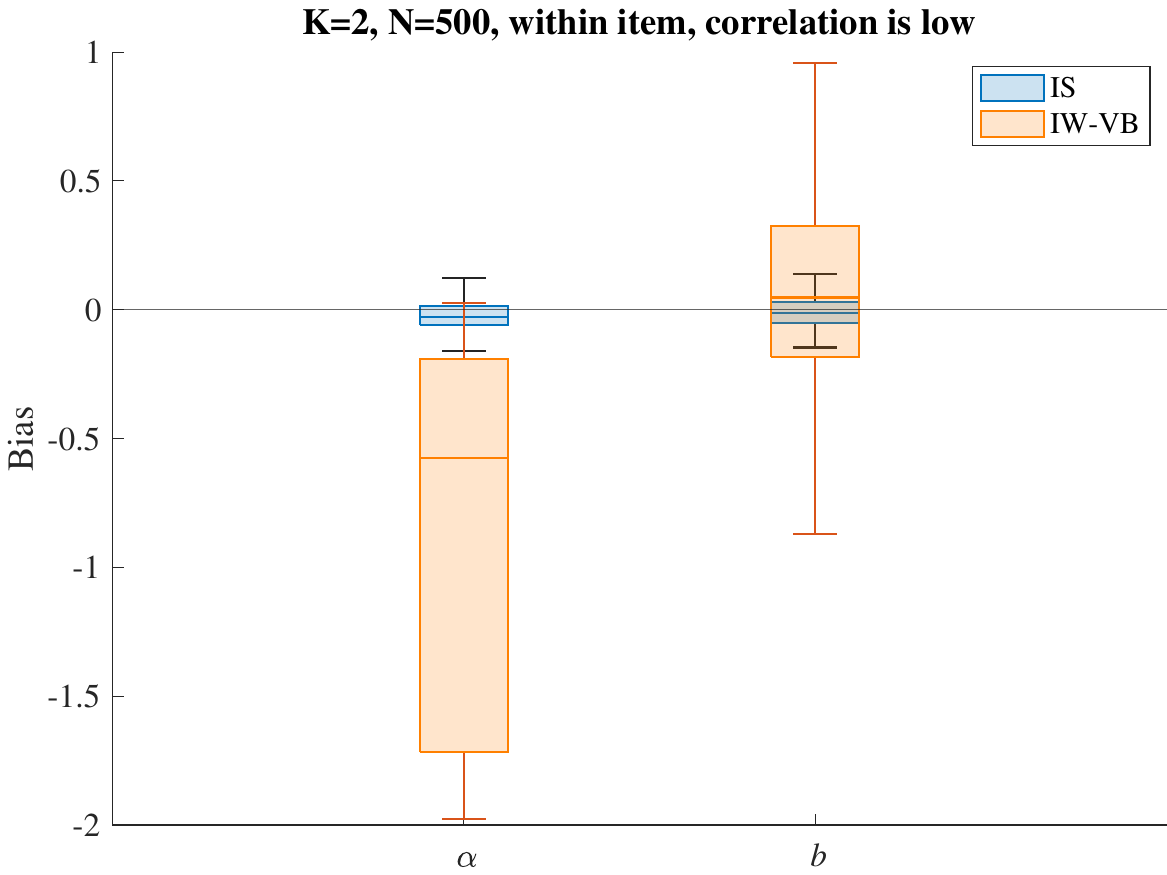}}
    \hspace{0.6in}
    \subfigure{
        \includegraphics[width=2.5in]{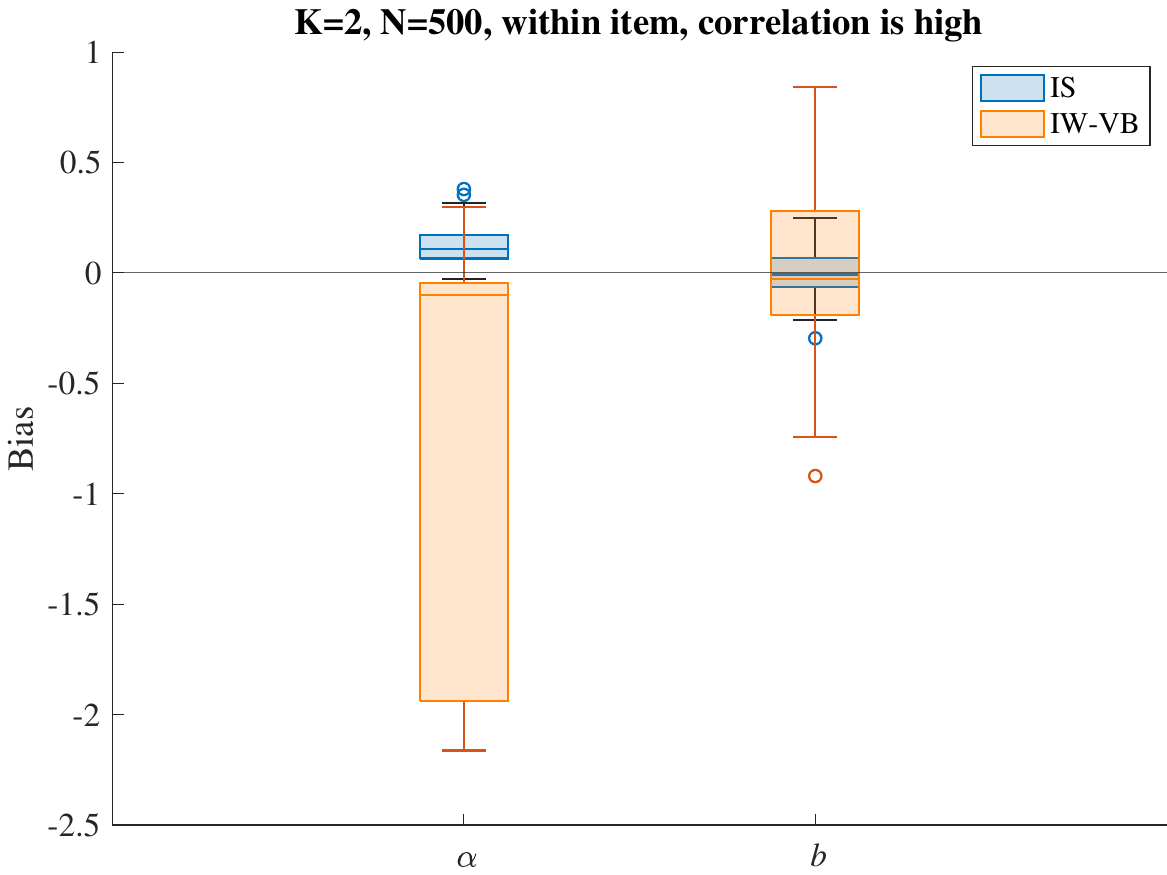}}
    \\
     \subfigure{
        \includegraphics[width=2.5in]{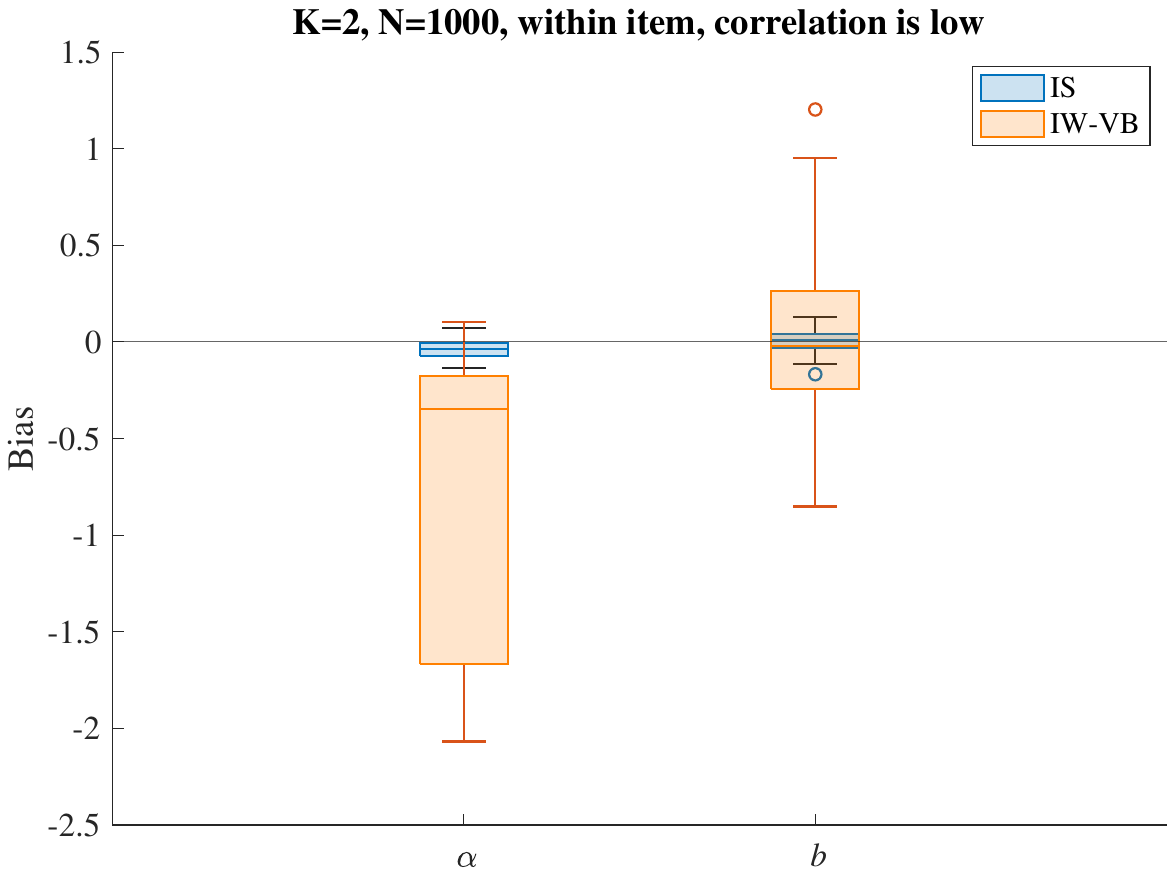}}
    \hspace{0.6in}
    \subfigure{
        \includegraphics[width=2.5in]{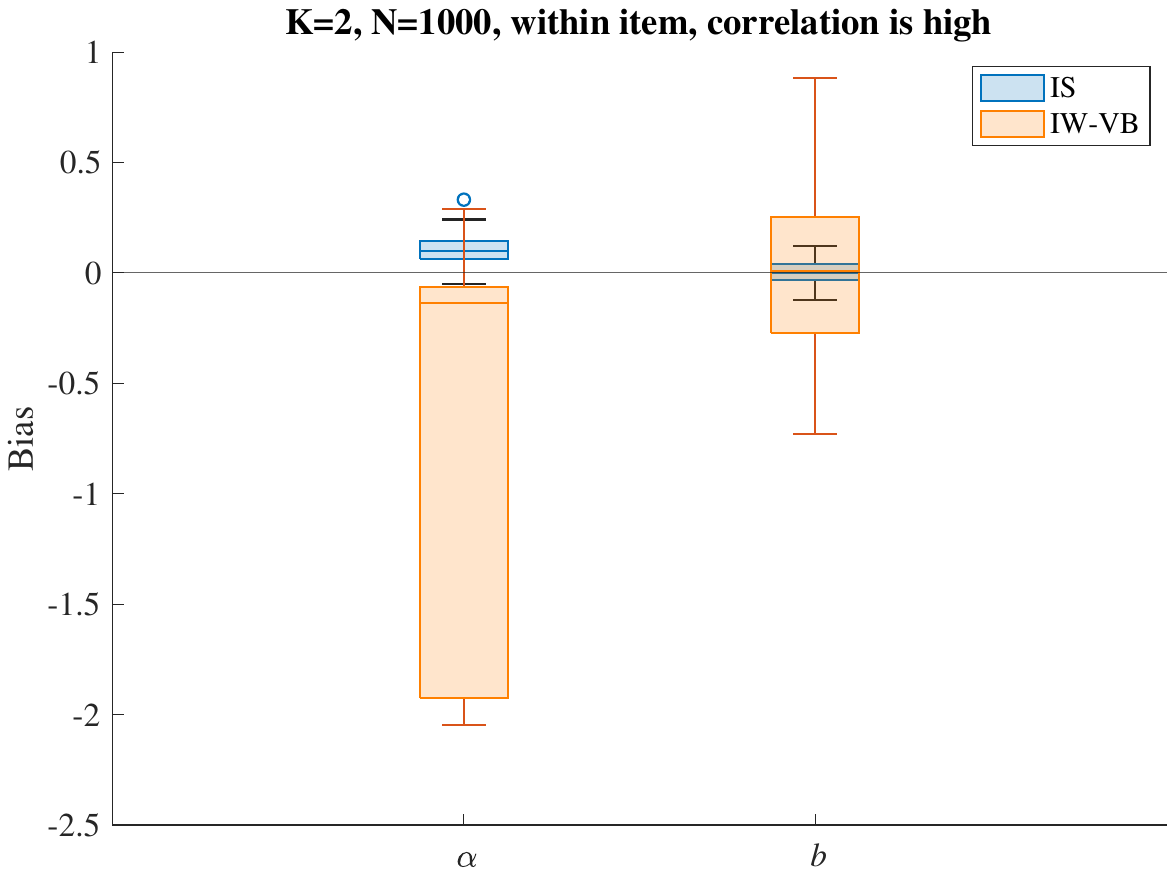}}
    \\
    \caption{Bias for $K=2$ within item under exploratory analysis   }
    \label{fig:bias-k2-within-explore compare vb}
\end{figure}

\newpage

\begin{figure}[ht!]
    \centering
    \subfigure{
        \includegraphics[width=2.5in]{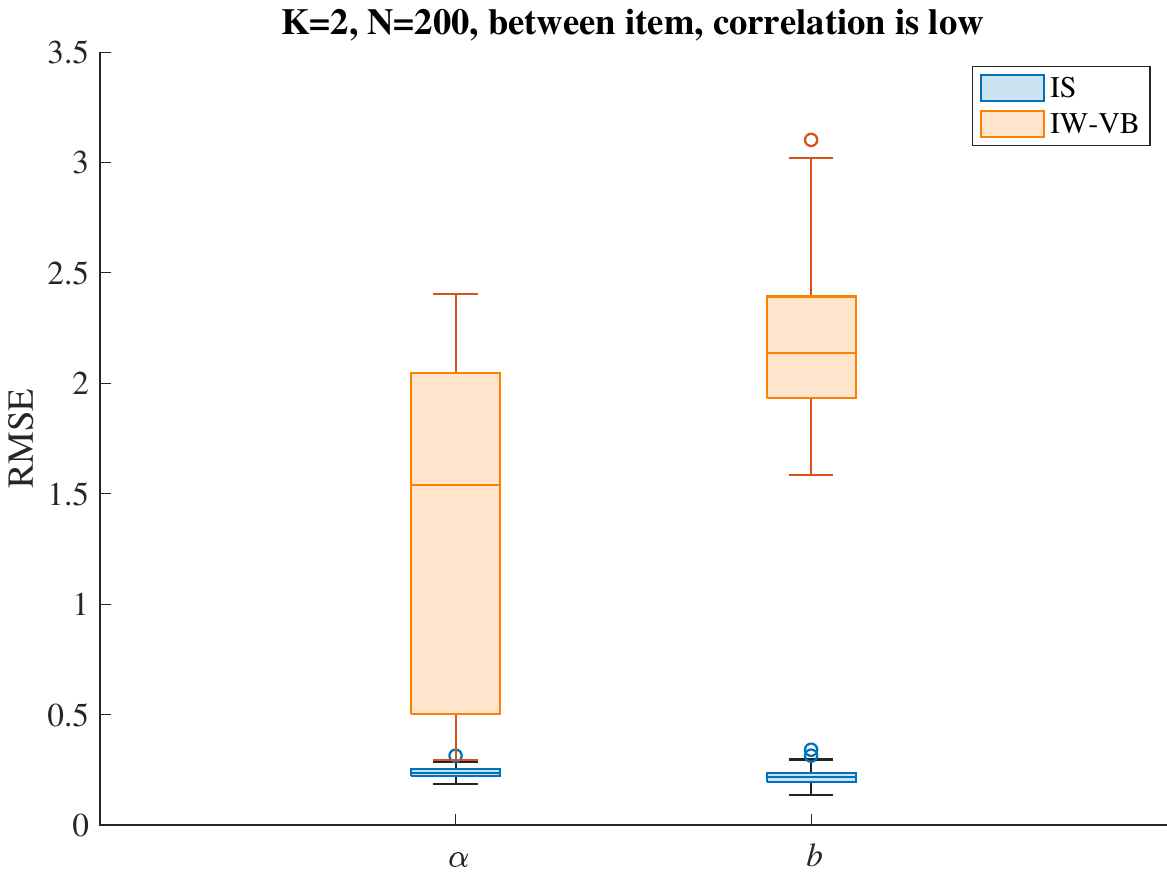}}
    \hspace{0.6in}
    \subfigure{
        \includegraphics[width=2.5in]{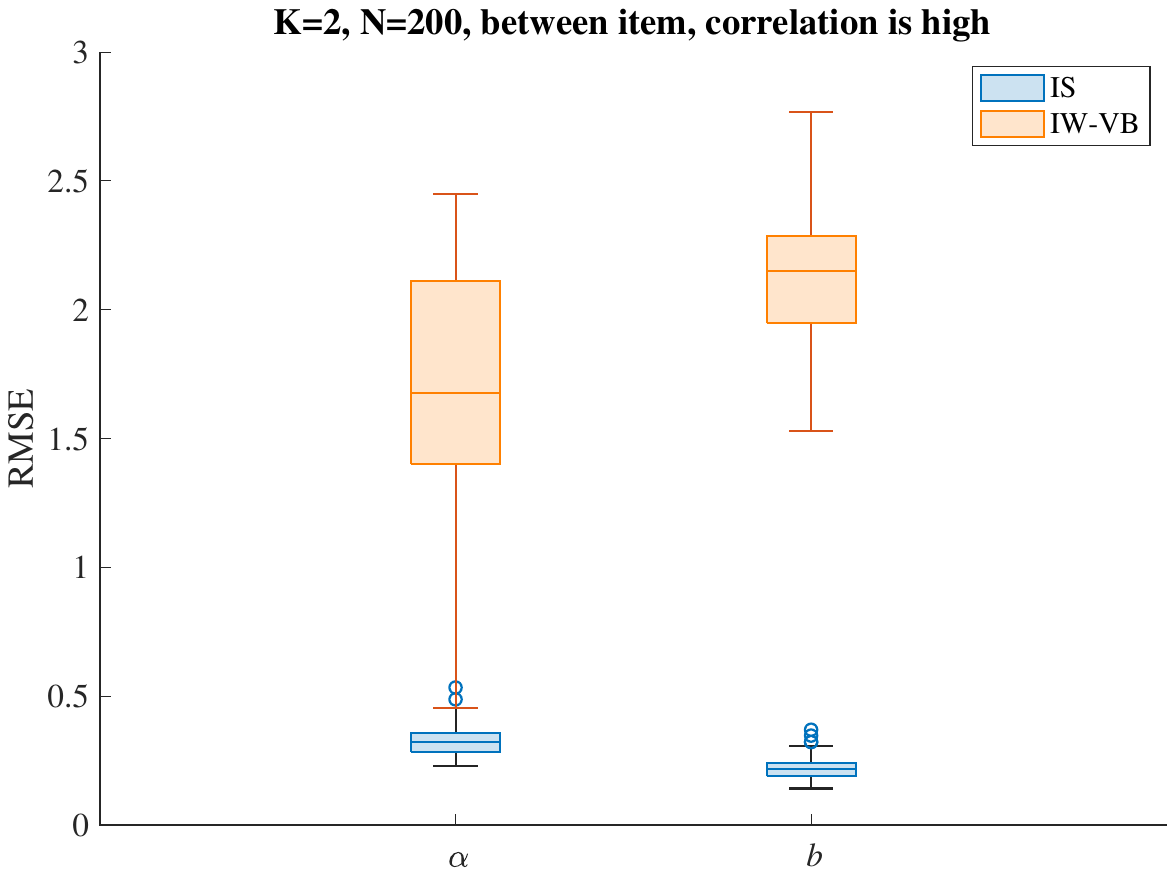}}
    \\
    \subfigure{
        \includegraphics[width=2.5in]{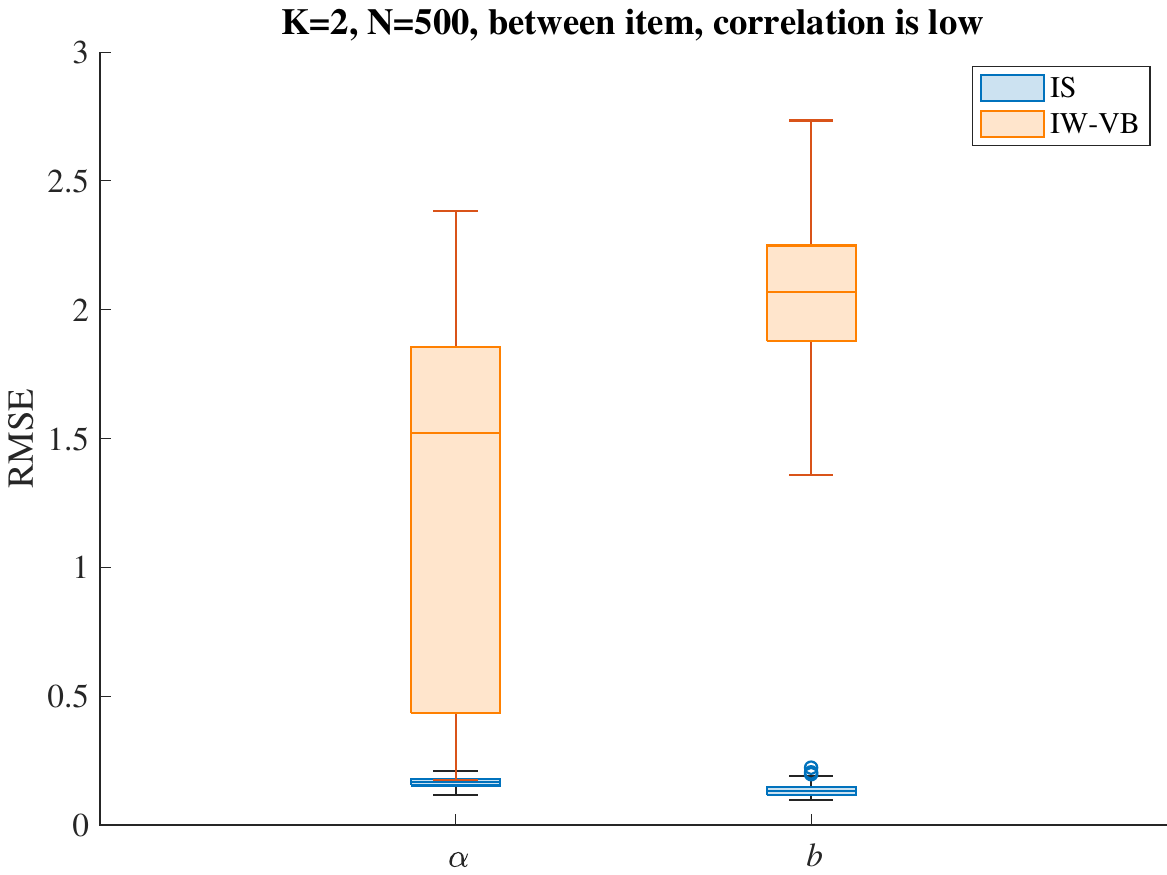}}
    \hspace{0.6in}
    \subfigure{
        \includegraphics[width=2.5in]{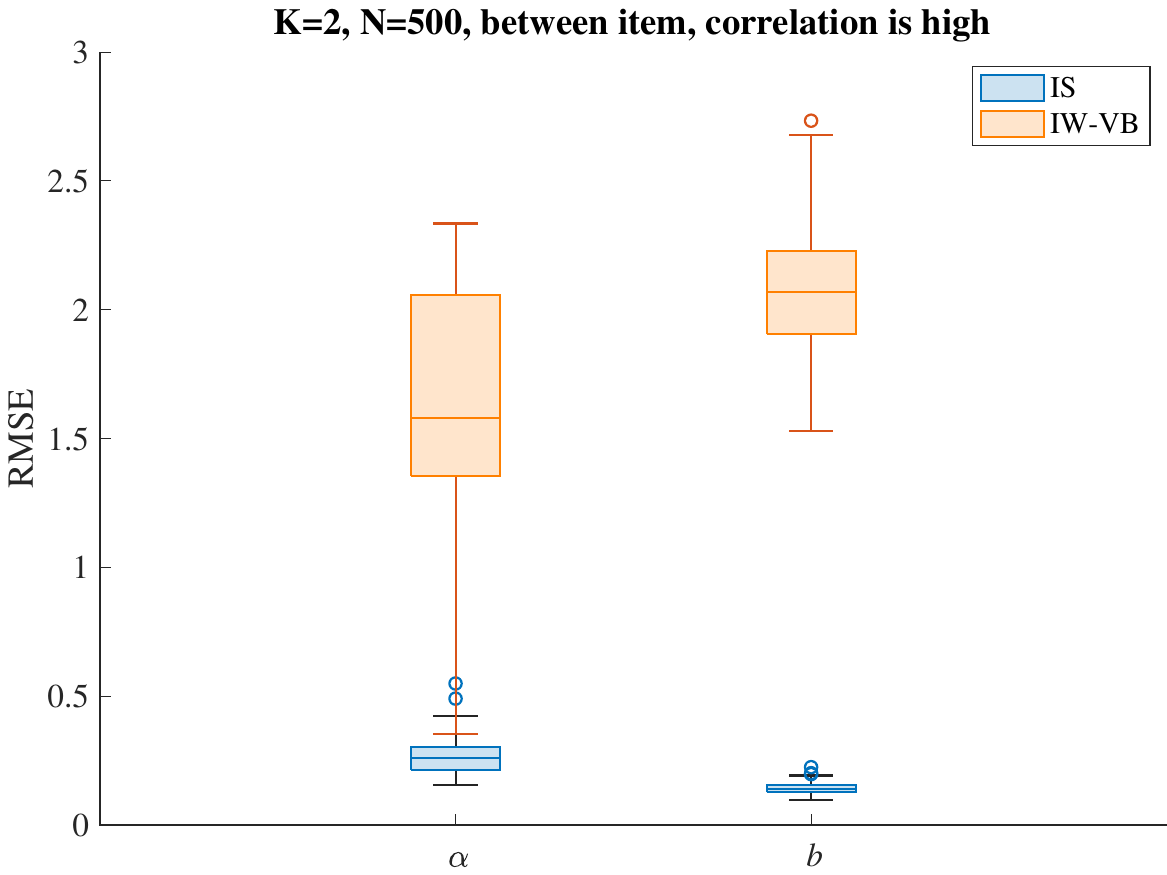}}
    \\
    \subfigure{
        \includegraphics[width=2.5in]{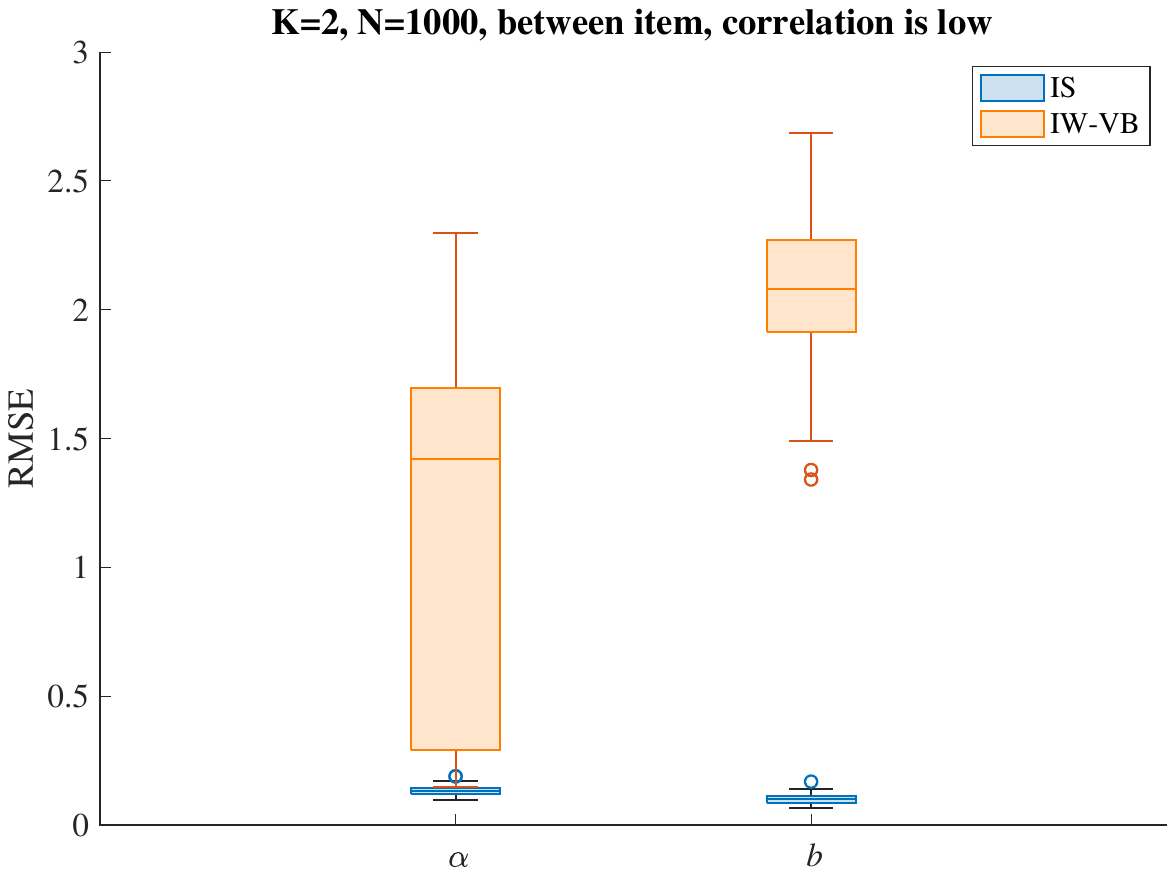}}
    \hspace{0.6in}
    \subfigure{
        \includegraphics[width=2.5in]{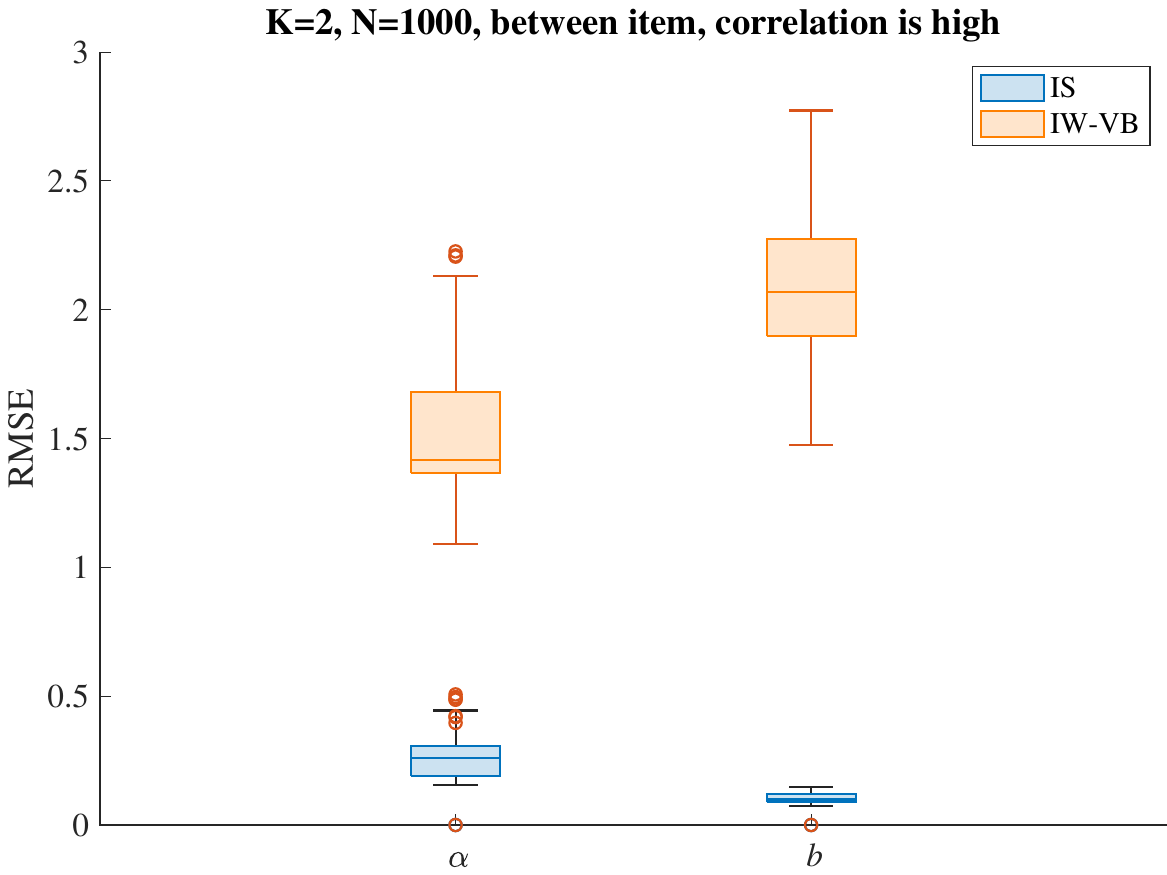}}
    \\
    \caption{RMSE for   $K=2$ between item under exploratory analysis
    }
    \label{fig:rmse-k2-between-explore compare vb}
\end{figure}

\newpage

\begin{figure}[ht!]
    \centering
    \subfigure{
        \includegraphics[width=2.5in]{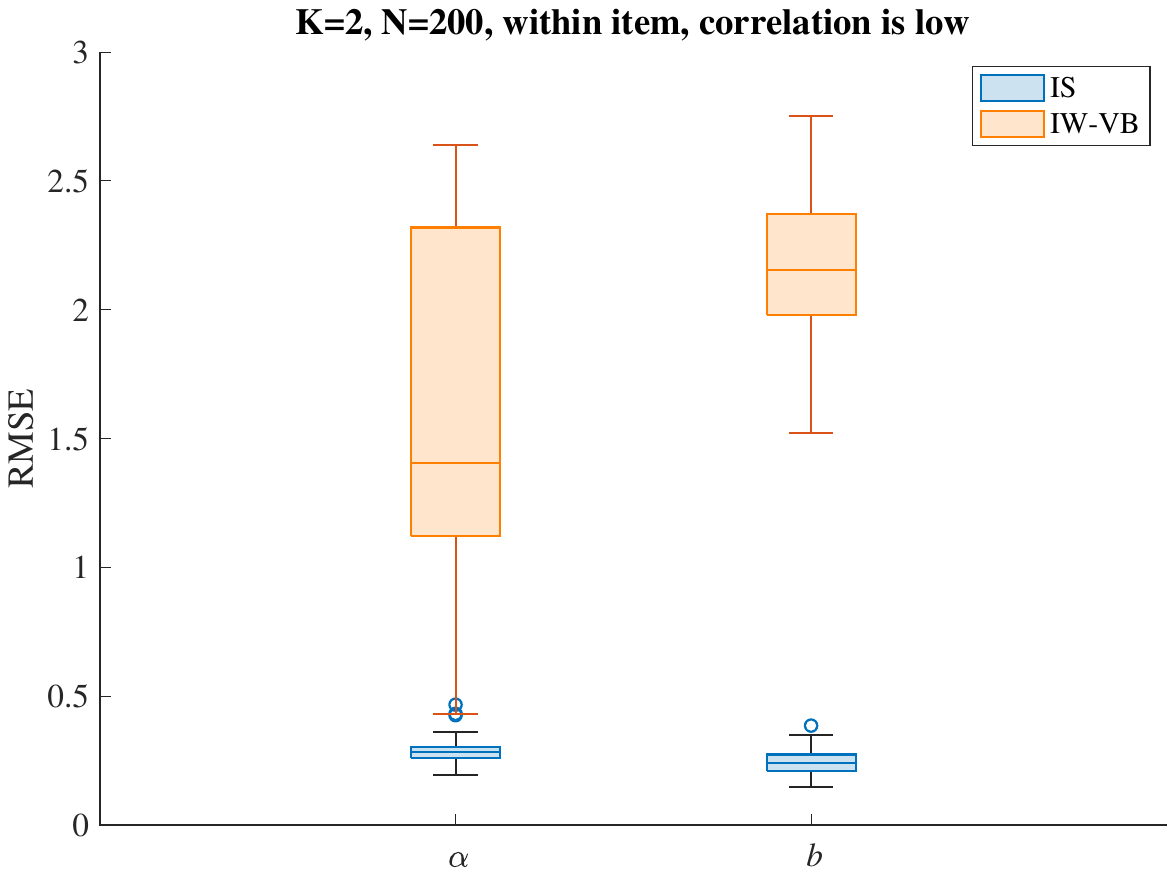}}
    \hspace{0.6in}
    \subfigure{
        \includegraphics[width=2.5in]{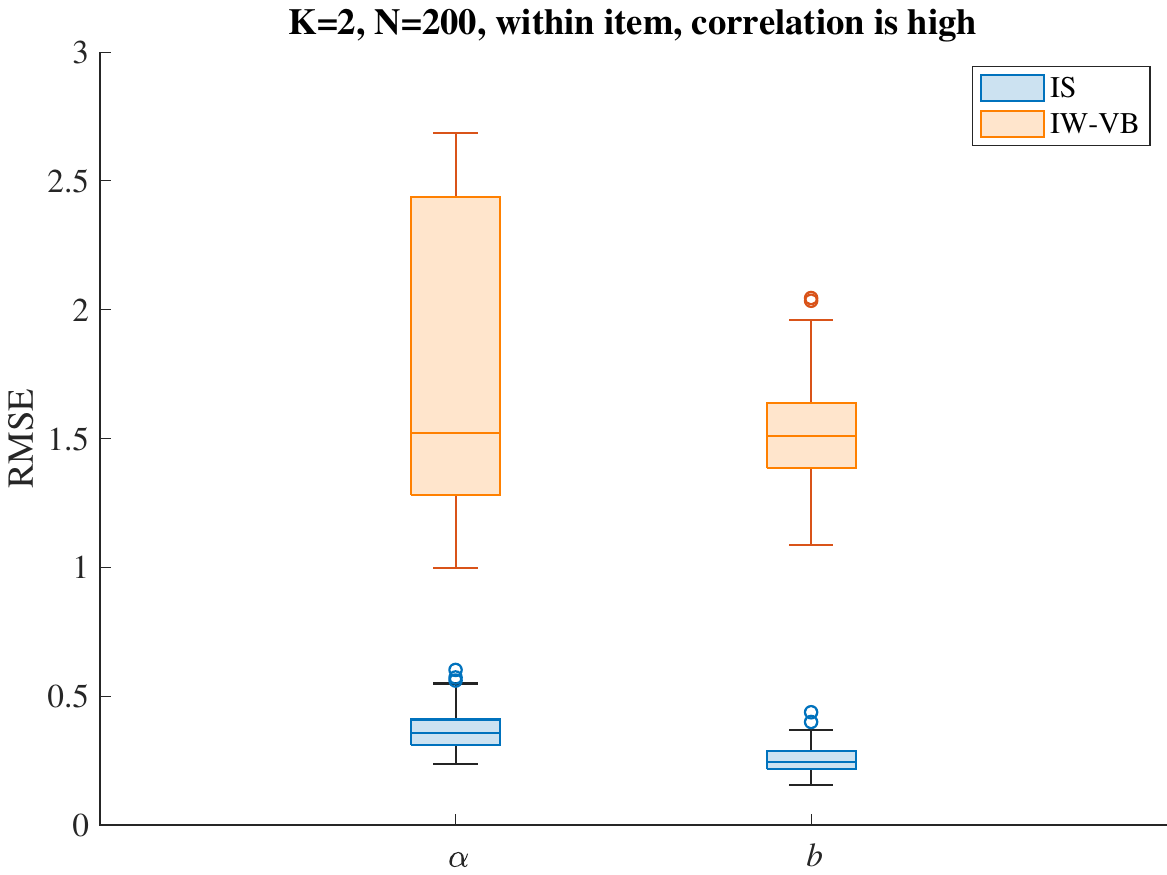}}
    \\
    \subfigure{
        \includegraphics[width=2.5in]{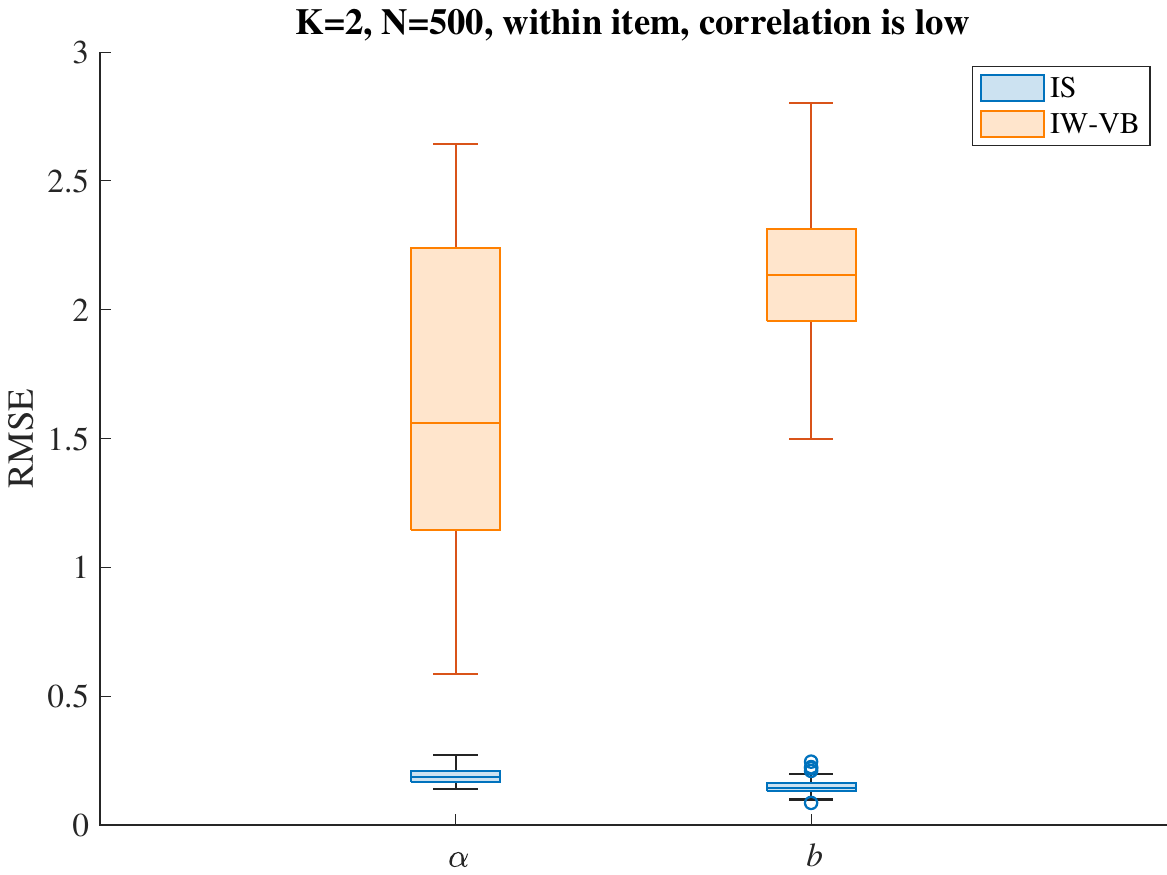}}
    \hspace{0.6in}
    \subfigure{
        \includegraphics[width=2.5in]{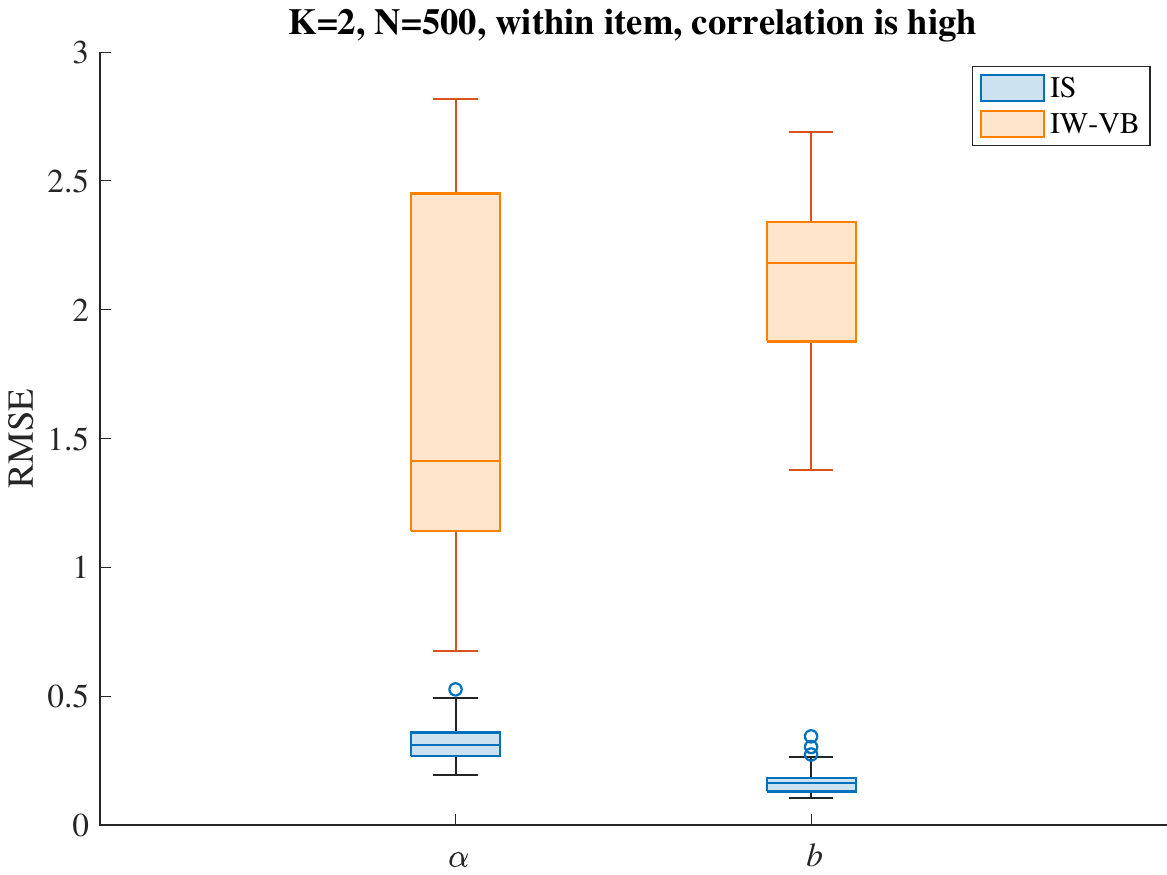}}
    \\
    \subfigure{
        \includegraphics[width=2.5in]{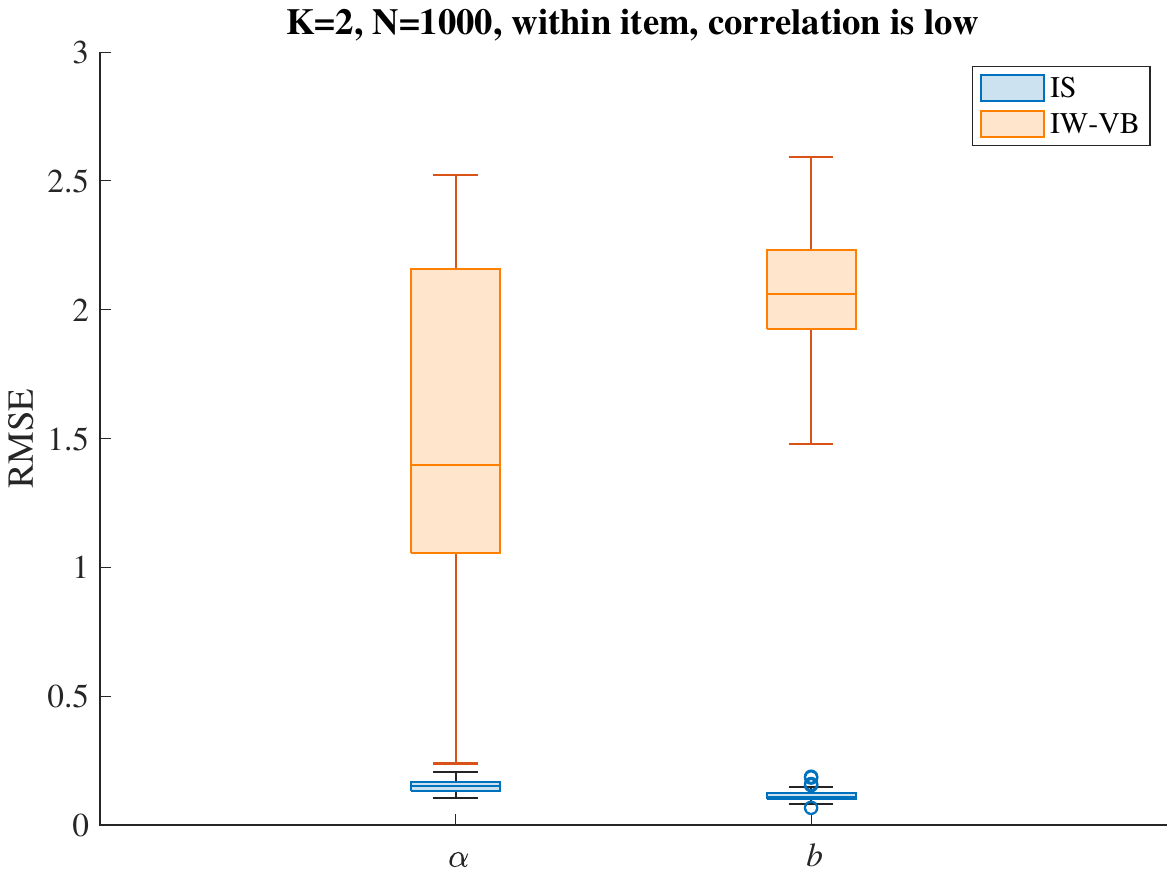}}
    \hspace{0.6in}
    \subfigure{
        \includegraphics[width=2.5in]{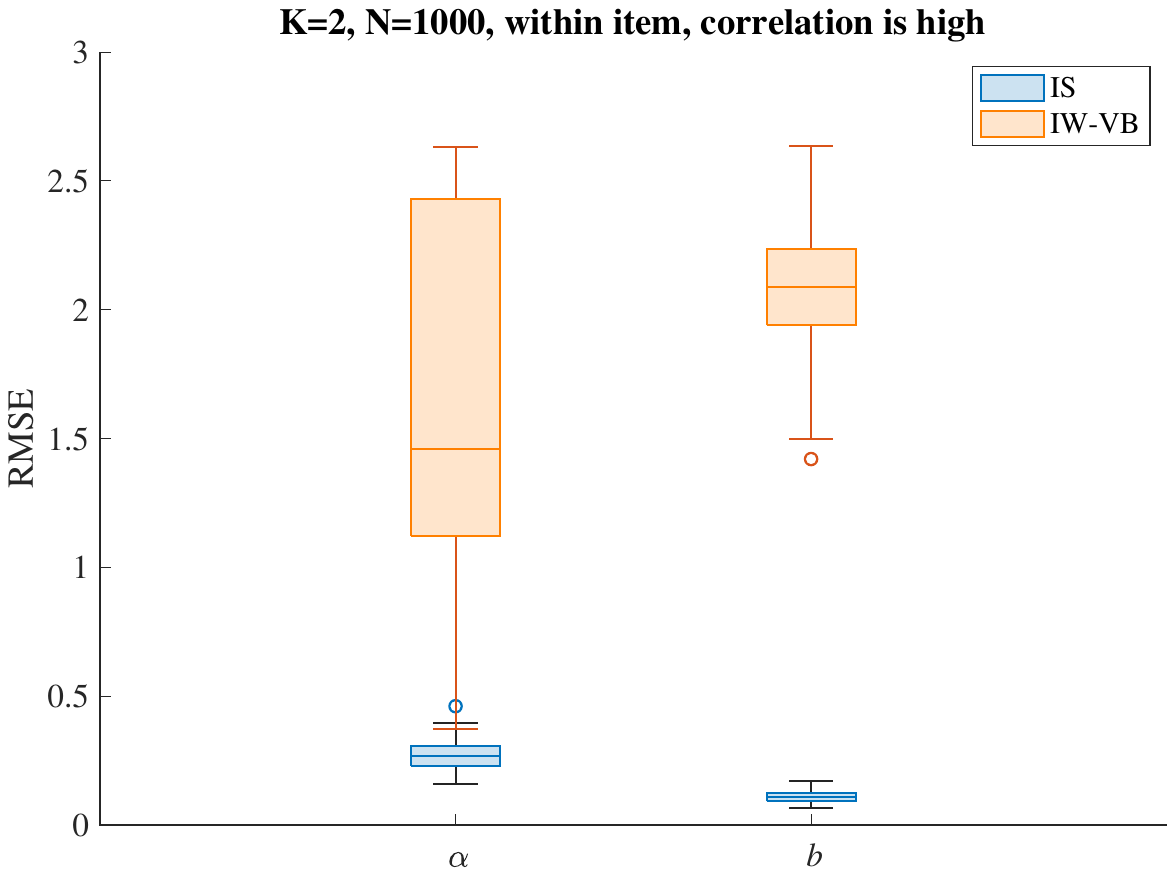}}
    \\
    \caption{RMSE for $K=2$ within item under exploratory analysis
    }
    \label{fig:rmse-k2-within-explore compare vb}
\end{figure}


\newpage 

\begin{figure}[ht!]
    \centering
    \subfigure{
        \includegraphics[width=2.5in]{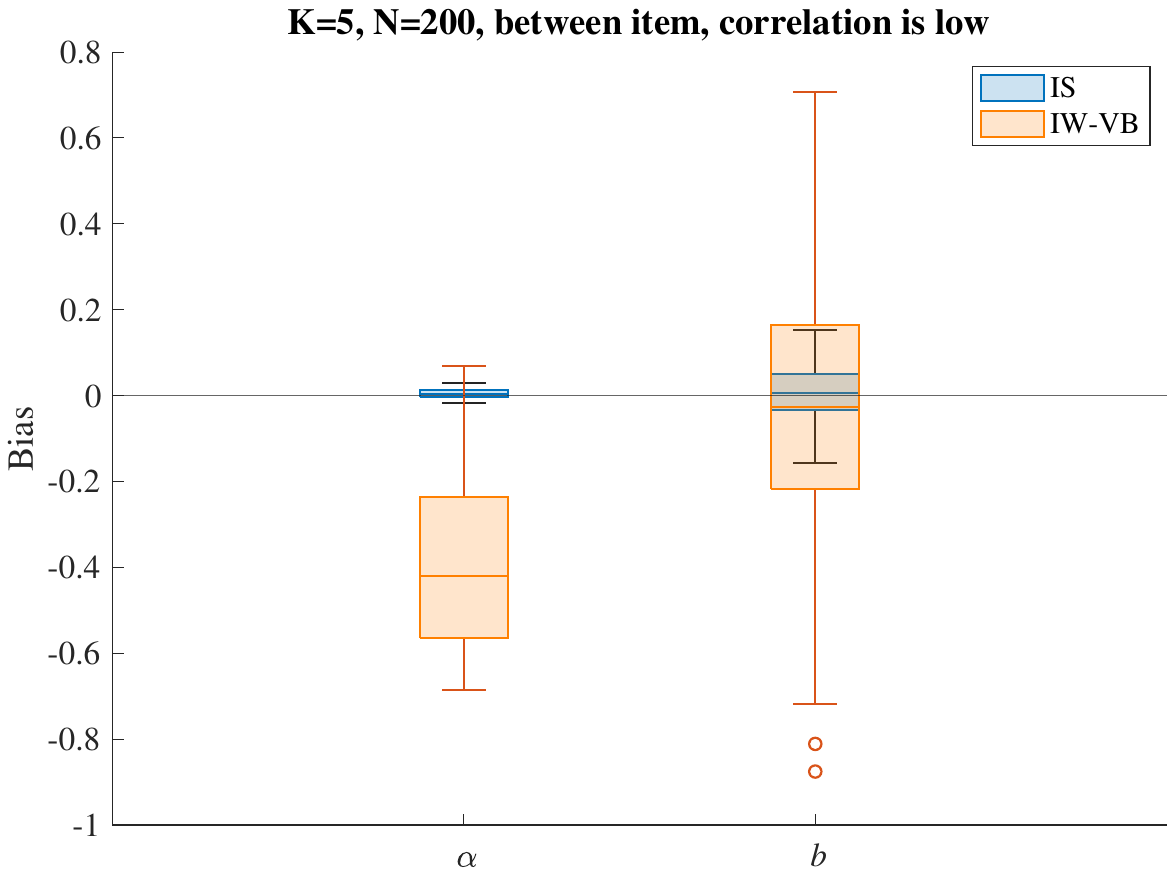}}
    \hspace{0.6in}
    \subfigure{
        \includegraphics[width=2.5in]{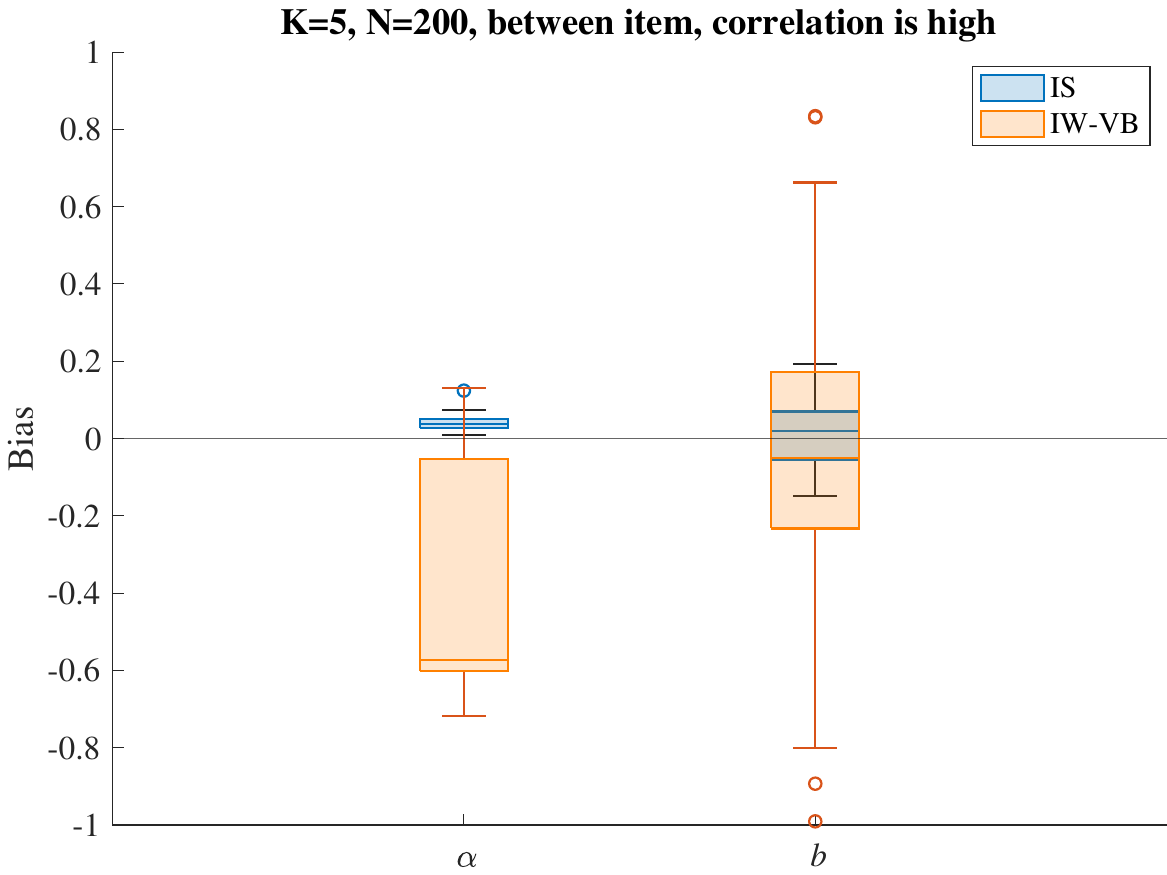}}
    \\
    \subfigure{
        \includegraphics[width=2.5in]{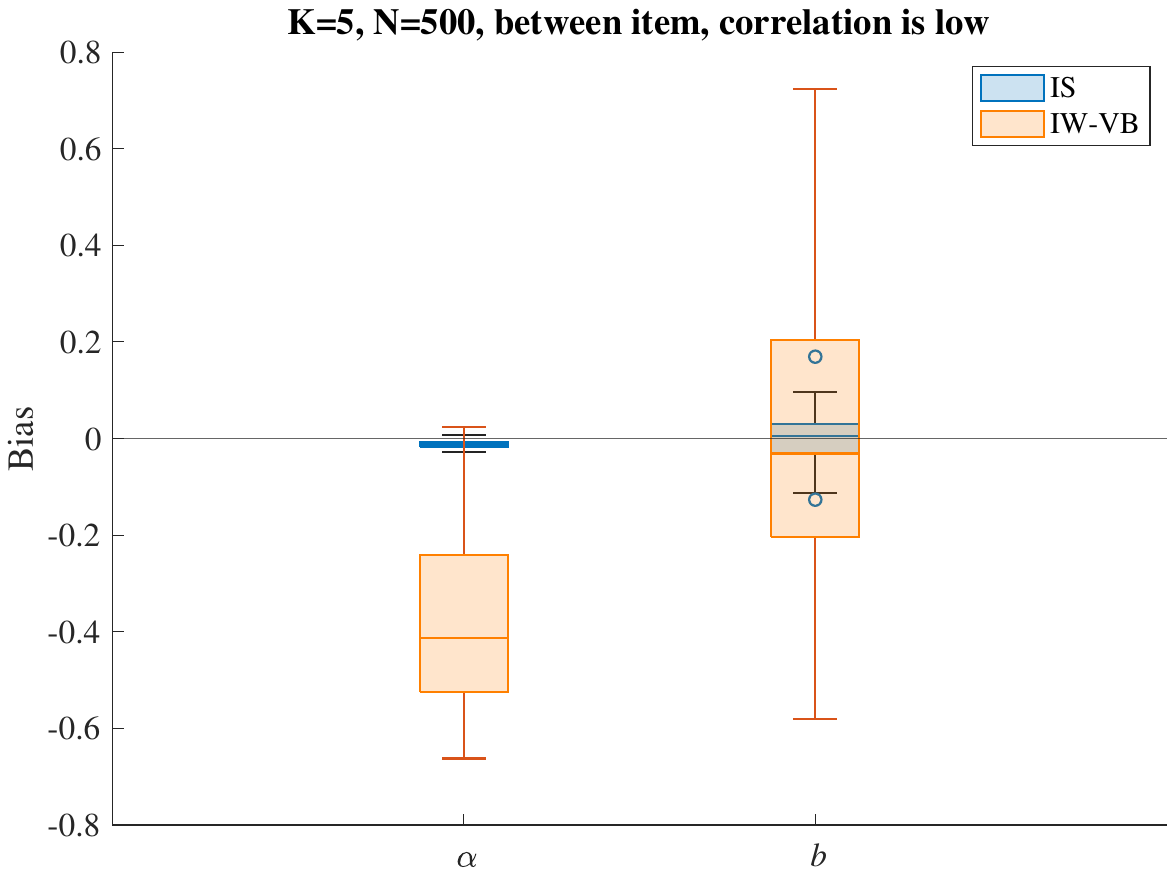}}
    \hspace{0.6in}
    \subfigure{
        \includegraphics[width=2.5in]{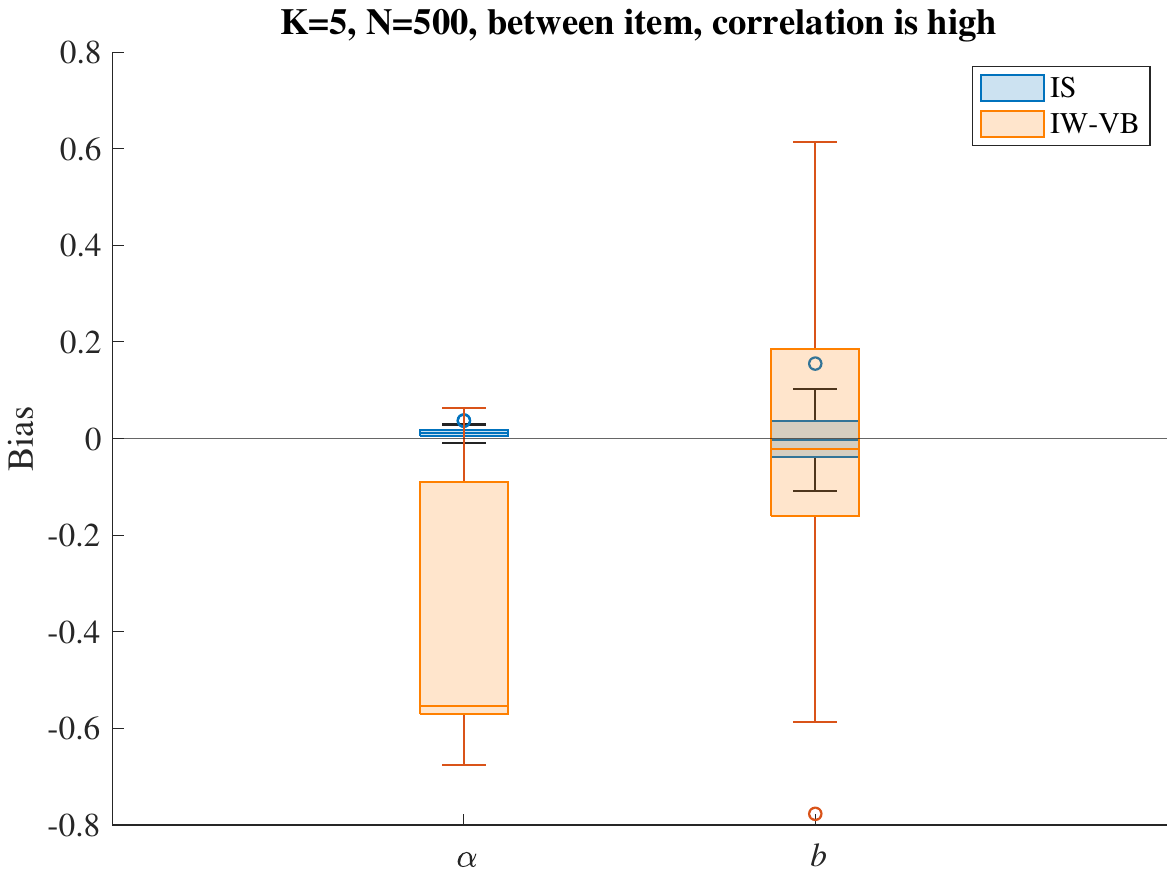}}
    \\
    \subfigure{
        \includegraphics[width=2.5in]{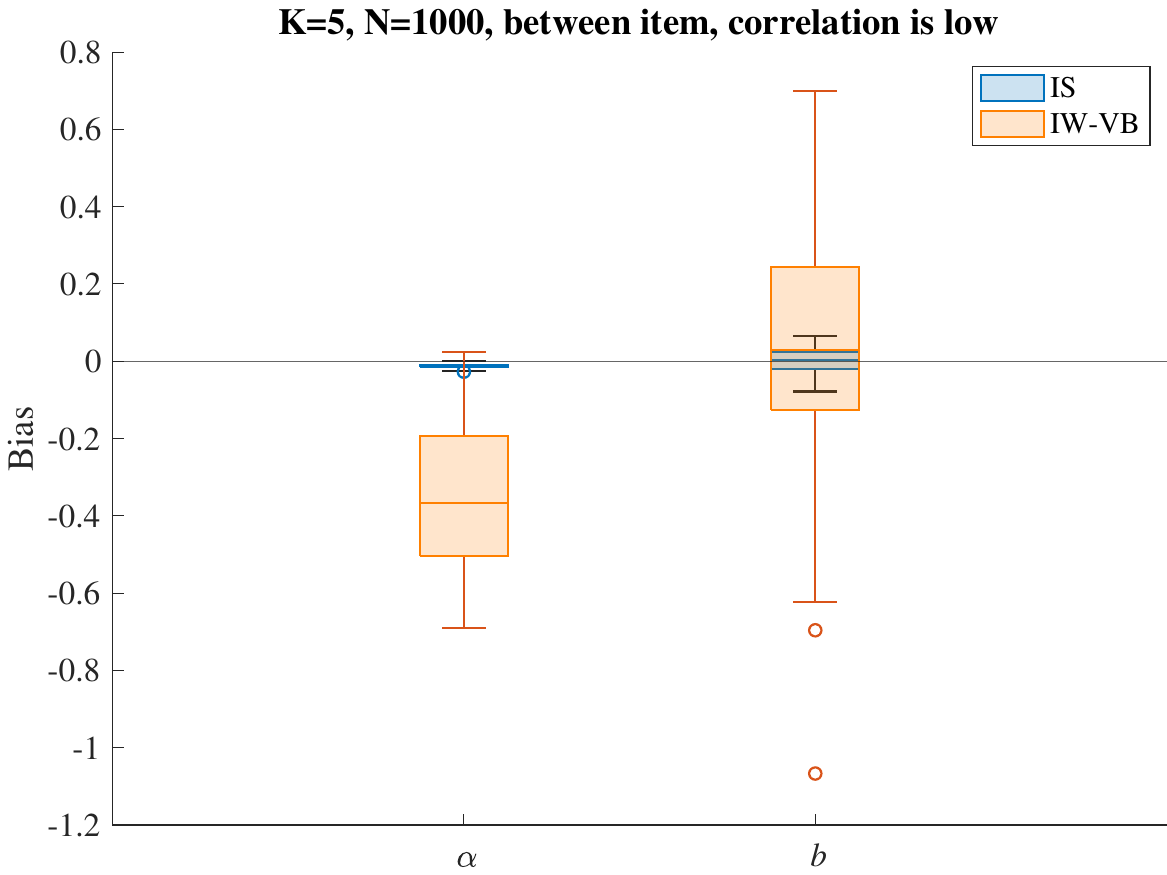}}
    \hspace{0.6in}
    \subfigure{
        \includegraphics[width=2.5in]{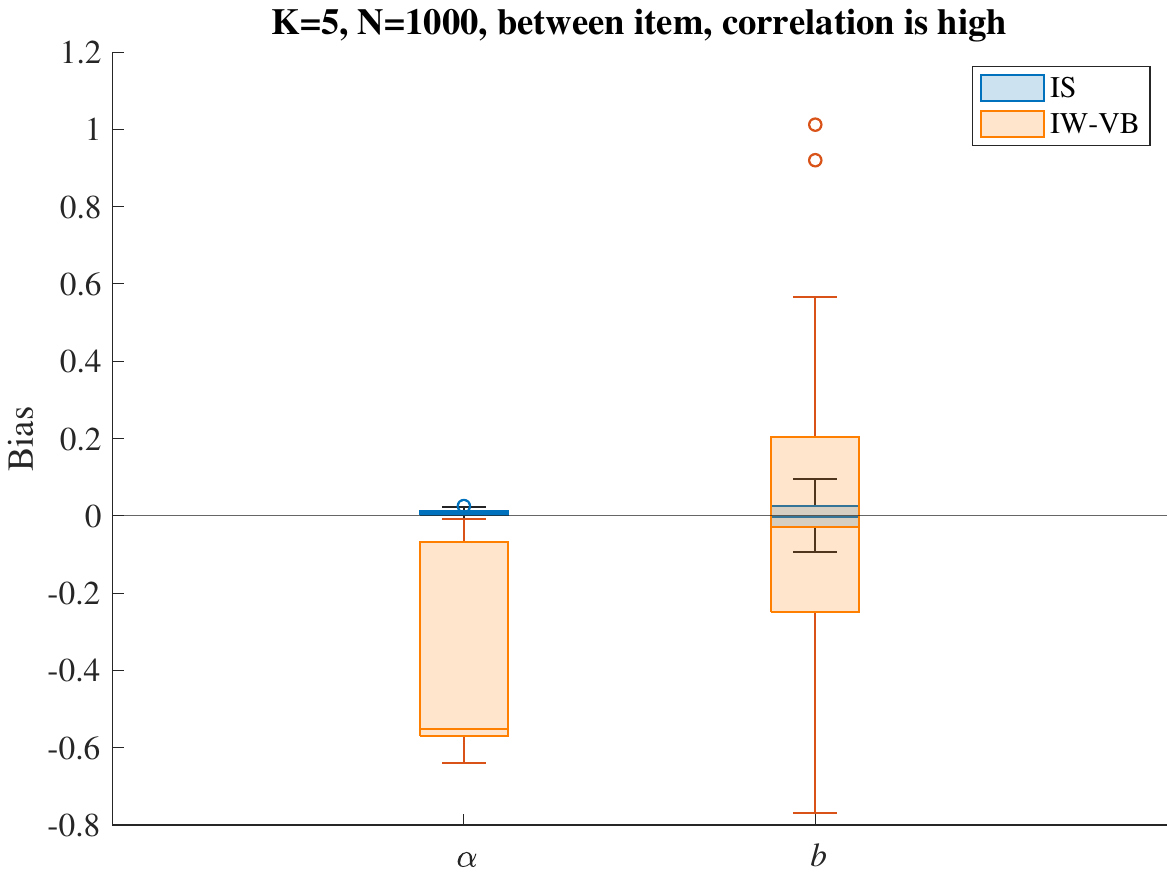}}
    \\
    \caption{Bias for   $K=5$ between item under exploratory analysis}
    \label{fig:bias-k5-between-explore compare vb}
\end{figure}

\newpage 

\begin{figure}[ht!]
    \centering
    \subfigure{
        \includegraphics[width=2.5in]{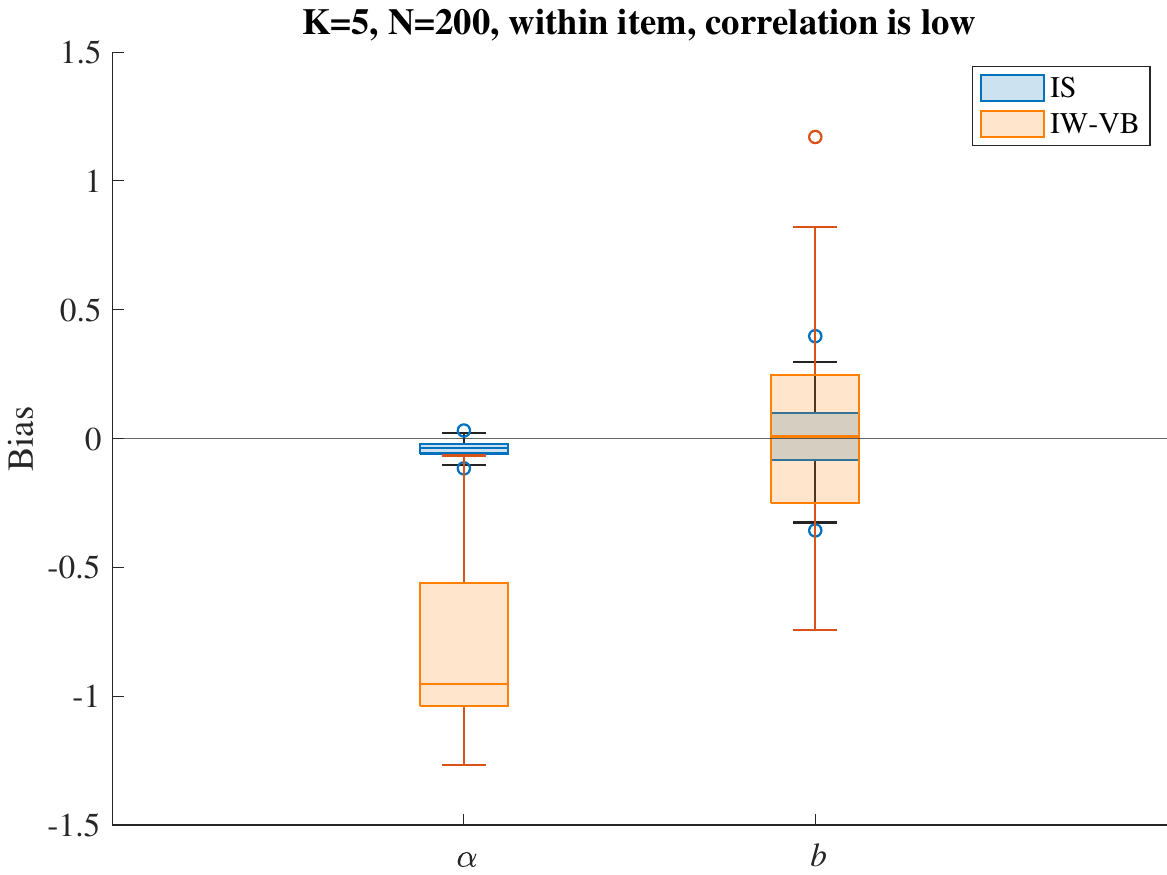}}
    \hspace{0.6in}
    \subfigure{
        \includegraphics[width=2.5in]{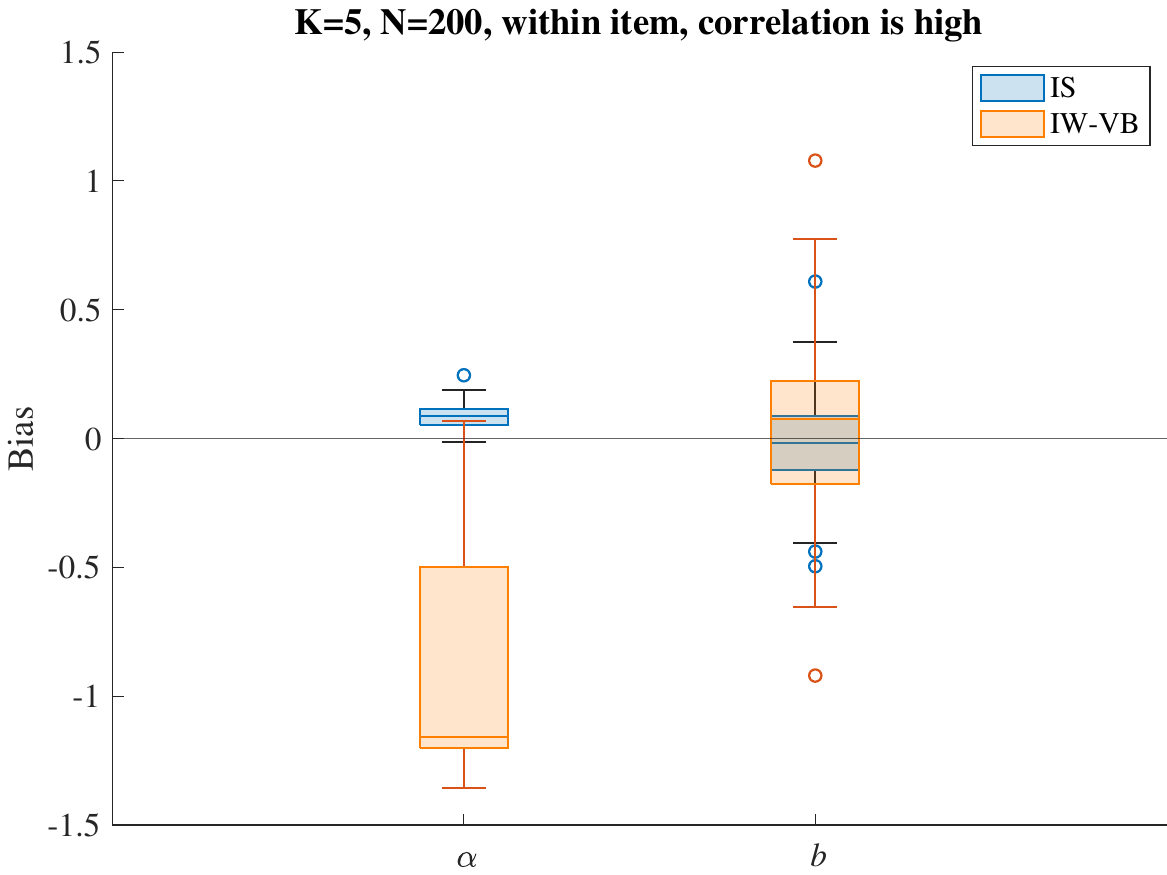}}
    \\
    \subfigure{
        \includegraphics[width=2.5in]{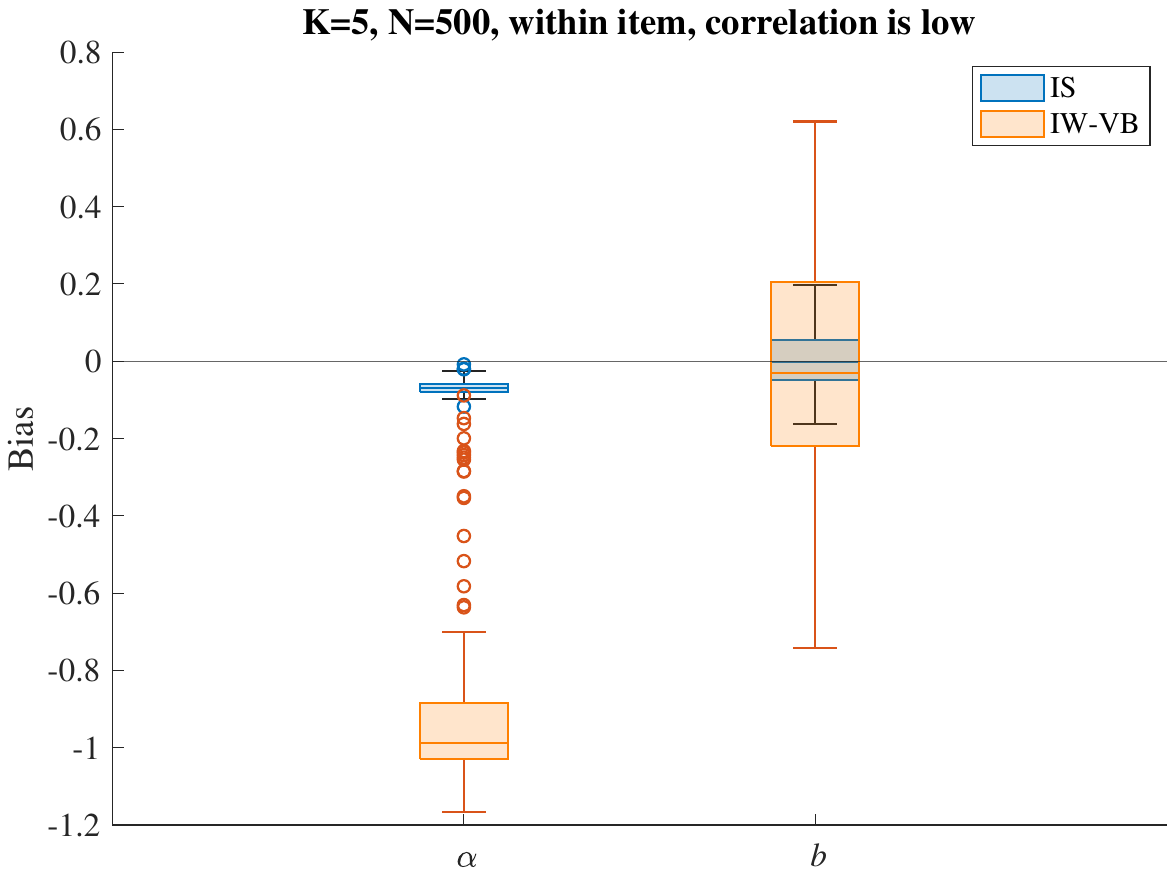}}
    \hspace{0.6in}
    \subfigure{
        \includegraphics[width=2.5in]{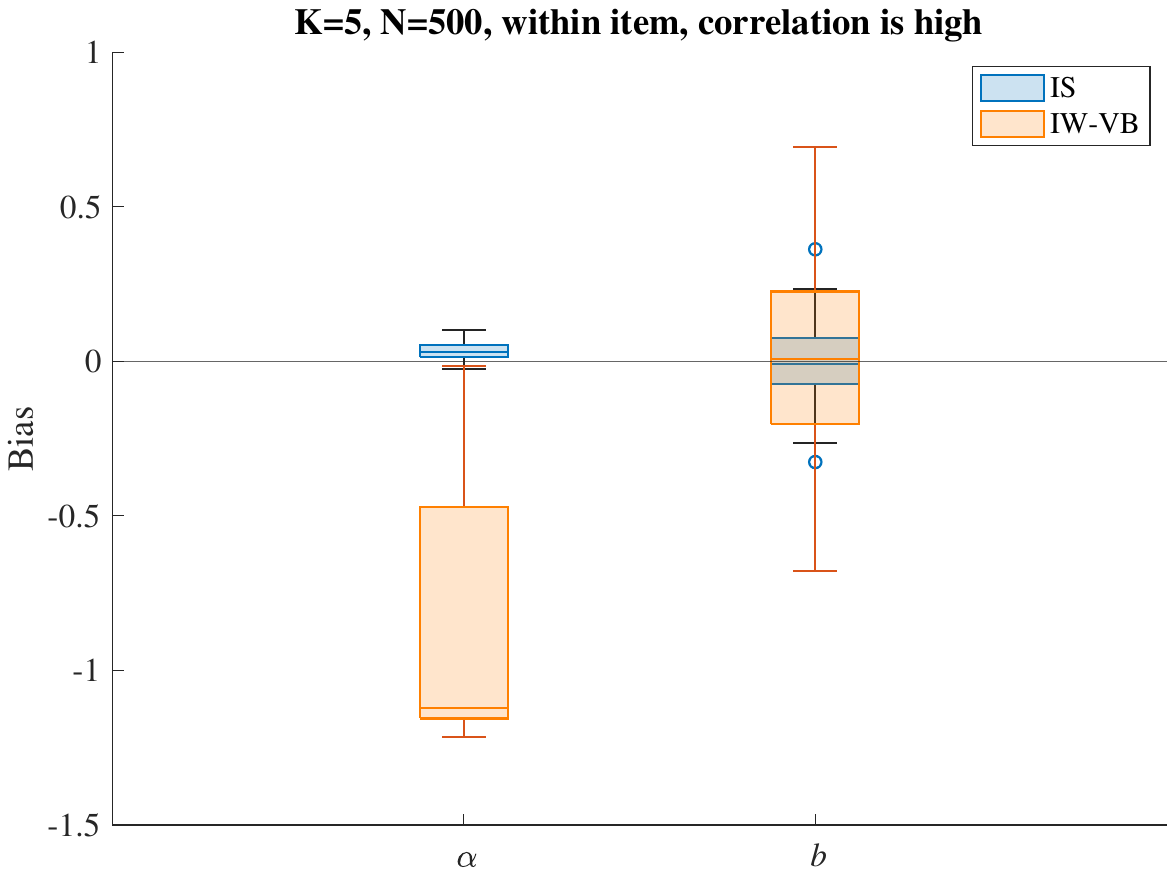}}
    \\
     \subfigure{
        \includegraphics[width=2.5in]{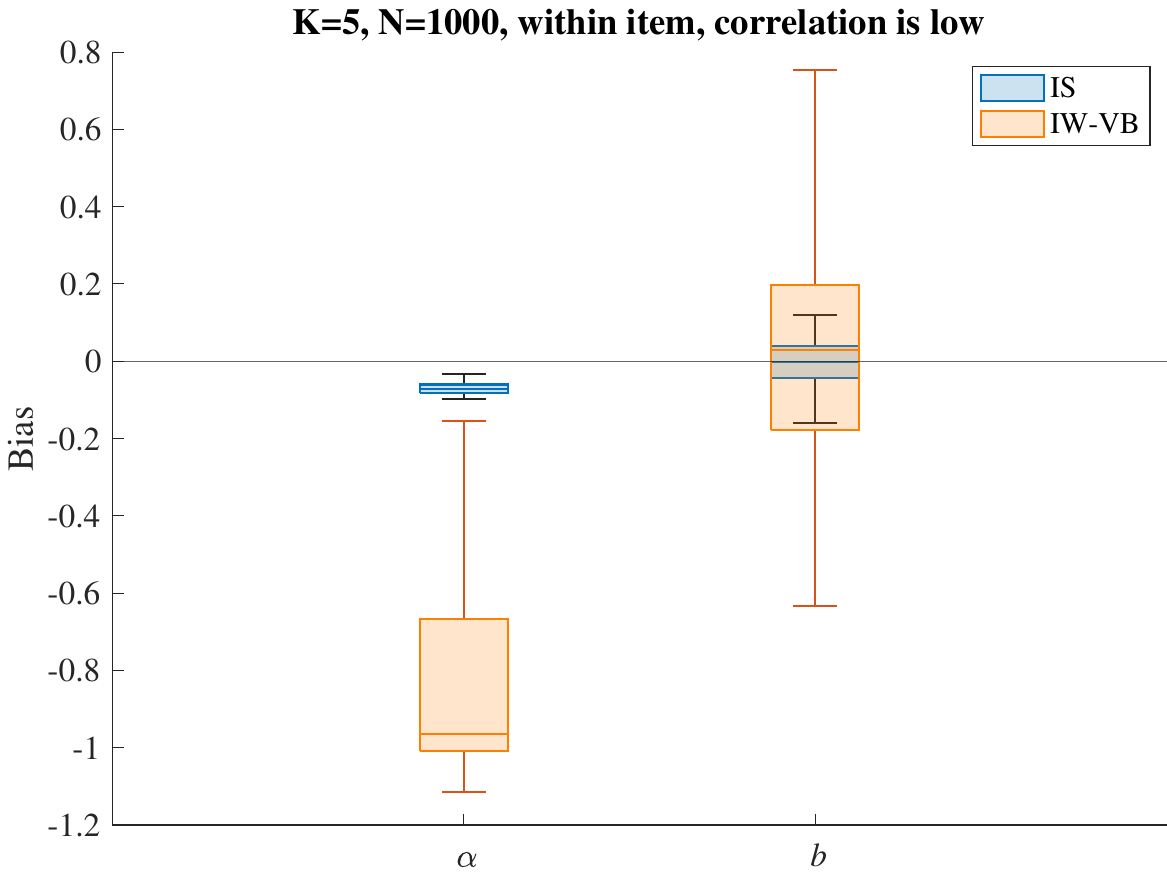}}
    \hspace{0.6in}
    \subfigure{
        \includegraphics[width=2.5in]{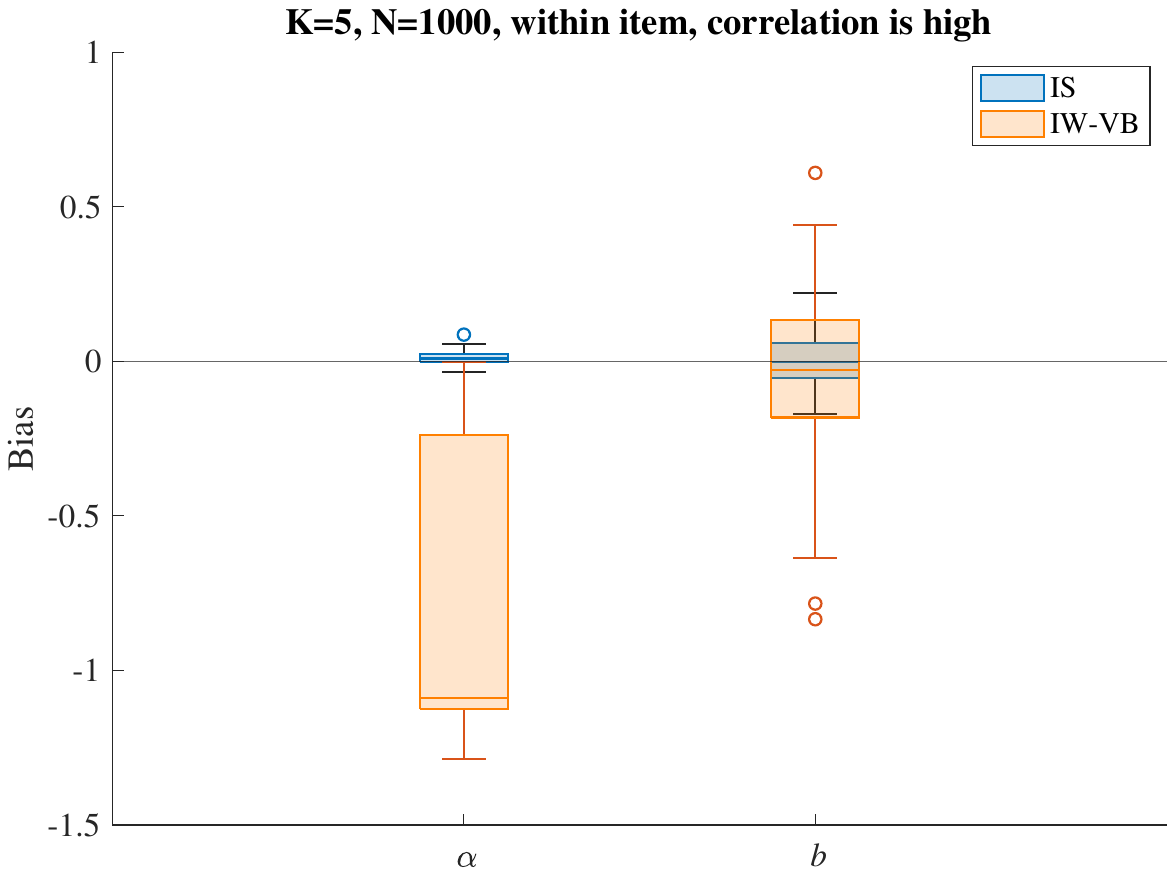}}
    \\
    \caption{Bias for   $K=5$ within item under exploratory analysis}
    \label{fig:bias-k5-within-explore compare vb}
\end{figure}

\newpage 

\begin{figure}[ht!]
    \centering
    \subfigure{
        \includegraphics[width=2.5in]{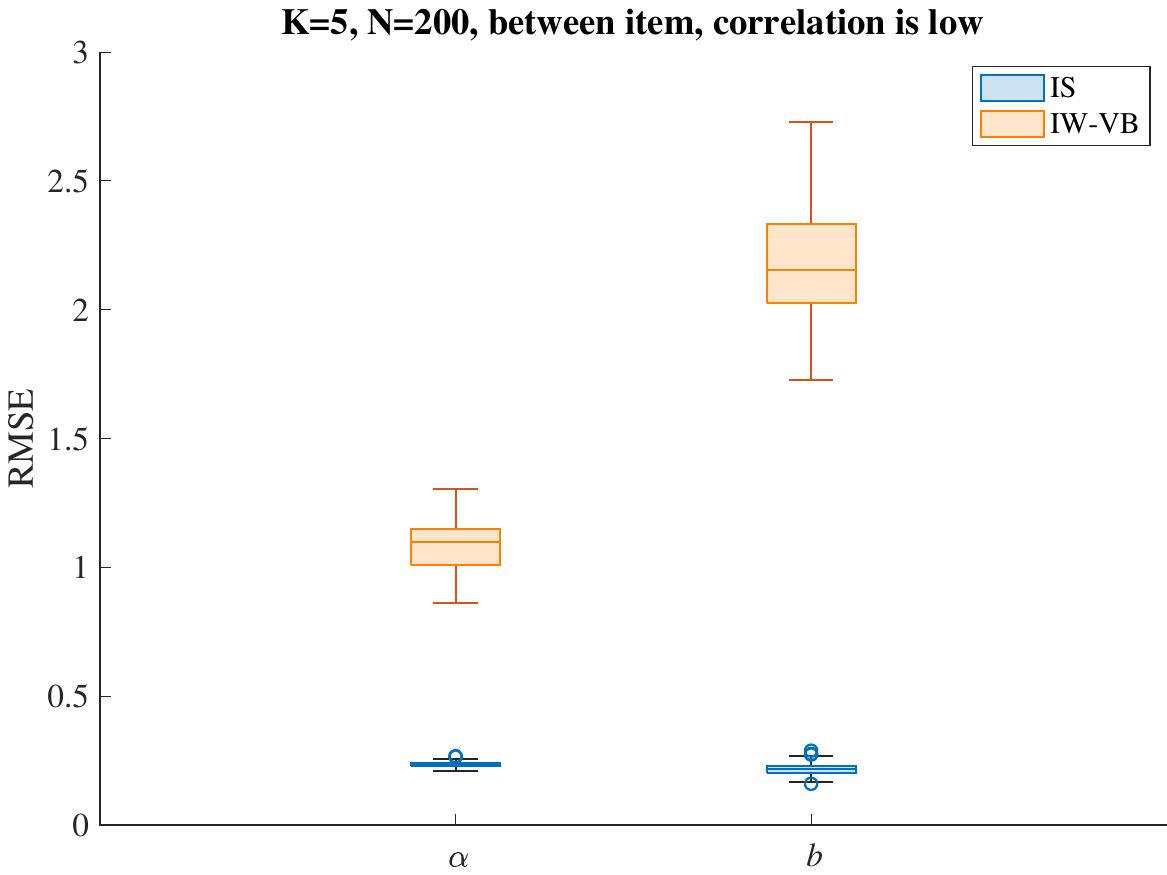}}
    \hspace{0.6in}
    \subfigure{
        \includegraphics[width=2.5in]{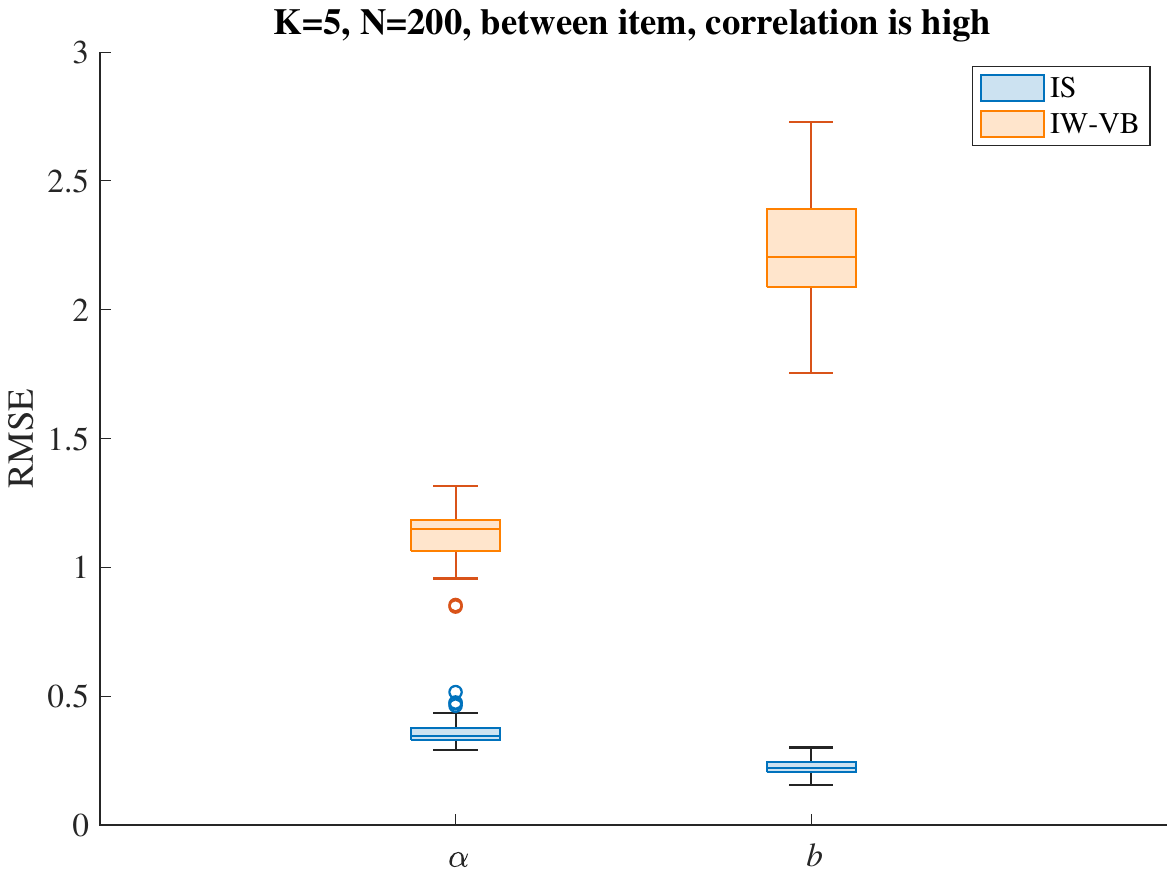}}
    \\
    \subfigure{
        \includegraphics[width=2.5in]{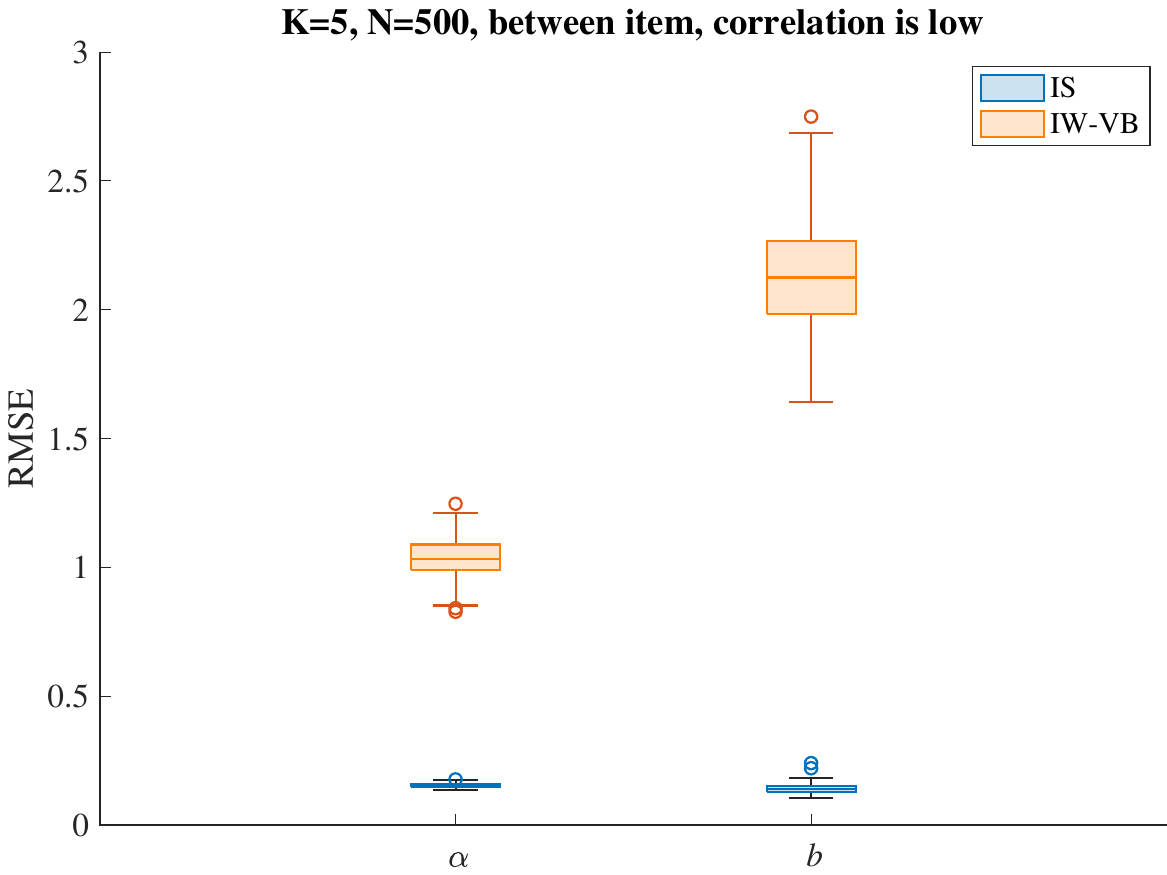}}
    \hspace{0.6in}
    \subfigure{
        \includegraphics[width=2.5in]{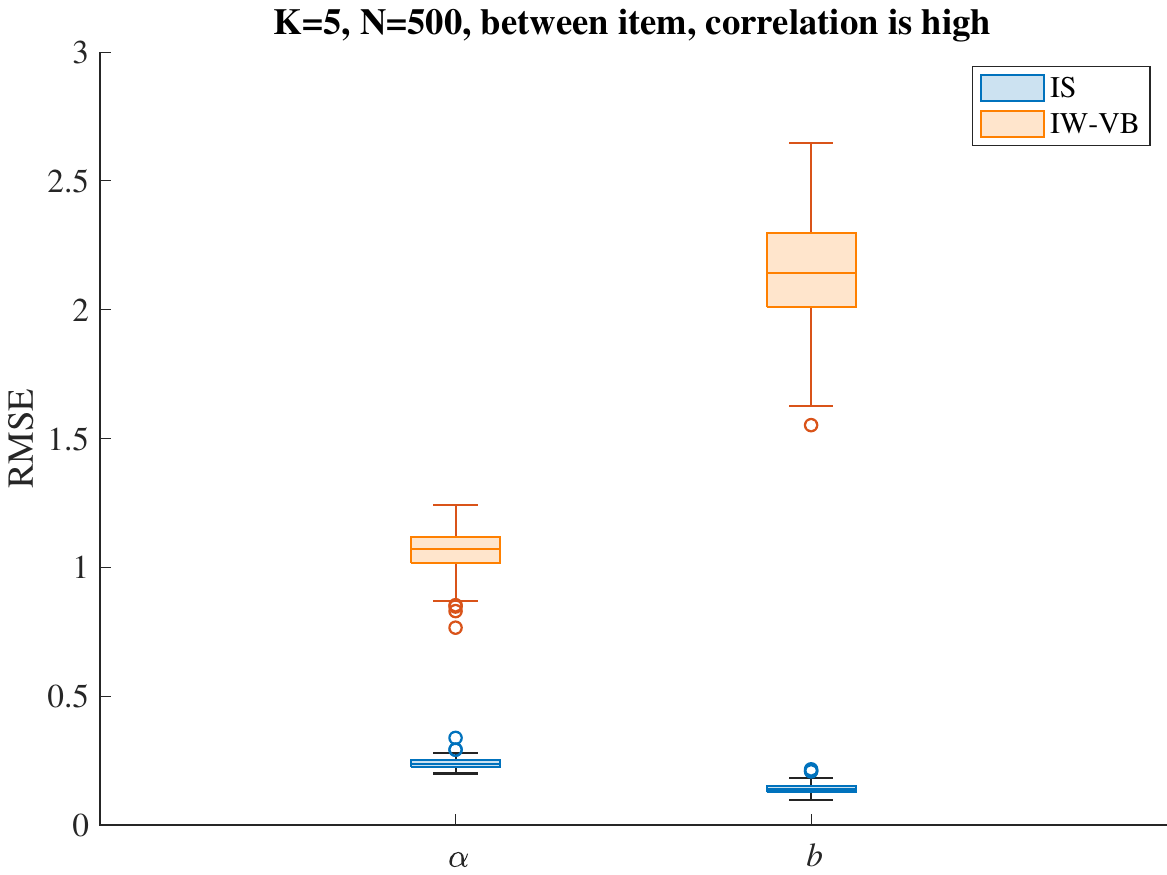}}
    \\
  \subfigure{
        \includegraphics[width=2.5in]{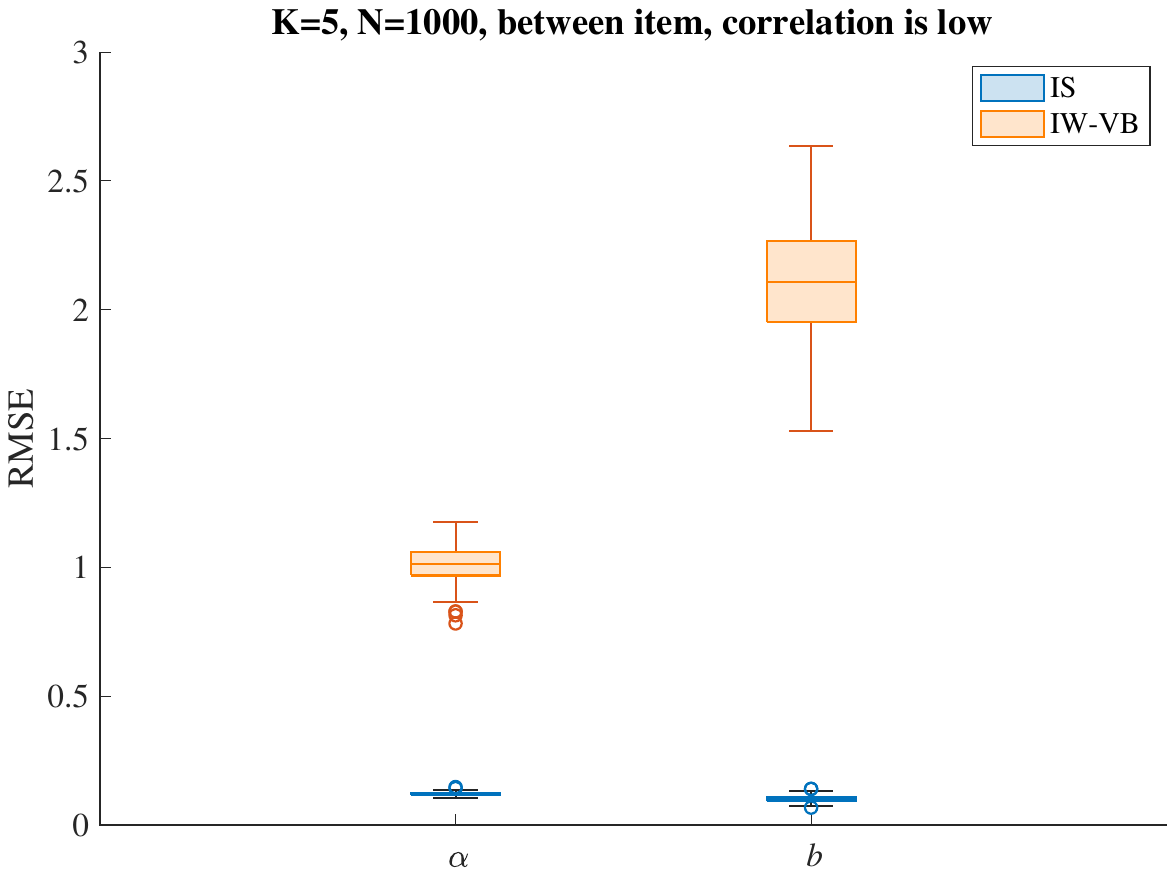}}
    \hspace{0.6in}
    \subfigure{
        \includegraphics[width=2.5in]{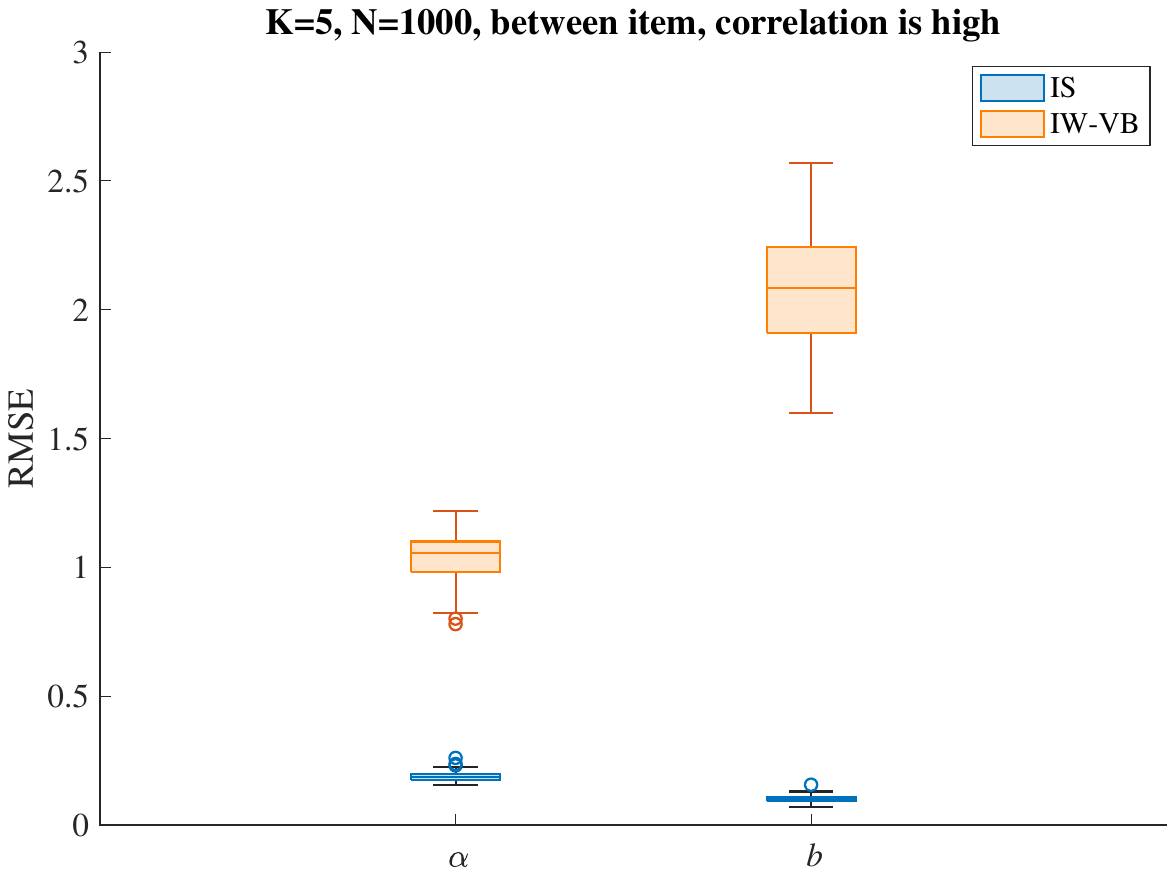}}
    \\
    \caption{RMSE for  $K=5$ between item under exploratory analysis
    }
    \label{fig:rmse-k5-between-explore compare vb}
\end{figure}

\newpage 

\begin{figure}[ht!]
    \centering
    \subfigure{
        \includegraphics[width=2.5in]{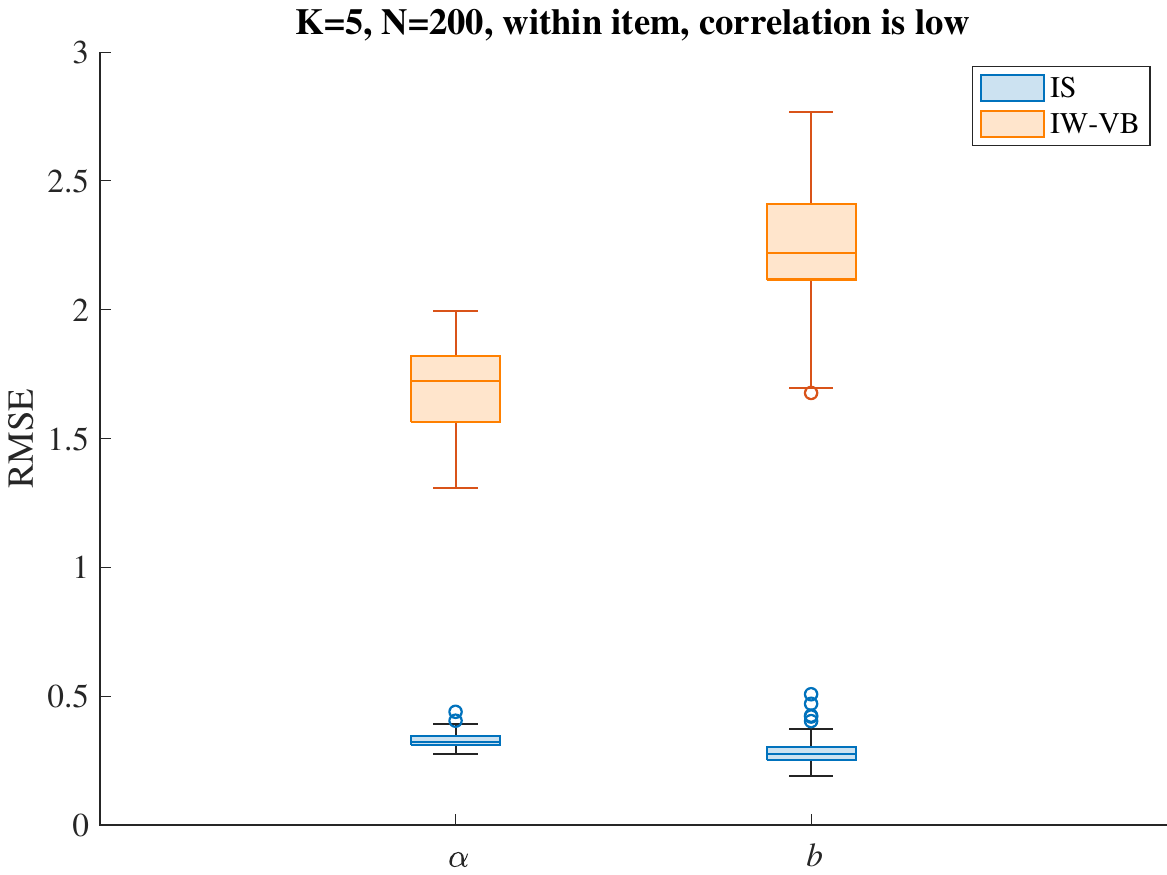}}
    \hspace{0.6in}
    \subfigure{
        \includegraphics[width=2.5in]{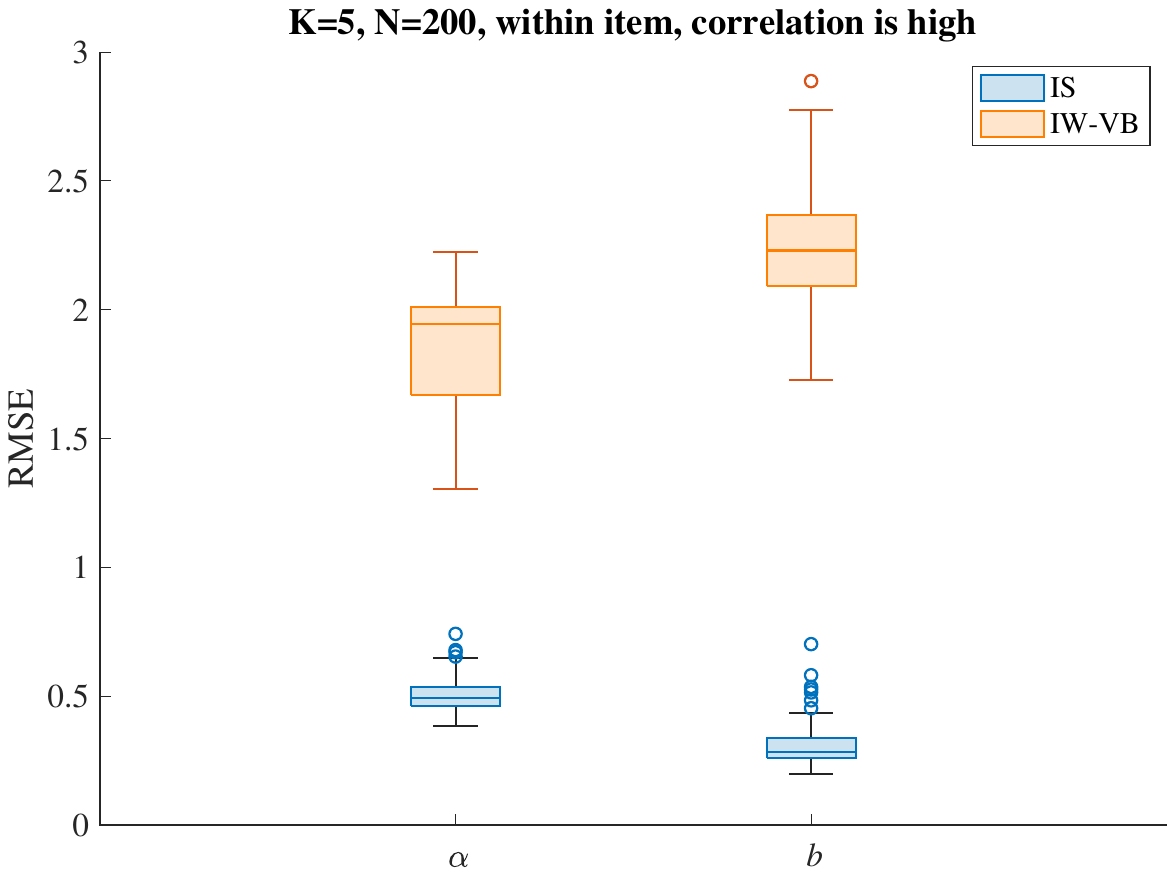}}
    \\
    \subfigure{
        \includegraphics[width=2.5in]{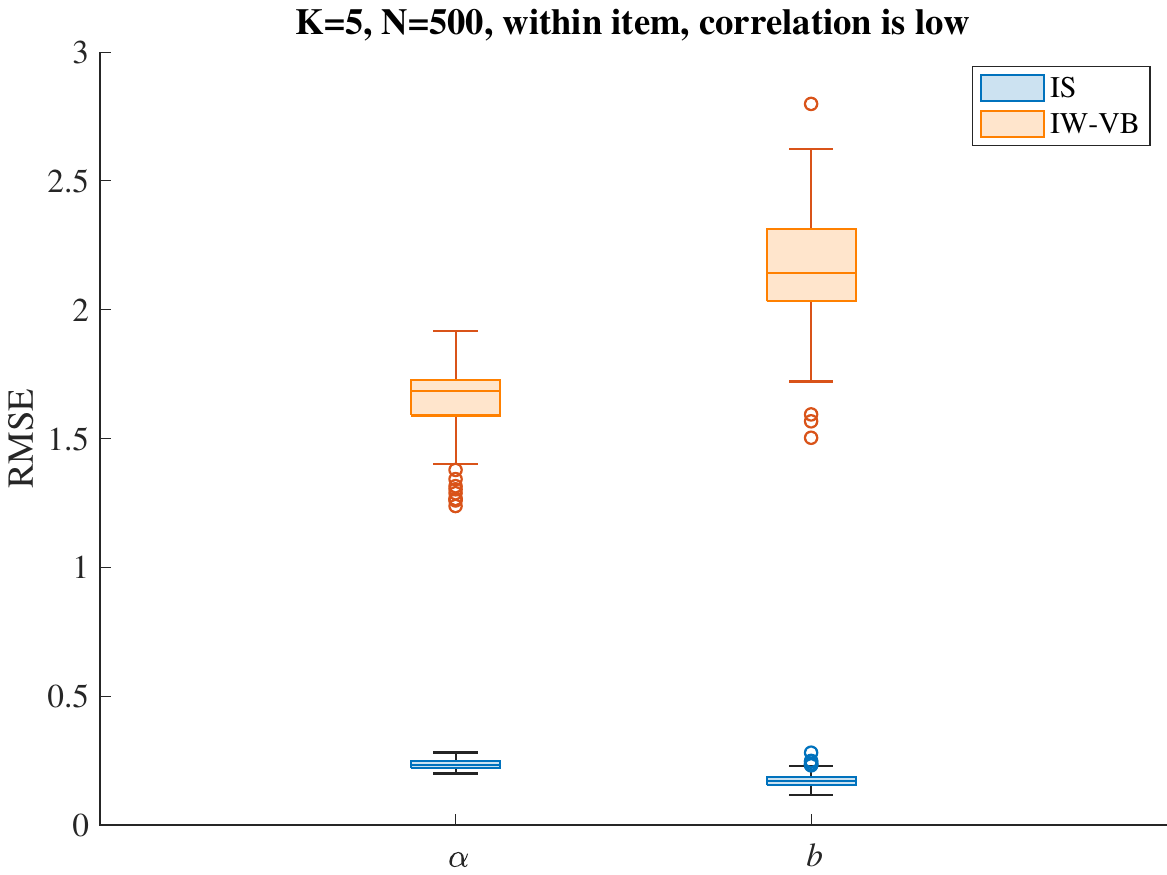}}
    \hspace{0.6in}
    \subfigure{
        \includegraphics[width=2.5in]{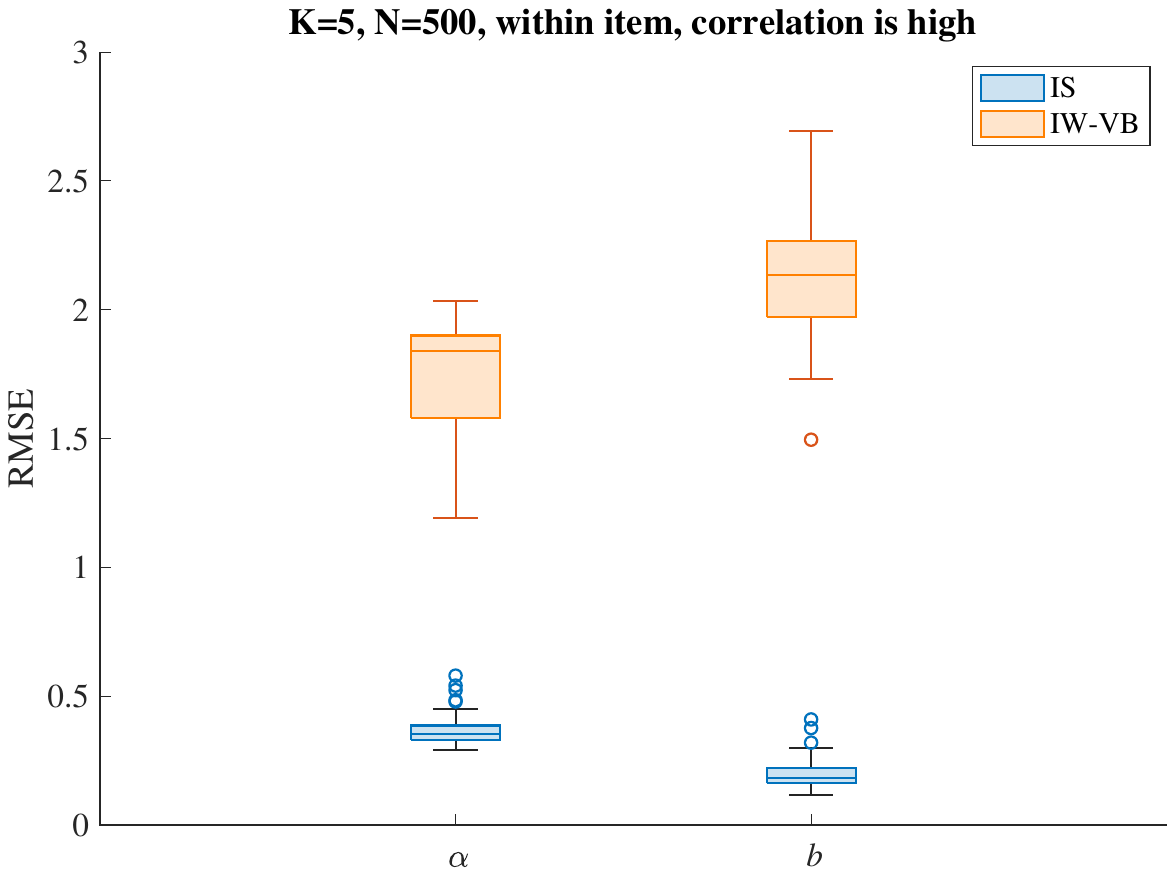}}
    \\
     \subfigure{
        \includegraphics[width=2.5in]{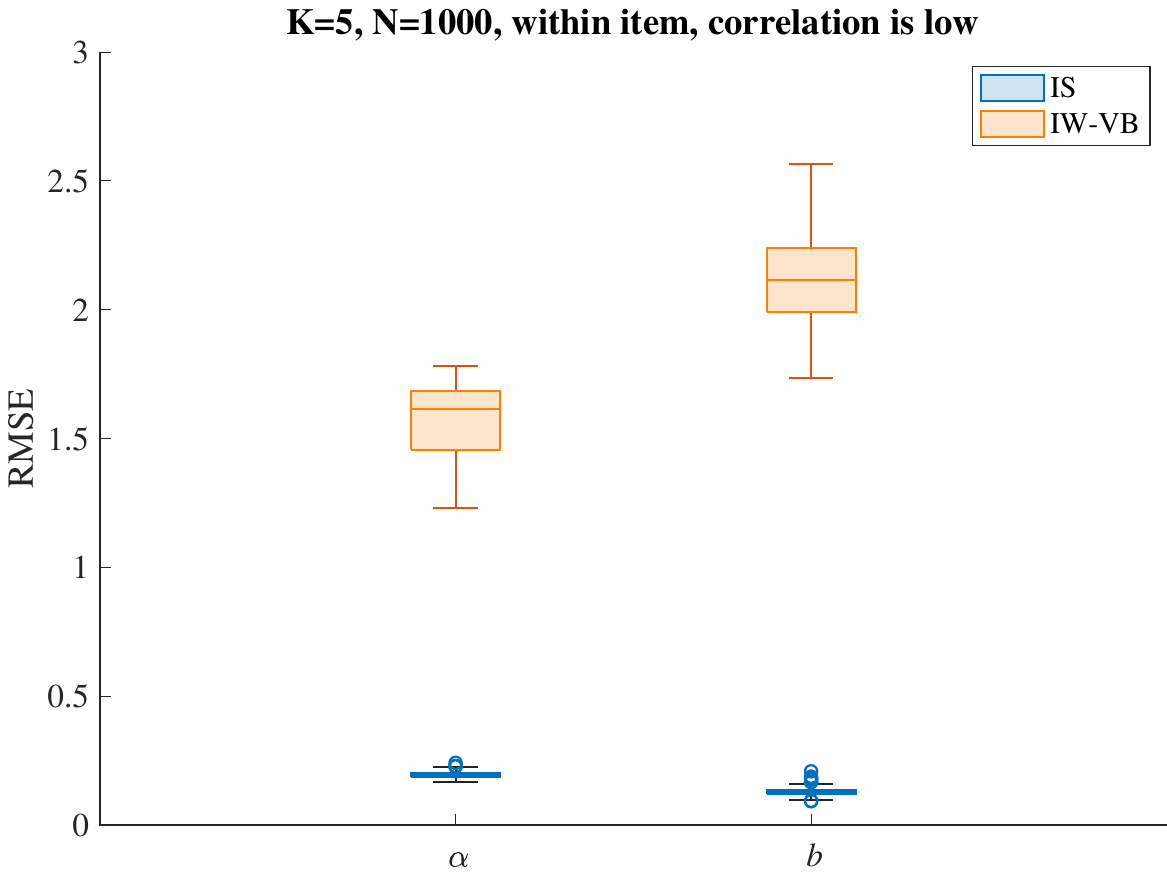}}
    \hspace{0.6in}
    \subfigure{
        \includegraphics[width=2.5in]{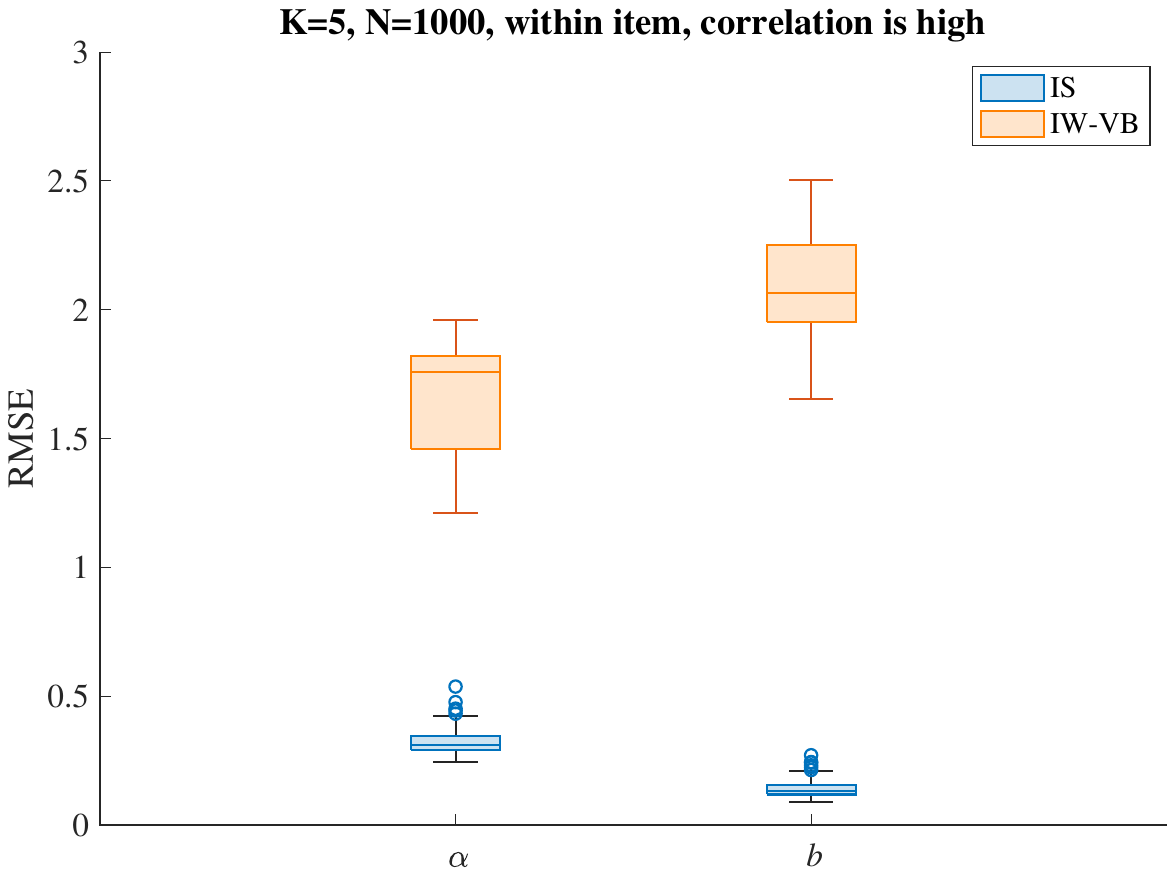}}
    \\
    \caption{RMSE for  $K=5$ within item under exploratory analysis
    }
    \label{fig:rmse-k5-within-explore compare vb}
\end{figure}

\newpage

\subsection{Comparing IW-GVEM with Joint Maximum Likelihood Method}
 The joint maximum likelihood (JML) estimator is a computationally efficient estimator with theoretical consistency established. It is proved in~\cite{chenlizhang} that JML estimator is consistent under high-dimensional settings and it outperforms the marginal maximum likelihood approaches in terms of computational costs. However, different from our IW-GVEM method, the latent abilities are treated as fixed effect parameters instead of random variables in JML method, which may constrain its performances in settings where latent factors are correlated. {\color{black}The JML estimation is also inconsistent in the setting when the number of items is fixed and the sample size grows to infinity. Because the number of parameters in the joint likelihood function grows to infinity, the standard theory for the maximum likelihood method cannot directly apply and the point estimation consistency for each item cannot be attained, which is known as Neyman-Scott phenomenon~\citep{neyman1948consistent}.}

Extensive simulation studies were conducted in~\cite{cho2021gaussian} to compare GVEM to JMLE method under the same simulation settings (sample sizes, within or between multidimensional structures, factor correlations, etc.) and using the same evaluation criteria (bias and RMSE) as in Section~\ref{sec-simulation}. Specifically, Figures 3 and 4 of~\cite{cho2021gaussian} compared the bias and RMSE of GVEM and JML and showed that GVEM has much lower bias and RMSE than JML across all settings. At certain challenging cases such as ``within item, correlation is high", JML estimator has even worse performances. This could be explained by that latent factors are fixed effects in JMLE whereas GVEM treats them as random effects with multivariate Gaussian distributions accounting for the correlations among factors. 
 
 As an improvement of GVEM method, our IW-GVEM method outperforms GVEM in confirmatory factor analysis and has overall comparable performances as GVEM in exploratory factor analysis, across all simulation settings. For a detailed comparison of the simulation results of IW-GVEM and GVEM, please refer to Section~\ref{sec-results}.
As our IW-GVEM is comparable to, if not better than, GVEM, the performance of our IW-GVEM is also better than JML under our simulation settings. 
}\\

\section{Additional Simulation Study}
\label{sec:additional simulation study}
{\color{black} In this section, we present finite-sample simulation studies to show that our proposed IW-GVEM greatly improves the ELBO from GVEM. For the purpose of illustration, we consider the four settings under $N = 200$ and $J= 30$: (1) within-item and low factor correlation; (2) between-item and low factor correlation; (3) within-item and high factor correlation; (4) between-item and high factor correlation. For each setting, we generate the ELBOs from the GVEM algorithm and importance-weighted ELBOs for different sample sizes $M = 5, 10, 50$, and $100$ at the importance sampling step over 100 replications. The calculated ELBOs are presented in Figure~\ref{fig:elbo-n200-j30}.
From Figure~\ref{fig:elbo-n200-j30}, we see that the importance sampling step leads to a tighter importance-weighted ELBO ($M = 5, 10, 50, 100$) than that of GVEM. As the sample $M$ in the importance sampling step increases, the ELBOs converge, which is consistent with theoretical results in Proposition~\ref{prop:converge}.

\begin{figure}[h!]
\centering    
    \subfigure{
        \includegraphics[width=2.7in]{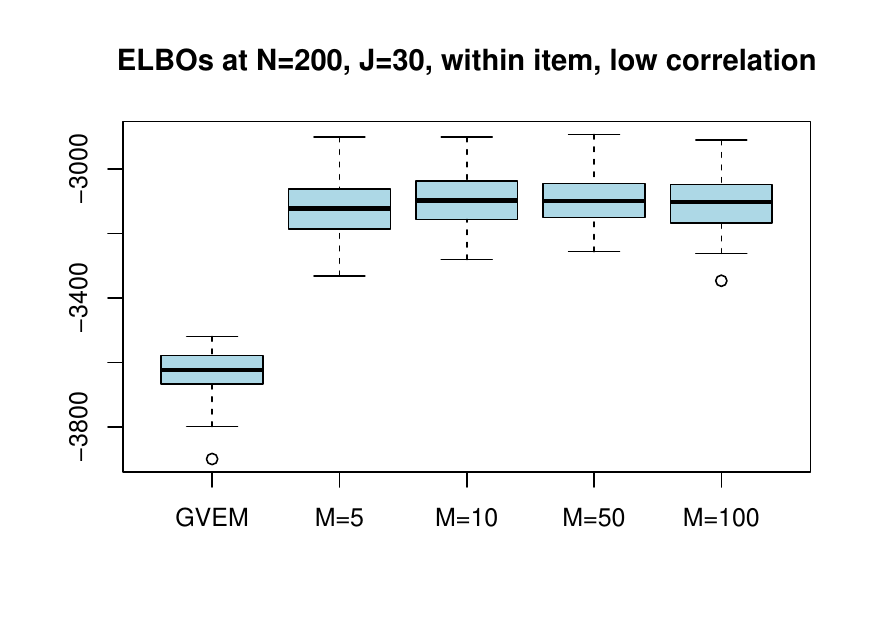}}\hspace{0.2in}
    \subfigure{
        \includegraphics[width=2.7in]{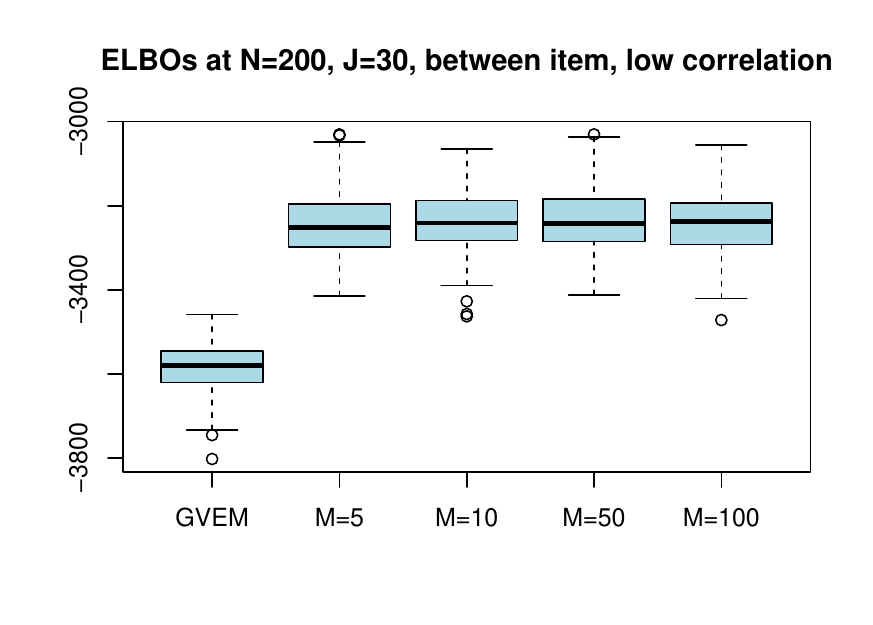}}
    \\
     \subfigure{
        \includegraphics[width=2.7in]{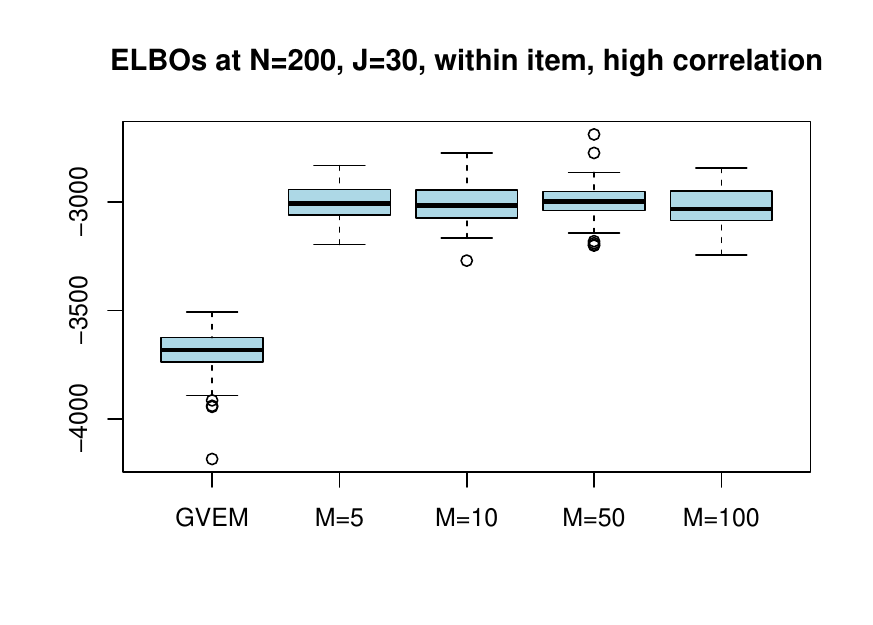}}\hspace{0.2in}
    \subfigure{
        \includegraphics[width=2.7in]{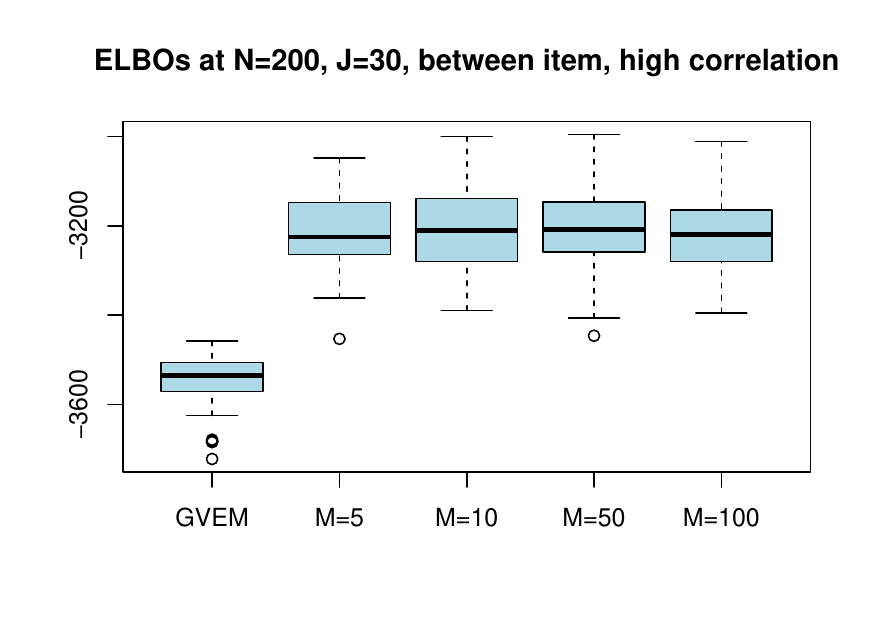}}
    \\
    \caption{Importance-weighted ELBO at $N = 200, J = 30$.}
    \label{fig:elbo-n200-j30}
\end{figure}
}

 \newpage
 
\bibliographystyle{apalike}
\bibliography{bibref}

\begin{thebibliography}{}

\bibitem[Albert, 1992]{albert1992bayesian}
Albert, J.~H. (1992).
\newblock Bayesian estimation of normal ogive item response curves using gibbs
  sampling.
\newblock {\em Journal of educational statistics}, 17(3):251--269.

\bibitem[Bates et~al., 2014]{bates2014fitting}
Bates, D., M{\"a}chler, M., Bolker, B., and Walker, S. (2014).
\newblock Fitting linear mixed-effects models using lme4.
\newblock {\em arXiv preprint arXiv:1406.5823}.

\bibitem[Bishop, 2006]{bishop2006pattern}
Bishop, C.~M. (2006).
\newblock {\em Pattern Recognition and Machine Learning}.
\newblock Springer.

\bibitem[Blei et~al., 2017]{blei2017variational}
Blei, D.~M., Kucukelbir, A., and McAuliffe, J.~D. (2017).
\newblock Variational inference: {A} review for statisticians.
\newblock {\em Journal of the American Statistical Association},
  112(518):859--877.

\bibitem[Bock and Aitkin, 1981]{bock1981marginal}
Bock, R.~D. and Aitkin, M. (1981).
\newblock Marginal maximum likelihood estimation of item parameters:
  Application of an em algorithm.
\newblock {\em Psychometrika}, 46(4):443--459.

\bibitem[Briggs and Wilson, 2003]{briggs2003introduction}
Briggs, D.~C. and Wilson, M. (2003).
\newblock An introduction to multidimensional measurement using rasch models.

\bibitem[Burda et~al., 2015]{burda2015importance}
Burda, Y., Grosse, R., and Salakhutdinov, R. (2015).
\newblock Importance weighted autoencoders.
\newblock {\em arXiv preprint arXiv:1509.00519}.

\bibitem[Cai, 2008]{caisem}
Cai, L. (2008).
\newblock Sem of another flavor: Two new applications of the supplemented em
  algorithm.
\newblock {\em British Journal of Mathematical and Statistical Psychology},
  61:309--329.

\bibitem[Cai, 2010a]{cai2010high}
Cai, L. (2010a).
\newblock High-dimensional exploratory item factor analysis by a
  metropolis--hastings robbins--monro algorithm.
\newblock {\em Psychometrika}, 75(1):33--57.

\bibitem[Cai, 2010b]{cai2010metropolis}
Cai, L. (2010b).
\newblock Metropolis-hastings robbins-monro algorithm for confirmatory item
  factor analysis.
\newblock {\em Journal of Educational and Behavioral Statistics},
  35(3):307--335.

\bibitem[Cai and Hansen, 2018]{cai2018improving}
Cai, L. and Hansen, M. (2018).
\newblock Improving educational assessment: multivariate statistical methods.
\newblock {\em Policy Insights from the Behavioral and Brain Sciences},
  5(1):19--24.

\bibitem[Cai et~al., 2011]{cai2011generalized}
Cai, L., Yang, J.~S., and Hansen, M. (2011).
\newblock Generalized full-information item bifactor analysis.
\newblock {\em Psychological methods}, 16(3):221.

\bibitem[Chen and Wang, 2021]{chenwang}
Chen, P. and Wang, C. (2021).
\newblock Using em algorithm for finite mixtures and reformed supplemented em
  for mirt calibration.
\newblock {\em Psychometrika}, 86:299--326.

\bibitem[Chen et~al., 2019]{chenlizhang}
Chen, Y., Li, X., and Zhang, S. (2019).
\newblock Joint maximum likelihood estimation for high-dimensional exploratory
  item factor analysis.
\newblock {\em Psychometrika}, 84(1):124--146.

\bibitem[Cho et~al., 2021]{cho2021gaussian}
Cho, A.~E., Wang, C., Zhang, X., and Xu, G. (2021).
\newblock Gaussian variational estimation for multidimensional item response
  theory.
\newblock {\em British Journal of Mathematical and Statistical Psychology},
  74:52--85.

\bibitem[Cho et~al., 2022]{cho2022qestimation}
Cho, A.~E., Xiao, J., Wang, C., and Xu, G. (2022).
\newblock Regularized variational estimation for exploratory item response
  theory.
\newblock {\em Psychometrika}, pages 1--29.

\bibitem[CRESST, 2017]{cresst2017English}
CRESST (2017).
\newblock English language proficiency assessment for the 21st century: Item
  analysis and calibration.

\bibitem[Curi et~al., 2019]{curi2019interpretable}
Curi, M., Converse, G.~A., Hajewski, J., and Oliveira, S. (2019).
\newblock Interpretable variational autoencoders for cognitive models.
\newblock In {\em 2019 international joint conference on neural networks
  (ijcnn)}, pages 1--8. IEEE.

\bibitem[Domke and Sheldon, 2018]{domke2018importance}
Domke, J. and Sheldon, D.~R. (2018).
\newblock Importance weighting and variational inference.
\newblock {\em Advances in neural information processing systems}, 31.

\bibitem[Gibbons and Hedeker, 1992]{gibbons1992full}
Gibbons, R.~D. and Hedeker, D.~R. (1992).
\newblock Full-information item bi-factor analysis.
\newblock {\em Psychometrika}, 57(3):423--436.

\bibitem[Hamilton et~al., 1995]{hamilton1995enhancing}
Hamilton, L.~S., Nussbaum, E.~M., Kupermintz, H., Kerkhoven, J.~I., and Snow,
  R.~E. (1995).
\newblock Enhancing the validity and usefulness of large-scale educational
  assessments: Ii. nels: 88 science achievement.
\newblock {\em American Educational Research Journal}, 32(3):555--581.

\bibitem[Hartig and H{\"o}hler, 2009]{hartig2009multidimensional}
Hartig, J. and H{\"o}hler, J. (2009).
\newblock Multidimensional irt models for the assessment of competencies.
\newblock {\em Studies in Educational Evaluation}, 35(2-3):57--63.

\bibitem[Hui et~al., 2017]{hui2017variational}
Hui, F.~K., Warton, D.~I., Ormerod, J.~T., Haapaniemi, V., and Taskinen, S.
  (2017).
\newblock Variational approximations for generalized linear latent variable
  models.
\newblock {\em Journal of Computational and Graphical Statistics},
  26(1):35--43.

\bibitem[Jeon et~al., 2017]{jeon2017variational}
Jeon, M., Rijmen, F., and Rabe-Hesketh, S. (2017).
\newblock A variational maximization--maximization algorithm for generalized
  linear mixed models with crossed random effects.
\newblock {\em Psychometrika}, 82(3):693--716.

\bibitem[Jordan, 2004]{jordan2004graphical}
Jordan, M.~I. (2004).
\newblock Graphical models.
\newblock {\em Statistical science}, 19(1):140--155.

\bibitem[Kingma and Ba, 2014]{kingma2014adam}
Kingma, D.~P. and Ba, J. (2014).
\newblock Adam: A method for stochastic optimization.
\newblock {\em arXiv preprint arXiv:1412.6980}.

\bibitem[Kupermintz et~al., 1995]{kupermintz1995dedication}
Kupermintz, H., Ennis, M.~M., Hamilton, L.~S., Talbert, J.~E., and Snow, R.~E.
  (1995).
\newblock In dedication: Leigh burstein: Enhancing the validity and usefulness
  of large-scale educational assessments: I. nels: 88 mathematics achievement.
\newblock {\em American Educational Research Journal}, 32(3):525--554.

\bibitem[Lindstrom and Bates, 1988]{lindstrom1988newton}
Lindstrom, M.~J. and Bates, D.~M. (1988).
\newblock Newton—raphson and em algorithms for linear mixed-effects models
  for repeated-measures data.
\newblock {\em Journal of the American Statistical Association},
  83(404):1014--1022.

\bibitem[Liu et~al., 2022]{liu2022estimating}
Liu, T., Wang, C., and Xu, G. (2022).
\newblock Estimating three- and four-parameter mirt models with
  importance-weighted sampling enhanced variational auto-encoder.
\newblock {\em Frontiers in Psychology}, 13.

\bibitem[McCulloch, 1997]{mcculloch1997maximum}
McCulloch, C.~E. (1997).
\newblock Maximum likelihood algorithms for generalized linear mixed models.
\newblock {\em Journal of the American statistical Association},
  92(437):162--170.

\bibitem[Natesan et~al., 2016]{natesan2016bayesian}
Natesan, P., Nandakumar, R., Minka, T., and Rubright, J.~D. (2016).
\newblock Bayesian prior choice in irt estimation using mcmc and variational
  bayes.
\newblock {\em Frontiers in psychology}, 7:1422.

\bibitem[Neyman and Scott, 1948]{neyman1948consistent}
Neyman, J. and Scott, E.~L. (1948).
\newblock Consistent estimates based on partially consistent observations.
\newblock {\em Econometrica: Journal of the Econometric Society}, pages 1--32.

\bibitem[OECD, 2003]{oecd2003pisa}
OECD, N. (2003).
\newblock The pisa 2003 assessment framework: Mathematics, reading, science and
  problem solving knowledge and skills.

\bibitem[Ormerod and Wand, 2010]{ormerod2010explaining}
Ormerod, J.~T. and Wand, M.~P. (2010).
\newblock Explaining variational approximations.
\newblock {\em The American Statistician}, 64(2):140--153.

\bibitem[Patz and Junker, 1999]{patz1999applications}
Patz, R.~J. and Junker, B.~W. (1999).
\newblock Applications and extensions of mcmc in irt: Multiple item types,
  missing data, and rated responses.
\newblock {\em Journal of educational and behavioral statistics},
  24(4):342--366.

\bibitem[Pinheiro and Bates, 1995]{pinheiro1995approximations}
Pinheiro, J.~C. and Bates, D.~M. (1995).
\newblock Approximations to the log-likelihood function in the nonlinear
  mixed-effects model.
\newblock {\em Journal of computational and Graphical Statistics}, 4(1):12--35.

\bibitem[Reckase, 2009]{reckase2009multidimensional}
Reckase, M.~D. (2009).
\newblock Multidimensional item response theory models.
\newblock In {\em Multidimensional item response theory}, pages 79--112.
  Springer.

\bibitem[Rijmen and Jeon, 2013]{rijmen2013fitting}
Rijmen, F. and Jeon, M. (2013).
\newblock Fitting an item response theory model with random item effects across
  groups by a variational approximation method.
\newblock {\em Annals of Operations Research}, 206(1):647--662.

\bibitem[Rijmen et~al., 2008]{rijmen2008latent}
Rijmen, F., Vansteelandt, K., and De~Boeck, P. (2008).
\newblock Latent class models for diary method data: Parameter estimation by
  local computations.
\newblock {\em Psychometrika}, 73(2):167--182.

\bibitem[Thissen, 2013]{thissen2013using}
Thissen, D. (2013).
\newblock Using the testlet response model as a shortcut to multidimensional
  item response theory subscore computation.
\newblock In {\em New developments in quantitative psychology}, pages 29--40.
  Springer.

\bibitem[Urban and Bauer, 2021]{urban2021deep}
Urban, C.~J. and Bauer, D.~J. (2021).
\newblock A deep learning algorithm for high-dimensional exploratory item
  factor analysis.
\newblock {\em psychometrika}, 86(1):1--29.

\bibitem[von Davier and Sinharay, 2010]{von2010stochastic}
von Davier, M. and Sinharay, S. (2010).
\newblock Stochastic approximation methods for latent regression item response
  models.
\newblock {\em Journal of Educational and Behavioral Statistics},
  35(2):174--193.

\bibitem[Wainer et~al., 2007]{wainer2007testlet}
Wainer, H., Bradlow, E.~T., and Wang, X. (2007).
\newblock {\em Testlet response theory and its applications}.
\newblock Cambridge University Press.

\bibitem[Wang and Xu, 2015]{wang2015mixture}
Wang, C. and Xu, G. (2015).
\newblock A mixture hierarchical model for response times and response
  accuracy.
\newblock {\em British Journal of Mathematical and Statistical Psychology},
  68(3):456--477.

\bibitem[Wu et~al., 2020]{wu2020variational}
Wu, M., Davis, R.~L., Domingue, B.~W., Piech, C., and Goodman, N. (2020).
\newblock Variational item response theory: Fast, accurate, and expressive.
\newblock {\em arXiv preprint arXiv:2002.00276}.

\bibitem[Yamaguchi and Okada, 2020a]{yamaguchi2020variational1}
Yamaguchi, K. and Okada, K. (2020a).
\newblock Variational {Bayes} inference algorithm for the saturated diagnostic
  classification model.
\newblock {\em psychometrika}, 85(4):973--995.

\bibitem[Yamaguchi and Okada, 2020b]{yamaguchi2020variational}
Yamaguchi, K. and Okada, K. (2020b).
\newblock Variational {Bayes} inference for the {DINA} model.
\newblock {\em Journal of Educational and Behavioral Statistics},
  45(5):569--597.

\bibitem[Zhang et~al., 2020a]{zhang2020note}
Zhang, H., Chen, Y., and Li, X. (2020a).
\newblock A note on exploratory item factor analysis by singular value
  decomposition.
\newblock {\em psychometrika}, 85:358--372.

\bibitem[Zhang et~al., 2020b]{zhang2020improved}
Zhang, S., Chen, Y., and Liu, Y. (2020b).
\newblock An improved stochastic em algorithm for large-scale full-information
  item factor analysis.
\newblock {\em British Journal of Mathematical and Statistical Psychology},
  73(1):44--71.

\end{thebibliography}

\end{document}